\documentclass{aa}

\usepackage{graphicx}
\usepackage{amsmath}
\usepackage{amssymb}
\usepackage{units}
\usepackage{placeins}  % for \FloatBarrier
\usepackage{natbib}
\bibpunct{(}{)}{;}{a}{}{,} % to follow the A&A style
\usepackage{tikz}
\usetikzlibrary{graphs, positioning, shapes.geometric, arrows.meta, shapes.multipart}
\usepackage{txfonts,textcomp}
\usepackage[allcolors=blue]{hyperref}
\begin{document}

\title{Multicomponent imaging of the \textit{Fermi} gamma-ray sky\\in the spatio-spectral domain\thanks{All reconstructions are released as data products at \url{https://doi.org/10.5281/zenodo.7970865}.}}

\author{
   L.\,I.~Scheel-Platz  \inst{1,2,3,4,5} \and
   J.~Knollm{\"u}ller \inst{1,2,6,7} \and
   P.~Arras \inst{1,8} \and
   P.~Frank \inst{1} \and
   M.~Reinecke \inst{1} \and
   D.~J{\"u}stel \inst{3,4,5} \and
   T.\,A.~En{\ss}lin \inst{1,2}
}

\institute{
  Max~Planck~Institute~for~Astrophysics,\
  Karl-Schwarzschild-Str.\,1, 85748~Garching, Germany\\
  \email{lplatz@mpa-garching.mpg.de, ensslin@mpa-garching.mpg.de}
\and
  Ludwig-Maximilians-Universität~München,\
  Geschwister-Scholl-Platz~1, 80539~Munich, Germany
\and
  Institute~of~Biological~and~Medical~Imaging,\
  Helmholtz~Zentrum~München,\
  Ingolstädter~Landstraße~1, 85764~Neuherberg, Germany
\and
  Institute~of~Computational~Biology,\
  Helmholtz~Zentrum~München,\
  Ingolstädter~Landstraße~1, 85764~Neuherberg, Germany
\and
  Technical~University~of~Munich;\
  School~of~Medicine,\
  Chair~of~Biological~Imaging~at~the~Central~Institute~for~Translational~Cancer~Research~(TranslaTUM),\
  Einsteinstraße~25, 81675~Munich, Germany
\and
  Technical~University~of~Munich;\
  School~of~Natural~Sciences,\
  Arcisstraße~21, 80333~Munich, Germany
\and
  Excellence~Cluster~ORIGINS,\
  Boltzmannstraße~2, 85748~Garching, Germany
\and
  Technical~University~of~Munich;\
  School~of~Computation,~Information~and~Technology,\
  Boltzmannstr.\,3, 85748~Garching, Germany
}

\date{Received 20 April 2022 / Accepted 05 August 2023}
 
\abstract{
% Context
The gamma-ray sky as seen by the Large Area Telescope (LAT) on board the \textit{Fermi} satellite
is a superposition of emissions from many processes.
%
%Because it allows the indirect study of high-energy astrophyical phenomena,
%gamma-ray observations as made by the LAT have been studied extensively
%and have led to many advances in our understanding of the high-energy dynamics in our Galaxy and beyond.
%
%The gamma-ray sky includes galactic diffuse foreground emissions
%(tracing matter, electromagnetic radiation, and cosmic ray (CR) populations within the Galaxy),
%Galactic and extragalactic point source (PS) and extended object emissions and an extragalactic background.
%
To study them, a rich toolkit of analysis methods for gamma-ray observations
has been developed,
most of which rely on emission templates to model foreground emissions.
%
%To make analyses robust to templating errors, methods to explicitly model template misfits
%and template-free deep learning based approaches have been published.
%
% Aims
%
Here, we aim to complement these methods
by presenting a template-free spatio-spectral imaging approach for the gamma-ray sky,
%
% Methods
%
based on a phenomenological modeling of its emission components.
It is formulated in a Bayesian variational inference framework
and allows a simultaneous reconstruction and decomposition of the sky into
multiple emission components,
enabled by a self-consistent inference of their spatial and spectral correlation structures.
Additionally, we formulated the extension of our imaging approach to template-informed imaging,
which includes adding emission templates to our component models
while retaining the ``data-drivenness'' of the reconstruction.
%
% Results
%
We demonstrate the performance of the presented approach on the ten-year \textit{Fermi} LAT data set.
With both template-free and template-informed imaging,
we achieve a high quality of fit and
show a good agreement of our diffuse emission reconstructions
with the current diffuse emission model published by the \textit{Fermi} Collaboration.
We quantitatively analyze the obtained data-driven reconstructions
and critically evaluate the performance of our models,
highlighting strengths, weaknesses, and potential improvements.
All reconstructions have been released as data products.
}

\keywords{
  Gamma rays: general --
  Gamma rays: ISM --
  methods: data analysis --
  methods: statistical
}

\maketitle

\section{Introduction} \label{sec:introduction}

Situated at the upper end of the electromagnetic spectrum,
gamma rays bear witness to highly energetic processes.
Because of their low interaction cross-section with most matter and radiation,
gamma rays in the GeV to TeV range provide a window into their generation sites,
which in other parts of the electromagnetic spectrum
is often obscured by matter in the line of sight.

Based on observations of the gamma radiation,
the processes and objects involved in creating it can be studied.
This includes the production, acceleration, and dissemination
of cosmic rays \citep[CRs;][]{grenier2015reviewCR,tibaldo2021reviewCR,liu2022reviewCR},
whose interaction with matter and radiation fields contribute
the majority of observed gamma rays
and potentially dark matter (DM),
whose properties can be constrained based on searches of
dark matter annihilation (DMA) emission \citep{bergstrom2000reviewDM,gianfranco2005reviewDM}.

The gamma-ray fluxes that reach Earth are a superposition
of emissions from various sources:
First, a strong Galactic foreground is produced by interactions of CRs
with the interstellar medium (ISM) of the Milky Way.
The dominant processes for this are inelastic collisions of CR protons with protons present in the thermal ISM, which create neutral pions that quickly decay into gamma rays,
bremsstrahlung produced by cosmic ray electrons (CRe-) and positrons (CRe+) in ionized gas,
and inverse Compton (IC) up-scattering of photons from the interstellar radiation
field (star light, thermal infrared, and microwave emission from dust grains)
and the cosmic microwave background by CRe- and CRe+.
These Galactic emissions trace the distribution of the respective
target and CR populations within the Galaxy, making them span large parts of the sky
and giving them characteristic spatial structures.

Second, there is a large population of very localized gamma-ray emitters,
appearing as point sources (PSs) from Earth, including blazars, pulsars, and supernova remnants
\citep{abdo2010points,2012ApJS..199...31N,2015ApJS..218...23A,abdollahi2020fermi,abdollahi2022_4FGL_DR2}.
Because these objects actively accelerate CRs, their gamma-ray spectra differ somewhat from the
spectra observed for the Galactic ISM emission.
Some of them, because of their relatively close proximity to Earth,
appear as extended objects in the sky \citep{lande2012extendedsources,ackermann2017extendedsources,ackermann2018extendedsources}.
Notable examples include the Vela, Crab, and Geminga pulsar wind nebulae.

Third, there is an isotropic background of gamma radiation from faint Galactic and
extragalactic sources (see \citealt{fornasa2015dgrb}, \citealt{ackermann2018grb}, and \citealt{roth2021grb}).
To contrast the first and third group from PSs, the former are usually
referred to as diffuse emissions.\\

% Subsect: The Fermi Large Area Telescope
Today's largest data set of gamma-ray observations is based on measurements
by the \textit{Fermi} Large Area Telescope \citep[LAT;][]{atwood2009fermilat}.
The LAT is an orbital pair-conversion telescope, sensitive in the \unit{MeV} to \unit{TeV} energy range,
that features a large instantaneous field of view (roughly \unit[20]{\%} of the sky)
and an angular resolution from about \unit[3]{degrees} at \unit[0.1]{GeV} to a few arcminutes at \unit{TeV} energies.
It detects individual gamma-ray photons and records
an estimate of their origin direction, energy, and arrival time.

Notably, \textit{Fermi} LAT observations led to the discovery of the so-called Fermi bubbles
\citep[FBs;][]{su2010bubbles,ackermann2014bubbles},
two symmetric large-scale emission structures located north and south of the Galactic center (GC),
and the Galactic center excess \citep[GCE;][]{goodenough2009gce,hooper2011gce},
a population of gamma rays originating from within a few degrees around the GC
that are not predicted by emission models based on observations in other energy bands.
The discovery of the GCE sparked a still ongoing debate about its origin\footnote{
\citet{calore2015gce,ajello2016gce,huang2016gce,bartels2016wavelet,macias2019gce,
ackermann2017gce,bartels2018gce,bartels2018gceskyfact,caron2018ml_gce,
leane2020enigmatic,buschmann2020shts,calore2021nptf,burns2021gce,
karwin2023diffuse,cholis2022return,mcdermott2022gce};
\citet[Review]{hooper2023gce_review};
\citet{caron2023gce}},
with the leading hypotheses being DMA emissions \citep{hooper2011gce}
and a dense population of faint millisecond pulsars \citep{abazajian2011gceMSPs}.
Together with the study of the extragalactic gamma-ray background,
the debate about the GCE origin led to
the development of a broad range of analysis methods for gamma-ray data sets,
all addressing the fundamental difficulty of attributing gamma-ray fluxes to
specific production processes.\\

% Subsect: Previous work
Here, we present an overview of established gamma-ray analysis approaches
to provide context for the method we introduce.
A fundamental technique employed in almost all analyses is emission template fitting.
For this, the spatial and/or spectral distribution of gamma-ray emission from different channels
is predicted based on other astrophysical observations, theoretical considerations,
or, in the case of only gamma-visible structures, preprocessed gamma-ray data.
The obtained templates are then used in parametric models of the gamma-ray sky,
which are fit to the observed data.
This way, even complex skies can be adequately expressed with only a few fit parameters
if correct templates of all contained emission components are available.
The emission from major channels can be predicted to a high degree of accuracy (see \citealt{ackermann2012diffuse}, \citealt{acero2016templates}, \citealt{buschmann2020shts}, and \citealt{karwin2023diffuse}
for details of the template creation).
Because of this success in modeling, most analyses make use of templates to
account for the emission from these channels.

In the study of extended and diffuse emissions, a second fundamental technique
(often combined with template fitting, but also used in other ways)
is the masking of known PSs.
For this, data bins within a chosen distance to known PSs are discarded.
This alleviates the need to model bright PS emissions
and reduces the sensitivity of the analysis to instrumental point spread mismodeling.
When using PS masking,
a trade-off between PS flux contamination and
data set size degradation needs to be made.

As laid out by \citet{storm2017skyfact},
likelihood-based gamma-ray analyses can be understood on a continuum from low to high parameter count,
representing different trade-offs between parameter space explorability,
statistical power, goodness of fit, and sensitivity to unexpected emission components.
On the low parameter count end of the continuum are pure template fitting analyses,
with only a few parameters per emission template (global or large-area scaling constants).
Many early publications of the \textit{Fermi} Collaboration are based on this approach (see for example
\citealt{abdo2010spectrum}, \citealt{abdo2010points}, \citealt{ackermann2012diffuse}, and \citealt{ackermann2012anisotropies}).

As emission structures not predicted by other observations were found, such as the FBs,
multiple methods for deriving data-driven templates for these structures were developed.
Examples include the work of
\citet{casandjian2015fermi}, which models emission from IC interactions of CRe- and CRe+ with the cosmic microwave background
via pixel-wise fits of an IC emission spectrum model to filtered and
point spread function (PSF) de-convolved residuals of the already templated foregrounds,
and post-processed using a spatial domain {2\textdegree} wavelet high-pass filter.
Similarly, \citet{acero2016templates} built models of ``extended excess emissions'' that include the FBs
and Loop 1 from wavelet-filtered baseline model residuals.
Based on the combination of conventional and data-driven templates,
the diffuse gamma-ray sky can be modeled to a high fidelity (see for example \citealt{calore2022template3d}).

Many publications assume the remaining misfits to point out the existence of true excess populations.
These are typically studied by extending the template model with additional parametric models of the excesses,
derived from the properties of the hypothesized excess origins.
The comparative quality of fit using different excess models
is then taken as evidence in favor of or against the respective origin hypotheses.
Many analyses of the GCE are based on this approach.

A notable family of techniques that follow this pattern are
photon count statistics methods,
first introduced for gamma rays by \citet{malyshevhogg2011}.
They exploit the difference in photon statistics expected
from truly diffuse emitters and dense but faint PS populations to infer properties
of sub-detection-threshold PSs,
for example their source count distribution (SCD).
\citet{lee2016nptf} and \citet{mishra2017nptfit}
implemented this concept under the name non-Poissonian template fitting (NPTF).
Independently, \citet{zechlin2016unveiling} formulated
methods based on photon count statistics  centered around
one-point probability distribution functions (1pPDFs),
which include pixel-dependent variations in the expected
photon statistics, for example from the instrument response.
The authors used this to study GCE signals, DM signals,
and the extragalactic isotropic background \citep{zechlin2018background}.
Analysis approaches of the photon count statistics family have been extended to other modalities,
including neutrino astronomy \citep{feyereisen2018icecube,aartsen2020icecube}.

A conceptual extension of 1pPDF methods to methods based on two-point correlation statistics  was shown by \citet{manconi2020testing}.
They derived expected angular power spectrum (APS) models
for different emission types, similar to the 1pPDF models.
Using the 1pPDF and APS models, they studied the extragalactic background using \textit{Fermi} LAT data and find good agreement between the two statistics.
\citet{baxter2022approximate} show with synthetic data how approximate Bayesian computation \citep{sisson2007abc,beaumont2009abc,blum2010abc},
a likelihood-free method, could be applied to 1pPDF studies of the extragalactic gamma-ray emissions.

A crucial limitation of template-based approaches is their dependence on accurate emission templates.
As has been pointed out on multiple occasions,
a mismodeling of the foreground emissions can severely bias template-based analyses.
An example of this is the disagreement between \citet{leane2020enigmatic},
who studied the ability of NPTF to recover an artificially injected DM signal and report NPTF to be biased toward neglecting it,
and \citet{buschmann2020shts}, who argue the observed bias is caused by insufficiently accurate foreground emission models.

To eliminate templating errors post creation,
approaches for optimizing existing templates in a data-driven way have been published.
\citet{buschmann2020shts} demonstrate a template optimization scheme in the spherical harmonic (SH) domain,
adjusting the normalization constant of all SH components of the templates individually
and reassembling updated versions of the templates from the scaled SH components afterward.
%For this, the templates are decomposed with a spherical harmonic transform (SHT) and the best-fit normalization for each $(\ell, m)$ combination is obtained.
%The templates are then reassembled from these rescaled SH components.

To mitigate biases introduced by template misfits, a move toward additional nuisance parameters that explicitly model the template misfits can be observed.
This includes various imaging techniques that derive data-driven reconstructions of the whole sky, as well as works toward full modifiability of templates.

Criticizing poor qualities of fit in previous template analyses,
\citet{storm2017skyfact} introduced the SkyFACT framework.
It enables the inclusion of full-resolution Gaussian template modification fields
into sky reconstructions,
making it a hybrid parameter estimation and imaging method.
Spatio-spectral distributions of emission components are modeled by the outer product
of spatial and spectral templates and (if desired) their respective modification fields.
They introduced a custom regularization term for the
modification fields that preserves the convexity of the optimization problem.
By consecutively adding modification fields to a template analysis of the \textit{Fermi} LAT data set,
the authors demonstrate the effect of transitioning from low to high parameter count analyses.
For the model parameter estimation, the authors employed a quasi-Newton method and estimated the errors of individual model parameters via the Fisher information matrix and a sparse Cholesky decomposition (see \citealt{storm2017skyfact} for details).
This work can also be used to derive optimized emission templates as demonstrated by \citet{calore2021nptf}, who obtained optimized emission templates with a SkyFACT fit of the LAT data and then performed a 1pPDF analysis of the residuals.

\citet{mishra2020gaussianprocess} introduced a variational inference framework that includes Gaussian process-based template modification fields and a neural network (NN) to
predict posterior distributions of other model parameters
depending on the modification field state.
The Gaussian processes are implemented using a
deep learning (DL) framework, making the implementation compatible with graphics processing units.
The authors demonstrate the potential of their method by analyzing the GC emission as captured by the \textit{Fermi} LAT.

\citet{selig2015fermi} demonstrate a template-free imaging of the gamma-ray sky
based on the D$^{3}$PO algorithm \citep{selig2015d3po}.
Here, the gamma-ray emissions are modeled freely,
with prior statistics of the logarithmic diffuse sky brightness encoded in
a Gaussian processes with a to-be-inferred statistically homogeneous
correlation structure, and PS fluxes implemented via spatial sparsity
enforcing priors.
This work performed the reconstructions for each energy band independently,
only making use of spatial correlations to guide the reconstruction.
\citet{pumpe2018d4po} developed the D$^{4}$PO algorithm
to additionally exploit spectral correlations for spatio-spectral imaging.

A benefit of these high parameter count reconstruction methods is their potential to have a higher sensitivity to unexpected emission components,
especially to small structures,
as they do not impose as strong a priori assumptions about the distribution of gamma-ray flux
as fixed template-based fits do.
However, this comes at the price of a higher vulnerability to instrument mismodeling
in data-driven analyses, where an accurate modeling of the instrument response is crucial,
and potentially a higher susceptibility to measurement noise.

A different line of inquiry involves wavelet-based methods.
\citet{schmitt2012wavelet} demonstrate an iterative de-noising and PSF de-convolution algorithm based on SH wavelets with synthetic LAT data.
\citet{bartels2016wavelet} performed a template-free comparison of GCE excess models
based on how well they predict wavelet coefficient statistics
of \textit{Fermi} data at length scales where no foreground structure is expected.
\citet{balaji2018wavelet} show a wavelet-based analysis
of photon counts unexplained by state-of-the-art emission templates
and derive data-driven maps of the emissions of
the FBs, the GCE, and the ``cocoons'' inside the FBs.

Finally, over the last few years, a number of promising DL-based gamma-ray
data analysis approaches have been developed, training NNs to
directly predict quantities of interest from binned photon counts.
The NNs are trained on synthetic data generated with existing emission templates
and various mixtures of additional emission structures,
tested on further synthetic data samples,
and then applied to the \textit{Fermi} LAT data.

\citet{caron2018ml_gce} demonstrate a convolutional NN
trained to predict the PS emission fraction of the GCE directly
from photon count data.
\citet{list2020ml_gce} demonstrate a graph convolutional NN
trained to predict unmixed gamma-ray emission maps for the GC region,
including a separation of smooth DM emissions and GCE PS emissions.
They report emission component recovery accuracies to the percent level
on a synthetic test data set
and provide estimates of the aleatoric uncertainty for each component.
\citet{list2021ml_gce} extended this work, predicting the SCD
of the identified GCE emissions in a nonparametric form.
Recently, \citet{mishra2022ml_sbi} published an orthogonal approach that
uses NNs to enable fast simulation-based inference (SBI).
They simultaneously trained a convolutional-NN-based feature extractor
and a normalizing-flow-based inference network
to predict posterior parameters of a generative (forward) gamma-ray sky model
from photon count data.
With this SBI pipeline, they were able produce posterior samples of the forward model parameters
conditioned on the \textit{Fermi} LAT data.\\

% Subsect: Imaging based on information field theory
In this article we build on methods developed in the context of information field theory
\citep[IFT;][]{2009PhRvD..80j5005E,2013AIPC.1553..184E,2019AnP...53170017E}
following \citet{selig2015fermi}, \citet{storm2017skyfact}, \citet{pumpe2018d4po},
and \citet{mishra2020gaussianprocess}.
We demonstrate a template-free spatio-spectral imaging approach for the gamma-ray sky
using a diffuse emission model, published and validated in \citet{arras2021comparison}
and \citet{arras_variable_2022},
and a PS emission model built following the same philosophy.
We show how the approach can be extended into a hybrid approach of imaging and template analyses.
For this, we include emission templates in our model
but retain the data-drivenness of the template-free imaging.
We call this hybrid approach ``template-informed imaging.''
We created two corresponding sky emission models (template-free and template-informed) and reconstructed the \textit{Fermi} LAT ten-year gamma-ray sky based on them.
To test the presented approaches,
we analyzed the spatio-spectral sky reconstructions obtained with the two models
in light of results from the literature.
Finally, we briefly discuss how the presented approach can be extended by or merged with existing emission modeling approaches.\\

%Subsect: Structure of the work
This paper is structured as follows:
In Sect.~\ref{sec:models-and-methods} we describe the models and methods used,
as well as our data selection.
Section~\ref{sec:results} showcases the results obtained with the presented imaging approach.
Therein, Sects.~\ref{sec:results-m1} and~\ref{sec:results-m2}
provide the sky maps obtained with the template-free and template-informed imaging,
as well as analyses of their features.
Section~\ref{sec:results-comparison} then compares the two obtained sky maps.
In Sect.~\ref{sec:discussion} we discuss our findings, highlighting their strengths and limitations.
We conclude in Sect.~\ref{sec:conclusions}.

The appendices are structured as follows:
Appendix~\ref{sec:appendix-generative-models} gives further details on the sky models.
Appendix~\ref{sec:appendix-instrument-response-model} describes how we implemented the instrument response model.
Appendix~\ref{sec:appendix-color-coding} details how we created the spatio-spectral sky map renderings contained in this work.
Finally, Appendix~\ref{sec:appendix-plots} contains supplementary figures.

\section{Models and methods} \label{sec:models-and-methods}
\subsection{General methods} \label{sec:methods-general}
In this paper we present a Bayesian imaging approach.
Bayesian inference assigns probability values $\mathcal{P}(s|d)$
to every possible configuration of a quantity of interest (``signal'')
$s$, given the data $d$. This assignment takes into account the
likelihood $\mathcal{P}(d|s)$, the probability of having obtained the
observed data, $d$, for a given $s$, and the prior $\mathcal{P}(s)$,
the probability of $s$ given pre-measurement knowledge, via Bayes'
theorem:
\begin{equation}
\mathcal{P}(s|d)=\frac{\mathcal{P}(d|s)\,\mathcal{P}(s)}{\mathcal{P}(d)}.
\end{equation}
The term $\mathcal{P}(d)=\int \mathrm{d}s\,\mathcal{P}(d|s)\,\mathcal{P}(s)$
(``evidence'') ensures proper normalization of the so-called posterior
probability distribution $\mathcal{P}(s|d)$.

In our case, the signal of interest is the time-averaged gamma-ray photon flux density of the sky $\Phi(x,E)$\footnote{
formally $\frac{\mathrm{d^4N}}{\mathrm{dt\,dA\,dE\,dx}}$
with units $\left(\mathrm{s}\:\mathrm{m}^{2}\:\mathrm{GeV}\:\mathrm{sr}\right)^{-1}$}
as a function of photon origin direction $x$ and energy $E$
in the energy interval of \unit[0.56--316]{GeV}
and the first 10 years of LAT operation.
$\Phi(x,E)$ is a continuous, strictly positive function in both variables.
We cannot numerically represent and reconstruct such general functions
without approximation,
so we derive a discretization-aware signal formulation related to $\Phi(x,E)$
in Sect.~\ref{sec:methods-general-signal-def}.

We performed a binned analysis, meaning that
the data with respect to which we estimate the posterior signal distribution
is a high-dimensional histogram of photons recorded by the LAT.
The LAT instrument response depends on photon properties such as the photon energy,
the incidence direction with respect to the instrument orientation $\psi=(\theta,\phi)$, 
and the location of the pair conversions within the instrument
\citep[\texttt{FRONT} or \texttt{BACK} conversions,][]{2012ApJS..203....4A}.
To include this effect in our model of the measurement,
the photon events were binned along the dimensions x, E, $\phi$, and event types,
and we model the response function for each histogram bin individually.
We assume the counts in the histogram to follow Poissonian statistics:
\begin{equation}
\mathcal{P}(d|\lambda) = \frac{\lambda^{d}}{d!}\,\mathrm{e}^{-\lambda}
\label{eq:data-poissonian-distributed}
,\end{equation}
with $\lambda$ being the histogram-bin photon rate predicted by
the signal $\mathrm{s}$ and our instrument response model $\mathrm{R}$:
$\lambda = \mathrm{R}(\mathrm{s})$.
Section~\ref{sec:methods-instrument-model} provides details of the instrument response modeling
and Sect.~\ref{sec:methods-dataset} details the event selection.

To make our template-free imaging robust against measurement noise,
we need to constrain the signal model.
In the Bayesian framework, constraints like this are naturally introduced
via the signal prior, which is based on our physical understanding of the signal.
In Sect.~\ref{sec:methods-formulation-of-prior-knowledge}
we describe how we express our prior knowledge on the reformulated signal quantity.
We implement the instrument, sky models and the reconstruction
using the numerical information field theory (NIFTy) framework
\citep{2013A&A...554A..26S,2019AnP...53100290S,2019ascl.soft03008A},
Version 8.4\footnote{\url{https://gitlab.mpcdf.mpg.de/ift/nifty}, commit \texttt{ebd57b33}}.
The signal priors are expressed in the form of generative models,
following \citet{knollmueller2018encoding}.

One significant challenge in performing inferences of high-dimensional parameter vectors,
as is necessary for the given imaging problem,
is the curse of dimensionality.
With rising parameter number, exploring the full parameter space in the inference quickly becomes infeasible.
In this work, we rely on metric Gaussian variational inference \citep[MGVI;][]{knollmueller2019mgvi},
a variational inference framework built for the high parameter number limit.
It allows us to iteratively optimize a set of posterior samples for models with millions of parameters,
from which posterior estimates of quantities of interest can be derived
and their variances can be inspected.

\subsection{Discretized signal definition} \label{sec:methods-general-signal-def}
In the process of binning the event data into count maps,
information on the flux structure within the spatial and spectral bins is lost.
As we want to make our analysis as data-driven as possible and correspondingly
do not want to impose strong assumptions on the structure of $\Phi(x,E)$ within the bins,
we do not try to reconstruct $\Phi(x,E)$, but instead infer its integrated counterpart,
\begin{equation}
I_\mathrm{ij} = \frac{1}{|\Omega_\mathrm{j}|} \int_{E_\mathrm{i}}^{E_\mathrm{i+1}} dE \int_{\Omega_\mathrm{j}} dx\;\Phi(x, E)
\label{eq:I-from-phi}
,\end{equation}
where $\Omega_{j}$ is the j-th sky pixels area (solid angle) and the energy bin boundaries are chosen equidistantly on a log-energy scale.
Structural information on bin-sized scales and above is retained in the binned count maps,
allowing us to do relatively unconstrained reconstructions of $I_\mathrm{ij}$.
The next section outlines what this means in practice.

One side effect of the binning into logarithmically equidistant energy bins
is a change in flux scaling with energy.
For logarithmically equidistant energy bins, the bin width is proportional to the base energy of the bins.
Because of this, $I_\mathrm{ij}$ effectively scales with energy like $E \cdot \Phi$
and spectral index estimates based on $I_\mathrm{ij}$ need to be adjusted by $1$
to compare with spectral index estimates for un-integrated fluxes, $\Phi$:
\begin{equation}
\alpha_{\,I} = \alpha_{\,\Phi} + 1.
\label{eq:spectral-index-adjustment}
\end{equation}

We pixelized the sky based on the HEALPix scheme introduced by \citet{gorski2005healpix}
and use an $\mathtt{nside}$ value of 128,
corresponding to a pixel size of approximately {0.46\textdegree}.
In the energy dimension, we pixelized the signal with a density of four bins per decade
between \unit[0.56]{GeV} and \unit[316]{GeV}.

\subsection{Formulation of prior knowledge and image regularization} \label{sec:methods-formulation-of-prior-knowledge}

\subsubsection{General modeling of the sky brightness} \label{sec:general-modeling-of-the-sky-brightness}

% ---- fig: data spectral plot ----
\begin{figure}
\resizebox{\hsize}{!}{\includegraphics{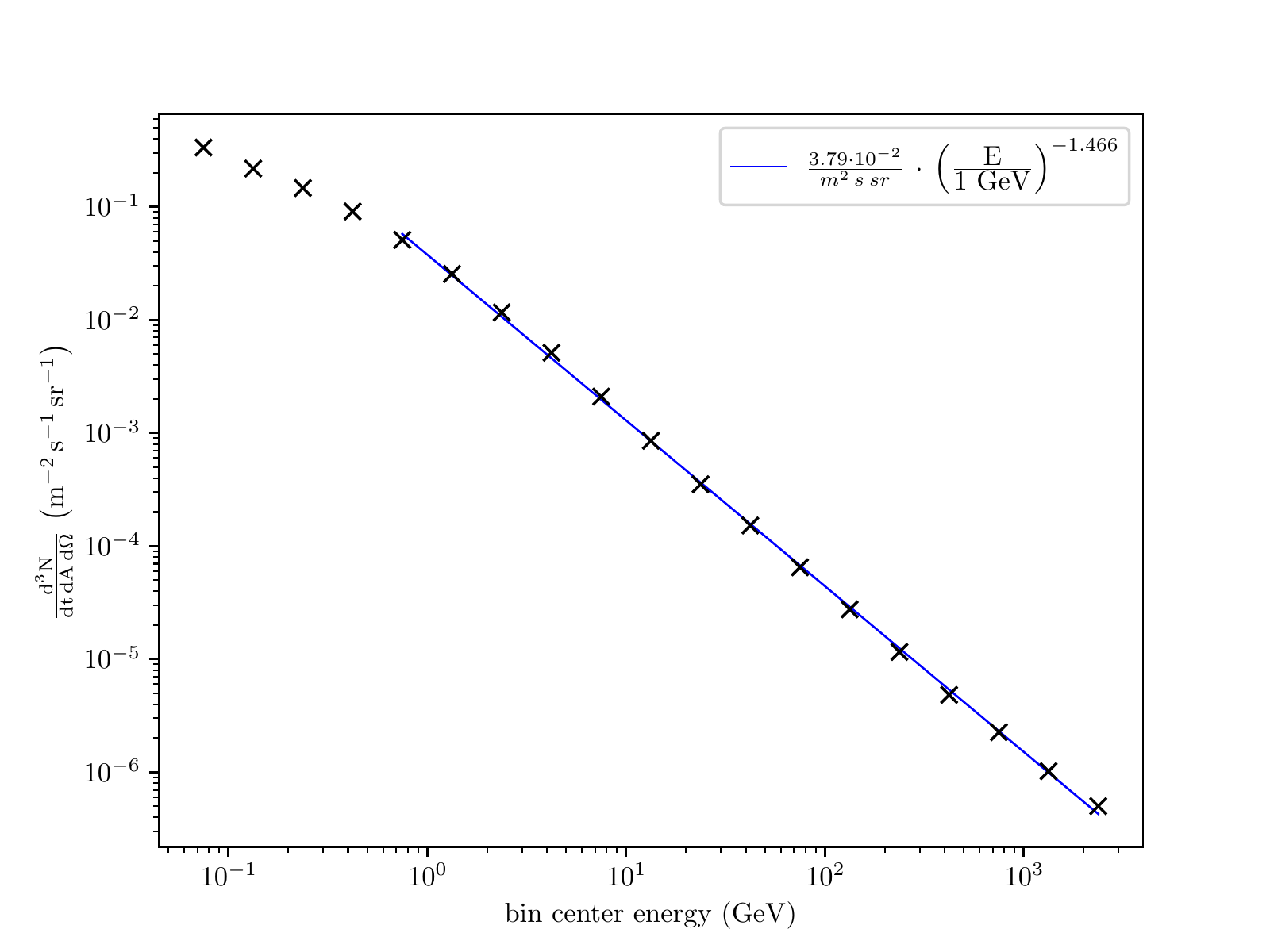}}
\caption{Sky-averaged estimate of $I_\mathrm{ij}$ on log-log scale based on the exposure- and effective-area-corrected photon count map.
Because we show fluxes integrated over logarithmically equidistant energy bins,
spectral index values are offset by $+1$ with respect to un-integrated energy spectra
(see Eqs.~\ref{eq:I-from-phi} and~\ref{eq:spectral-index-adjustment} for details).}
\label{fig:data-exposure-corrected-spectral-plot}
\end{figure}

Both $\Phi(x,E)$ and $I_\mathrm{ij}$ are strictly positive and vary over many orders of magnitude.
A rudimentary estimate of $I_\mathrm{ij}$ can be obtained by
dividing the data histogram by the exposure time and instrumental sensitivity for each data bin.
Figure~\ref{fig:data-exposure-corrected-spectral-plot} shows this quantity averaged over the spatial (x) and incidence direction ($\theta$) domains.
We observe it to follow a power law with a spectral index of $-1.466$ in the \unit[0.56--316]{GeV} energy range studied in this paper.
We note that this is the spectral index per logarithmic energy.
The corresponding spectral index per energy is $-2.466$.

Motivated by this, we model the bin-integrated sky flux as
\begin{equation}
I_\mathrm{ij} = I_{0} \cdot \left(\frac{E_i}{E_0}\right)^\alpha \cdot {10}^{\,\tau_\mathrm{ij}}
\label{eq:I-basic-model}
,\end{equation}
with 
\begin{equation}
\tau_\mathrm{ij}=\log_{10}\left(\frac{I_\mathrm{ij}}{I_\mathrm{0}\cdot(E_\mathrm{i}/E_\mathrm{0})^{\alpha}}\right)\label{eq:tau-general-definition}
\end{equation}
giving the variations in the sky with respect to a power-law spectrum with spectral index $\alpha$
on a decadic logarithmic scale. The employed value of the reference flux scale $I_\mathrm{0}$
and the reference energy $E_\mathrm{0}$ are provided in the following section.

This parameterization is convenient as it allows us to jointly model
the spatial variations in the integrated sky flux and
its spectral deviations from power-law spectra, both of which are expected to take values over many orders of magnitude,
on linear scales.
The modeling via logarithmic flux variations allows us
to accommodate this.

The sky exhibits correlations of the logarithmic fluxes over the spatial
and spectral domains, respectively,
$\langle\tau_\mathrm{ij}\,\tau_\mathrm{i'j'}\rangle_{(\tau)}$,
with $\langle f(\tau)\rangle_{(\tau)}:=\int \mathrm{d}\tau\,\mathcal{P}(\tau)\,f(\tau)$
being the expectation value of some function $f$ over the probability
distribution on $\tau$ as indicated by the index.
In the spectral direction the correlation structures emerge from high-energy astroparticle
processes, while the spatial correlations are related to macroscopic
events, for example the temporal evolution of our Galaxy.
We therefore assume these correlations to be separable in the spatial and energy
direction. A priori, we also do not want to single out any directions
or energy scale, so we additionally assume statistical homogeneity
with respect to locations and log-energies $y_\mathrm{i}=\log_{10}\left(E_\mathrm{i}/E_\mathrm{0}\right)$,
$\langle\tau_\mathrm{ij}\,\tau_\mathrm{i'j'}\rangle_{(\tau)}=C(|x_\mathrm{j}-x_\mathrm{j'}|)\,D(|y_\mathrm{i}-y_\mathrm{j'}|)$,
where we changed from the energy $E$ to the log-energy coordinate $y$.
By using the log-energy coordinate $y$ and the assumption of homogeneous
spectral correlations, we assume that the to be reconstructed energy spectra behave
similarly over several decades in energy.

Spatial homogeneity of the correlations of $\tau$ expresses the assumption
that the relative contrast of emission is structured similarly everywhere
on the sky. This is not true in general, as for example point-like
sources and diffuse emission certainly differ morphologically.
Similarly, hadronic and leptonic emission should exhibit different
structures due to the differently structured target densities, those of the nuclei
and photons, respectively. Consequently, we model the total gamma-ray
flux as a superposition of several emission processes $I_\mathrm{ij}^\mathrm{\:comp}$
 (in the following ``components''), always including point-like emission and
 one or more diffuse emission components.
Each follows the functional form outlined in Eq.~\ref{eq:I-basic-model} and
has its own correlation structure:
\begin{equation}
\langle\tau_\mathrm{ij}^\mathrm{\:c}\tau_\mathrm{i'j'}^\mathrm{\:c'}\rangle_{(\tau)}
= \delta_\mathrm{c,\,c'}\,C^\mathrm{\:c}(|x_\mathrm{j}-x_\mathrm{j'}|)\,D^\mathrm{\:c}(|y_\mathrm{i}-y_\mathrm{i'}|).
\end{equation}
A priori, we assumed each component to be uncorrelated with the others.
Here $C^\mathrm{\:c}(\Delta x)$ is the spatial correlation function of component c
and $D^\mathrm{\:c}(\Delta y)$ the spectral one.
The Kronecker-delta $\delta_\mathrm{c,\,c'}$ encodes a priori
independence between the different components.
This separability of the spatial and spectral correlations
expresses the assumption that at different spatial locations of an
emission component one expects similar, but not identical, spectral
structures and vice versa.

The corresponding models of $\tau^\mathrm{\:c}$ were built based on
the Gaussian correlated field model introduced by \citet[][Sect.~3.4]{arras2021comparison}
and \citet{arras_variable_2022}.
It allows us to translate prior knowledge on the correlation functions $C^\mathrm{\:c}$
and $D^\mathrm{\:c}$ into hierarchical generative models of $\tau^\mathrm{\:c}$.
During reconstruction, a self-consistent pair of correlation functions and $\tau$s was
inferred, providing a data-informed regularization of the inferred component sky maps
without a priori imposing specific spatio-spectral flux shapes or flux intensity scales.
Appendix~\ref{sec:appendix-generative-models} provides details of the generative modeling.

Flux from point-like sources can be modeled in this way as well.
Assuming their location and brightness to be independent,
their spatial correlation function is a delta function,
$C^\mathrm{\:point}(\Delta x)=\delta(\Delta x)$.
Each sky pixel harbors the photon flux contributed by all PSs situated within it.
Since the set of PSs comprises a large variety of different physical
objects, the energy spectra and total brightness values found for the pixels are expected to
strongly differ across the sky.
To accommodate this, we model the PS energy spectra differently
from the diffuse energy spectra.
Similarly to the diffuse energy spectra, we model them as following power-law spectra
with variations on logarithmic scales.
However, each pixel has an individual power-law spectral index $\alpha^\mathrm{\:point}_\mathrm{\;j}$
and the logarithmic deviations from pure power-law spectra are modeled for each pixel separately with a
Gaussian process along the spectral dimension $\epsilon_\mathrm{ij}$,
which we parameterized such that it naturally models sharp spectral features in the PS energy spectra.

Furthermore, the total brightness distribution function of the PS pixels $\mathcal{P}(I^\mathrm{\:point}_\mathrm{tot,\,j})$
is modeled as following a power-law distribution
in pixel brightness $I^\mathrm{\:point}_\mathrm{tot,\,j}$ with a fixed power-law index of $-2.5$ and an exponential cutoff at low brightness values
for normalizability.
This corresponds to the assumption of a uniform source distribution
in a steady-state and statically homogeneous Euclidean universe \citep{ryle1950,mills1952}.
This is a simplifying assumption and will bias the found source distribution, but was chosen
for simplicity.

The PS component is thus modeled as
\begin{equation}
I^\mathrm{\:point}_\mathrm{ij} = I^\mathrm{\:point}_\mathrm{tot,\:j} \cdot \frac{f_\mathrm{spec,\:ij}}{\sum_\mathrm{i} f_\mathrm{spec,\:ij}};
\;\;f_\mathrm{spec,\:ij} = {10}^{\:\alpha_\mathrm{j} \cdot y_\mathrm{i} \,+\, \epsilon_\mathrm{ij}}
\label{eq:I-point-source-basic}
.\end{equation}

We chose the prior brightness mean to be two orders of magnitude below the expected
diffuse emission brightness mean,
to avoid introducing an isotropic background via the PS component.
This effectively makes the PS brightness prior sparsity-enforcing,
as PSs need to deviate by many standard deviations from the prior mean
to contribute significant amounts of flux.
This removes the degeneracy between the diffuse and PS components,
which in principle could both account for the full sky flux
(as we have a PS in every pixel).
By biasing the PSs toward insignificant contributions we can ensure
PS flux is only reconstructed when strongly suggested by the data.
Section~\ref{sec:discussion} provides a discussion of potential effects of our choices regarding
the PS component prior.

To summarize, we assume the log-flux correlation function of the components
to be separable in spatial and log-energy direction. This does not
imply the sky components themselves to be separable into spatial and
spectral component functions, but favors such a separability unless the data requests nonseparable
spatio-spectral structures.
This is suboptimal for distinct extended objects
\citep{lande2012extendedsources,ackermann2017extendedsources,ackermann2018extendedsources}
that are neither well represented by the PS emission field nor correctly characterized
by the spectra and spatial properties of any of the all-sky diffuse emission fields.
Image reconstruction artifacts and misclassifications with respect to the assumed sky component classes
are expected to happen for such objects in the current approach.
Still, the majority of gamma-ray emission in the studied energy interval is expected to
fit these models and is efficiently parameterized by them, as the reconstruction results show.
The following subsection defines a template-free gamma-ray sky model, M1,
based on the given concepts.

\subsubsection{Template-free imaging model, M1} \label{sec:methods-m1}

Model M1 for the template-free reconstruction consists of a single diffuse emission component
$I^\mathrm{\:diff}$ and a PS component $I^\mathrm{\:point}$ as described in the previous section.
Appendix~\ref{sec:appendix-generative-models} provides details of the M1 sky model,
including a table of prior parameter values used. Here, we want to highlight a few especially important
model parameters that most strongly determine the characteristics of the modeled components.

First, both the PS and diffuse models contain power laws along the energy dimension,
for which a priori spectral indices need to be defined.
Because we model flux integrated over logarithmically equidistant energy bins $I_\mathrm{ij}$,
but in literature spectral indices are usually defined for un-integrated fluxes $\Phi$,
we adjusted spectral index values obtained from literature according to Eq.~\ref{eq:spectral-index-adjustment}.
In case of the diffuse component $I^\mathrm{\:diff}$,
we set the adjusted a priori spectral index to $\alpha^\mathrm{\:diff}=-1.466$,
the value obtained from the estimate of $I_\mathrm{ij}$ based on
the exposure- and effective-area-corrected data (see Fig.~\ref{fig:data-exposure-corrected-spectral-plot}).
Here we used the assumption that the gamma-ray sky is dominated by diffuse emission,
such that we can take the average spectral index observed for the estimate of $I_\mathrm{ij}$
as the prior mean for the diffuse component spectral index without modification.

For the PS component, the spectral index prior $\mathcal{P}(\alpha^\mathrm{\:point}_\mathrm{j})$
was set according to \citet[Fig.~16]{abdollahi2020fermi},
from which we estimated a mean PS spectral index of $\alpha_\Phi = -2.25$
and a spectral index standard deviation of $0.30$.
Adjusting for the fact that we model flux integrated over logarithmic energy bins,
we set the adjusted a priori mean spectral index of the PS pixels to
$\mu_\mathrm{ap} := \langle\alpha^\mathrm{\:point}_\mathrm{j}\rangle_\mathrm{prior}=-1.25$
and the prior spectral standard deviation to $\sigma_\mathrm{ap}=0.30$ (unchanged).

Secondly, we needed to specify how much the overall brightness of the diffuse component can vary
with respect to the reference scale $I^\mathrm{\:diff}_\mathrm{0}$.
This model property is controlled by the prior on the mean of $\tau^\mathrm{\:diff}$,
for which we chose a zero-centered Gaussian distribution with a standard deviation of $0.5$.
This corresponds to a log10-normal prior on the best-fit diffuse brightness scale $I^\mathrm{\:diff}_\mathrm{0,\,bf}$,
and allows our reconstruction to globally scale up or down the diffuse flux by a factor of 3 within one prior
standard deviation and a factor of 10 with two prior standard deviations.
This gives the reconstruction considerable freedom to correct a suboptimally chosen reference brightness scale $I^\mathrm{\:diff}_\mathrm{0}$.

Third, we needed to specify by how many orders of magnitude $\tau^\mathrm{\:diff}$ should be able to locally modify
the flux density of $I^\mathrm{\:diff}$.
This property of the model is controlled by the prior standard deviation of the fluctuations in $\tau^\mathrm{\:diff}$  with respect to its mean.
We chose a prior mean of $0.75$ for this parameter, meaning the reconstruction can locally change
the $I^\mathrm{\:diff}$ flux density by $0.75$ orders of magnitude within one prior standard deviation
and by $1.5$ orders of magnitude within two prior standard deviations.
Put another way, this assumes the spatial fluctuations of $I^\mathrm{\:diff}$
lie on the order of three orders of magnitude.

Lastly, the correlated field model employed for $\tau^\mathrm{\:diff}$ contains
priors for the spatial and spectral correlation structure of the fluctuation field $\tau^\mathrm{\:diff}$.
This serves to encode our prior knowledge on the spatial and spectral correlation structure
of $\tau^\mathrm{\:diff}$ and $I^\mathrm{\:diff}$.
Appendix~\ref{sec:appendix-generative-models} provides a more mathematical explanation of this model component,
but for the sake of completeness, we give a short summary here.
The spatial correlation structure model in $\tau^\mathrm{\:diff}$ assumes a priori that the spatial power spectrum follows
a power law in angular harmonic mode scale $C_\mathrm{\ell} \propto \ell^{\:\beta}$.
For the prior distribution of the power-law index, $\beta$, we chose a Gaussian distribution $\mathcal{N} \left(\beta\:|-3,\,0.25\right)$.
This corresponds to assuming that the spatial structure of $I^\mathrm{\:diff}$,
which needs to take up small-scale hydrogen density correlated features and larger scale smooth emission structures alike,
is smoother than integrated Brownian noise (Wiener process).
The energy dimension correlation structure model in $\tau^\mathrm{\:diff}$ assumes a priori that the spectral dimension power spectrum follows
a power law in harmonic energy mode scale $D_\mathrm{q} \propto \mathrm{q}^{\:\gamma}$.
For the prior distribution of the power-law index, $\gamma$, we chose a Gaussian distribution $\mathcal{N} \left(\gamma\:|-3.5,\,0.25\right)$.
This corresponds to a priori smoother structures in the energy spectrum dimension than in the spatial domain,
as is expected for most gamma-ray production processes.

Appendix~\ref{sec:appendix-generative-models} provides details on the correlated field model employed here and contains a table of all prior parameters.
Section~\ref{sec:results-m1} gives our results of the template-free imaging run.

\subsubsection{From imaging to template-informed imaging} \label{sec:methods-template-informed-imaging}

In principle, an arbitrarily complex sky can be modeled in the described way if a sufficient number of 
flexible sky components is included, one for each kind of emission.
However, in practice, if one allowed a flexible component for every known emission mechanism,
the inference problem would get too degenerate to solve, as the model degrees of freedom would not be sufficiently constrained
by the limited information available in the data.
This degeneracy can be lifted by adding informative templates to the models,
giving a priori spatial or spectral structure to some emission components.

For example, a spatial emission template $I^\mathrm{\:T} (x)$ can be included in a component model
by modifying Eq.~\ref{eq:I-basic-model} as follows:
\begin{equation}
I^\mathrm{\:c}_\mathrm{ij} = I^\mathrm{\:c}_{0} \cdot \left(\frac{E_i}{E_0}\right)^\alpha \cdot {10}^{\,\tau^\mathrm{\:c}_\mathrm{ij}}
                   \cdot \nicefrac{I^\mathrm{\:T}_\mathrm{\:j}}{\langle I^\mathrm{\:T}_\mathrm{j} \rangle_\mathrm{geom}}
\label{eq:I-dust-model}
,\end{equation}
where $\langle \cdot \rangle_\mathrm{geom}$ is the geometric mean function
and $I^\mathrm{\:T}_\mathrm{j}$ is the pixel surface integrated counterpart of $I^\mathrm{\:T} (x)$.
This imprints the template into the prior mean of the component.
Now, instead of modeling variations with respect to an a priori spatially uniform component brightness,
 $\tau^\mathrm{\:c}$ models spatio-spectral variations with respect to the template.

By choosing how much freedom we give $\tau^\mathrm{\:c}$ to make the component deviate
from the prior mean set by the template,
we can choose how data-driven the reconstruction of the component should be.
As long as at least one flexible component remains, the total reconstructed flux density
can be expected to be data-driven,
as the flexible component can take up all flux not adopted by the template-informed components.
The flexibility with respect to their templates of the remaining component picks a trade-off
between the degeneracy of the problem and the data-drivenness of the component reconstructions.

We see this approach as an extension of the works presented by \citet{storm2017skyfact}
and \citet{mishra2020gaussianprocess}, discussed in the introduction,
exploring the extreme end of the template fitting -- imaging continuum.

\subsubsection{Template-informed imaging model, M2} \label{sec:methods-m2}

To demonstrate the potential of our proposed template-informed imaging approach,
we show a gamma-ray sky reconstruction based on a template-informed model, M2.
Similar to the template-free model M1, it contains a PS component $I^\mathrm{\:point}$
and a template-free diffuse component $I^\mathrm{\:nd}$.
Additionally, it contains a template-informed diffuse component $I^\mathrm{\:dust}$
that we created as described in Eq.~\ref{eq:I-dust-model}.

As the most prominent diffuse Galactic foreground component is emission from hadronic interactions,
we employed a template informing the reconstruction about potential interaction sites for it.
A natural choice for the template would be one of the hadronic emission templates derived
for recent template-fit-based publications.
However, to demonstrate the ability of our approach to correct major template errors, we used
the \textit{Planck} 545\ GHz thermal dust emission map
\citep[in the following called the ``thermal dust map'';][]{planck2015thermaldust} as the template.
It is morphologically similar to the soft-spectrum structures visible in Fig.~\ref{fig:data-exposure-corrected-mf-plot}
and in our template-free diffuse reconstructions (see Sect.~\ref{sec:results-m1}),
and expected to trace the ISM density.
However, to challenge our method, we explicitly did not apply the usually applied corrections to our template,
such as hydrogen column density correction and CR transport effect modeling.
The fluctuation field $\tau^\mathrm{\:dust}$ thus was required to learn the necessary corrections to the template.

The second, template-free diffuse component can take up the diffuse emission from processes
whose target population distributions are uncorrelated with the template, unveiling them in the process.
We give it the suffix ``\texttt{nd},'' which stands for ``non-dust diffuse emission.''

Most model parameters were adapted from model M1, but some needed to be changed to account for the
newly introduced diffuse component.
First, the reference brightness scales for the two diffuse components $I^\mathrm{\:dust}$ and $I^\mathrm{\:nd}$
were set each to half the value used for the M1 diffuse component,
producing an a priori even split of diffuse flux contribution by both diffuse components,
which, however, does not force the a posteriori results to exhibit the same split.

Second, the prior energy spectrum spectral indices of the two components were tailored to the gamma-ray production processes modeled by them.
The dust map-informed component is expected to almost exclusively take up hadronic emissions,
for which we expected steeper energy spectra than for the mixture of diffuse emissions reconstructed in the M1 diffuse component.
We set the a priori spectral index of the template-informed component to $\alpha^\mathrm{\:dust}=-1.65$.
The template-free diffuse component was expected to take up
most leptonic emission, which has a flatter spectrum than the
mixture of diffuse emissions.
Accordingly, we set the a priori spectral index of the template-free component to $\alpha^\mathrm{\:nd}=-1.25$.

Third, as the template already traces fine details of the spatial source distribution for the hadronic emission,
the fluctuation field $\tau^\mathrm{\:dust}$ does not need to model them,
opposed to $\tau^\mathrm{\:diff}$ in M1.
We therefore set the correlation structure prior of $\tau^\mathrm{\:dust}$ to prefer smoother spatial structures
than $\tau^\mathrm{\:diff}$.

Appendix~\ref{sec:appendix-generative-models} contains the full set of prior parameters for the model M2.
Section~\ref{sec:results-m2} gives the results of our template-informed imaging run,
including a map of the corrections with respect to the thermal dust map that were determined.

\subsection{Instrument response model} \label{sec:methods-instrument-model}

In this section, we describe the model of the instrument
response $\mathrm{R}$ used in this work.
The LAT collects photon flux $\Phi (x,y)$ coming from sky
directions $x$ and log energies $y$. The incidence directions of the
collected photons are reconstructed in detector coordinates,
but because of physical and manufacturing limitations\footnote{
Such as the finite path length through the instrument of electron-positron
pairs created by low-energy (<100 MeV) gamma rays or the finite spacing of
charge tracker strips.}
they cannot
be reconstructed perfectly by the instrument \citep{2012ApJS..203....4A}.
The PSF describes the corresponding
spreading of reconstructed incidence directions.
Formally it is the conditional probability $\mathrm{PSF}(x'|x,y)=\mathcal{P}(x'|x,y)$
of a photon entering the instrument from direction x and with energy
y to end up as being classified to stem from $x'$ within the detector coordinates.
Thus, the detector plane flux is 
\begin{equation}
J(x',y)=\int \mathrm{d}x\,\mathrm{PSF}(x'|x,y)\,I(x,y).
\end{equation}
Analogously, the log-energy $y'$, with which the photon is registered,
can deviate from its true log-energy $y$.
This is described by an
energy-dispersion function (EDF),  $\mathrm{EDF}(y'|y)=\mathcal{P}(y'|y)$,
which we here assume to be independent of the sky photon direction.
The number density of photons ending up at detector coordinates $(x',y')$
is therefore 
\begin{equation}
J'(x',y')=\int \mathrm{d}y\,\mathrm{EDF}(y'|x',y)\,J(x',y).
\end{equation}
In practice, the instrument response functions (IRFs; consisting of PSF, EDF, and detector sensitivity)
all depend on the photon incidence direction
with respect to the instrument orientation $\psi=(\theta,\phi)$, the photon energy,
and the location of the pair conversions within the instrument
\citep[\texttt{FRONT} or \texttt{BACK} conversions,][]{2012ApJS..203....4A}.
As we did not model the photon flux $\Phi(x,y)$ directly, but only its bin-integrated
counterpart $I_\mathrm{ij}$, we represented the PSF and EDF by tensors,
\begin{equation}
J_\mathrm{vw'i'j'} = \sum_\mathrm{i,j} \mathrm{EDF}^\mathrm{\:v}_\mathrm{i'i} \cdot \mathrm{PSF}^\mathrm{\:i'vw'}_\mathrm{j'j} \cdot I_\mathrm{ij},
\end{equation}
that simulate the effect of the respective IRF on the integrated sky
and include the dependence on the event type, $\mathrm{v} \in \{\mathtt{FRONT},\mathtt{BACK}\}$,
the estimated photon incidence direction with respect to the instrument main axis, $\mathrm{w'}$ (binned),
and the estimated photon energy bin $\mathrm{i'}$.
In our model
the detector coordinate photon flux $J'(x',y')$ is further weighted with the product of the effective area
of the instrument, $\mathrm{A}_\mathrm{eff}(x, y)$, and the exposure, $\mathrm{T}_\mathrm{exp}(x)$.
For the discrete analog, we again modeled this with tensors.
The expected number of photons in the combined data bin $\mathrm{vw'i'j'}$ is then
\begin{equation}
\lambda_\mathrm{\,vw'i'j'} = \mathrm{T}_\mathrm{exp}^\mathrm{\:w'j'} \cdot \mathrm{A}_\mathrm{eff}^\mathrm{\:vw'i'} \cdot |\Omega_\mathrm{j}| \cdot  J_\mathrm{vw'i'j'}
.\end{equation}
The observed photon number in the respective data bin is assumed to originate from a Poisson process
with $\lambda_\mathrm{\,vw'i'j'}$ as its expectation value:
\begin{eqnarray}
d_\mathrm{vw'i'j'} & \sim & \mathcal{P}(d_\mathrm{\,vw'i'j'}|\lambda_\mathrm{\,vw'i'j'})
,\end{eqnarray}
where $d_\mathrm{\,vw'i'j'}$ are the entries of the data vector,
which carry the detected number of photons with event type $v$,
apparent incidence directions with respect to detector $\cos(\theta) \in\left(\cos(\theta_\mathrm{w'}),\cos(\theta_\mathrm{w'+1})\right]$,
apparent incidence directions $x_\mathrm{j'}\in\Omega_{j}$, and
apparent energies $y_\mathrm{i'}\in[y_\mathrm{i'},y_\mathrm{i'+1})$.

The \textit{Fermi} Collaboration provides renditions of the IRFs dependent
on all relevant variables, which we made full use of in the creation of our model tensors,
except for the Fisheye ($\phi$) correction, which we omitted for the sake of computational feasibility\footnote{
Further information about the IRFs is available at
\url{https://fermi.gsfc.nasa.gov/ssc/data/analysis/documentation/Cicerone/Cicerone_LAT_IRFs/index.html}.}.
Appendix \ref{sec:appendix-instrument-response-model} provides implementation details for the PSF and EDF tensors.

\subsection{Data set} \label{sec:methods-dataset}

As mentioned in the introduction,
the LAT records individual gamma-ray photon interactions.
Because the instrument is also sensitive to charged CRs,
it employs on-board event filtering to reject CR events.
To handle the remaining CR event contamination,
the \textit{Fermi} Collaboration performs extensive post-processing of the
recorded events and provides multiple data cuts.
Since the start of operation of the LAT,
the post-processing pipeline was updated
multiple times based on the most recent understanding of the instrument.
The current major data release by the \textit{Fermi} Collaboration, pass 8,
and its newest update, release 3 (\texttt{P8R3})
are described in \citet{atwood2013pass} and \citet{bruel2018pass}.
This includes a clustering of gamma-ray detection events
into classes of varying background contamination,
tuned to accommodate the requirements of different applications.

For our analysis, we use the \texttt{\detokenize{P8R3_SOURCE}} class,
which was optimized to provide a background rejection appropriate
for PS and extended object analyses, and
which is the most permissive data cut provided in the standard
data releases.\footnote{
The LAT data products overview published at \url{https://fermi.gsfc.nasa.gov/ssc/data/analysis/documentation/Cicerone/Cicerone_Data/LAT_DP.html}
provides more information on the \texttt{P8R3} data classes.}
We utilized data from mission weeks 9--511 and 514--599, taken as
weekly \texttt{photon} and \texttt{spacecraft} files
from the FTP servers linked on the \textit{Fermi} data access site\footnote{\url{https://fermi.gsfc.nasa.gov/ssc/data/access/}}.
Further, we restricted the photon energies to the \unit[1--316]{GeV} range
and to incidence directions with respect to the detector $\theta \leq$~{45.57\textdegree}.
The LAT recorded $2.4 \cdot {10}^8$~\texttt{\detokenize{P8R3_SOURCE}}
class gamma-ray events compatible with these selection criteria,
which averages to slightly below \unit[1]{event per second}.

As introduced in Sect.~\ref{sec:methods-general},
we created a 5D histogram of photon properties $d_\mathrm{\,vw'i'j'}$
from the included events
as the data with respect to which we reconstructed $I_\mathrm{ij}$.
The histogram dimensions and corresponding binning strategies are as follows.
The photon origin directions were binned according
to the HEALPix pixelization \citep{gorski2005healpix} with an $\mathtt{nside}$ of 128.
The photon energy values were binned to four logarithmic bins per decade.
The incidence directions with respect to the instrument $\theta$ were binned to three bins
equidistant in $\cos(\theta)$.
The conversion location classification by the instrument (\texttt{FRONT} or \texttt{BACK}) was retained as a separate dimension.
This results in a total of $1.18\cdot{10}^7$ data bins.

\begin{figure}
\begin{minipage}[t]{\hsize}%
\includegraphics[width=1\textwidth]{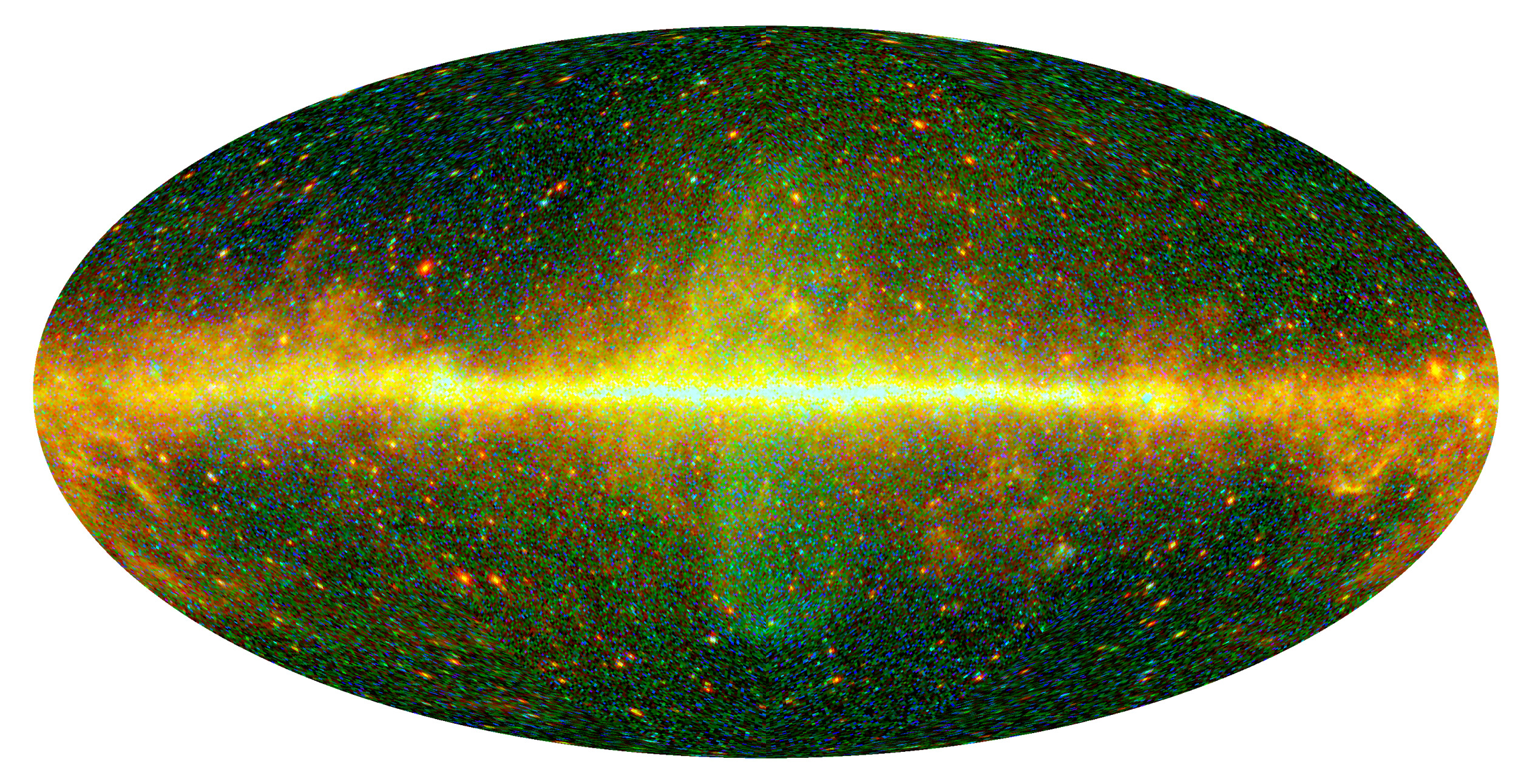}%
\end{minipage}
\begin{minipage}[t]{\hsize}%
\vspace{0.0cm}
\includegraphics[width=1\textwidth]{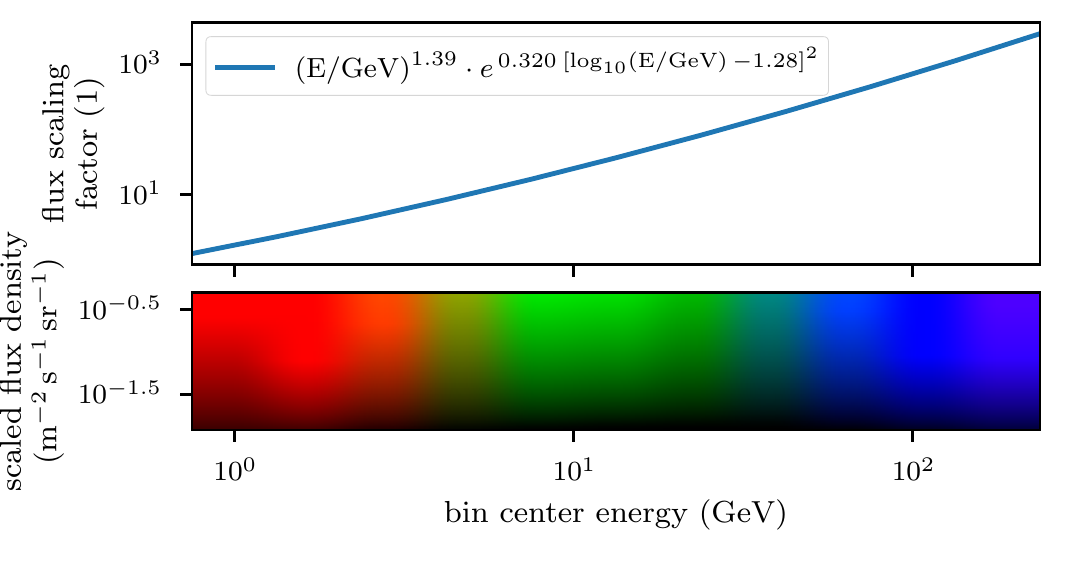}%
\end{minipage}
\caption{
Introduction to the spatio-spectral plotting and corresponding data plot.
\textbf{Top:} Exposure- and effective-area-corrected photon count observed
by the \textit{Fermi} LAT in the \unit[1--316]{GeV} range within the selected time frame.
The photon energies are color coded with red for the lowest and blue for the highest energies.
To compensate for the flux intensity change within the shown energy range,
we scaled the fluxes with an energy-dependent factor before plotting.
This presentation is optimized to show spectral variations.
Appendix~\ref{sec:appendix-color-coding} details the employed processing steps.
For reading flux values at specific energies, the reader is referred to the public release
of data products at the Zenodo data repository linked in the title footnote.
This and all further sky maps are based on the HEALPix pixelization \citep{gorski2005healpix}
with an $\mathtt{nside}$ of 128 and employ the Mollweide projection.
\textbf{Middle}: Energy-dependent flux scaling factor used for the spatio-spectral plotting.
\textbf{Bottom}: Mapping of photon energies and scaled flux densities to perceived colors
used in the spatio-spectral plotting.
}
\label{fig:data-exposure-corrected-mf-plot}
\end{figure}

As discussed in Sect.~\ref{sec:methods-formulation-of-prior-knowledge},
we can estimate the time-averaged spatio-spectral gamma-ray sky based
on the data histogram by dividing it by the exposure time and instrument sensitivity in each data bin
and aggregating the resulting flux estimates for all event types and incidence directions with respect to the instrument:
\begin{equation}
\:\mathrm{\tilde{I}}_\mathrm{i'j'} = \sum_\mathrm{vw'} \: d_\mathrm{\,vw'i'j'} \,/\, \left(\mathrm{T}_\mathrm{exp}^\mathrm{\:w'j'} \cdot \mathrm{A}_\mathrm{eff}^\mathrm{\:vw'i'}\right).
\end{equation}
Figure~\ref{fig:data-exposure-corrected-mf-plot} (top panel) shows $\mathrm{\tilde{I}}_\mathrm{i'j'}$
for the data set we use in this work
and highlights the high fidelity of the \textit{Fermi} LAT data,
which made it one of the driving instruments for gamma astronomy over the last decade.

% ---- fig: data plot

\section{Results} \label{sec:results}

In the following, we present the results of applying the two imaging models M1 and M2
to \textit{Fermi} LAT data
and perform analyses to connect our findings with results from the literature.
First, we look at the results of both models independently,
and then perform comparative analyses.

\subsection{Template-free spatio-spectral imaging via M1} \label{sec:results-m1}

% ---- fig: m1 sky maps ----
\begin{figure*}
\centering
\noindent
\begin{minipage}[t]{1\textwidth}%
\includegraphics[width=1\textwidth]{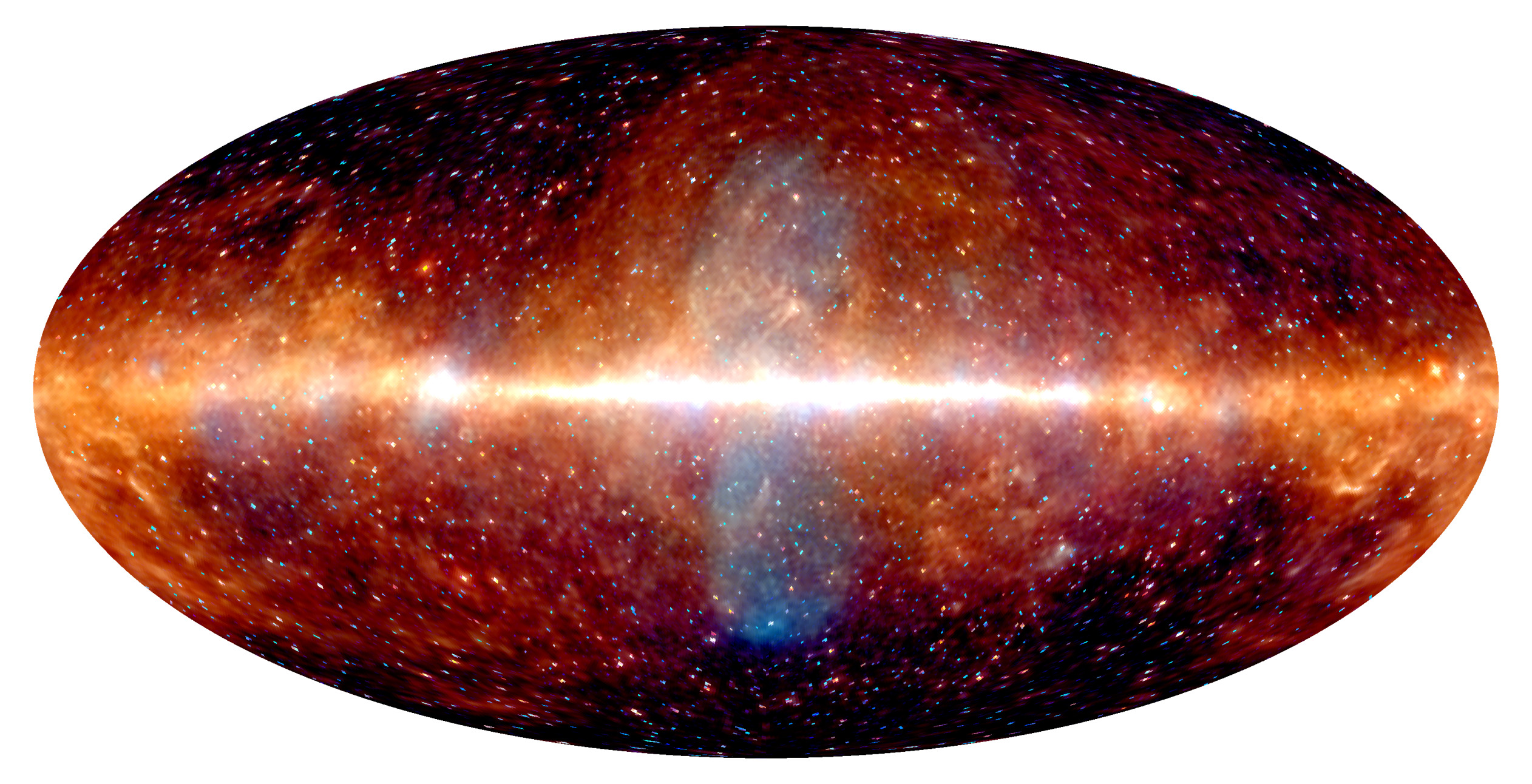}%
\end{minipage}

\begin{minipage}[t]{0.49\textwidth}%
\includegraphics[width=1\textwidth]{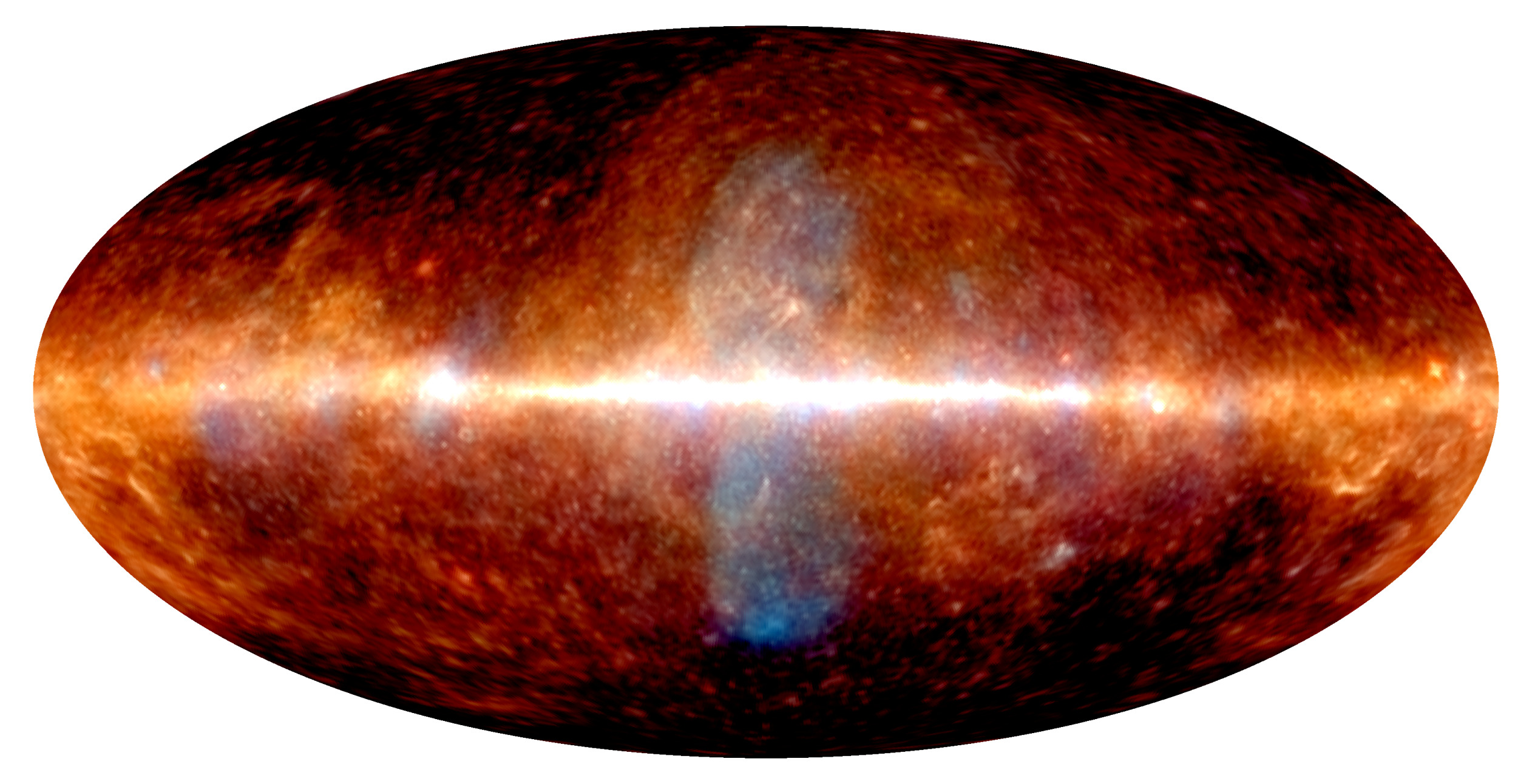}%
\end{minipage}\ %
\begin{minipage}[t]{0.49\textwidth}%
\includegraphics[width=1\textwidth]{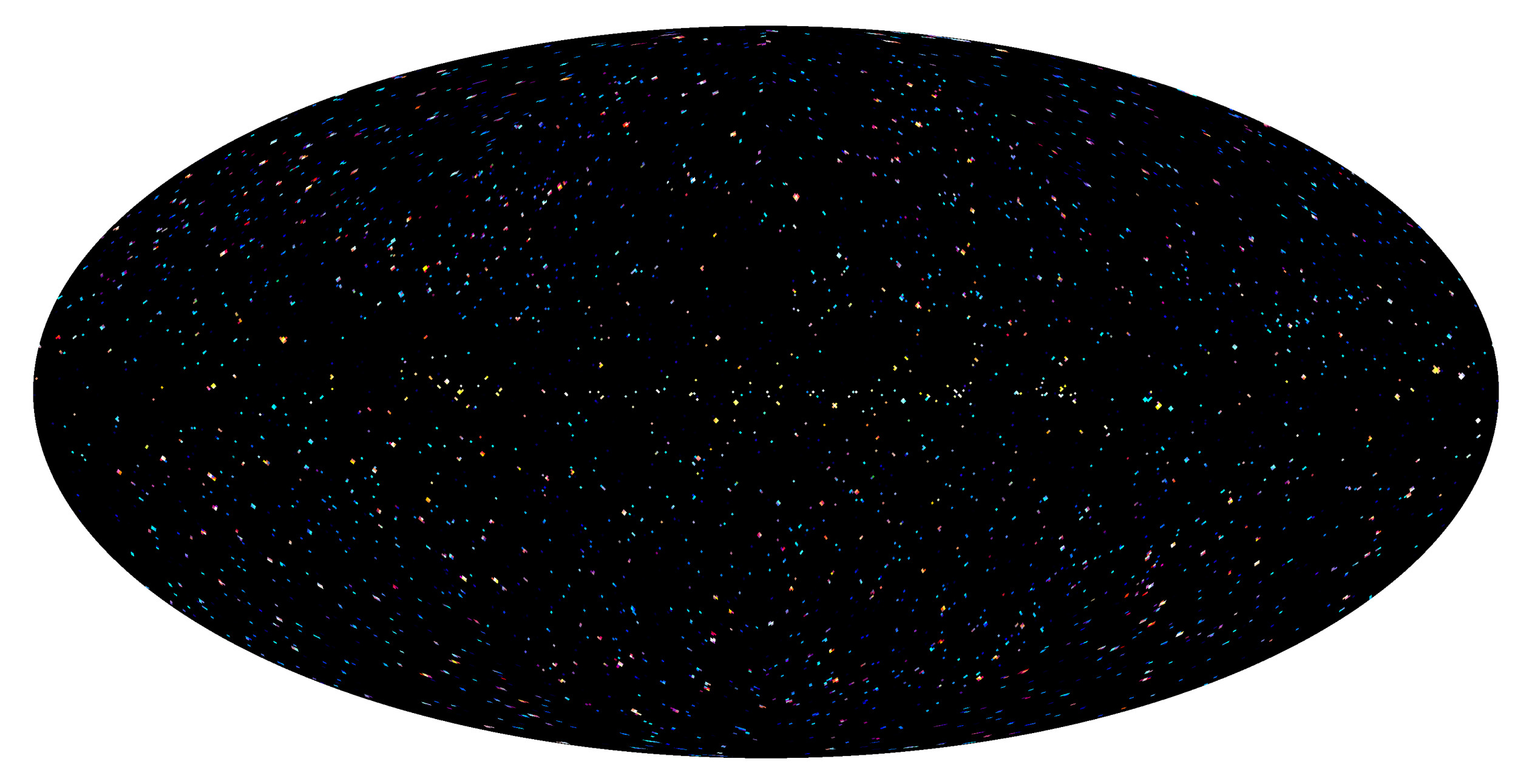}%
\end{minipage}

\begin{minipage}[t]{0.33\textwidth}%
\includegraphics[width=1\textwidth]{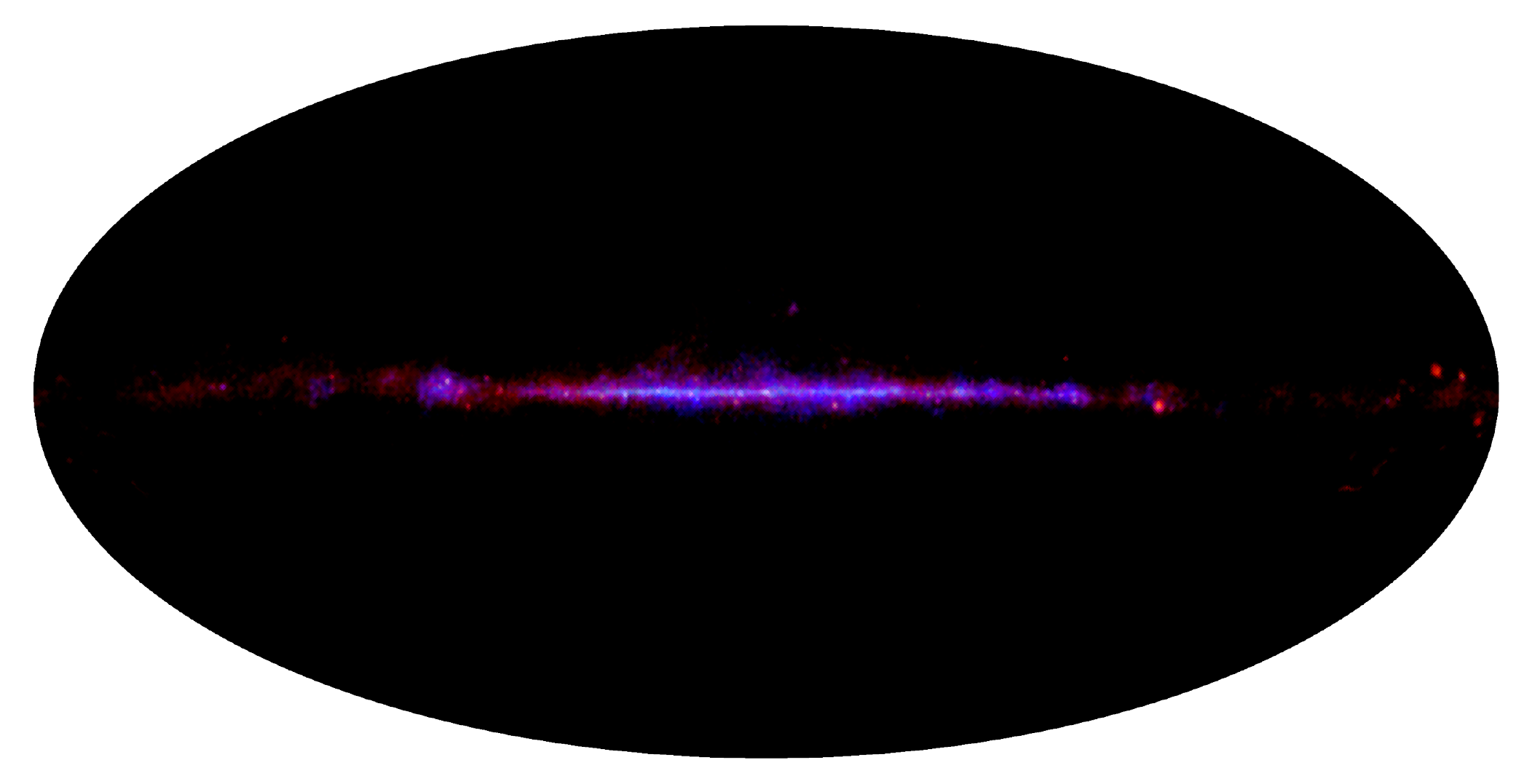}%
\end{minipage}%
\begin{minipage}[t]{0.33\textwidth}%
\includegraphics[width=1\textwidth]{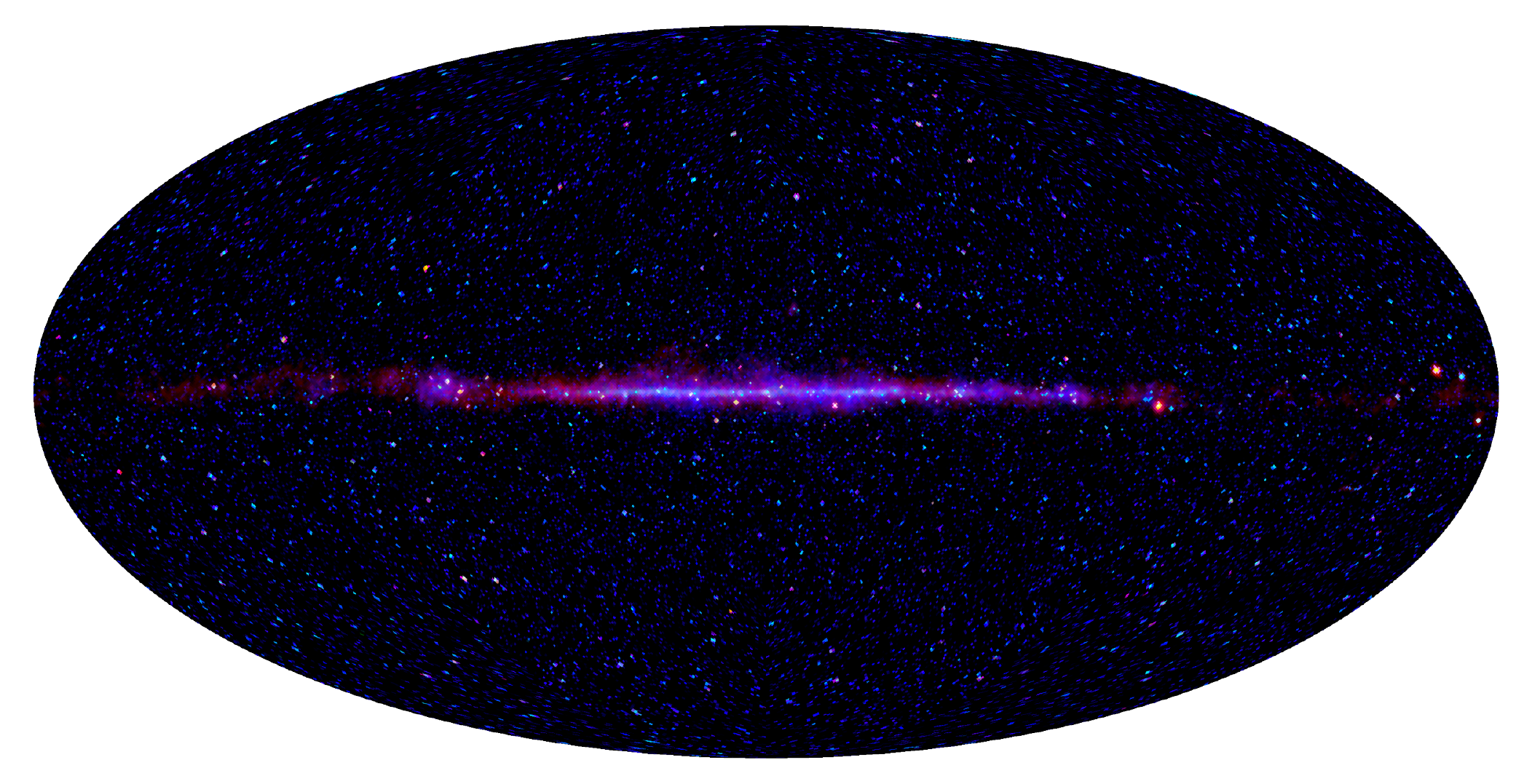}%
\end{minipage}%
\begin{minipage}[t]{0.33\textwidth}%
\includegraphics[width=1\textwidth]{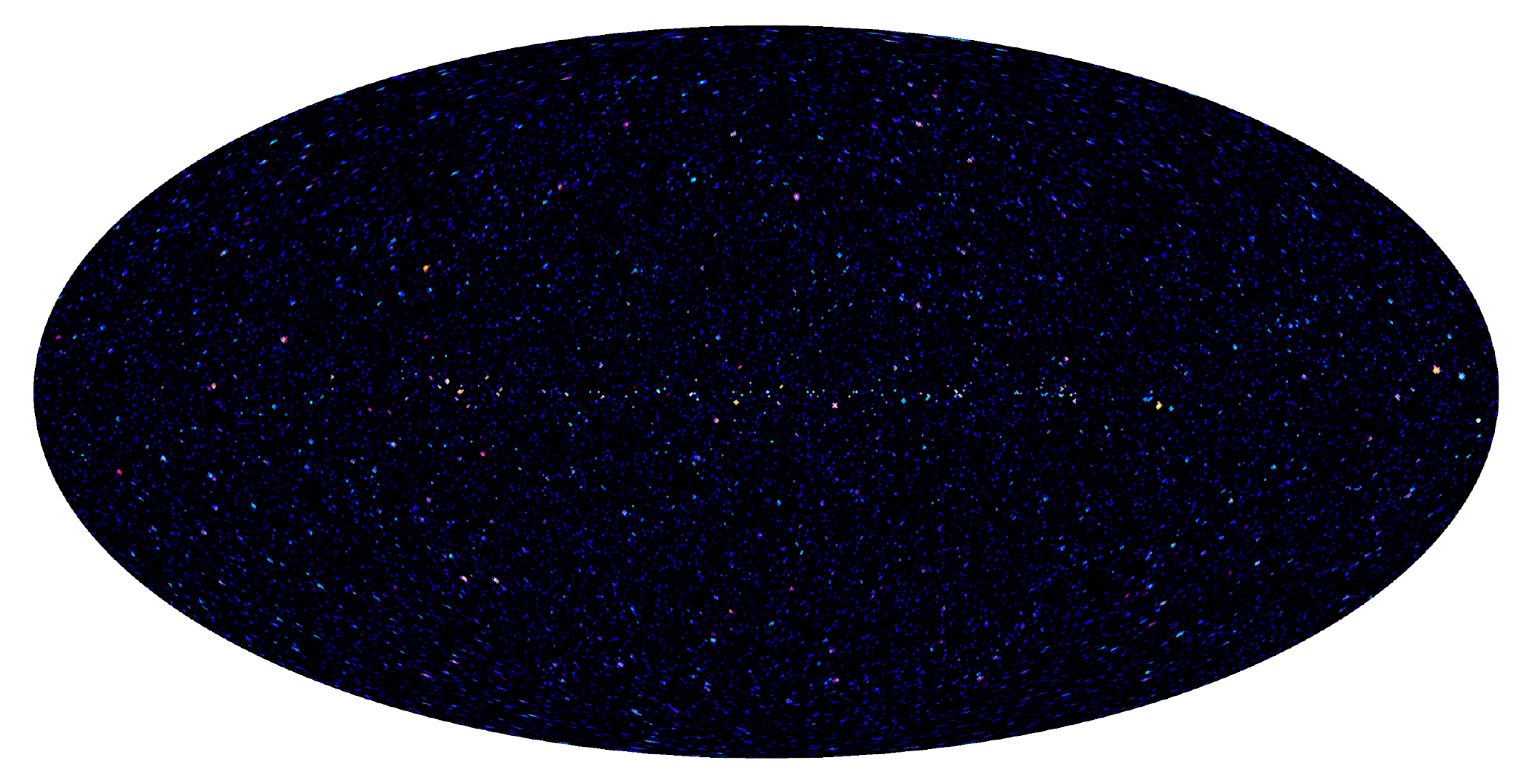}%
\end{minipage}

\caption{Results of the template-free reconstruction based on model M1.
The figure uses the spectral domain color mapping introduced in the caption
of Fig.~\ref{fig:data-exposure-corrected-mf-plot} and Appendix~\ref{sec:appendix-color-coding}.
It highlights spectral variations in the reconstructed skies.
The color scale is provided in the bottom panel of Fig.~\ref{fig:data-exposure-corrected-mf-plot}.
\textbf{First row}: Reconstructed gamma-ray sky.
\textbf{Second row}: Separated diffuse emission (left) and PS sky (right).
\textbf{Third row}: Absolute uncertainties of the diffuse component (left),
the full gamma-ray sky (middle), and the PS component (right).
The maps follow a Mollweide projection.
Single energy bin maps of the diffuse emission component are provided in Fig.~\ref{fig:m1-sf-plots}.}
\label{fig:m1-sky-maps}
\end{figure*}

The spatio-spectral gamma-ray sky according to our template-free model M1
and its decomposition in diffuse and point-like flux components
are shown in Fig.~\ref{fig:m1-sky-maps}.
With respect to the exposure- and effective-area-corrected photon count map
shown in Fig.~\ref{fig:data-exposure-corrected-mf-plot},
the shot noise and point spread introduced by the measurement process
are significantly reduced.

The reconstructed maps provide a detailed view of the gamma-ray sky,
including spectral variations in the reconstructed diffuse emission and the PSs.
The FBs are visible as gray and blue structures north and south of the GC,
behind a strong foreground of hadronic emission, appearing orange in the maps.
Visually comparing the Fig.~\ref{fig:m1-sky-maps} all-sky maps
with the D$^{3}$PO single-energy-band-based imaging results \citep{selig2015fermi},
the methodological progenitor of this work,
we recognize a similar sky morphology, but also a near elimination of
image artifacts associated with the bright Galactic disk exhibited
by the D$^{3}$P0 reconstructions.

% ---- fig: m1 residual histograms ----
\begin{figure*}
\centering
\noindent
\includegraphics[width=1\textwidth]{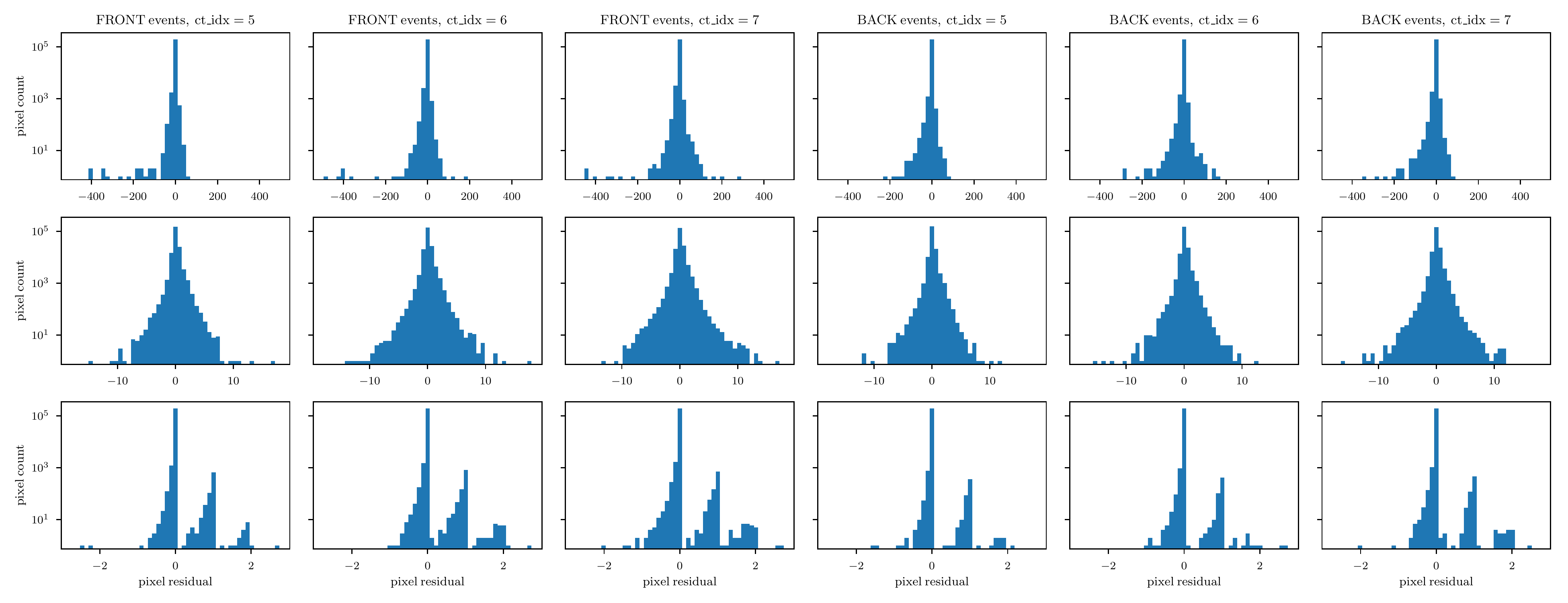}
\caption{Selected residual histograms for the M1 reconstruction.
The rows show all data bins for the energies
\unit[1.00--1.78]{GeV} (\textbf{top}), \unit[10.0--17.8]{GeV} (\textbf{middle}), and \unit[178--316]{GeV} (\textbf{bottom}).
Fig.~\ref{fig:data-histograms} shows the distribution of photon counts in the corresponding data bins.
Pixel counts are shown on a logarithmic scale.}
\label{fig:m1-residual-histograms}
\end{figure*}

% ---- fig: m1 residual maps ----
\begin{figure*}
\centering
\noindent
\begin{minipage}[t]{0.33\textwidth}%
\includegraphics[width=1\textwidth]{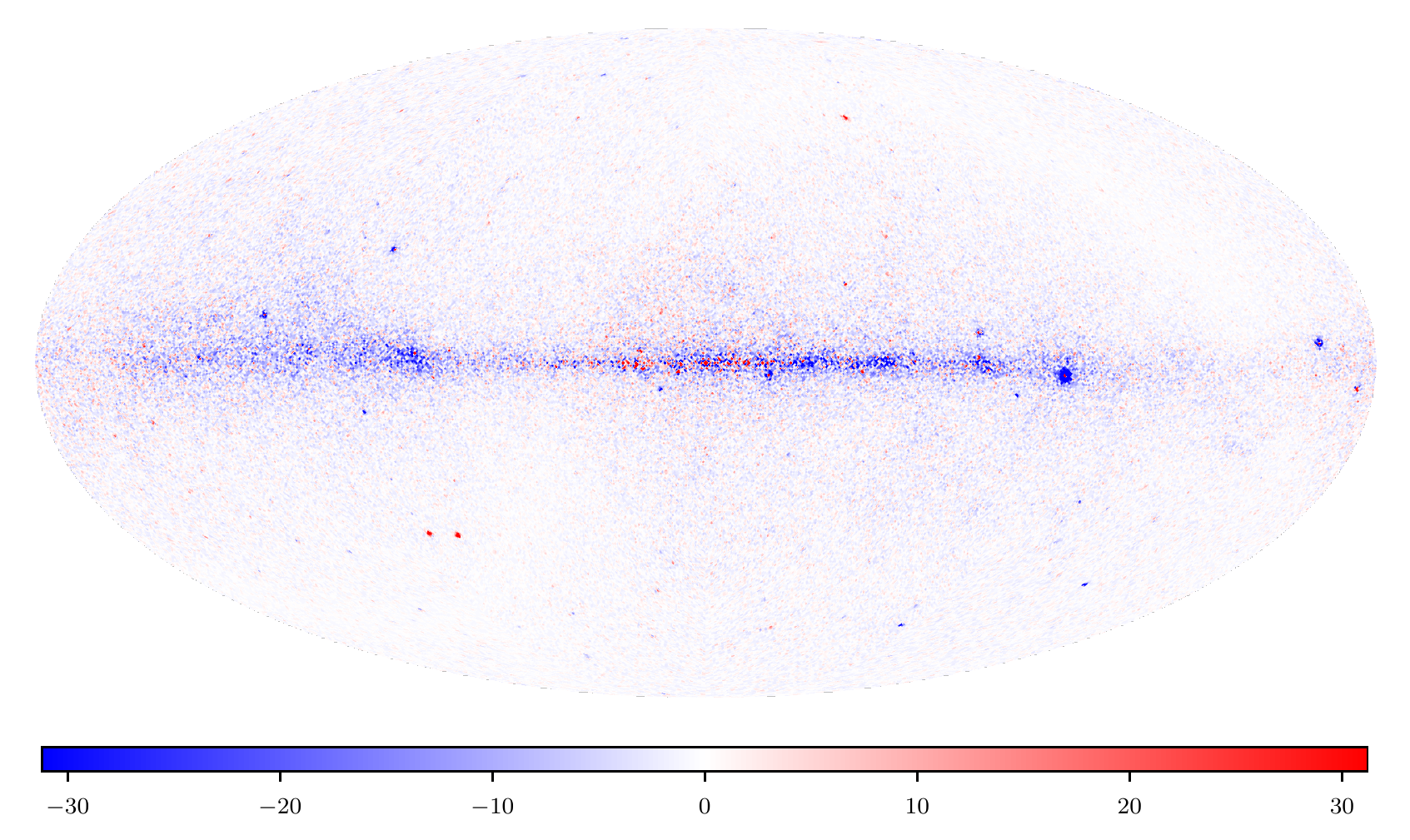}%
\end{minipage}%
\begin{minipage}[t]{0.33\textwidth}%
\includegraphics[width=1\textwidth]{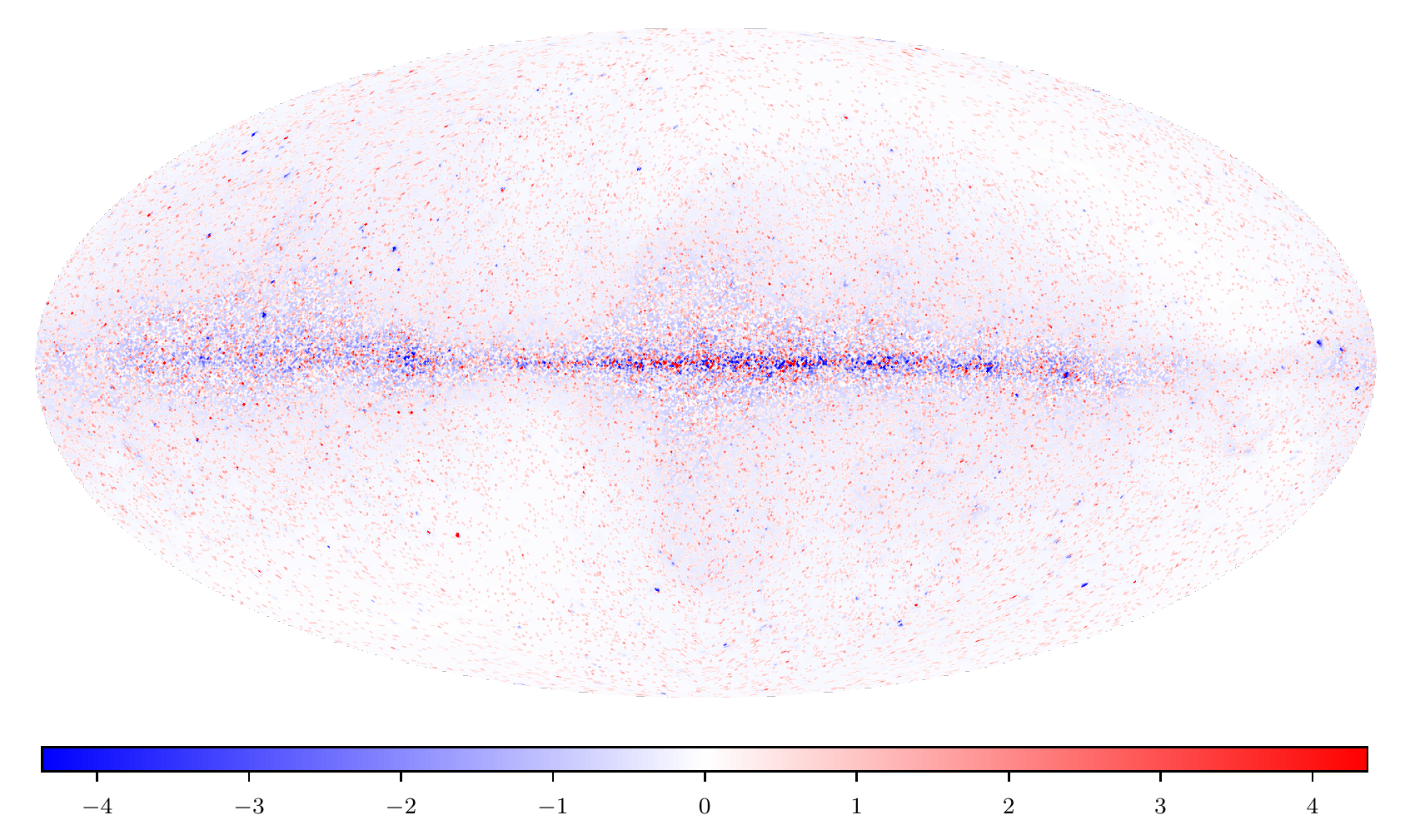}%
\end{minipage}%
\begin{minipage}[t]{0.33\textwidth}%
\includegraphics[width=1\textwidth]{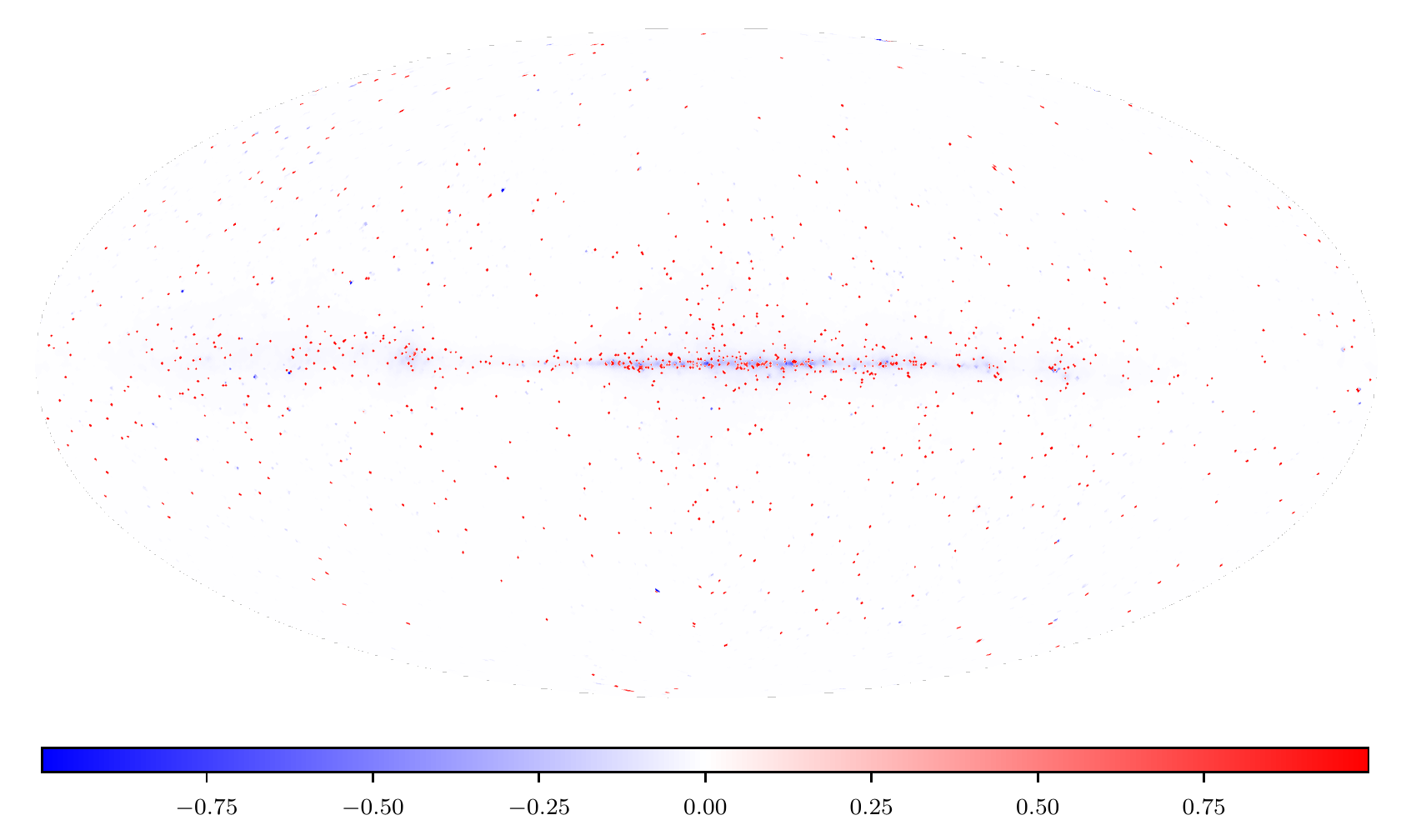}%
\end{minipage}

\begin{minipage}[t]{0.33\textwidth}%
\includegraphics[width=1\textwidth]{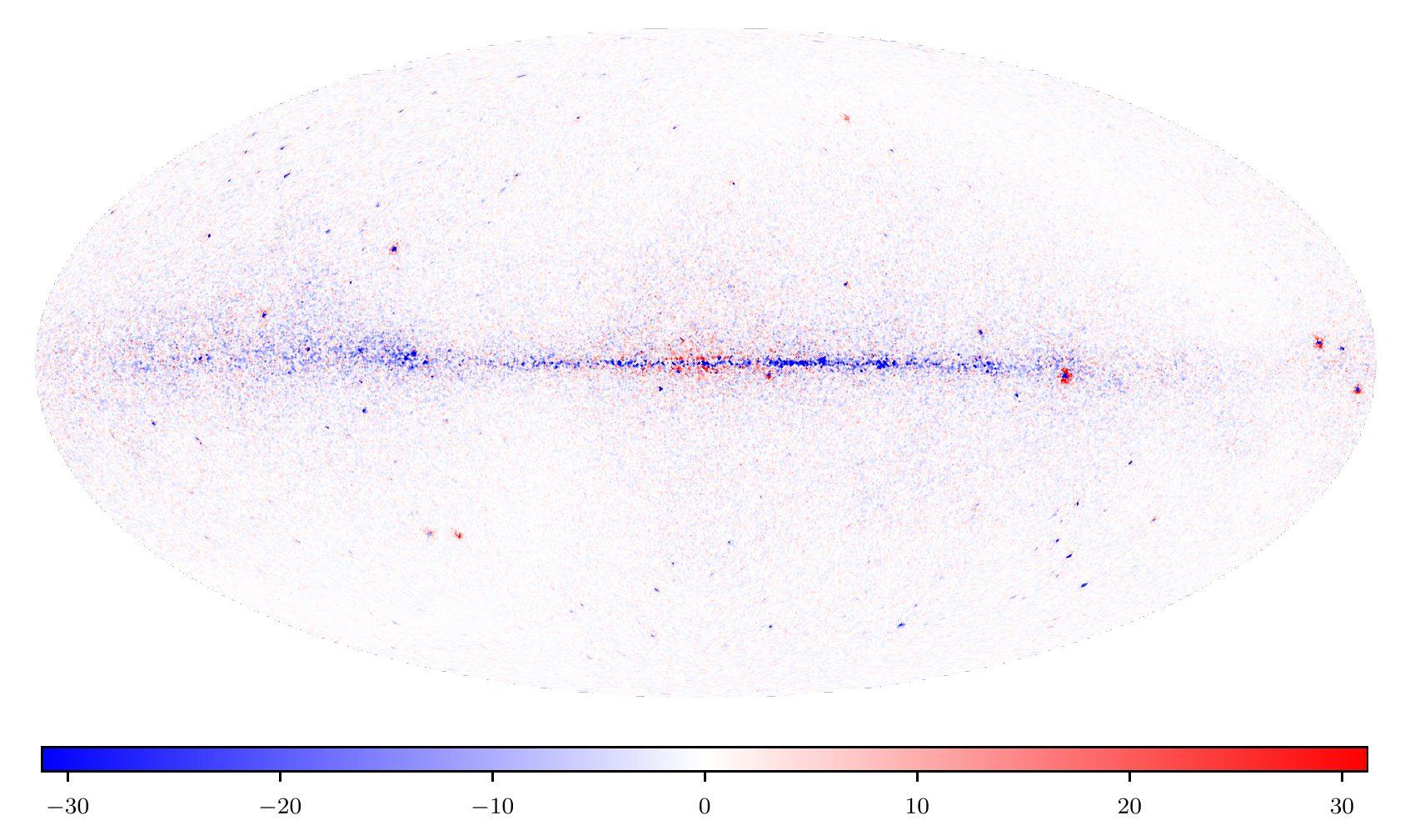}%
\end{minipage}%
\begin{minipage}[t]{0.33\textwidth}%
\includegraphics[width=1\textwidth]{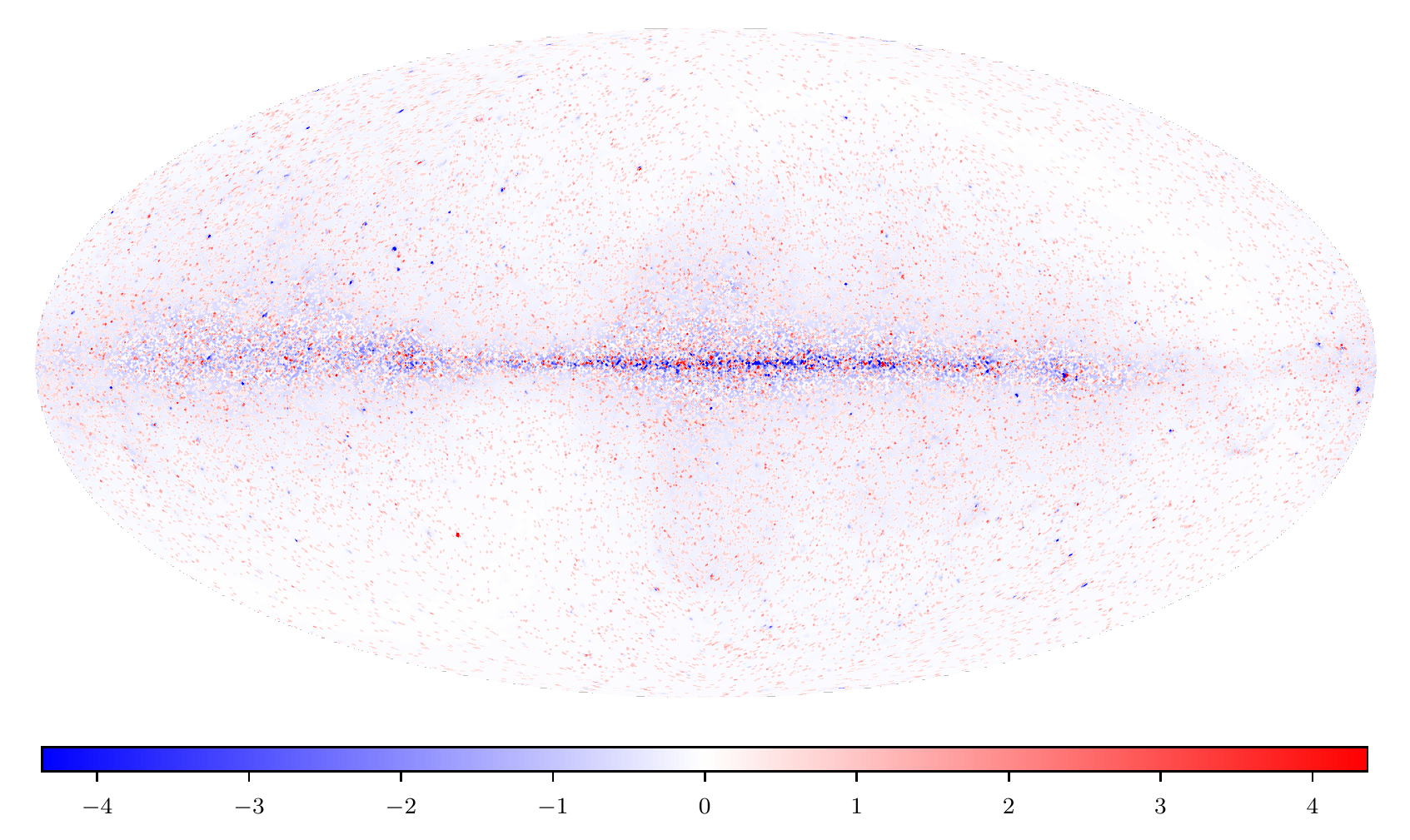}%
\end{minipage}%
\begin{minipage}[t]{0.33\textwidth}%
\includegraphics[width=1\textwidth]{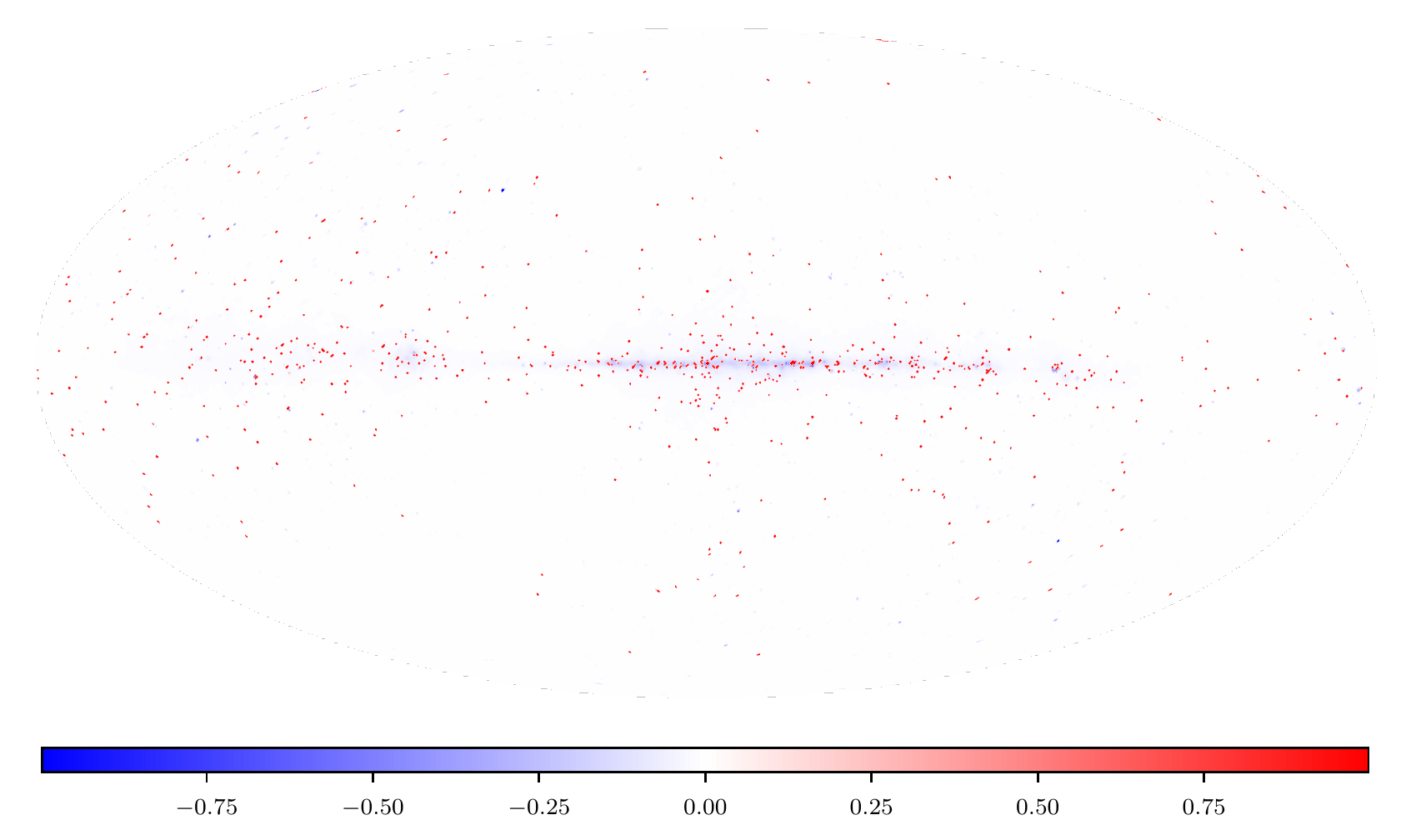}%
\end{minipage}
\caption{Selected residual maps for the M1 reconstruction, shown for the data bin with $\cos(\theta_\mathrm{w'}) \in (0.9,1.0]$
($\mathrm{ct\_idx}=7$ in Fig.~\ref{fig:m1-residual-histograms}).
\textbf{Top}: \texttt{FRONT} events.
\textbf{Bottom}: \texttt{BACK} events.
The energy bins are the same as in Fig.~\ref{fig:m1-residual-histograms} but shown from left to right:
\unit[1.00--1.78]{GeV} (\textbf{left}), \unit[10.0--17.8]{GeV} (\textbf{middle}), and \unit[178--316]{GeV} (\textbf{right}).
The maps are clipped to the 99.9th percentile of the residual amplitude in their respective energy bin.
%For the \unit[1.00--1.78]{GeV} bin, the colorbar limits corresponds to
%the 97.5th (\texttt{FRONT}) and 98.3th percentile (\texttt{BACK})
%of photon counts per pixel.
For comparison, Fig.~\ref{fig:data-histograms} shows the distribution of photon counts in the selected data bins.
}
\label{fig:m1-residual-maps}
\end{figure*}

To evaluate the quality of fit, we calculated reduced $\chi^2$ statistics for the data bins,
wherein we used our model's flux prediction as the expected residual variance in each spatio-spectral voxel.
For \texttt{FRONT} events, this yields an average reduced $\chi^2$ statistic of $1.1$, while for
back events we find an average value of $0.9$.

Figure~\ref{fig:m1-residual-histograms} shows residual histograms of selected data bins,
with representative energy bins chosen from the reconstructed energy range.
Residuals were calculated as $\,d_\mathrm{\,vw'i'j'} - \lambda_\mathrm{\,vw'i'j'} (I_\mathrm{ij})$.
The residual distributions are consistent within the individual energy bins,
but show strongly varying widths across the energy domain.
This is expected, as with increasing photon energy, the photon count strongly decreases
(see Fig.~\ref{fig:data-exposure-corrected-spectral-plot}).
The Poisson count distribution (Eq.~\ref{eq:data-poissonian-distributed}) assumed for the data
predicts residual variances equal to the expected photon count.
Correspondingly, the likelihood permitted much larger residuals at high observed photon counts
than at low observed photon counts.
In the highest energy bin, the discrete nature of the photon count becomes apparent,
and our model predicts count values slightly
above full integer count values, leading to the stepped appearance of the
\unit[178--316]{GeV} residual histograms.
This means our reconstructions in the highest spectral bins are partially
driven by counts in the medium and low energy bins via spectral extrapolation based on the a priori spectral continuity
built into our models.

Figure~\ref{fig:m1-residual-maps} shows spatial maps of residuals in selected data bins,
corresponding to the third and sixth row in Fig.~\ref{fig:m1-residual-histograms}.
The first column (\unit[1.00--1.75]{GeV}) shows an interesting pattern:
in the locations of bright PSs, both the \texttt{FRONT} (top) and \texttt{BACK} (bottom)
maps show large residuals, but with opposing signs.
Where in the \texttt{FRONT} event residual map the central pixel of these PS driven residuals
is positive and the surrounding pixel residuals are negative,
the inverse pattern can be observed in the \texttt{BACK} event residual map.
This indicates a PSF mismodeling, 
where for \texttt{FRONT} events the true point spread is weaker than modeled,
while for \texttt{BACK} events, the true point spread is stronger than modeled.
We believe this is also the reason for the mismatch in reduced $\chi^2$ statistics
between \texttt{FRONT} and \texttt{BACK} events observed.
We note that the Galactic ridge shows strong residuals in all shown maps.
This again is related to the observation that residual variance is expected to be proportional
to the photon flux, meaning high flux regions such as the Galactic ridge are expected to have
larger residuals than the remainder of the sky.
Besides this, the residual maps and histograms show a slight systematic mismodeling
of the FBs in the shown medium energy bin (\unit[10.0--17.8]{GeV}),
although only on the order of a fraction of a count.
We conclude that besides the apparent PSF mismodeling, a good quality of fit was achieved.

% ---- fig: m1 vs fermi template ----
\begin{figure*}
\centering
\noindent
\begin{minipage}[t]{0.65\textwidth}%
\includegraphics[width=1\textwidth]{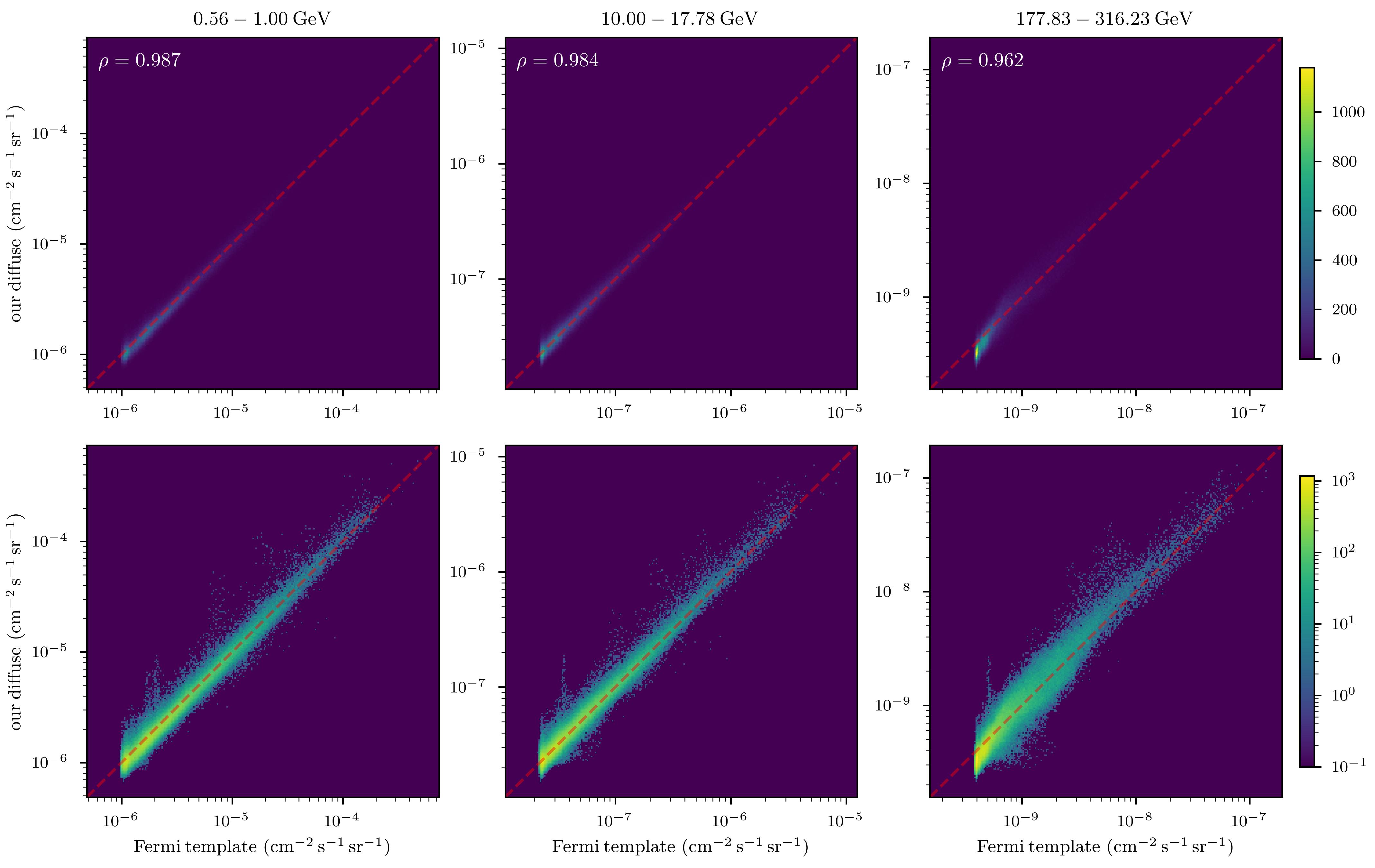}%
\vspace{0.2cm}
\end{minipage}\ %
\begin{minipage}[t]{0.25\textwidth}%
\vspace{-165pt}
\includegraphics[width=1\textwidth]{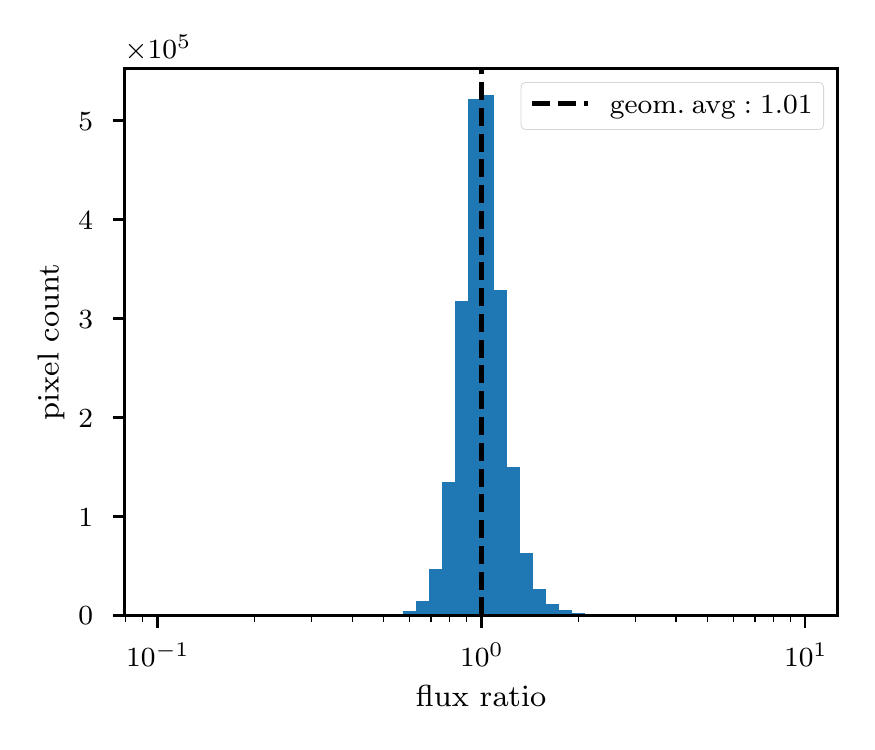}
\end{minipage}
\includegraphics[width=0.32\textwidth]{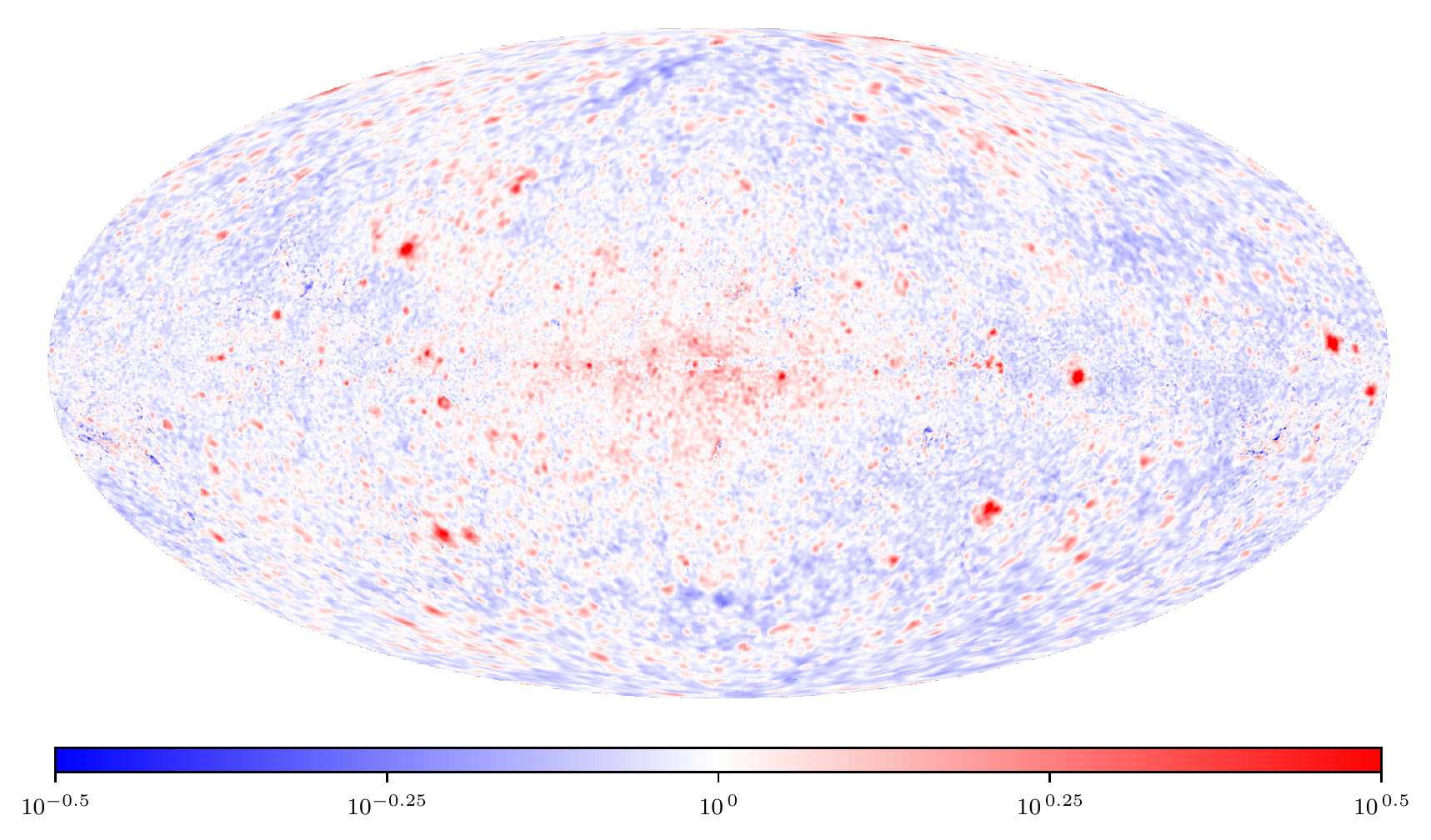}
\includegraphics[width=0.32\textwidth]{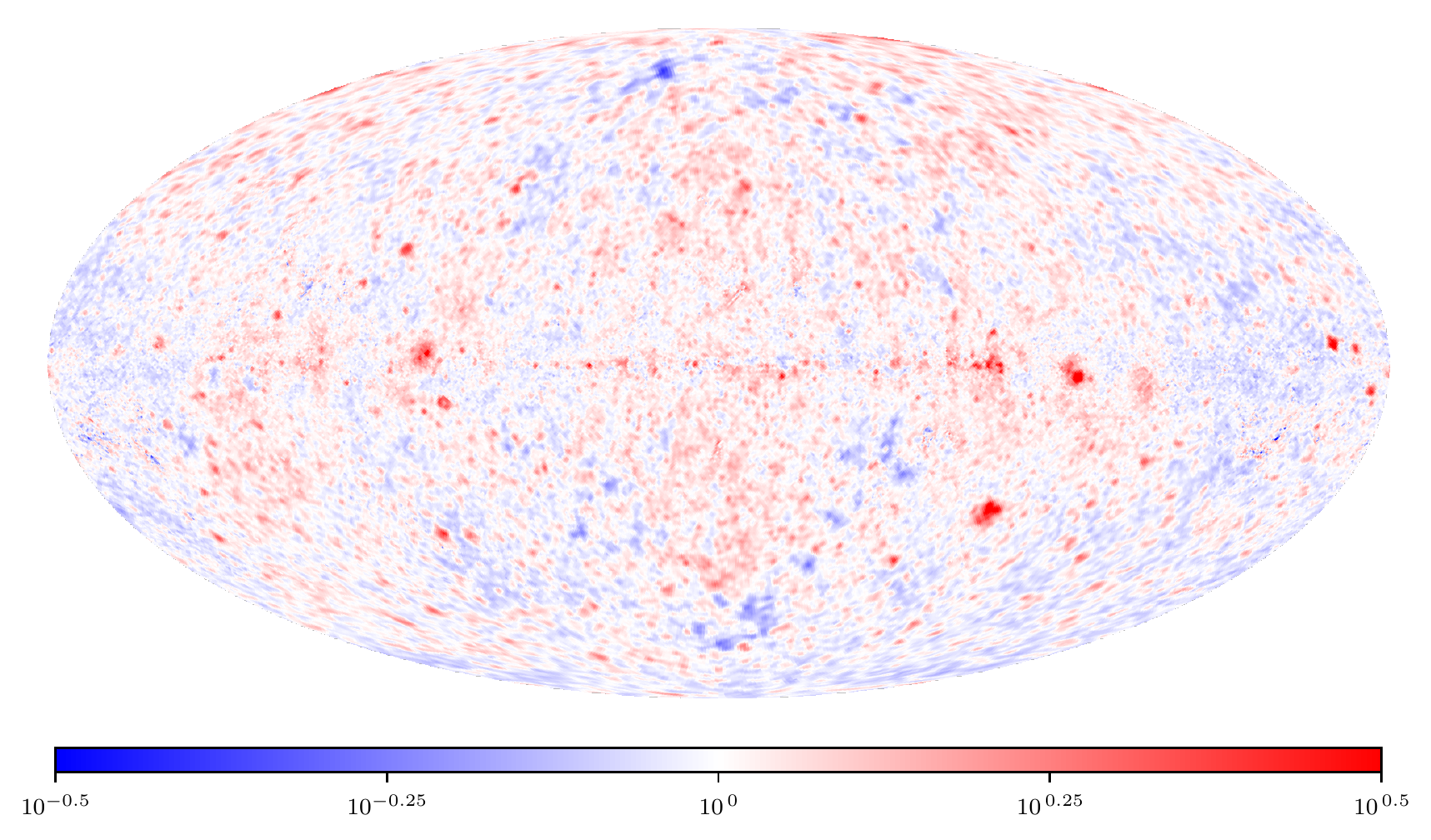}
\includegraphics[width=0.32\textwidth]{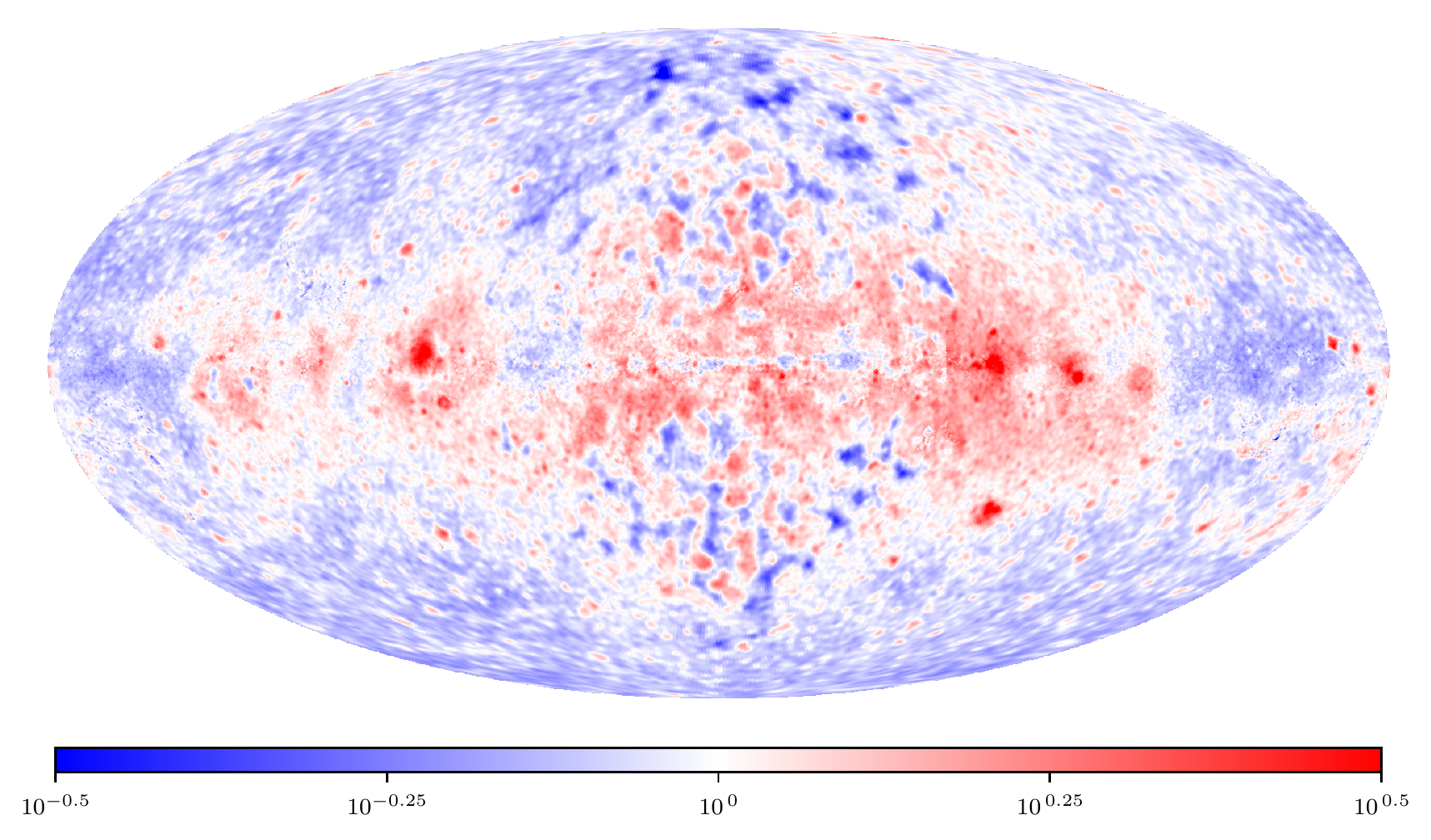}
\caption{Comparison of the M1 diffuse reconstruction with the diffuse emission templates published by the \textit{Fermi} Collaboration
(diffuse foreground: \texttt{gll\_iem\_v07}, isotropic background: \texttt{iso\_P8R3\_SOURCE\_V3\_v1}).
\textbf{Top Left}: 2D histograms of the diffuse flux found by our M1 reconstruction vs. the diffuse emission templates for the energy bins
\unit[0.56--1.00]{GeV}, \unit[10.0--17.8]{GeV}, and \unit[178--316]{GeV}.
The histogram bins are logarithmically spaced in both fluxes.
The top row shows the histogram counts with a linear color scale,
while the bottom row shows them with a logarithmic color scale.
The Pearson correlation between the template and our reconstruction on log-log scale, $\rho$, within the shown energy bins
is given in the top column panels.
The red dashed line marks perfect agreement.
\textbf{Top Right}: Histogram of flux ratios (our diffuse reconstruction divided by
the corresponding voxel values given by the template) for all spatio-spectral bins.
Flux ratios are binned and displayed on a logarithmic scale.
\textbf{Bottom row}: Flux ratio maps for the \unit[0.56--1.00]{GeV}, \unit[10.0--17.8]{GeV},
and \unit[178--316]{GeV} energy bins with a logarithmic color scale.
Numbers larger than $10^0$ ($=1$) indicate we reconstructed more flux than the template predicts, and vice versa.}
\label{fig:m1-vs-fermi-template}
\end{figure*}

To demonstrate our model is able to capture the full dynamic range of the diffuse gamma-ray sky,
Fig~\ref{fig:m1-vs-fermi-template}
shows a comparison of the diffuse reconstruction via model M1 with
the diffuse emission templates\footnote{\
Diffuse foreground: \texttt{gll\_iem\_v07}, isotropic background: \texttt{iso\_P8R3\_SOURCE\_V3\_v1}.
The templates specify differential flux values $\Phi(x,E)$.
To make them comparable to the $I_\mathrm{ij}$ values we reconstructed,
we integrated the published differential flux values over the spectral and spatial bins
as prescribed by Eq.~\ref{eq:I-from-phi}.
We implemented this using numerical quadrature,
and taking into account the power-law nature of gamma-ray emissions,
linear interpolation of the templates on $\log\Phi$, $\log E$, and $x$ scales.
The $I_\mathrm{ij}$ values resulting from the spatio-spectral integration
were then used as the $I^\mathrm{\:diff}_\mathrm{ij}$ prediction of the templates
in the comparison with our reconstructions.
}
developed by the \textit{Fermi} Collaboration
in preparation of the fourth \textit{Fermi} point source catalog \citep[4FGL;][]{abdollahi2020fermi}.
Overall, we observe a strong agreement of our diffuse reconstruction with the template,
with increasing deviation in the high-energy limit.
As the flux ratio histogram in the top-right panel of Fig~\ref{fig:m1-vs-fermi-template} shows,
the majority of spatio-spectral voxels have flux ratios in the interval $[0.8, 1.25]$
and are distributed symmetrically around
the geometric mean of flux ratios, $1.01$, which is very close to unity.
We observe a geometric standard deviation of 17.5 percent.

The 2D histograms in the top-left panel of Fig~\ref{fig:m1-vs-fermi-template} show
our reconstructions follow a linear relationship with the template on log-log scales,
with an average Pearson correlation coefficient of $\left<\rho\right>=0.98$.
Toward lower flux intensities in each energy bin, the correlation weakens slightly,
which is visible in the 2D histograms.
In the \unit[178--316]{GeV} energy bin, we observe a decreased correlation ($rho = 0.96$) with the templates.
This stems from flux deviations in pixels with medium to low flux
(with reference to the flux values found in this energy bin).
In this low flux regime, the \textit{Fermi} templates predict more flux than we have reconstructed.
This is driven by the isotropic background template, which imprints itself
as a hard lower cutoff for the template flux values.
The apparent flux cutoff is three times higher than the lowest value we reconstruct in this energy bin.
A similar cutoff is also present in the \unit[0.56--1.00]{GeV} and \unit[10.0--17.8]{GeV} bins,
but it differs only by $30\%$ from the lowest values we find in our reconstruction.

The bottom row of Fig~\ref{fig:m1-vs-fermi-template} displays
flux ratio maps between our diffuse reconstruction and the templates,
calculated as $\mathrm{I}^\mathrm{\:diff}/\mathrm{I}^\mathrm{fermi\:templates}$,
for the same energy bins as used in the 2D histograms.
Areas of flux over-assignment with respect to the templates are shown in red,
while areas of under-assignment are shown in blue.
Over-assignment is most prominent in the locations of extended sources,
which our diffuse reconstruction includes but whose emissions are not included in the templates.
In the highest energy bin shown (\unit[178--316]{GeV})
the deviation between our reconstruction and the template is markedly stronger
than in the two other energy bins, consistent with the observations above.
In an elliptic region centered on the GC,
reaching to high Galactic latitudes ($|b| \lesssim$~{80\textdegree})
and longitudes of $|\ell| \lesssim$~{60\textdegree},
we observe a strong but irregular pattern of deviations with a characteristic length scale of
approximately {5\textdegree}.
East and west of this centered on the Galactic plane are regions of slight flux over-prediction
with respect to the template, visible in red and reaching to Galactic latitudes of $|b| \lesssim$~{30\textdegree}.
At large Galactic latitudes and in the Galactic anticenter,
we find on average approximately 20 percent less flux than the templates assume.

% ---- fig: res power spectra ----
\begin{figure*}
\centering
\noindent
\begin{minipage}[t]{0.49\textwidth}%
\includegraphics[width=1\textwidth]{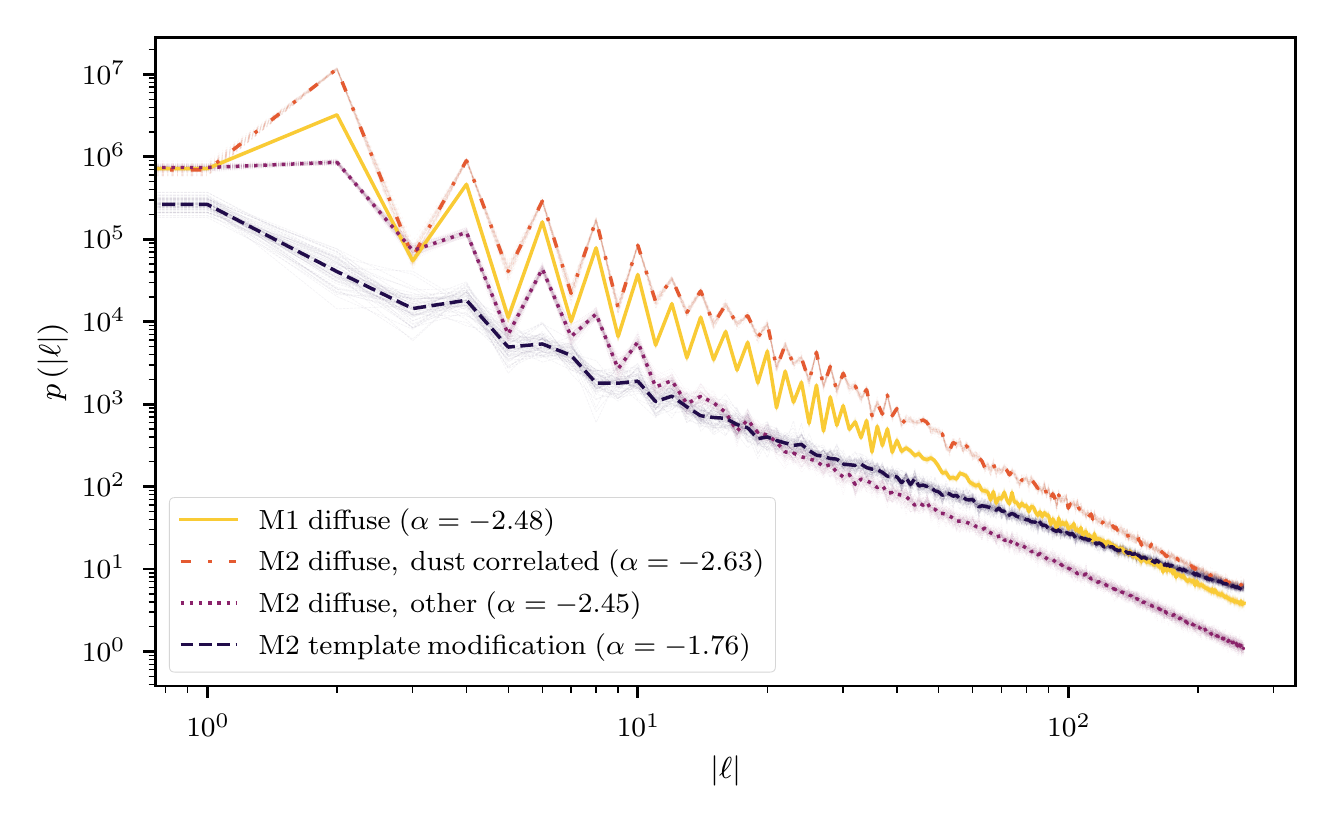}%
\end{minipage}%
\begin{minipage}[t]{0.49\textwidth}%
\includegraphics[width=1\textwidth]{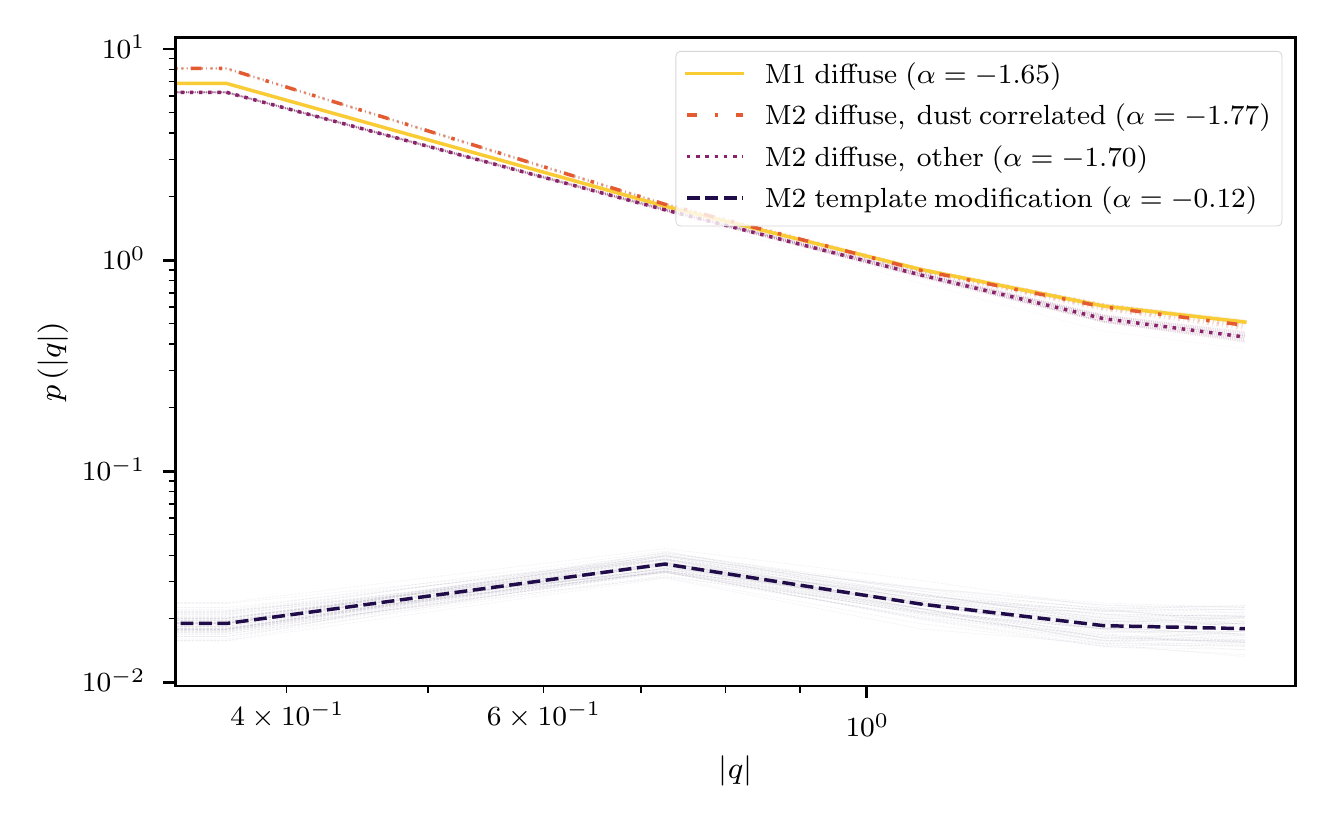}%
\end{minipage}

\begin{minipage}[t]{0.49\textwidth}%
\includegraphics[width=1\textwidth]{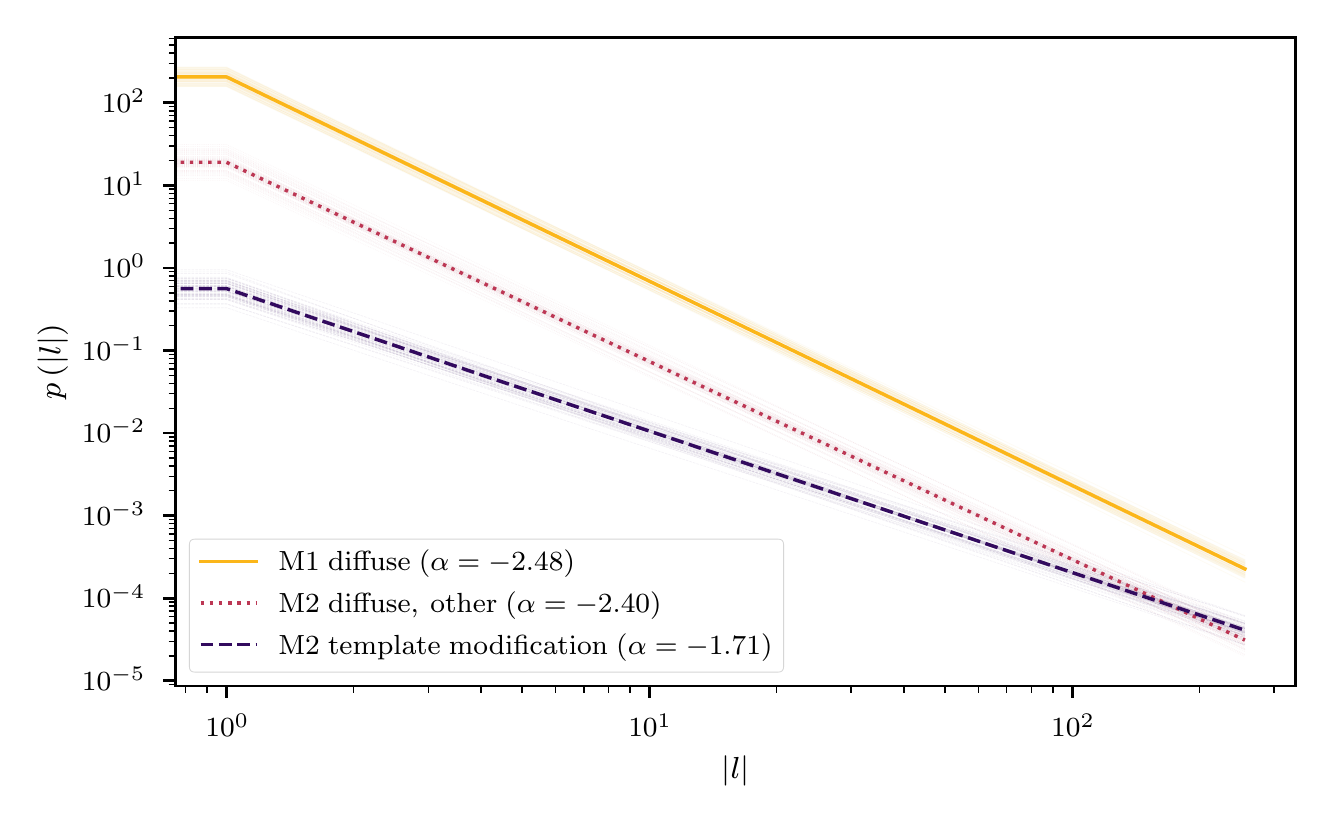}%
\end{minipage}%
\begin{minipage}[t]{0.49\textwidth}%
\includegraphics[width=1\textwidth]{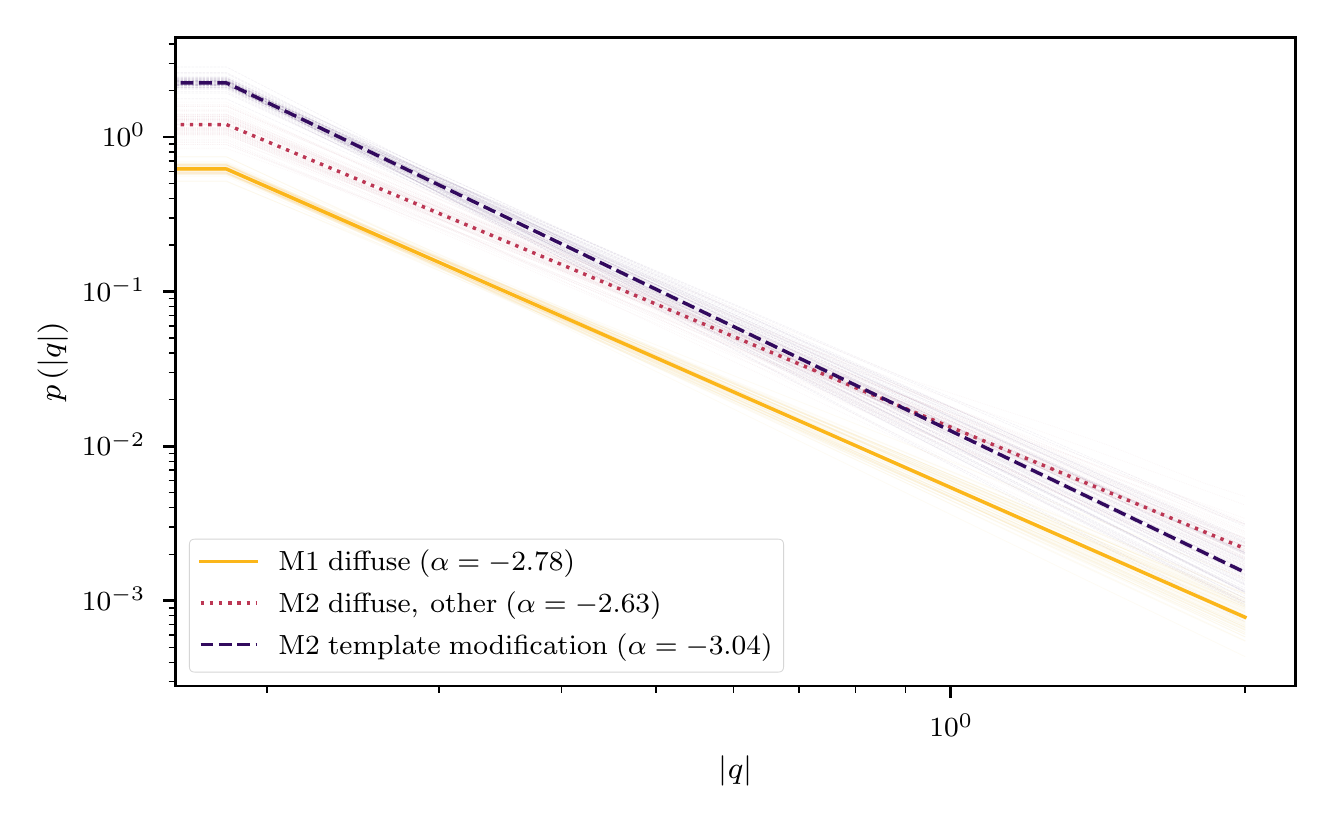}%
\end{minipage}
\caption{
Power spectrum results of the two reconstruction runs.
\textbf{Top row}: Empirical power spectra calculated directly from the reconstructed sky components (on log10 scale).
\textbf{Bottom row}: Posterior power spectrum models of the correlated field models representing the $\tau^\mathrm{\:c}$.
\textbf{Left column}: Angular power spectra. \textbf{Right column}: Energy spectrum power spectra.
Bold lines show the geometric posterior sample means, while thin lines show individual posterior samples.
All power spectra are plotted on log-log scale.
}
\label{fig:res-power-spectra}
\end{figure*}

A significant quantity in our modeling of the sky are the spatial (angular) and spectral correlation power spectra (CPSs).
As detailed in Sect.~\ref{sec:methods-formulation-of-prior-knowledge},
the generative models for the $\tau^\mathrm{\:c}$ fields (here $\tau^\mathrm{\:diff}$)
contain CPS models that provide self-consistent regularization for them
and allow us to formulate prior knowledge of their spatial and spectral correlation structures 
$C^\mathrm{\:c}(\Delta x)$ and $D^\mathrm{\:c}(\Delta y)$.
Figure~\ref{fig:res-power-spectra} shows both the empirical CPSs of the reconstructed component maps (upper row)
and the internal CPS models used in the correlated fields $\tau^\mathrm{\:c}$ (bottom row).
For the M1-based diffuse reconstruction,
we find a (posterior mean) empirical APS index of $-2.48$ and a (posterior mean) empirical
energy spectrum power spectrum (EPS) index of $-1.65$.
The empirical APS shows a zigzag pattern in the lower even and odd multipoles,
which is a spectral imprint of the bright Galactic disk, exciting even angular modes stronger than odd ones.
This pattern wanes toward smaller angular scales, as expected, as
small angular scale structures exist in almost all regions of the sky.

% ---- fig: m1 diffuse spectra

%\textbf{Merge}:
%The spatio-spectral imaging allows us to investigate a data-driven
%spectrum of the diffuse flux at all sky locations.
%Figure \textbf{XX} shows such spectra for the Galactic equator and a great circle
%orthogonal to this.
%On the first glance, these are largely similarly falling power-law-like spectra.
%However, the spectra taken from the great circle in the Galactic plane show a steeper slope (average: )
%than the spectra taken on the great circle perpendicular to it (average: ),
%hinting at different production processes underlying the observed fluxes. \textbf{FIXME: gen numbers}.
%\textbf{/merge}

% ---- fig: m1 diffuse spectral index map ----
\begin{figure}
\resizebox{\hsize}{!}{\includegraphics{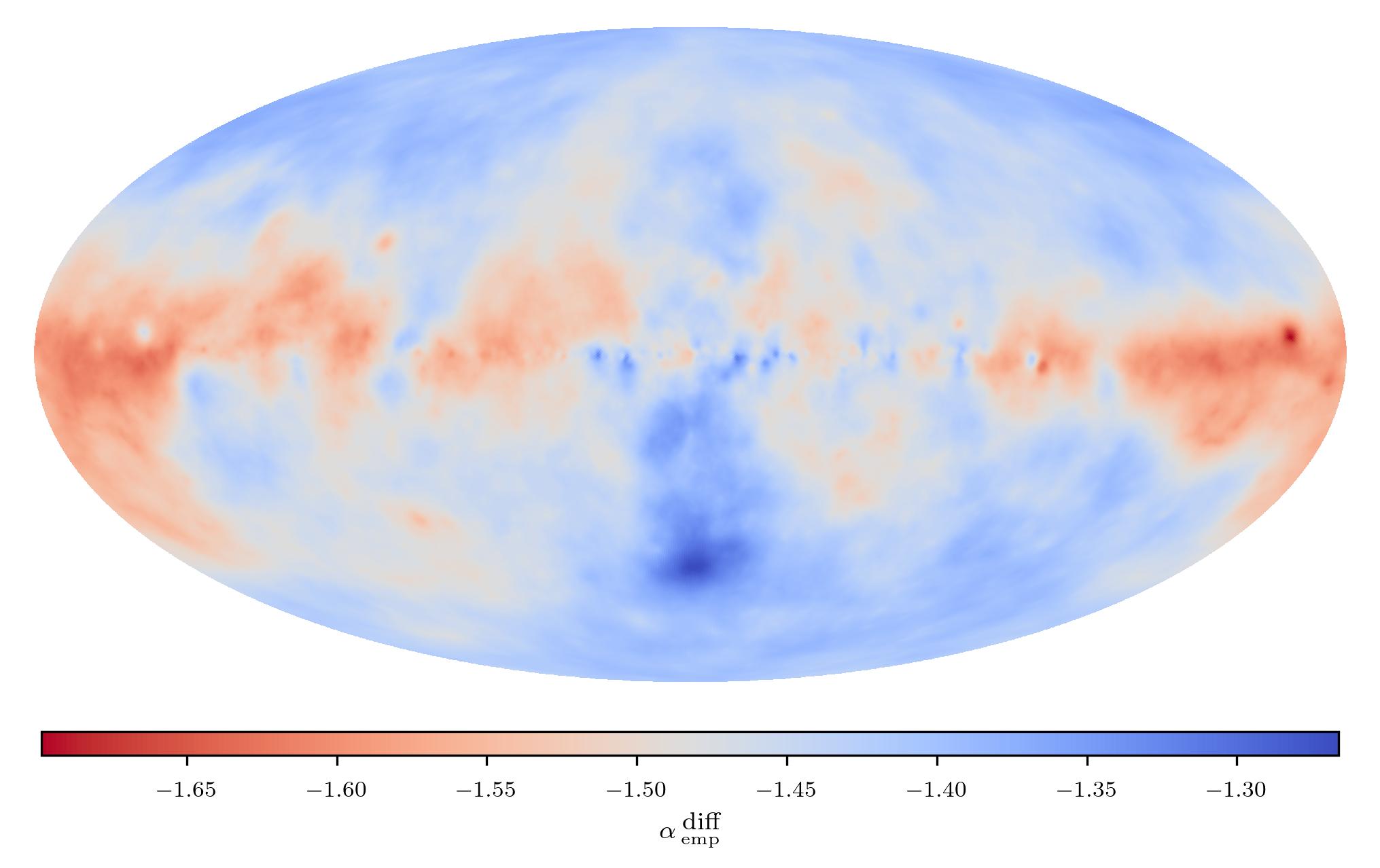}}
\caption{Empirical M1 diffuse spectral index map posterior mean, obtained by power-law fits to the diffuse emission component samples provided by the M1 reconstruction
and subsequent averaging.
Because we reconstruct fluxes integrated over logarithmically equidistant energy bins,
spectral index values are offset by $+1$ with respect to un-integrated energy spectra
(see Eqs.~\ref{eq:I-from-phi} and~\ref{eq:spectral-index-adjustment} in Sect.~\ref{sec:methods-general} for details).}
\label{fig:m1-diffuse-spectral-index-map}
\end{figure}

The spatio-spectral imaging allows us to investigate a data-driven
spectrum of the diffuse flux at all sky locations.
To complement the qualitative spectral overview provided by the Fig.~\ref{fig:m1-sky-maps} spatio-spectral maps,
we performed a pixel-wise power-law fit to the reconstructed diffuse sky energy spectra,
resulting in the empirical spectral index map shown in Fig.~\ref{fig:m1-diffuse-spectral-index-map}.
We remind the reader that we have reconstructed fluxes integrated within logarithmically equidistant energy bins,
which exhibit energy spectral indices offset by $+1$ compared to nonintegrated fluxes
(see Eq.~\ref{eq:spectral-index-adjustment}).
The empirical spectral index of the diffuse emission shown in Fig.~\ref{fig:m1-diffuse-spectral-index-map}
varies slowly across the sky,
with the exception of a few small-scale structures discussed below.
In regions dominated by neutral pion decay emission originating from the dense ISM,
empirical spectral indices $\alpha^\mathrm{\:diff}_\mathrm{\,emp}$ in the range $\left[-1.65, -1.50\right]$
can be observed, appearing in red shades in the figure.
In the region inhabited by the FBs, flatter spectra were found, with empirical spectral indices
in the range of $\left[-1.45, -1.26\right]$, with a significant spectral hardening toward the tip of the
southern bubble, appearing in blue.
In regions where the Galactic foregrounds are weak, a spectral index around $-1.40$ is observed.
In the outer Galaxy, the Galactic plane is permeated by regions
with spectral indices between $-1.45$ and $-1.40$,
suggestive of outflows of relativistic plasma from the Galactic disk.
In the Fig.~\ref{fig:m1-sky-maps}, these are visible as white veils above the remaining emission.
Regarding small-scale structures, we note multiple strongly localized hard spectrum regions
in the Galactic disk, close to the GC,
with a spectral indices in the $\left[-1.35, -1.26\right]$ range.
The region surrounding the Geminga pulsar shows an average spectral index of $-1.70$, making it the region with the
lowest spectral index in this map.

% ---- fig: m1 ps source count distribution ----
\begin{figure}
\resizebox{\hsize}{!}{\includegraphics{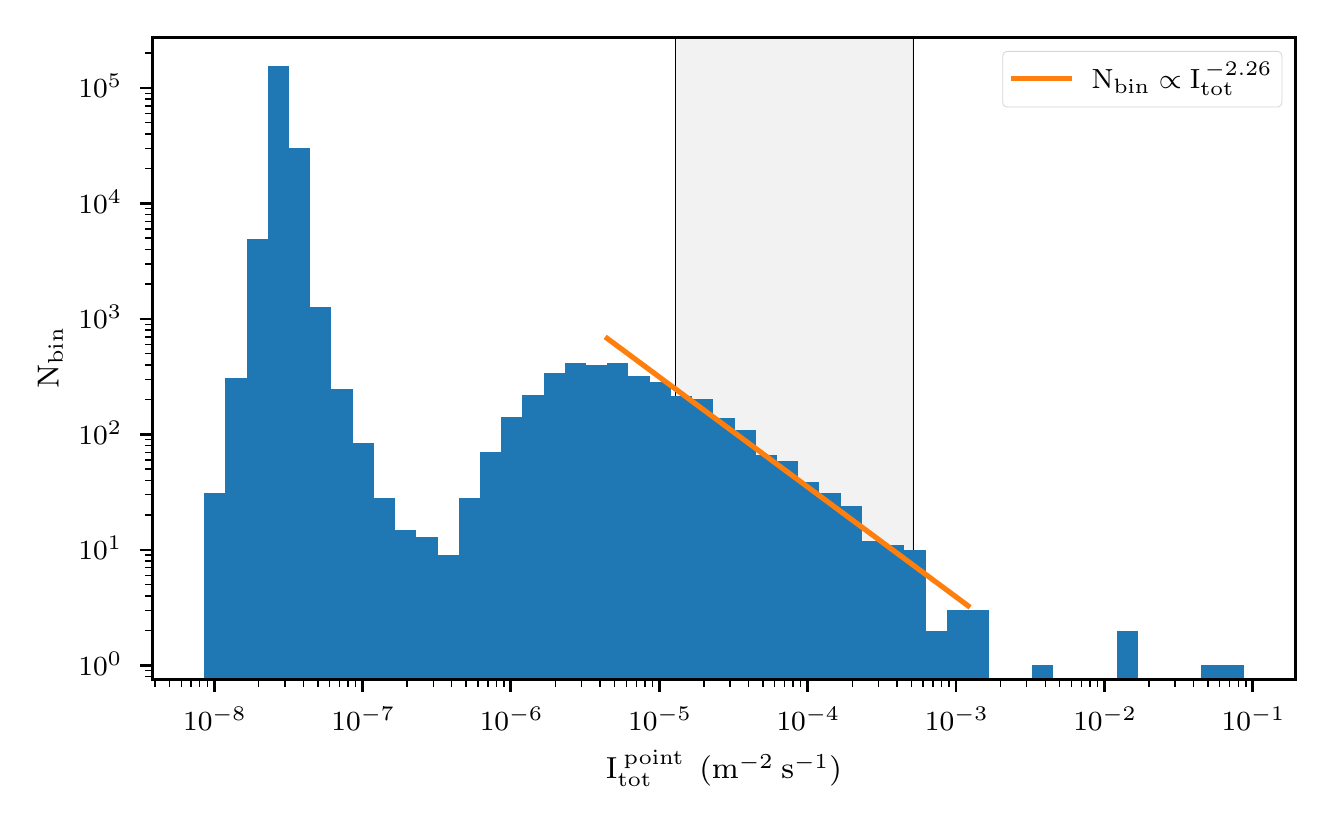}}
\caption{PS pixel count histogram for the M1 reconstruction on log-log scale.
The orange line shows a power-law fit to the brightness distribution in the brightness range highlighted in gray.
Flux values below \unit[$5\cdot{10}^{-7}$]{$\mathrm{m}^{-2}\,\mathrm{s}^{-1}$}
should be regarded as non-detections.
The distribution function in that area is prior-driven.}
\label{fig:m1-ps-source-count-distribution}
\end{figure}

We now turn to the analysis of the PS results made with the M1 model.
As described in Sect.~\ref{sec:methods-formulation-of-prior-knowledge},
we used a fixed PS brightness prior
that follows a power law with index $-2.5$ for large brightness values
and has an exponential cutoff for low brightness values,
with the prior mean set to lie two orders of magnitude below the expected
diffuse emission brightness mean.
Figure~\ref{fig:m1-ps-source-count-distribution} shows the posterior PS pixel
brightness distribution found with model M1.
In it, three distinct populations of PS pixels can be seen:

First, the majority of PS pixels exhibit a posterior mean brightness at the low end of the brightness scale
between ${10}^{-8}$ and \unit[${10}^{-7}$]{$\mathrm{m}^{-2}\,\mathrm{s}^{-1}$}.
These are PSs that were not ``switched on'' during the reconstruction,
leaving their posterior brightness close to the prior mean.
They should be regarded as non-detections of PS flux.

Second, between ${10}^{-6}$ and \unit[${10}^{-3}$]{$\mathrm{m}^{-2}\,\mathrm{s}^{-1}$},
lies a population of PS pixels for which the data requested significant flux contributions.
Their distribution matches the shape found for PS distributions in other works; \citet[Fig.~15]{abdollahi2020fermi} provide the distributions found in the first to fourth source catalogs
(1FGL -- 4FGL) published by the \textit{Fermi} Collaboration.
The number of sources in this group falls off quickly below a pixel brightness of \unit[$2 \cdot {10}^{-6}$]{$\mathrm{m}^{-2}\,\mathrm{s}^{-1}$},
giving the effective detection limit of our analysis.
We perform a power-law fit to the found distribution in the range of
${10}^{-5}$ to \unit[$5 \cdot {10}^{-4}$]{$\mathrm{m}^{-2}\,\mathrm{s}^{-1}$},
where the distribution follows a stable slope on log-log scales.
We find a power law with index $-2.26$,
which is close enough to the prior assumption of $-2.5$
that we do not expect strong biases in the estimated
flux values within that range.

Third, a few isolated sources exhibit brightness values above \unit[${10}^{-3}$]{$\mathrm{m}^{-2}\,\mathrm{s}^{-1}$}.
These correspond to the brightest sources found, including the Vela and Geminga pulsar.
We discuss these findings in the context of previous works in Sect.~\ref{sec:discussion}.

% ---- fig: m1 brightest ps spectra (spec, map)
\begin{figure*}
%\centering
\noindent
\begin{minipage}[t]{0.50\textwidth}% # 0.40
\includegraphics[width=1\textwidth]{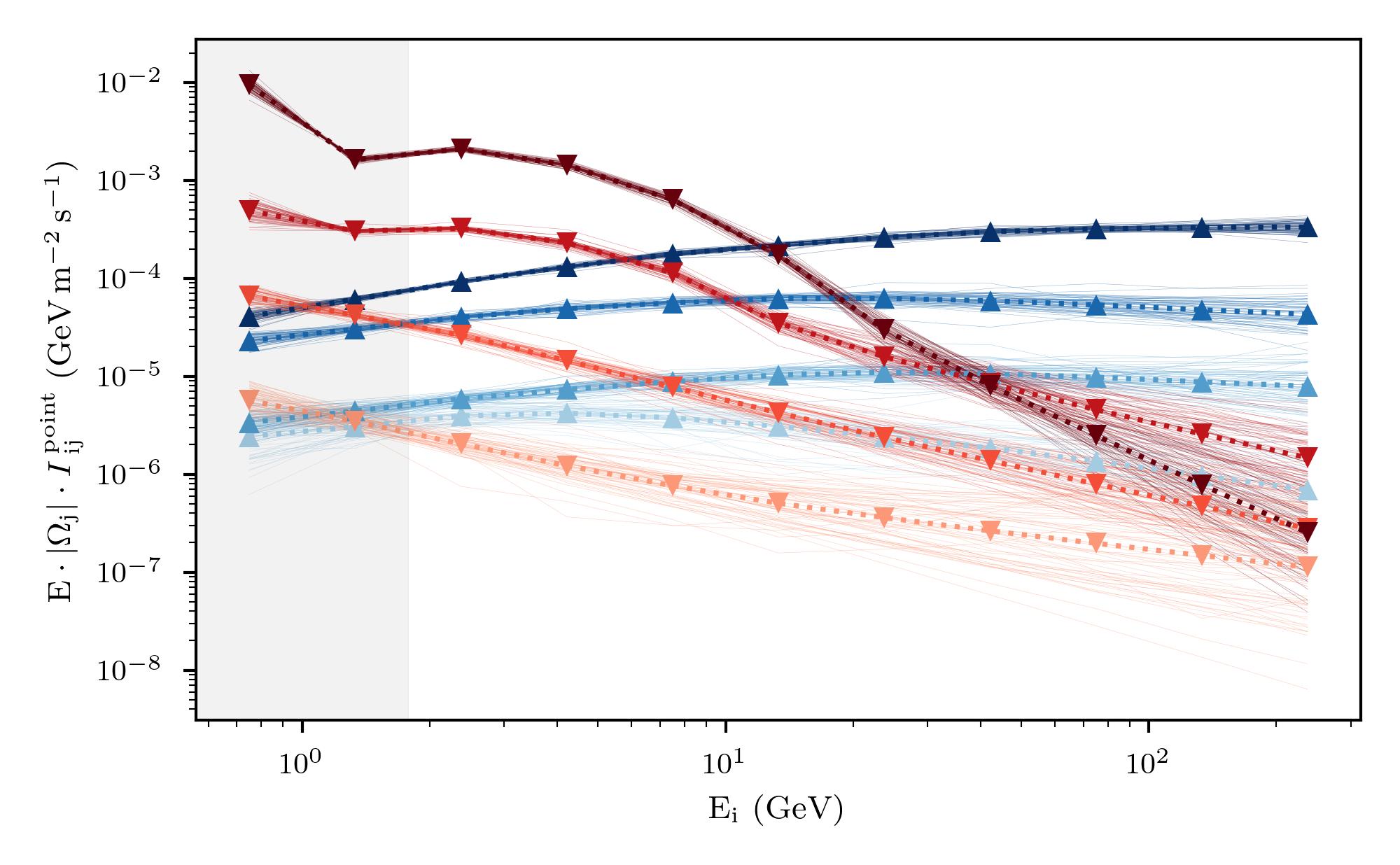}%
\end{minipage}%
\hskip30pt
\begin{minipage}[t]{0.36\textwidth}% #0.24
\vspace{-140pt}
\hskip5pt\includegraphics[width=1\textwidth]{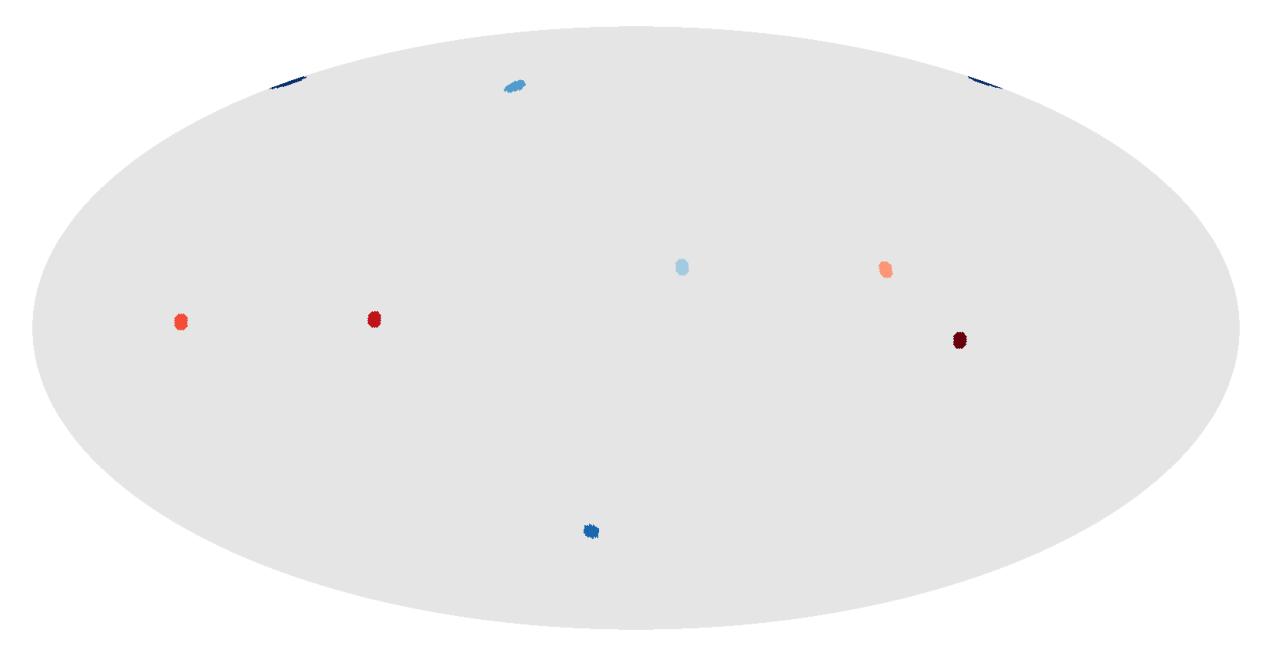}%
\end{minipage}
\caption{
Reconstructed energy spectra of selected PS pixels based on model M1.
The pixels were selected by sorting them by their geometric mean flux
in the \unit[1.00-1.77]{GeV} (red, down-pointing triangles)
and \unit[100-177]{GeV} energy bins (blue, up-pointing triangles)
and then choosing the brightest, 10th, 100th, and 1000th brightest source in each of the two energy bins.
\textbf{Right}: Positions of the selected PSs.
\textbf{Left}: Corresponding energy spectra, matched by color.
The triangle markers display the posterior geometric mean flux reconstructed for the sources in each energy bin,
while the thin continuous lines show the posterior spectrum samples associated with sources.
The spectra are plotted on log-log scale.}
\label{fig:m1-ps-spectra}
\end{figure*}

Continuing our analysis of the M1 PS results,
we turn to the full reconstructed energy spectra of PSs.
Figure~\ref{fig:m1-ps-spectra} shows this for a few PSs,
selected as the brightest, 10th, 100th, and 1000th brightest source
in the \unit[1.00-1.77]{GeV} and \unit[100-177]{GeV} energy bins.

The reconstructed spectra deviate from pure power laws in energy,
which would appear as straight lines in the graph.
The spectrum of the brightest source pixel in the \unit[1.00-1.77]{GeV} energy bin
shows a spectral hump characteristic of hadronic emission
and flattens to a straight line above \unit[40]{GeV}.
The same is true of the 10th brightest source in this bin,
although its spectrum already flattens above \unit[1]{GeV}.
The sources have soft spectra with spectral indices below $-1.5$
and all lie within the Galactic plane (except for the faintest).
The PSs selected in the \unit[100-177]{GeV} energy bin
all show comparatively straight energy spectra with a slight uniform bend to them.
They all have harder spectra with spectral indices close to $-1.0$
and lie at high Galactic latitudes (except for the faintest, again).

As seen by the posterior sample variability in the graph,
the posterior uncertainty generally increases toward higher energies
and with decreasing source brightness.
This is expected, as discussed above, because of low photon count
at high energies and thus less informative data,
and because of low intensity sources being ``covered'' by stronger diffuse emission,
making their spectra less constrained by data than those of brighter sources.

Figure~\ref{fig:m1-ps-spectra} also reveals that the
spectral deconvolution is facing a problem at lower energies,
which becomes most apparent for the steeper spectrum sources.
Because of the large EDF spread at low energies\footnote{
See \url{https://fermi.gsfc.nasa.gov/ssc/data/analysis/documentation/Pass8_edisp_usage.html} for a visualization
of the EDFs energy dependence.},
the data corresponding to the lowest reconstructed energy bin
is contaminated by photons with true energies below its low energy boundary.
Since the model has no representation of this,
we excluded the data of the lowest reconstructed energy bin from our likelihood.
This has the downside of making the lowest reconstructed energy bin
under-informed compared to the higher energy bins.
As it is only informed by photon counts from the second-lowest energy bin,
errors in the EDF model or its numerical representation were imprinted onto
the low energy end of the reconstruction without mitigation.
We therefore recommend to exclude the lowest energy bin in analyses
of the reconstruction results.

% ---- fig: m1 ps spectral index analyses (map, distribution)
\begin{figure*}
\centering
\noindent
\begin{minipage}[t]{0.49\textwidth}%
\includegraphics[width=1\textwidth]{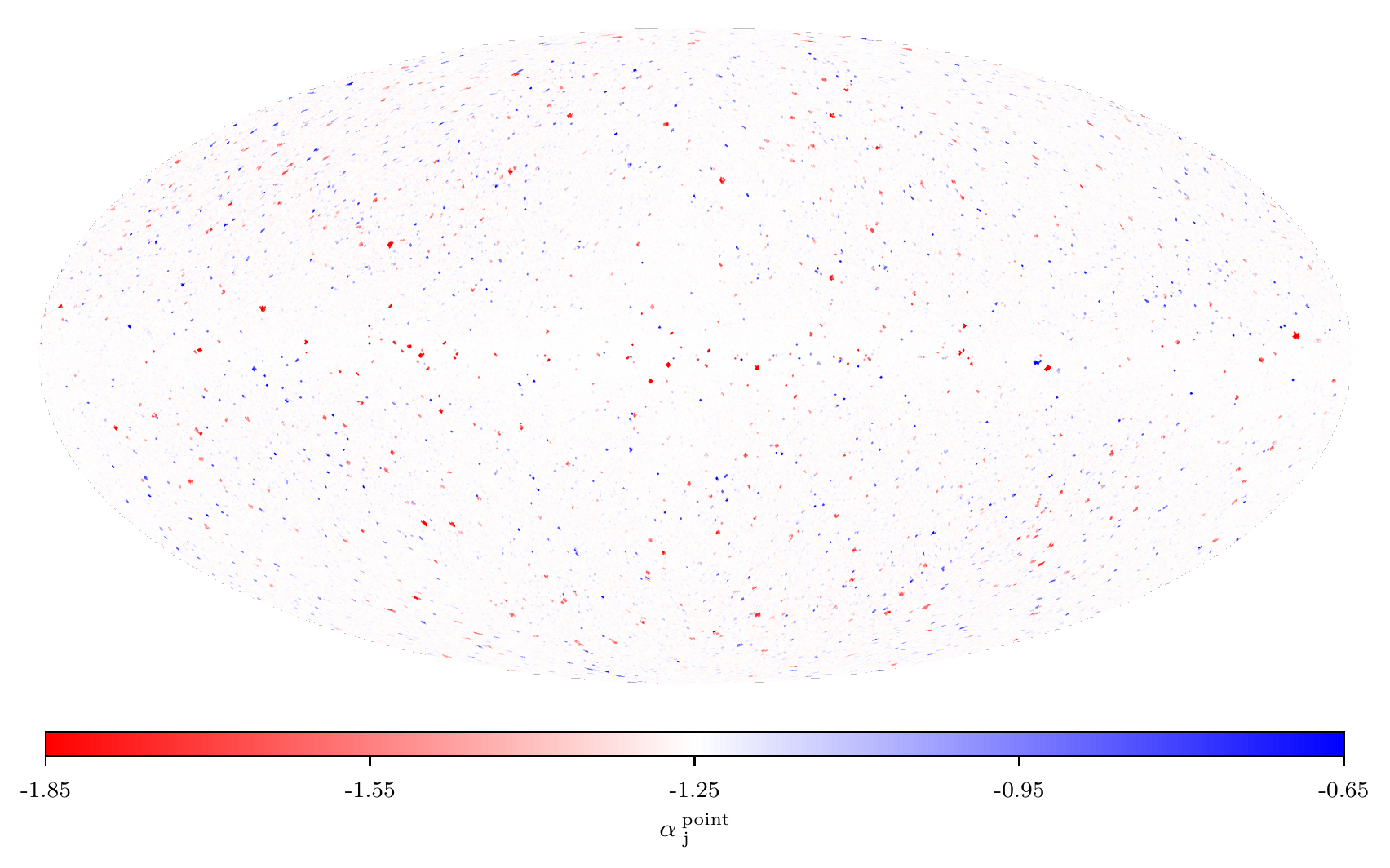}%
\end{minipage}%
\begin{minipage}[t]{0.49\textwidth}%
\includegraphics[width=1\textwidth]{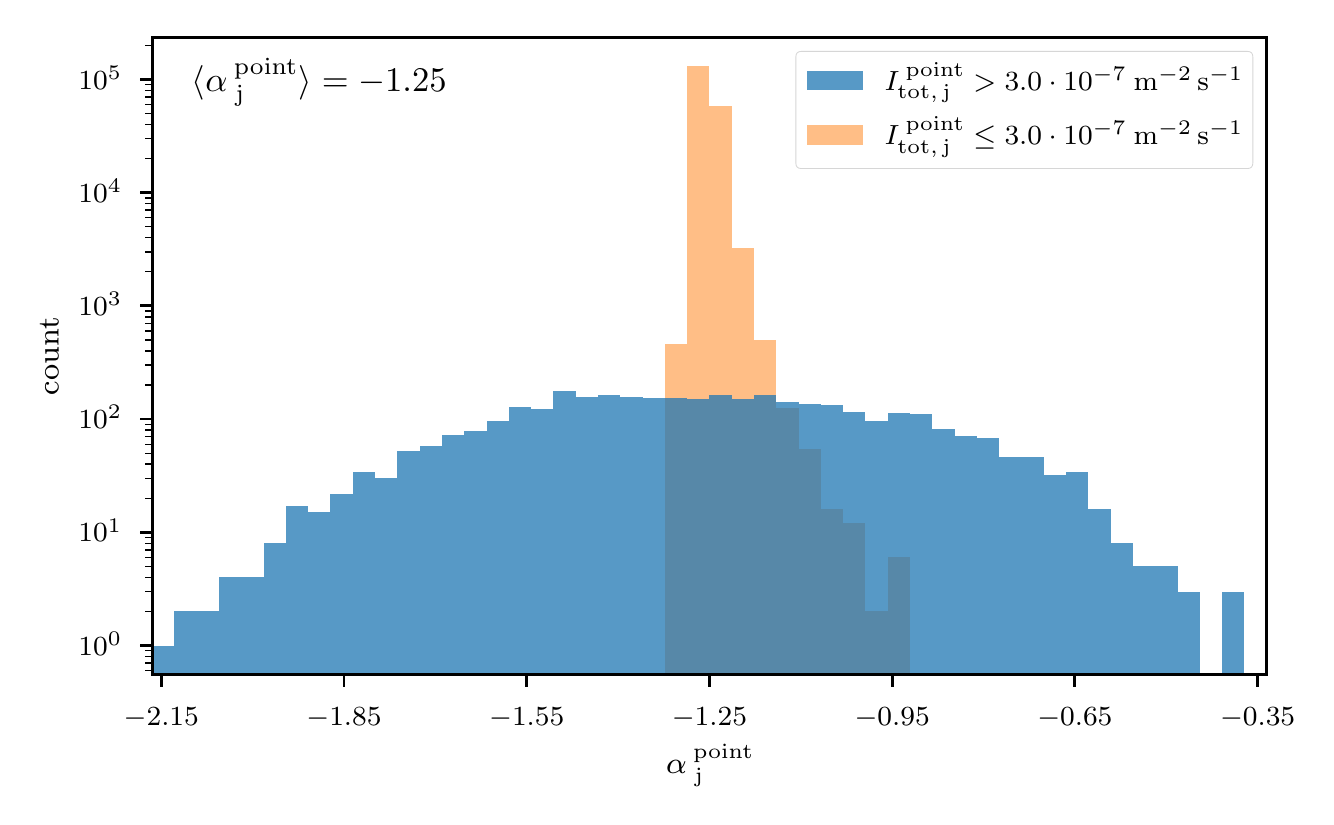}%
\end{minipage}
\caption{M1 PS spectral index $\alpha^\mathrm{\:point}_\mathrm{j}$ posterior means.
We note that the spectral indices have an offset of $+1$ introduced by the logarithmic binning of the energy dimension (see Eq.~\ref{eq:spectral-index-adjustment}).
\textbf{Left}: Map of spectral indices recovered for the PS pixels.
The color scale is centered on the prior mean, $-1.25$, and
ranges from two prior standard deviations above to two standard deviations below the prior mean.
\textbf{Right:} Histogram of posterior PS pixel spectral index means
for switched~on (\textbf{blue}) and switched~off PS pixels (\textbf{orange}).
The x-axis is centered on the prior mean, with the ticks indicating prior
standard deviation sized steps from the prior mean.
The y-axis shows counts on a logarithmic scale.}
\label{fig:m1-ps-spectral-indices}
\end{figure*}

Taking a broader view at the whole population of reconstructed PSs,
Fig.~\ref{fig:m1-ps-spectral-indices} shows the posterior mean of all sources' spectral indices.
As can be seen both from the left and right panel of the figure,
most PS pixels do not deviate from the prior mean.
This is mainly because of ``switched~off'' source pixels,
which strongly clusters around the spectral index prior mean of $-1.25$ (see the orange histogram in the right panel of Fig.~\ref{fig:m1-ps-spectral-indices}).
In contrast, the population of switched~on PS pixels
shows a broad distribution of posterior spectral indices
ranging from $-2.15$ to $-0.35$ (see the blue histogram in the right panel of Fig.~\ref{fig:m1-ps-spectral-indices}).
It extends markedly beyond the limits observed for the diffuse emission ($-1.70$ to $-1.26$).
For all source pixels, individual posterior mean spectral indices can be read off
in the left panel of Fig.~\ref{fig:m1-ps-spectral-indices}.
Noteworthy sources include the Vela pulsar, for whose containing pixel we find a soft spectrum ($\alpha < -1.85$),
and Vela-X, for whose containing pixel we find a hard spectrum ($\alpha > -0.95$).

% ---- fig: scatterplot dust m1 diffuse
\begin{figure}
\resizebox{\hsize}{!}{\includegraphics{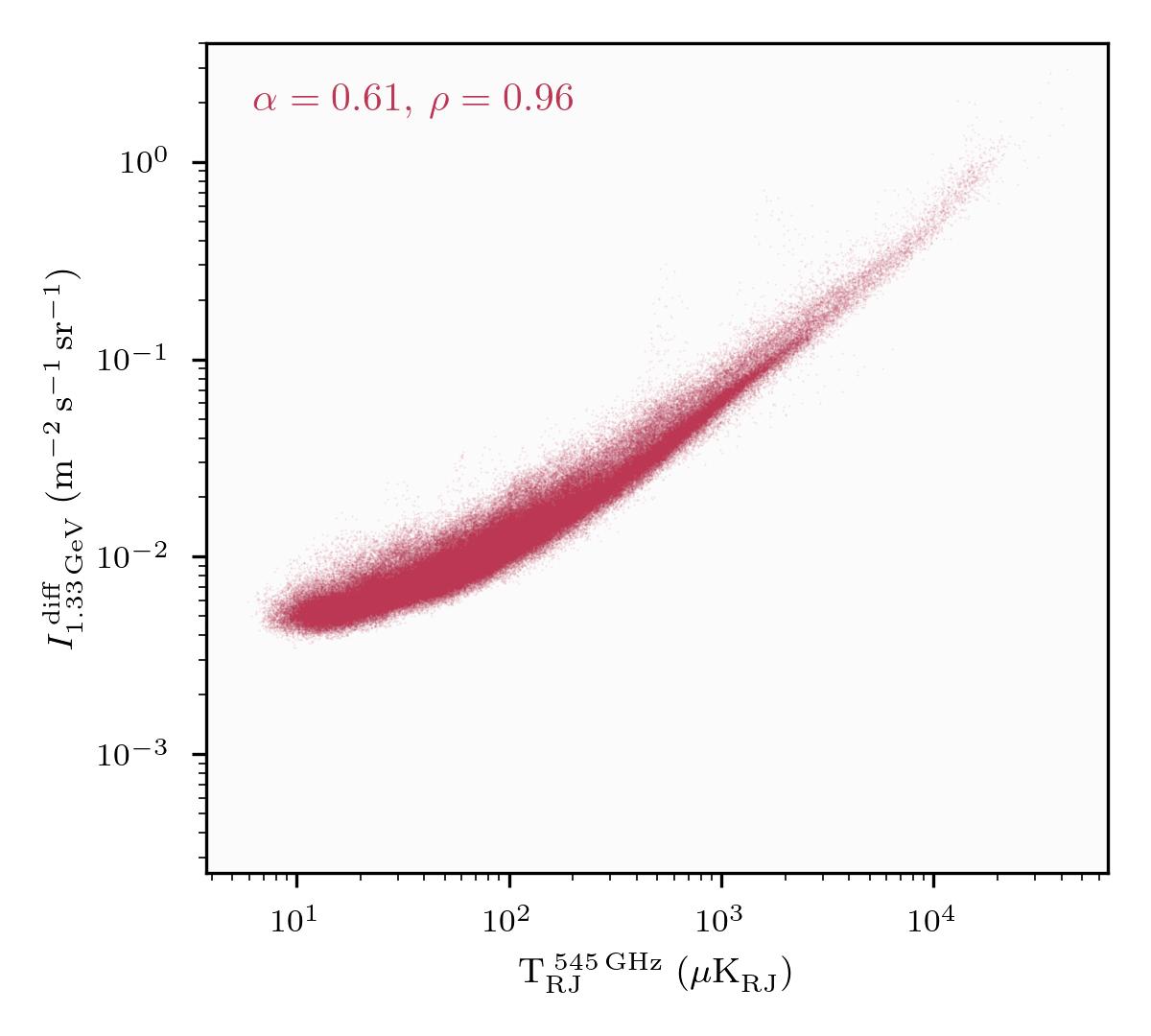}}
\caption{Scatter plot of the M1 diffuse flux density values in the \unit[1.00-1.77]{GeV} energy bin
against the \textit{Planck} \unit[545]{GHz} thermal dust emission map values.
Each pixel is represented by a dot,
with the x-axis showing the brightness temperature $\mathrm{T}_\mathrm{RJ}^{\:545\,\mathrm{GHz}}$
measured by \textit{Planck} for this pixel
and the y-axis showing the corresponding flux density value found by our reconstruction.
Both quantities are shown on a logarithmic scale.
The text in the upper left corner shows the slope of a linear fit to the dots, $\alpha$,
and the Pearson correlation coefficient between the two maps on log-log scale, $\rho$.}
\label{fig:m1-diffuse-scatterplot-dust}
\end{figure}

Closing the results section of the template-free model M1,
Fig.~\ref{fig:m1-diffuse-scatterplot-dust} shows a scatter plot
of the diffuse reconstruction based on M1 against
the \textit{Planck} thermal dust emission map used as a template in model M2.
Above diffuse gamma-ray fluxes of \unit[$3\cdot{10}^{-2}$]{$\left(\mathrm{m}^2\,\mathrm{s}\,\mathrm{sr}\right)^{-1}$}
we observe a close to linear scaling of the two quantities on log-log scale,
corroborated by a high Pearson correlation coefficient of $\rho = 0.96$.
A linear fit between the logarithmic pixel values of the two maps yields a best-fit slope of $\alpha = 0.61$.
Below the stated flux limit, the correlation weakens,
as other dust-independent emission processes begin contributing significant proportions of the diffuse emission.
These emissions are unveiled in the template-informed reconstruction, where dust-correlated emission was taken up by the
template-informed diffuse component and other flux is modeled with the second, free, diffuse component.
The results of this are presented in the following section.

\subsection{Template-informed spatio-spectral imaging via M2} \label{sec:results-m2}

% ---- fig: m2 sky maps
\begin{figure*}
\centering
\noindent\begin{minipage}[t]{1\textwidth}%
\includegraphics[width=1\textwidth]{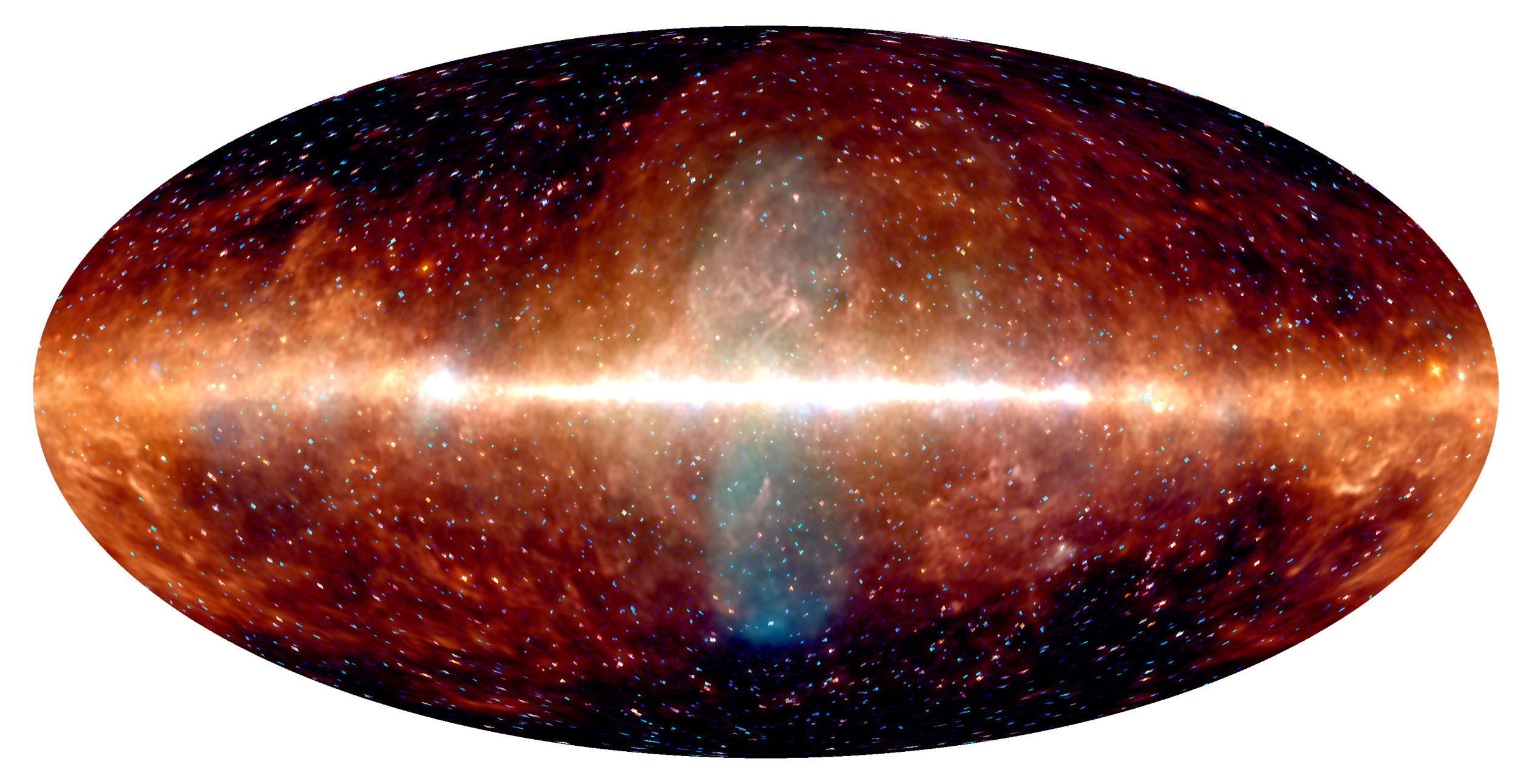}%
\end{minipage}

\begin{minipage}[t]{0.49\textwidth}%
\includegraphics[width=1\textwidth]{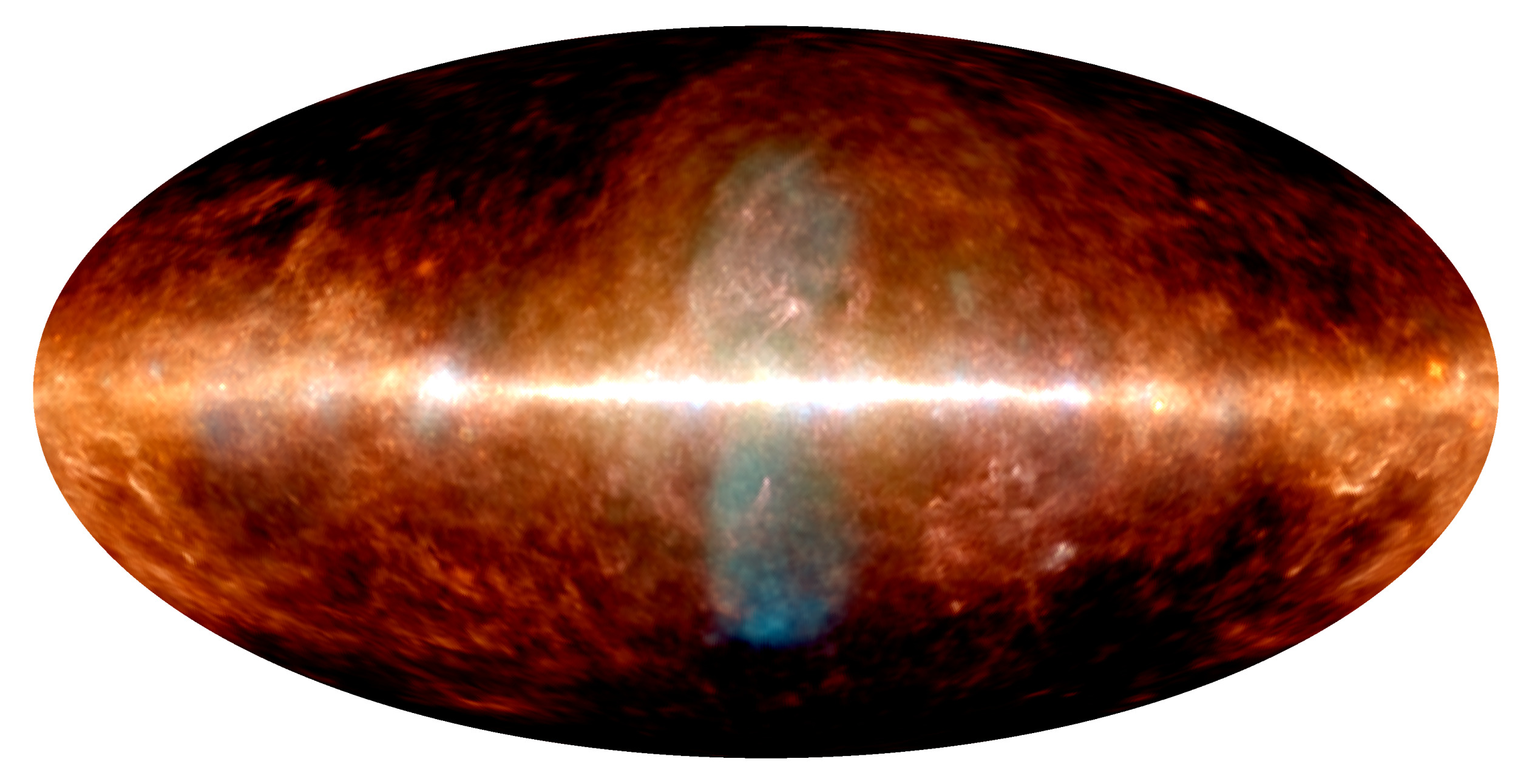}%
\end{minipage}\ %
\begin{minipage}[t]{0.49\textwidth}%
\includegraphics[width=1\textwidth]{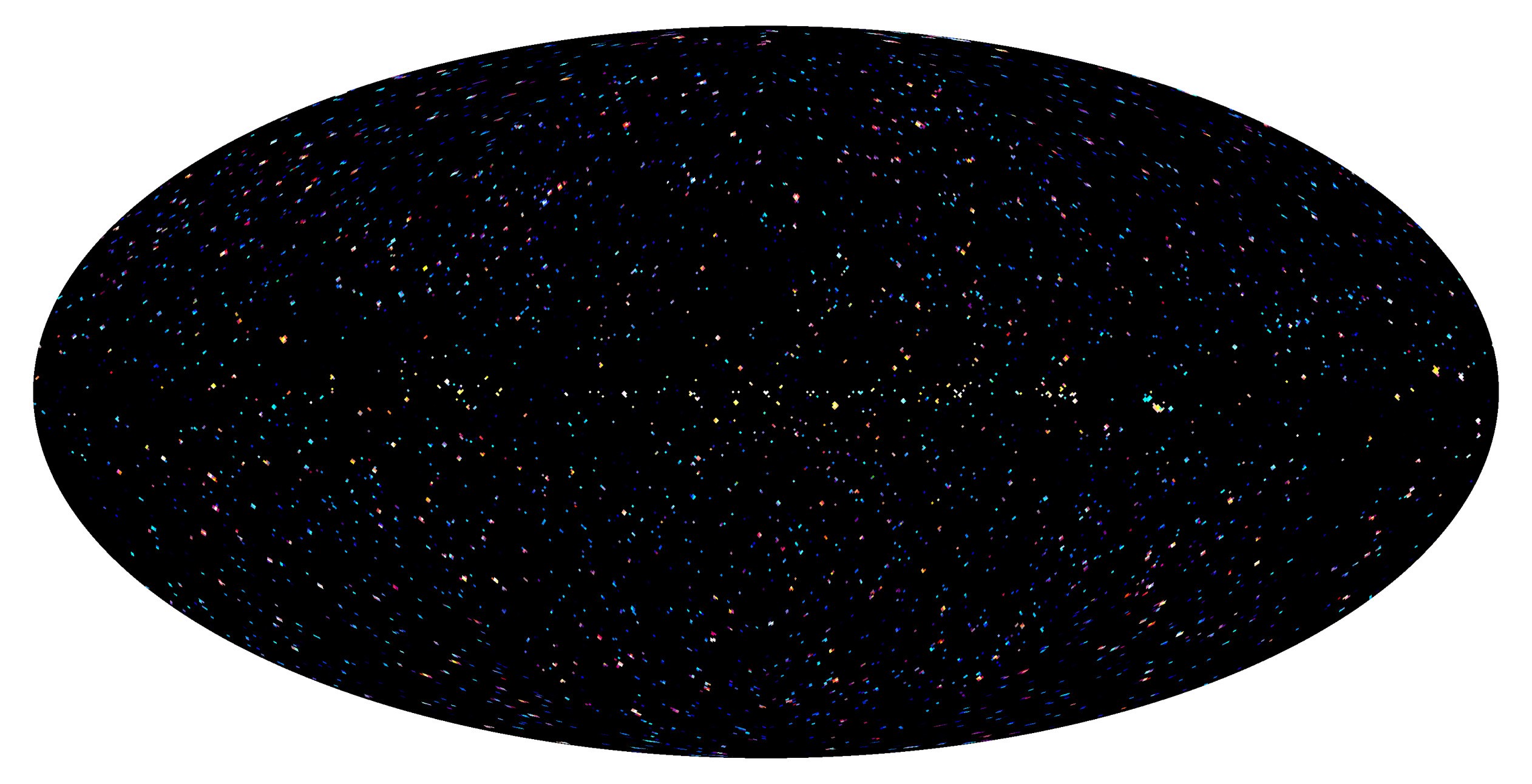}%
\end{minipage}

\begin{minipage}[t]{0.49\textwidth}%
\includegraphics[width=1\textwidth]{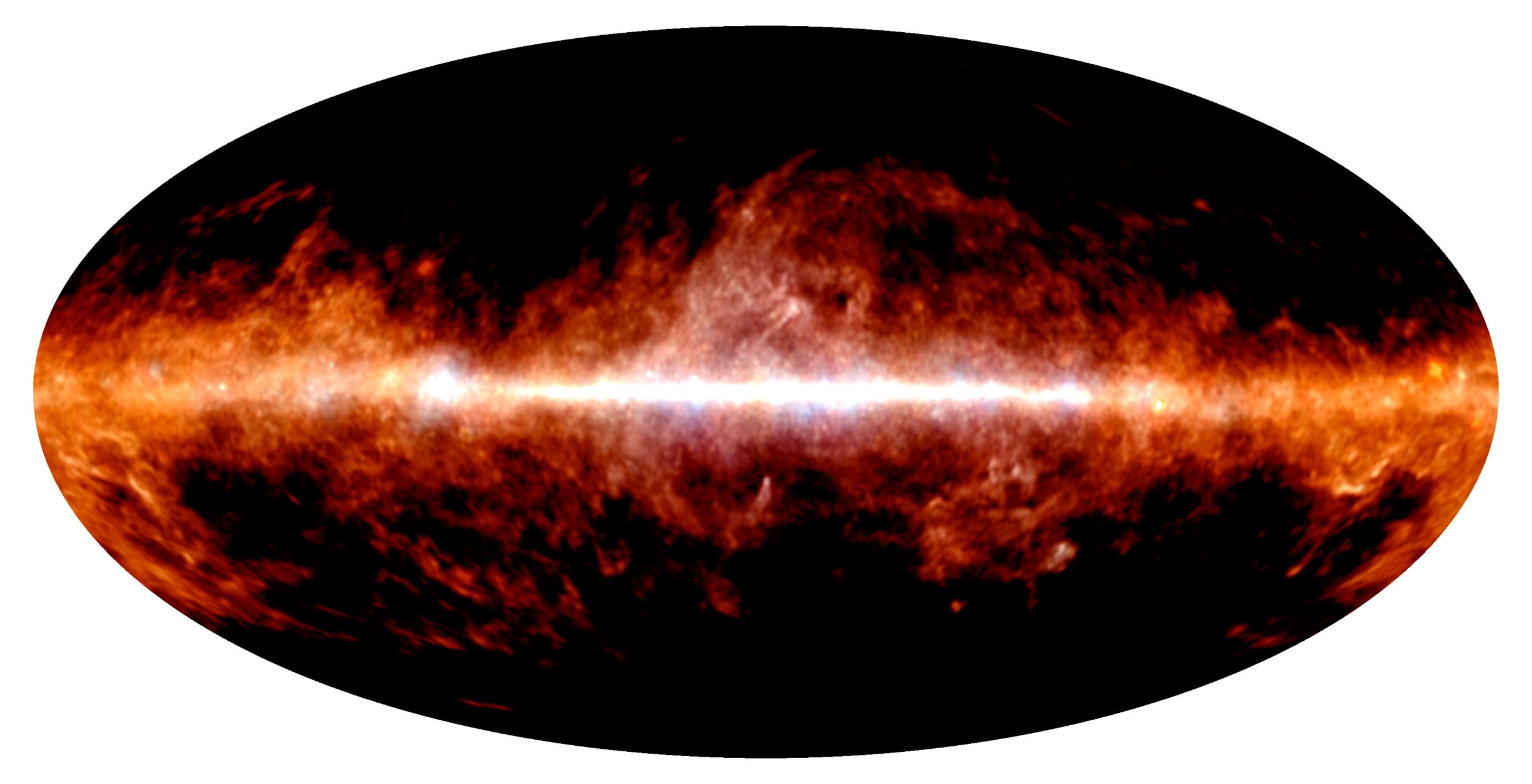}%
\end{minipage}\ %
\begin{minipage}[t]{0.49\textwidth}%
\includegraphics[width=1\textwidth]{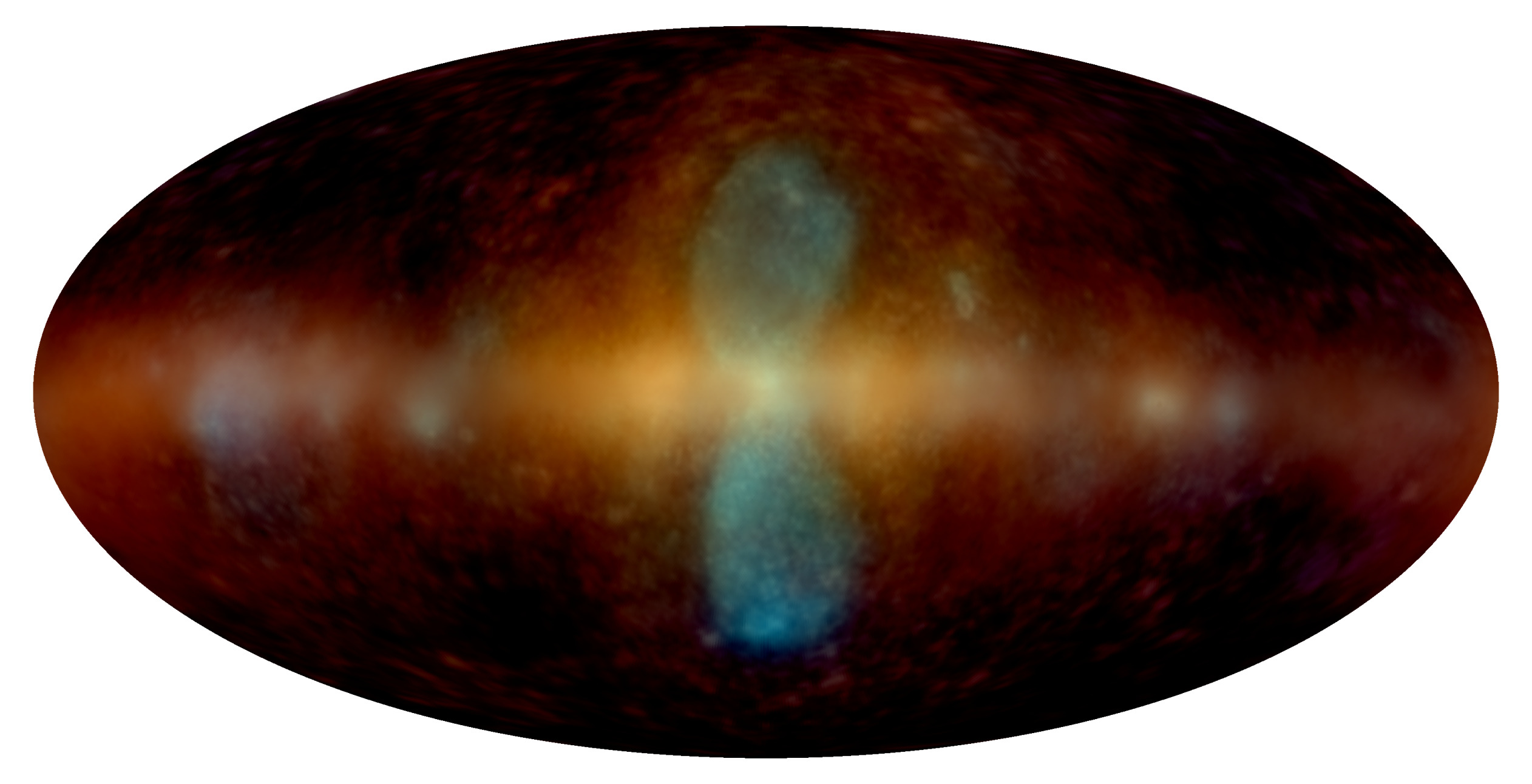}%
\end{minipage}

\begin{minipage}[t]{0.33\textwidth}%
\includegraphics[width=1\textwidth]{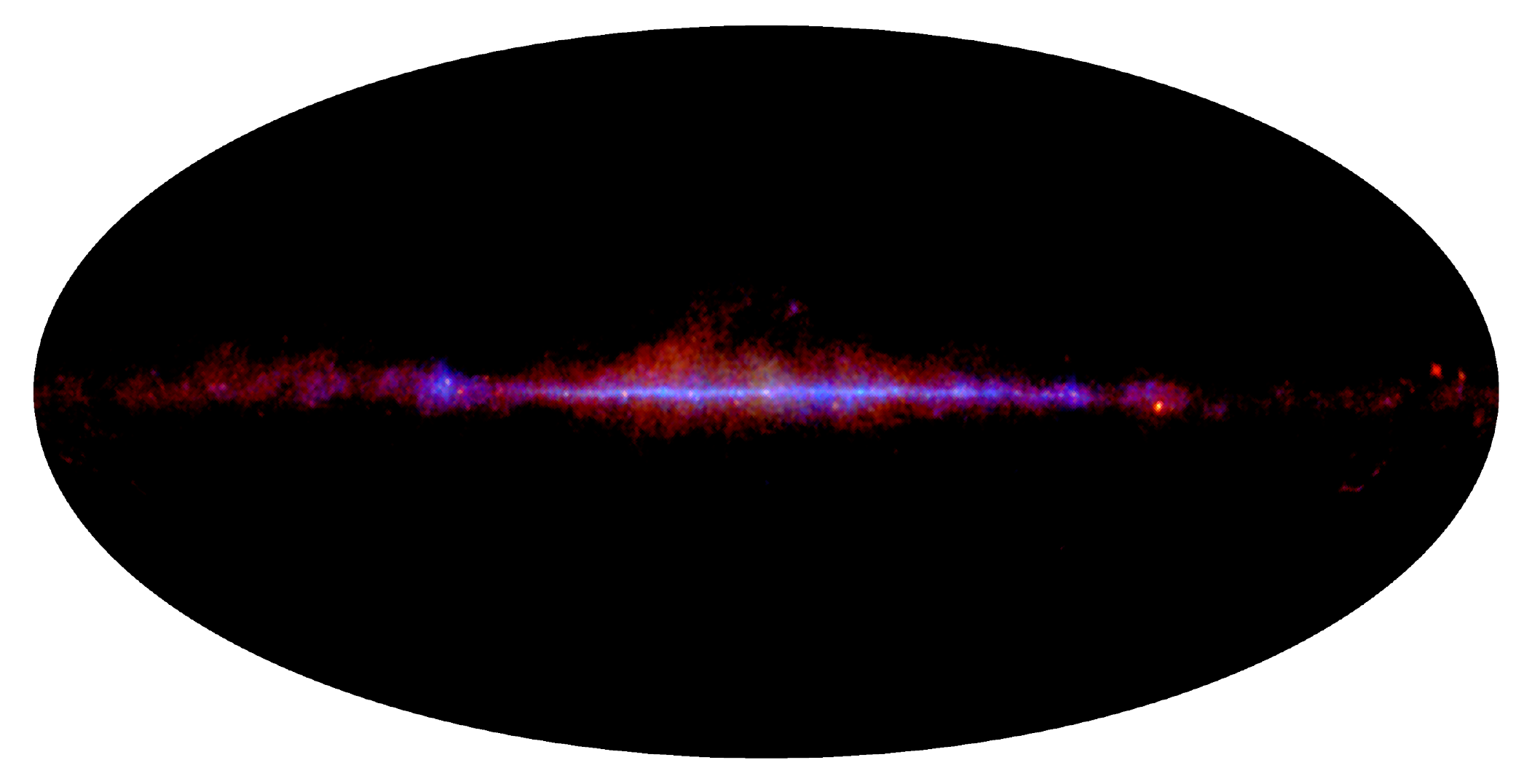}%
\end{minipage}%
\begin{minipage}[t]{0.33\textwidth}%
\includegraphics[width=1\textwidth]{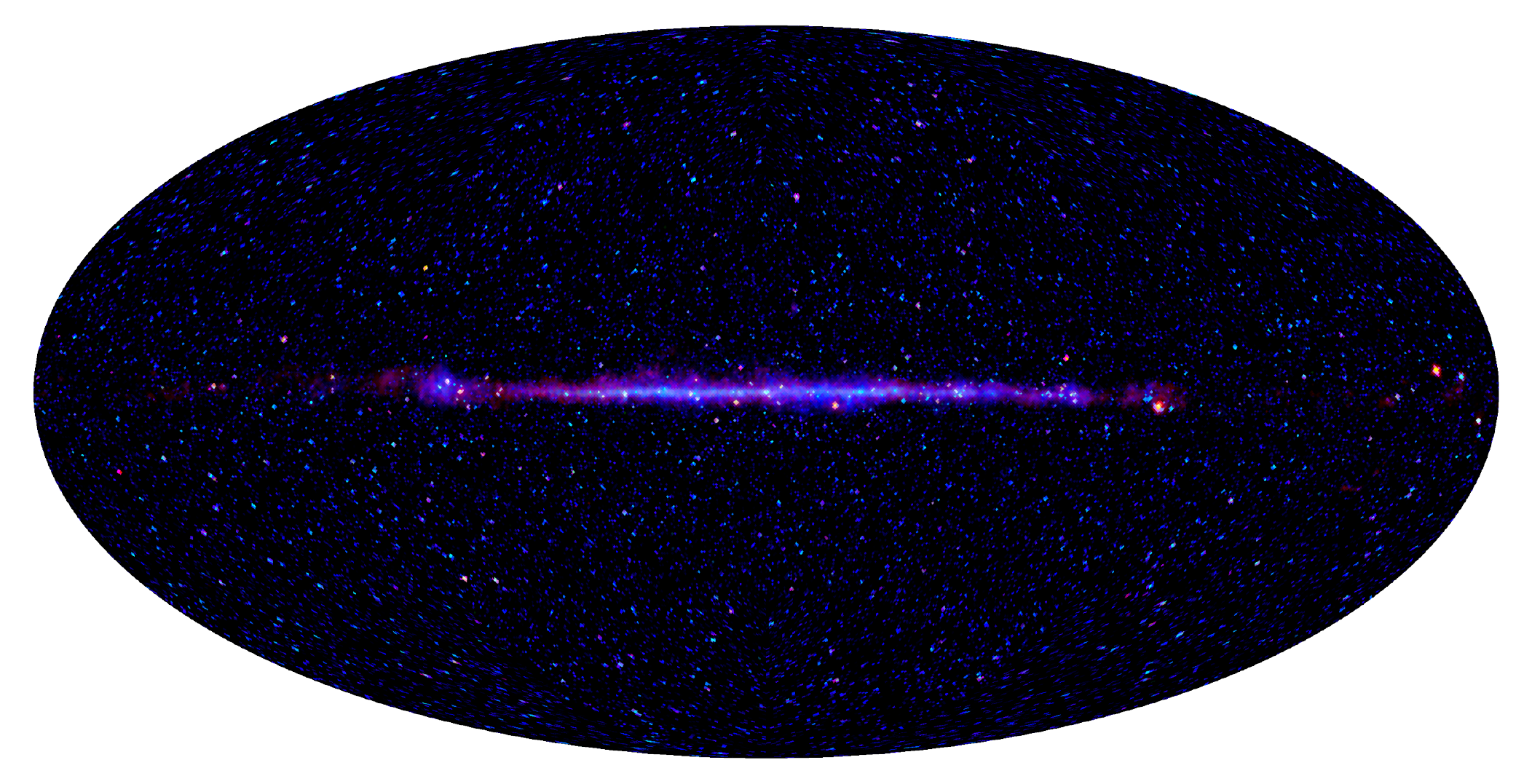}%
\end{minipage}%
\begin{minipage}[t]{0.33\textwidth}%
\includegraphics[width=1\textwidth]{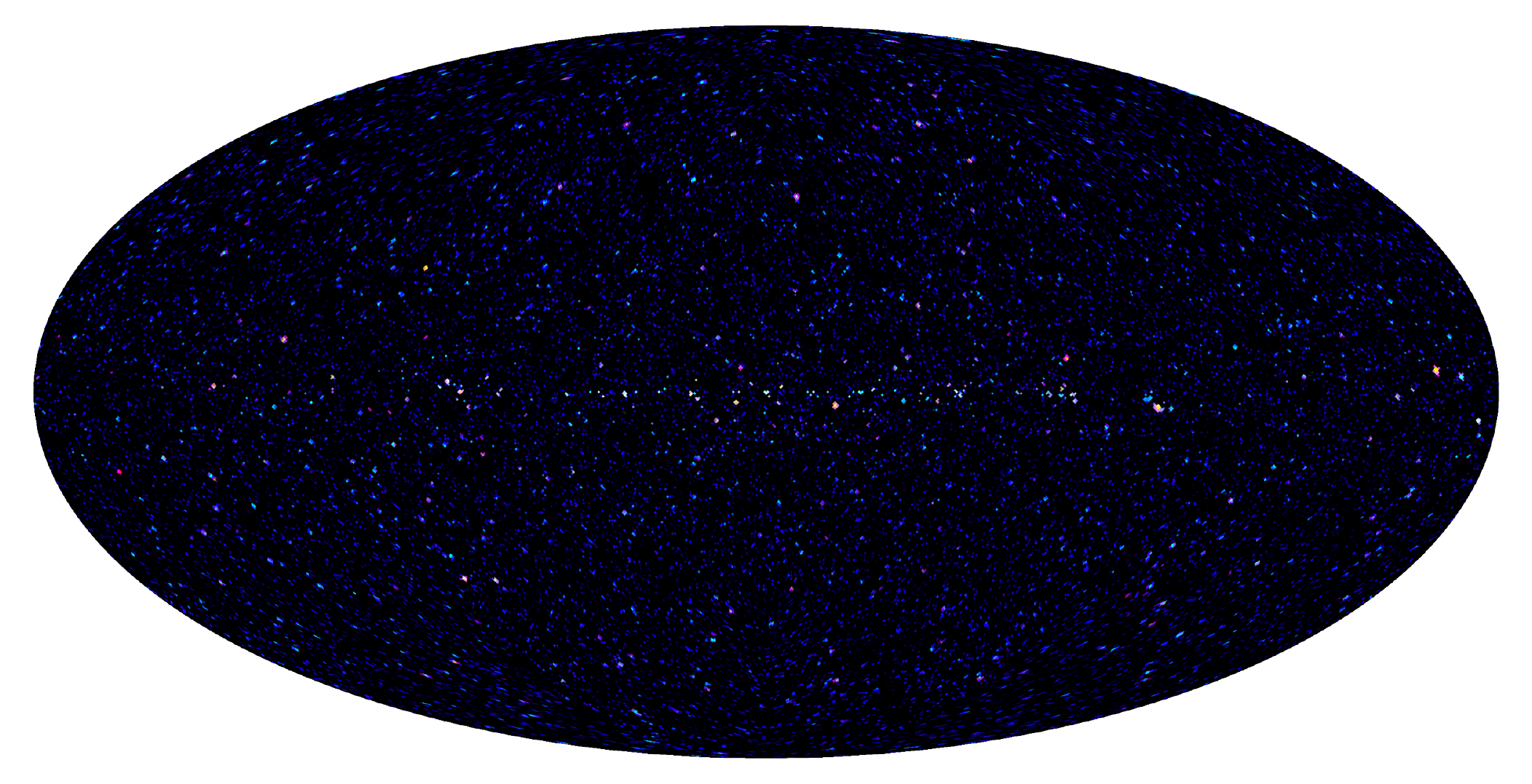}%
\end{minipage}

\caption{Results of the template-informed reconstruction based on model M2.
The figure uses the spectral domain color mapping introduced in the caption of Fig.~\ref{fig:data-exposure-corrected-mf-plot}.
The color scale is provided in the bottom panel of Fig.~\ref{fig:data-exposure-corrected-mf-plot}.
\textbf{First row}: Reconstructed gamma-ray sky.
\textbf{Second row}: Separated diffuse emission sum (left) and PS emissions (right).
A rendering of the diffuse emission sum without color saturation enhancement is show in Fig.~\ref{fig:m2-mf-plot-desaturated}.
\textbf{Third row}: Dust-correlated diffuse emissions (left) and other diffuse emissions (right).
\textbf{Fourth row}: Posterior standard deviation of the sum of the diffuse emissions (left),
the full sky (middle), and the PS emissions (right).}
\label{fig:m2-sky-maps}
\end{figure*}

The spatio-spectral gamma-ray sky reconstruction according to our template-informed model M2
is shown in Fig.~\ref{fig:m2-sky-maps}.
Additional to the decomposition into point-like and diffuse emission,
the M2 model allows us to further decompose the diffuse emission into a thermal dust emission correlated component
and other diffuse emissions.
Maps of the two diffuse components are shown in the third row of Fig.~\ref{fig:m2-sky-maps}.
Unveiled from the dust-correlated emission, the $I^\mathrm{\:nd}$ map (right)
shows all dust-independent diffuse emission.
First, this includes the FBs, as expected.
Second, there appears a bright soft spectrum emission structure in the inner Galaxy,
left and right of the FBs and symmetric around the north-south axis of the Galaxy.
It traces the shape of the FBs to Galactic latitudes of $|\mathrm{b}| <$~{30\textdegree}
at a longitudinal distance of approximately {10\textdegree} on the map and has a smooth appearance.
Third, along the Galactic plane in the western hemisphere of the sky,
there is a faint soft spectrum emission structure, also of smooth appearance.
Lastly, there is a structure of small clumps of emission surrounding the FBs
and extending to high Galactic latitudes in the northern hemisphere.
Structures similar to this are also visible in the southern bubble,
even in the M1 diffuse sky maps (Fig.~\ref{fig:m1-sky-maps}).
These might be part of the larger collection of small-structured emissions just mentioned
or correspond to a internal structure of the bubble emissions.

% ---- fig: m2 sf diffuse plots
\begin{figure*}
\centering
\noindent
\begin{minipage}[t]{0.32\textwidth}%
\includegraphics[width=1\textwidth]{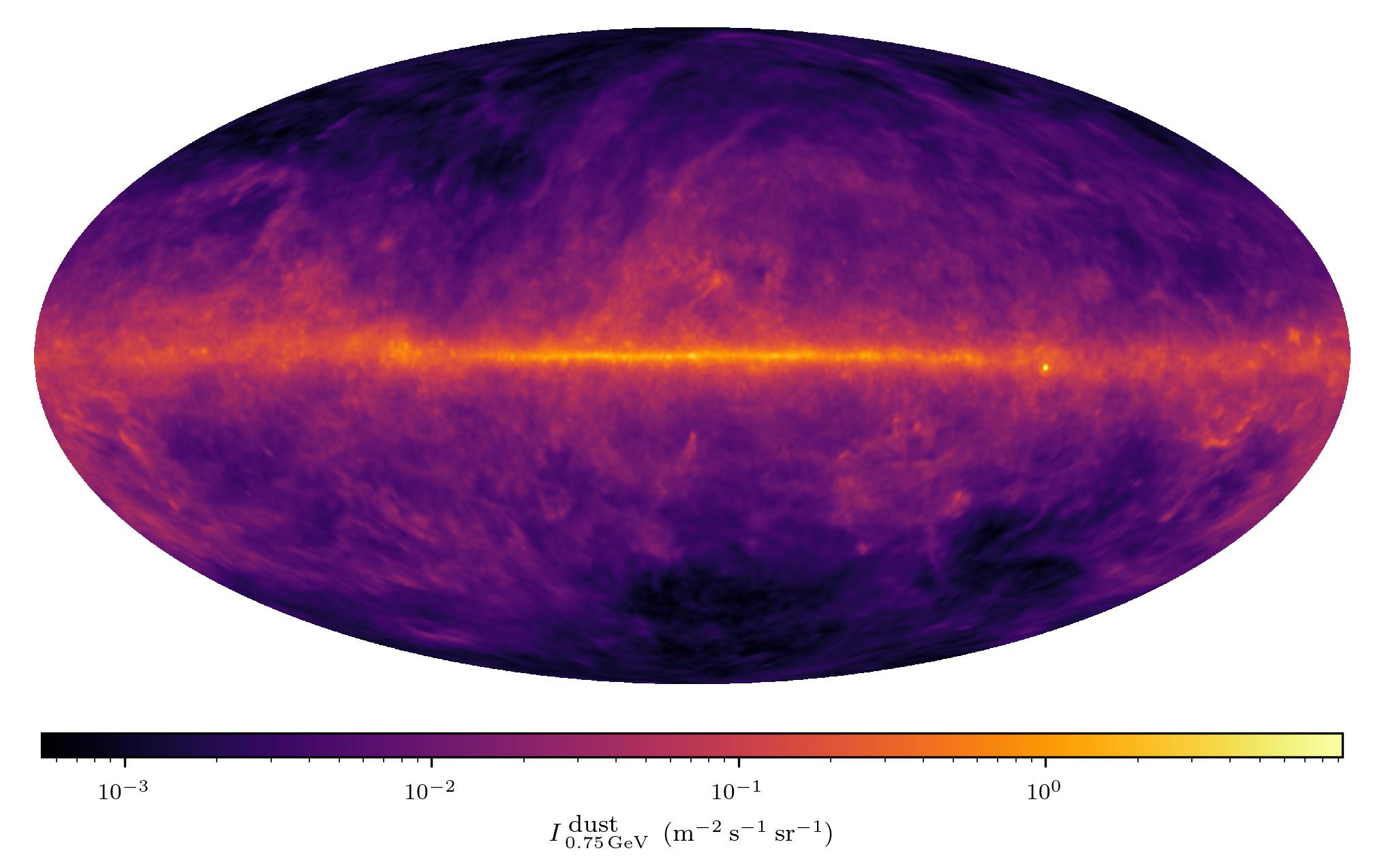}%
\end{minipage}\ %
\begin{minipage}[t]{0.32\textwidth}%
\includegraphics[width=1\textwidth]{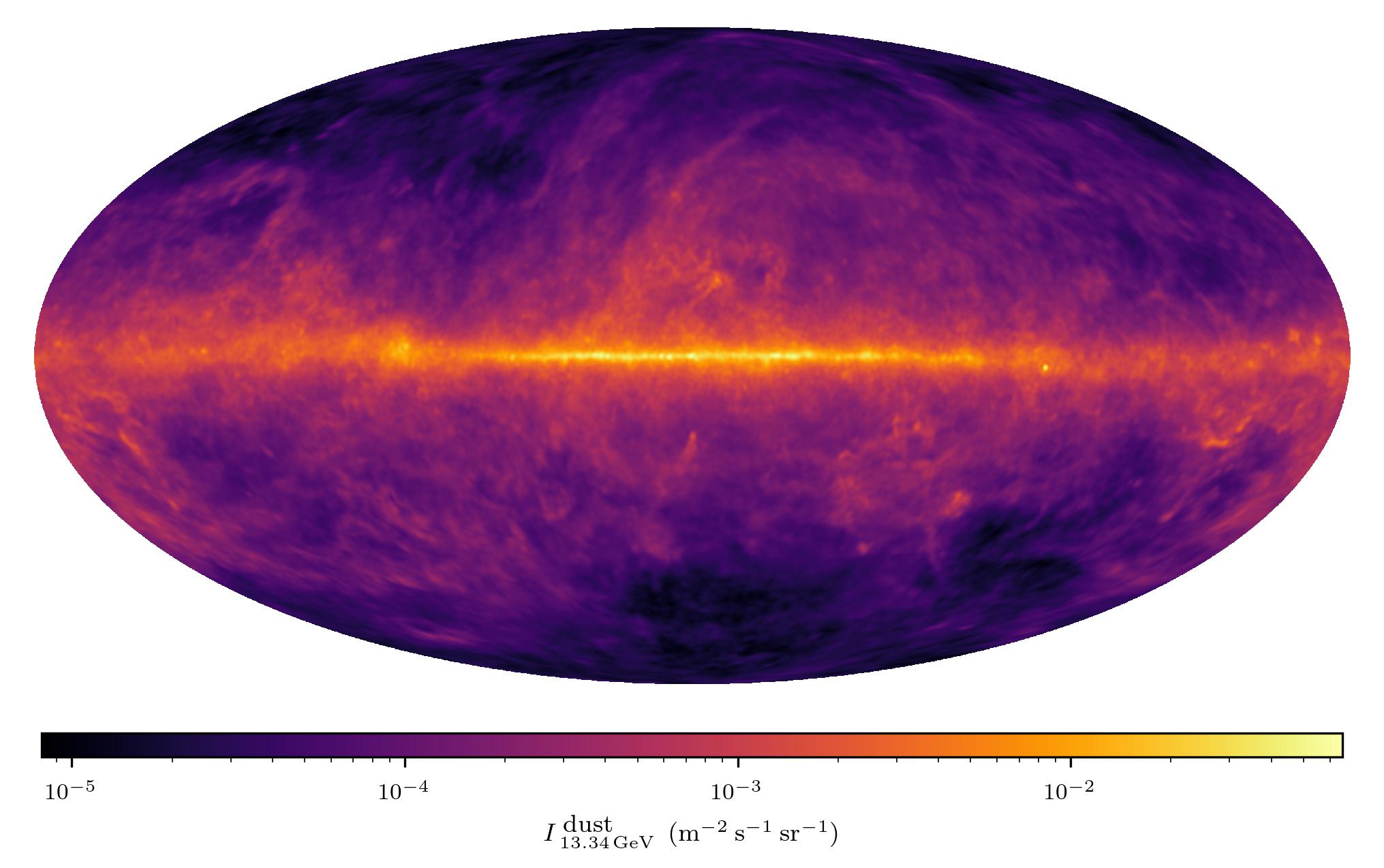}
\end{minipage}\ %
\begin{minipage}[t]{0.32\textwidth}%
\includegraphics[width=1\textwidth]{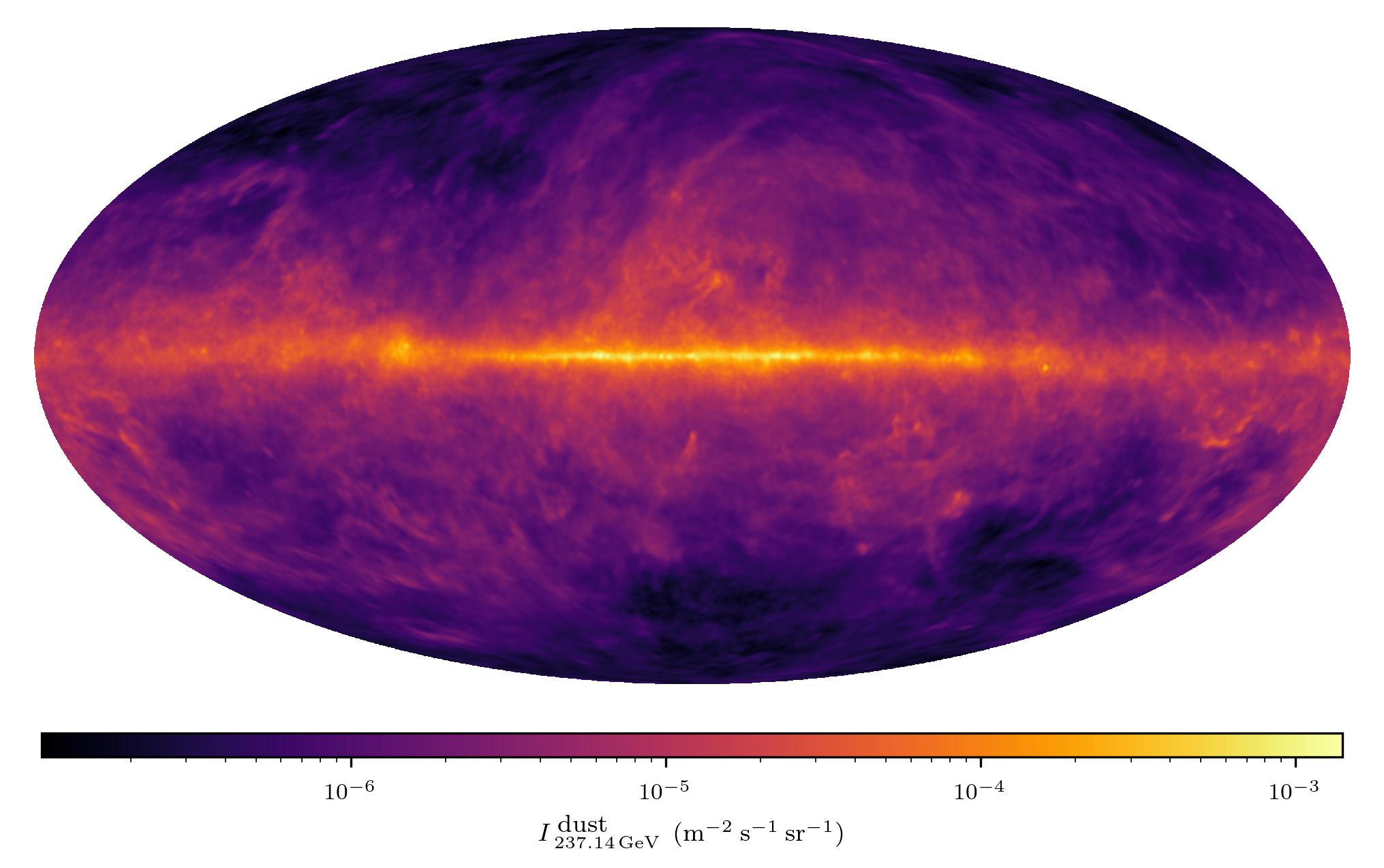}
\end{minipage}
\begin{minipage}[t]{0.32\textwidth}%
\includegraphics[width=1\textwidth]{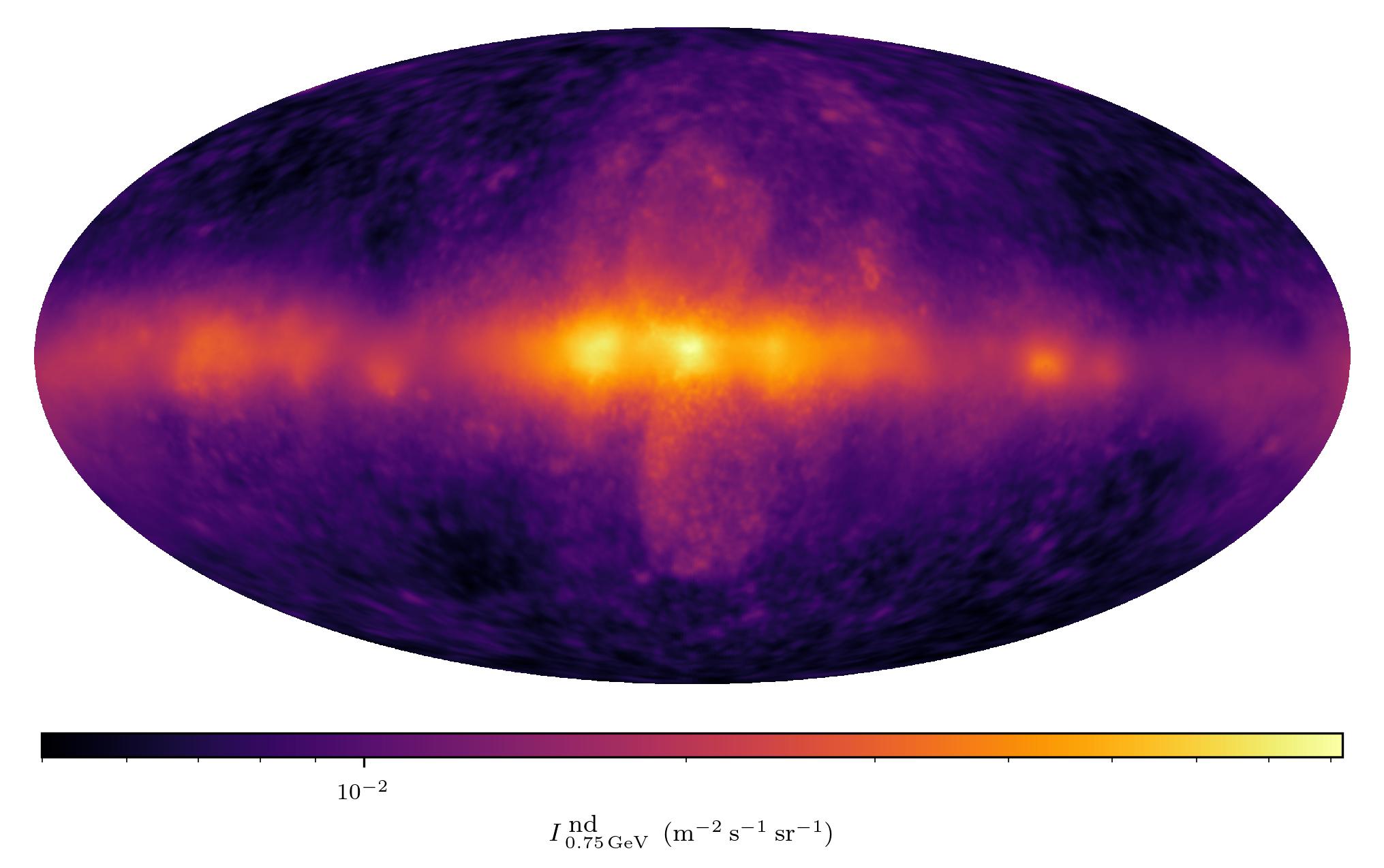}%
\end{minipage}\ %
\begin{minipage}[t]{0.32\textwidth}%
\includegraphics[width=1\textwidth]{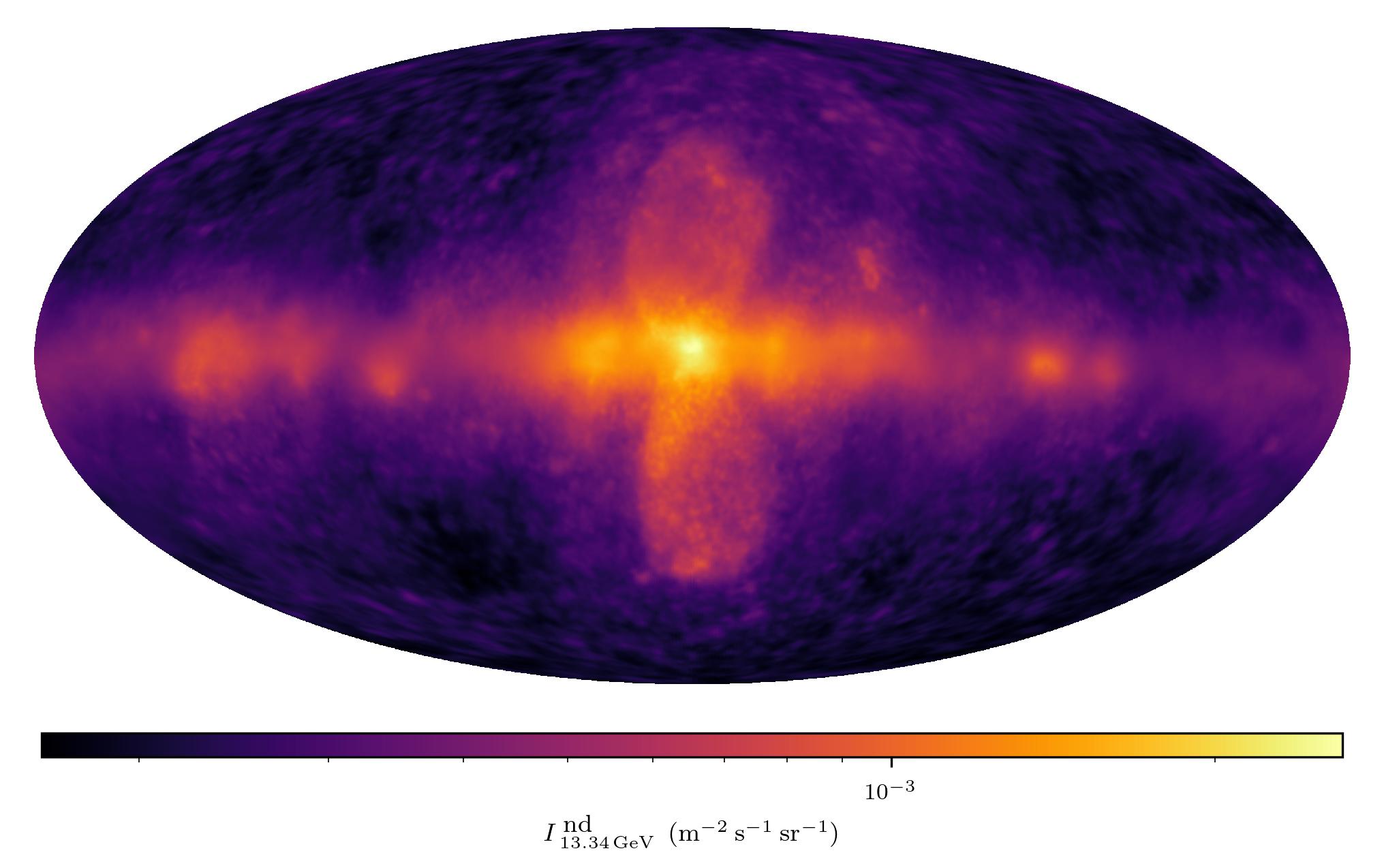}
\end{minipage}\ %
\begin{minipage}[t]{0.32\textwidth}%
\includegraphics[width=1\textwidth]{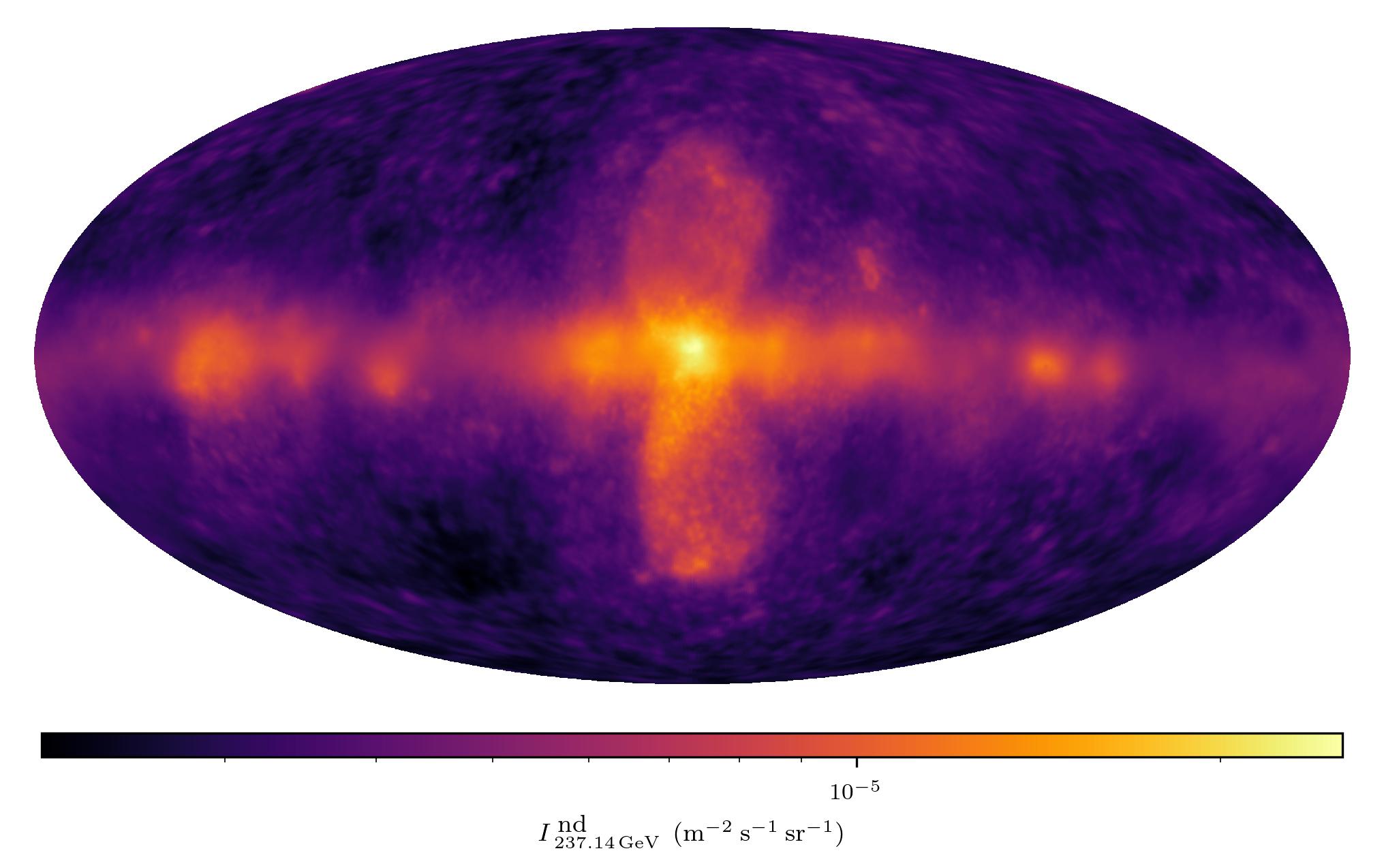}
\end{minipage}
\caption{Single energy plots of the M2 diffuse components with a logarithmic color scale.
\textbf{Top row}: Dust-correlated (template-informed) diffuse emission.
\textbf{Bottom row}: A priori dust-independent (free) diffuse emission.
Energy bins: \unit[0.56--1.00]{GeV} (\textbf{left}), \unit[10.0--17.8]{GeV} (\textbf{middle}),
and \unit[178--316]{GeV} (\textbf{right}).
All panels have individual brightness scales and visualize the full dynamic range
of their respective maps.}
\label{fig:m2-sf-plots}
\end{figure*}

Figure~\ref{fig:m2-sf-plots} shows single energy bin plots of the two M2 diffuse component reconstructions.
From the $I^\mathrm{\:dust}$ panels in the upper row, the large dynamic range of the dust-correlated
emission reconstruction becomes apparent, which is induced by the thermal dust emission template.
In the $I^\mathrm{\:nd}$ maps, the constellation of emission clumps surrounding the FBs
can be seen to be part of a more continuous outer bubble structure,
reminiscent of the X-ray arcs observed by eRosita \citep{predehl_2020_erosita_bubbles}.

% ---- fig: scatterplot dust m2 diffuse
\begin{figure}
\resizebox{\hsize}{!}{\includegraphics{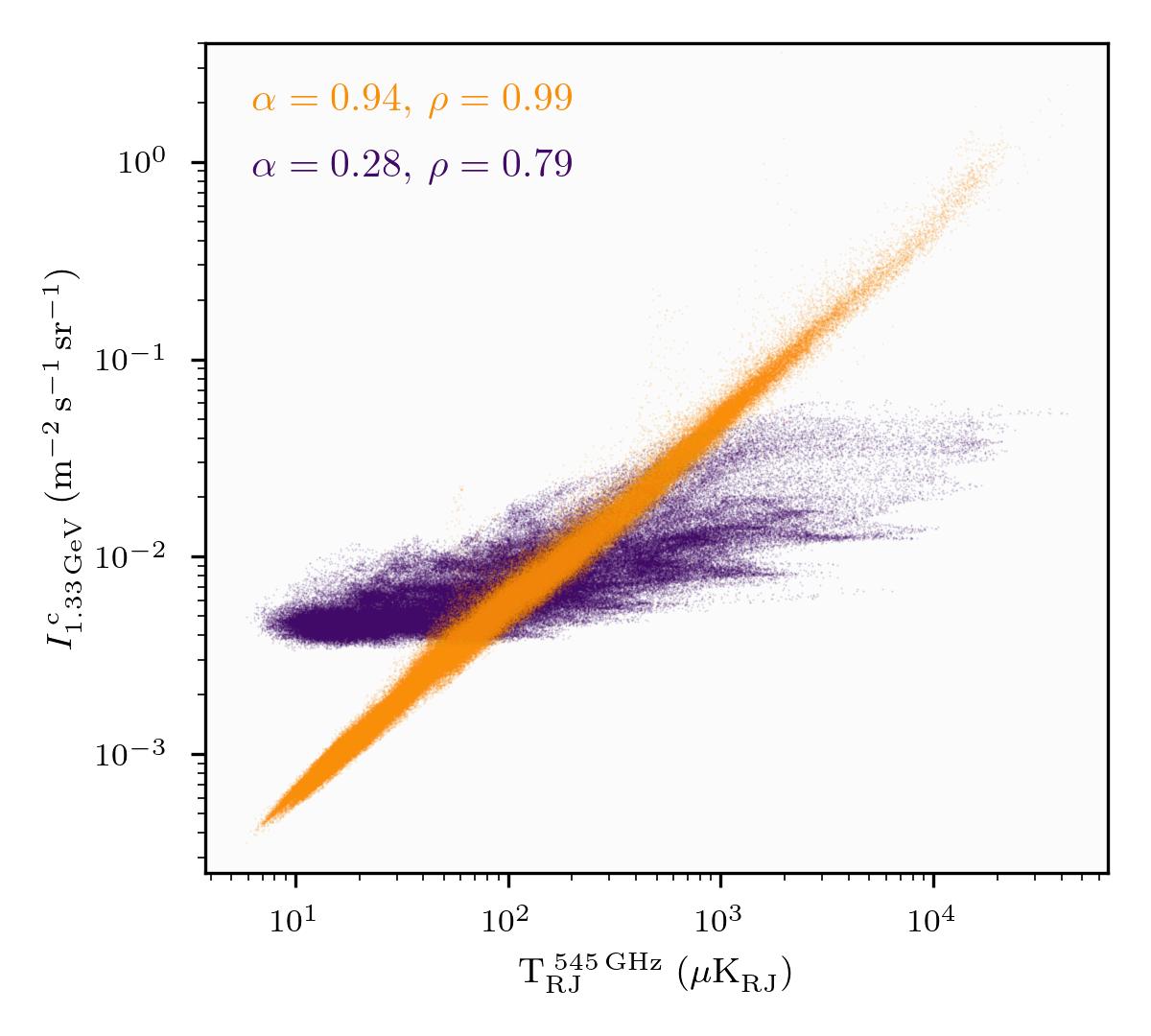}}
\caption{Scatter plot of the M2 diffuse flux density values in the \unit[1.00-1.77]{GeV} energy bin
against the \textit{Planck} \unit[545]{GHz} thermal dust emission map values.
Each pixel is represented by a dot,
with the x-axis showing the brightness temperature $\mathrm{T}_\mathrm{RJ}^{\:545\,\mathrm{GHz}}$
measured by \textit{Planck} for this pixel
and the y-axis showing the corresponding flux density value of our reconstruction.
Both axes have a logarithmic scale.
The points for the dust template informed diffuse component $I^\mathrm{\:dust}$ are shown in orange,
while the points for the a priori dust-independent diffuse component $I^\mathrm{\:nd}$ are shown in purple.
The text in the upper left corner shows the slope of a linear fit to the dots, $\alpha$,
and the Pearson correlation coefficient between the thermal dust map and our reconstructions on log-log scale, $\rho$.}
\label{fig:m2-diffuse-scatterplot-dust}
\end{figure}

So far, we have assumed that the template-free diffuse emission component $I^\mathrm{\:nd}$
reconstructs dust-independent emissions.
To test whether this assumption holds true,
Fig.~\ref{fig:m2-diffuse-scatterplot-dust} shows a scatter plot
of the M2 diffuse gamma-ray maps $I^\mathrm{\:dust}$ and $I^\mathrm{\:nd}$
with the \textit{Planck} thermal brightness map.
$I^\mathrm{\:dust}$ shows a strong linear relationship with the thermal brightness map on log-log scale ($\alpha=0.94$),
while $I^\mathrm{\:nd}$ only has a weak linear relationship with it ($\alpha=0.28$).
This indicates a good unmixing of the emission components.
The same pattern,
albeit less pronounced, is observed in the Pearson correlation coefficients on log-log scale
with $\rho = 0.99$ for $I^\mathrm{\:dust}$ and $\rho = 0.79$ for $I^\mathrm{\:nd}$.
The strong Pearson correlation of $I^\mathrm{\:nd}$ with the dust map
indicates $I^\mathrm{\:nd}$ is dominated by ISM-tracing gamma-ray emissions.

% ---- fig: m2 dust modification map
\begin{figure}
\resizebox{\hsize}{!}{\includegraphics{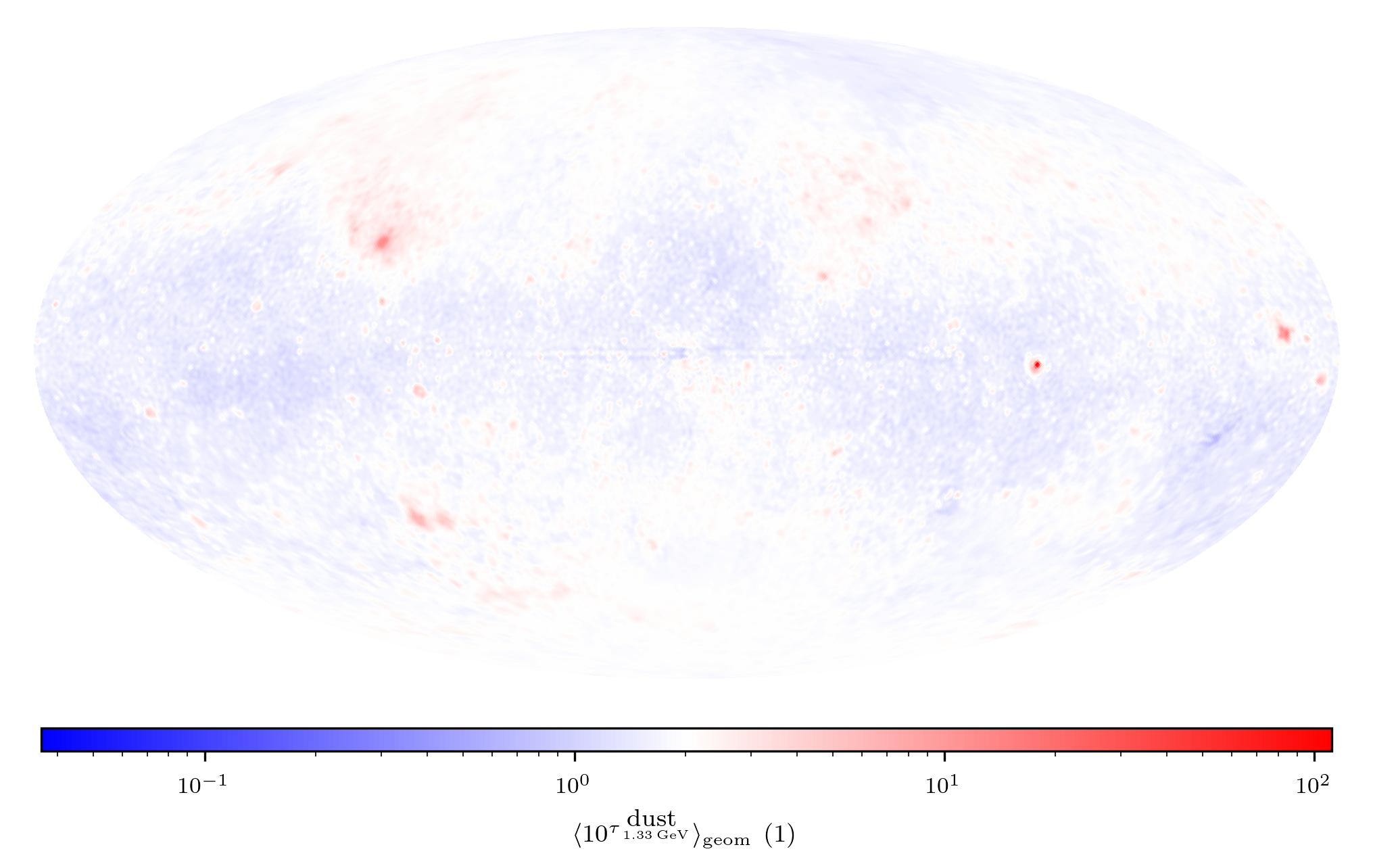}}
\caption{M2 multiplicative thermal dust template modification field
for the \unit[1.00-1.77]{GeV} energy bin (posterior geometric mean) on a logarithmic color scale.
Fig.~\ref{fig:m2-dust-modification-field-all-e} shows the modification field maps for two additional energy bins and corresponding posterior uncertainty maps.}
\label{fig:m2-dust-modification-field-low-e}
\end{figure}

Figure~\ref{fig:m2-dust-modification-field-low-e} shows the dust template modification field
$\tau^\mathrm{\:dust}$ in the energy bin from 1.00 to \unit[1.77]{GeV}.
It shows a median value of 2.0, indicating the brightness reference scale $I^\mathrm{\:dust}_\mathrm{0}$
was chosen too low by that factor.
The strongest modifications can be seen in the regions of
the Vela pulsar and nebula, the Geminga pulsar, the Crab pulsar and nebula,
PSR J1836+5925, and the blazar 3C 454.3.
All these are very bright PSs or extended objects,
here contaminating the dust-correlated emission map.
Besides this, the dust modification map of the energy bin only shows very little structure,
with most values lying in the range of 1.0 -- 3.0.
One notable structure in the map is a sharp line along the Galactic plane,
with modification field values around $2.0$ on the disk and slightly below $1.0$ directly next to it.
This is further evidence of a PSF mismodeling,
necessitating this unphysical correction to the dust-correlated Galactic disk emission.

% ---- fig: m2 diffuse component sepectral index maps -> spectral hardening
\begin{figure*}
\centering
\noindent
\begin{minipage}[t]{0.49\textwidth}%
\includegraphics[width=1\textwidth]{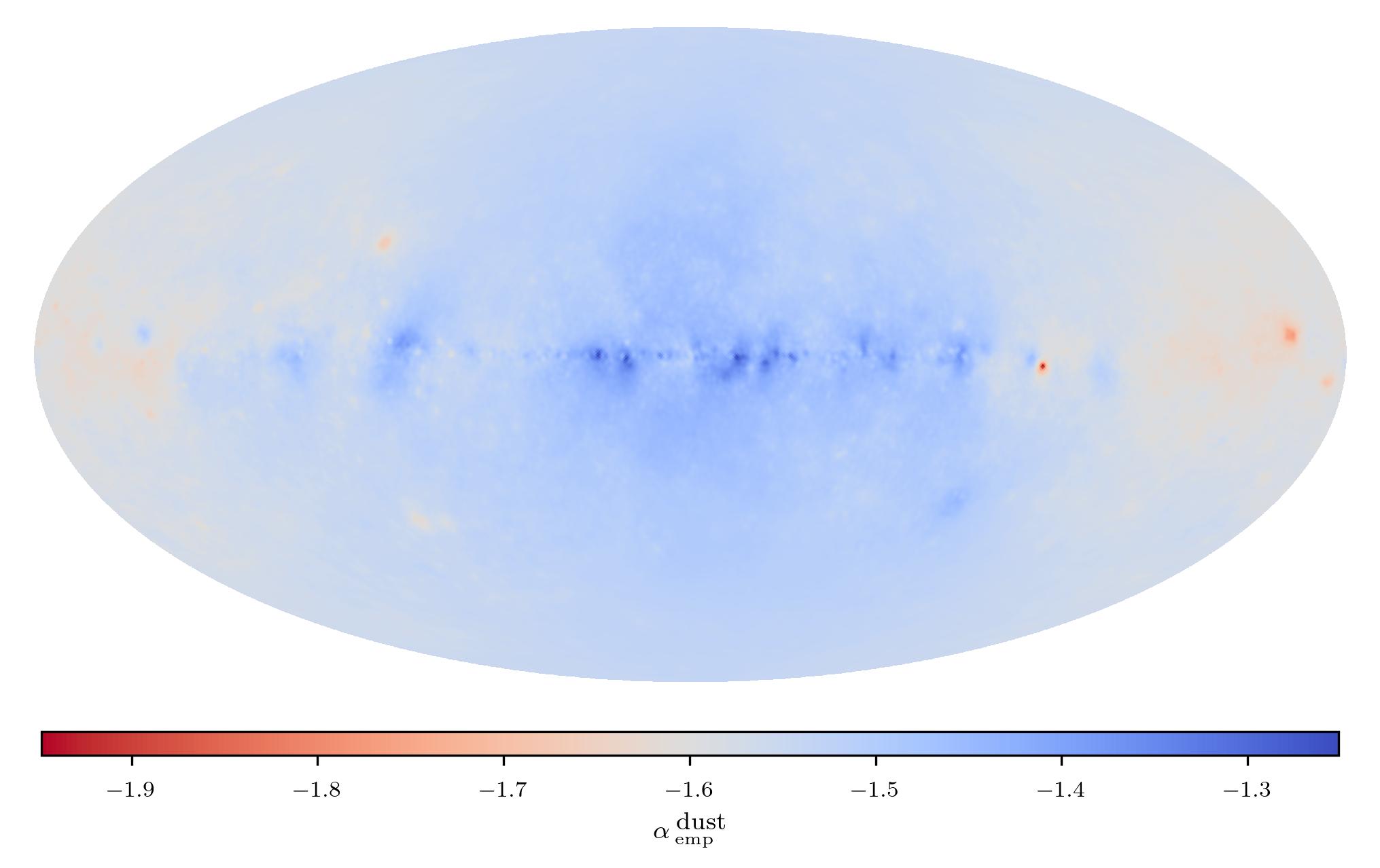}%
\end{minipage}%
\begin{minipage}[t]{0.49\textwidth}%
\includegraphics[width=1\textwidth]{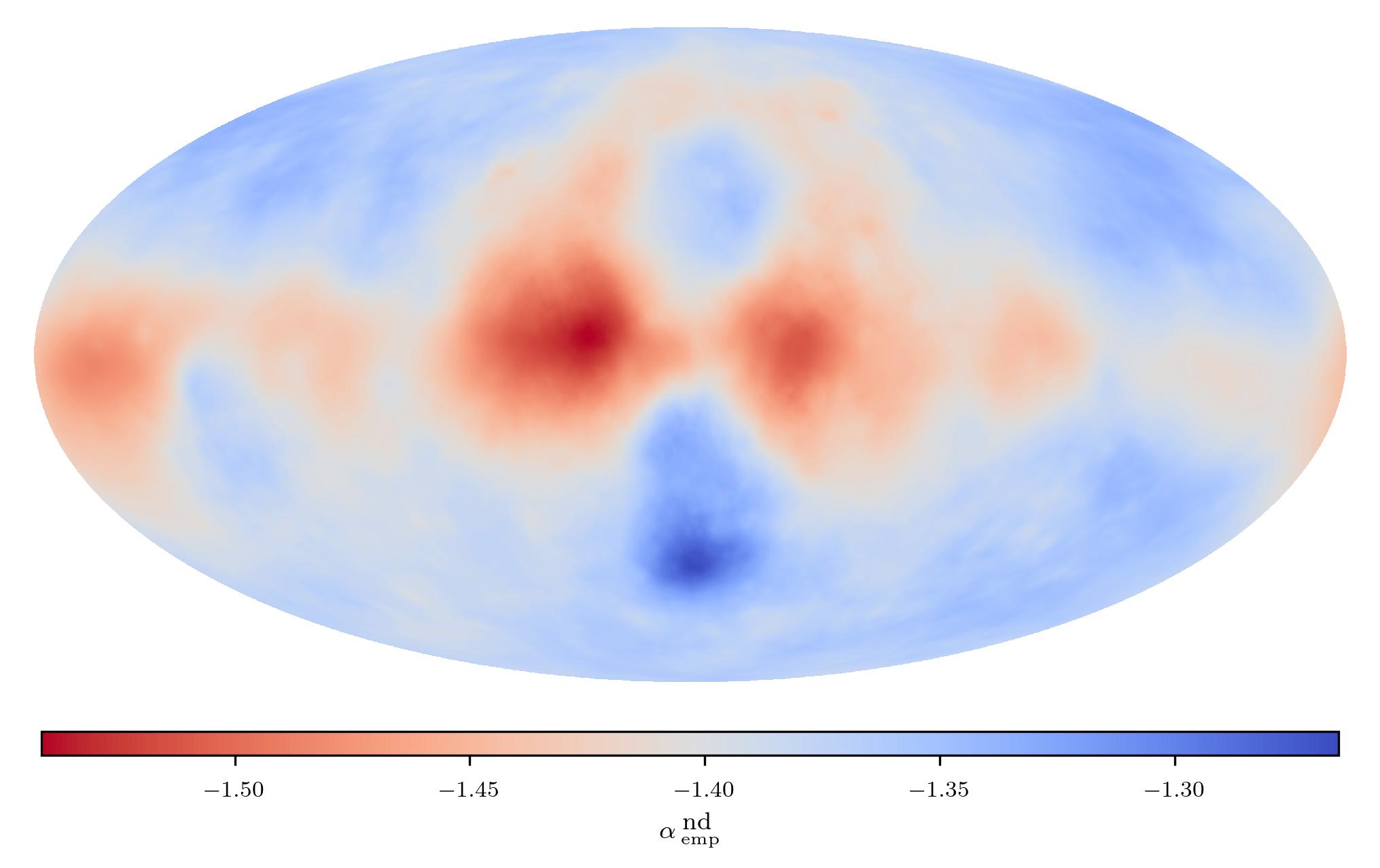}%
\end{minipage}
\caption{Maps of the M2 diffuse component empirical spectral index posterior means
obtained by power-law fits to the diffuse emission component samples provided by the M2 reconstruction and subsequent averaging.
\textbf{Left}: Map of the dust-template informed diffuse component pixel-wise spectral indices.
\textbf{Right}: Map of the template-free diffuse component pixel-wise spectral indices.}
\label{fig:m2-diffuse-spectral-index-maps}
\end{figure*}

Figure~\ref{fig:m2-diffuse-spectral-index-maps} shows empirical spectral index maps
for the two diffuse components of M2.
We obtained the values via power-law fits along the energy dimension of the diffuse maps
and a subsequent averaging over posterior samples.

The left panel displays the spectral index map for the template-informed diffuse component.
It shows a progressive spectral hardening toward the GC,
from spectral indices of $-1.7$ in the Galactic anticenter to spectral indices of
$-1.4$ near the GC ($-1.45$ when excluding the region occupied by the FBs).
Only a few small-scale features are present.
Among them are the region around extended objects already identified as
deviating strongly from the dust template in Fig.~\ref{fig:m2-dust-modification-field-low-e}.
For those objects, a deviation from the spectral indices of the surrounding regions is expected,
as they represent contamination in this map.
More interestingly, the small-scale flat spectrum structures in the Galactic ridge already observed in
Fig.~\ref{fig:m1-diffuse-spectral-index-map}, can now be seen with more clarity.
They have no counterpart in the dust modification field (see Fig.~\ref{fig:m2-dust-modification-field-low-e}),
so we believe these structures to instead originate from a change in the local CR spectrum,
making them indicative of hard spectrum CR injections in these locations.

The right panel of Fig.~\ref{fig:m2-diffuse-spectral-index-maps} displays the
spectral index map for the template-free diffuse component $I^\mathrm{\:nd}$.
It shows no small-scale structures, but has notable large-scale features.
First, the spectral imprint of the FB emission is visible in blue, with spectral indices
ranging from $-1.375$ to the highest observed value of $-1.26$.
The bubbles are surrounded by a steep spectrum region symmetric around the GC,
already observed in the spatio-spectral maps.
It has spectral indices ranging from $-1.4$ down to the lowest observed value of $-1.54$.
Additionally, in the Galactic plane but far from the GC
lie areas of mildly steeper than average spectral indices, which roughly trace the dense ISM,
but with a much smoother morphology.
The remaining regions show slightly flatter than average spectra with spectral
indices ranging from $-1.4$ to $-1.35$.

The properties of the two recovered diffuse emission components can also be studied by their
CPSs as displayed in Fig.~\ref{fig:res-power-spectra}.
The empirical APS of the template-informed component $I^\mathrm{\:dust}$
qualitatively follows the empirical APS of the M1 diffuse component.
This is expected, as the M1 reconstruction is dominated by the dust-correlated emissions
now taken up by $I^\mathrm{\:dust}$.
However, it has a slightly steeper (posterior mean) power spectrum index of $-2.63$.
The empirical APS of the M2 template-free diffuse component $I^\mathrm{\:nd}$
also shows a zigzag pattern induced by the bright Galactic disk,
but this vanishes for angular modes above $|\ell| > 10$.
Before this threshold, the $I^\mathrm{\:nd}$ APS shows a steeper slope than the $I^\mathrm{\:dust}$ APS,
while above it, they equalize in slope.
The (posterior mean) APS index of $I^\mathrm{\:nd}$ is found to be $-2.45$.
The APS of the dust modification field $\tau^\mathrm{\:dust}$ shows an overall
flatter spectrum, with an empirical APS index of $-1.76$.
This corresponds to a high degree of small-scale corrections to the emission template.
Finally, the EPS indices for both components are very similar to each other
($-1.77$ for $I^\mathrm{\:dust}$ and $-1.70$ for $I^\mathrm{\:nd}$)
and slightly lower than the value found for the M1 diffuse component ($-1.65$).
The dust modification field shows a very flat EPS, with an empirical EPS index of $-0.12$.
It has a peak at the harmonic log-energy scale $|q| = 0.1$,
pointing to a characteristic modulation scale of the $I^\mathrm{\:dust}$ energy spectra.

For an analysis of the PS results,
we refer to the analysis done for M1 in Sect.~\ref{sec:results-m1},
as the results do not differ qualitatively.
Figure~\ref{fig:m2-ps-source-count-distribution} shows the PS pixel brightness count distribution for M2.
Similarly, the residual maps and histograms of the M2 reconstruction do not show qualitative differences
to those of the M1 reconstruction.
We therefore do not discuss them here, but include them in Appendix~\ref{sec:appendix-plots}.
The M2 residuals have (to the stated precision) identical $\chi^2$ statistics as the M1 residuals.

% PS spectra of M2
% Compare stddev with M1 PS spectra. Larger?

\subsection{Comparison of the imaging results} \label{sec:results-comparison}

In this section we compare the results of imaging the gamma-ray sky
with the two presented models and evaluate them for consistency.
%

% ---- fig: energy spectra sky integrated

% ---- fig: m1 vs m2
\begin{figure*}
\centering
\noindent
\begin{minipage}[t]{0.33\textwidth}%
\includegraphics[width=1\textwidth]{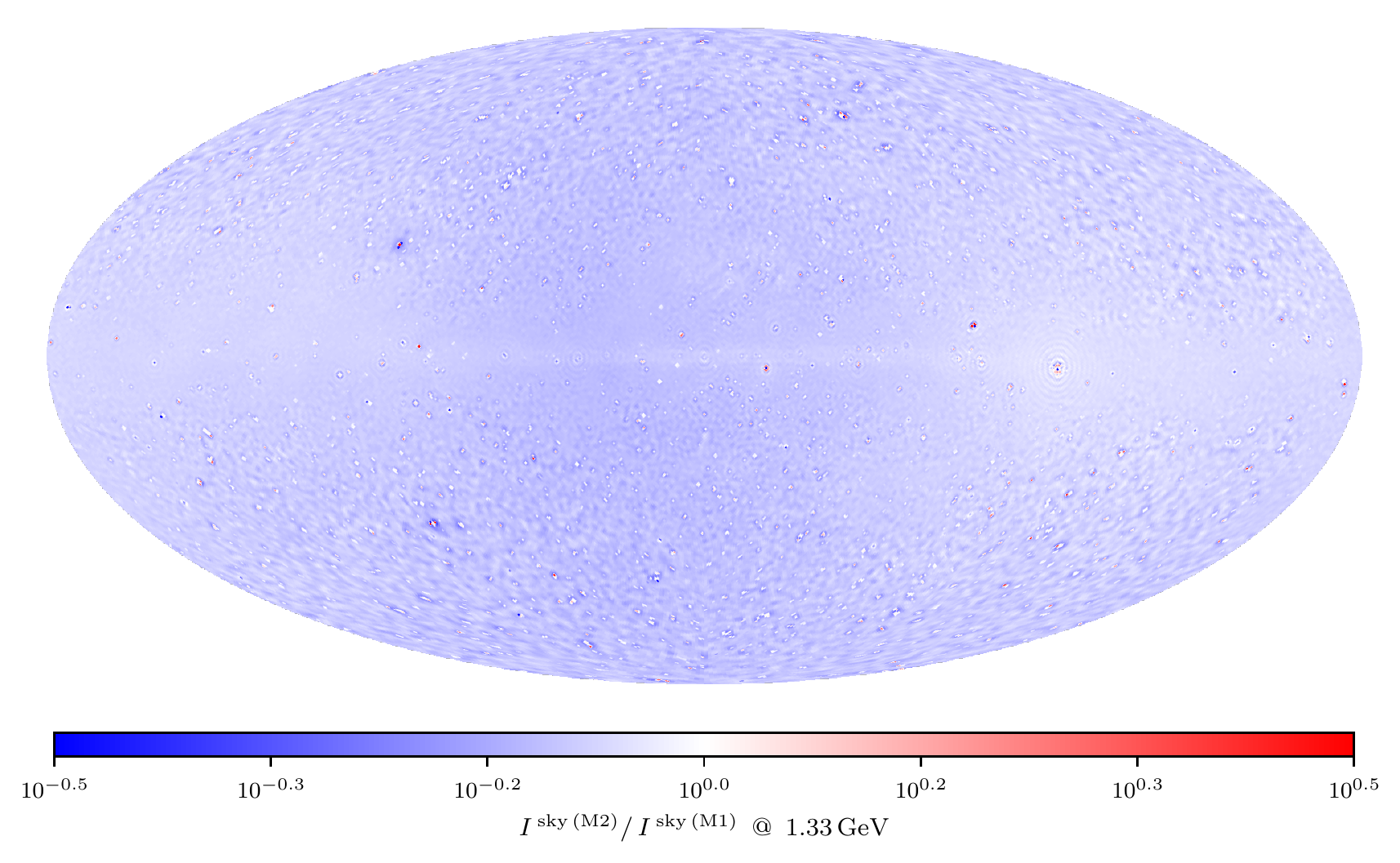}%
\end{minipage}%
\begin{minipage}[t]{0.33\textwidth}%
\includegraphics[width=1\textwidth]{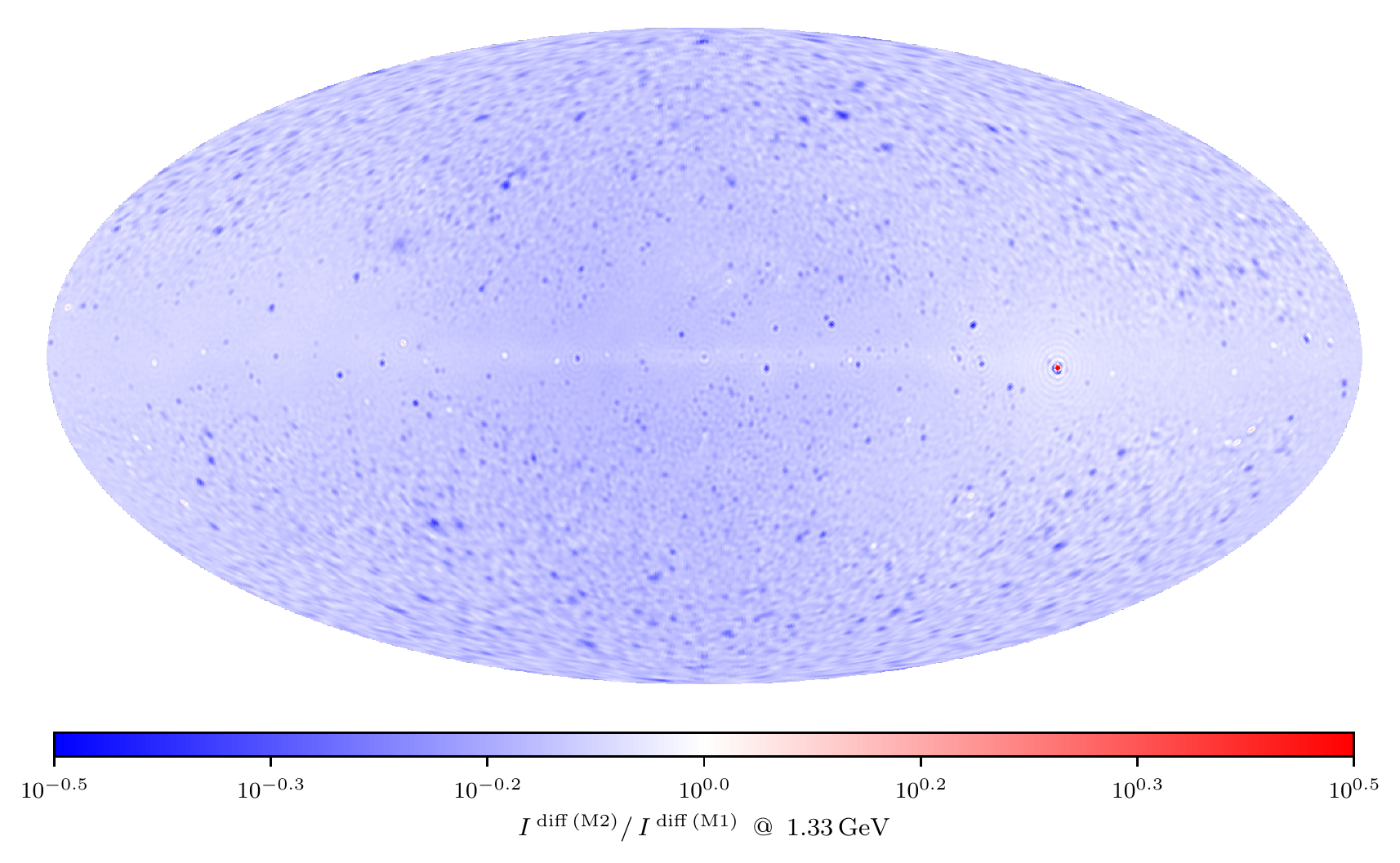}%
\end{minipage}%
\begin{minipage}[t]{0.33\textwidth}%
\includegraphics[width=1\textwidth]{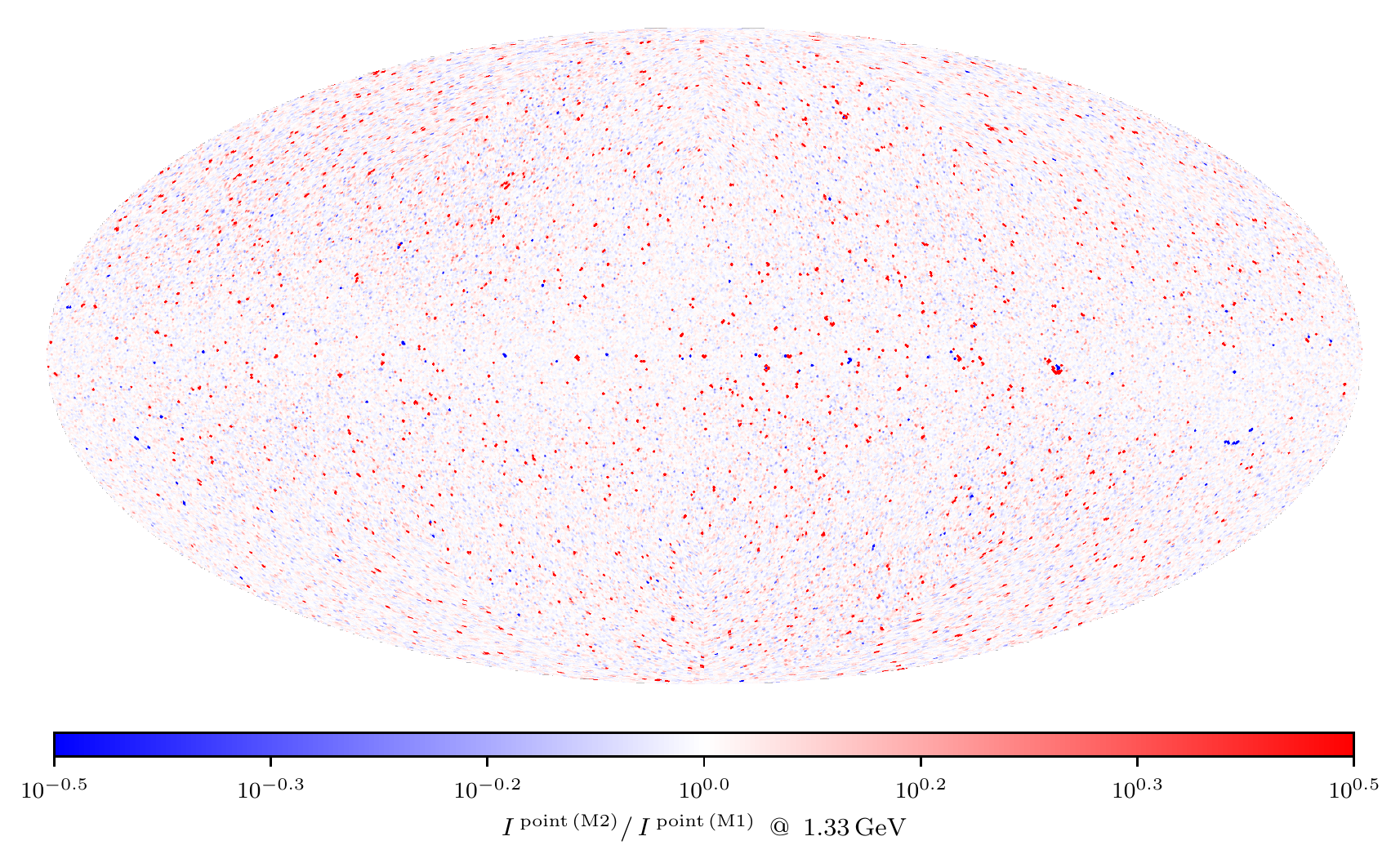}%
\end{minipage}
\begin{minipage}[t]{0.33\textwidth}%
\includegraphics[width=1\textwidth]{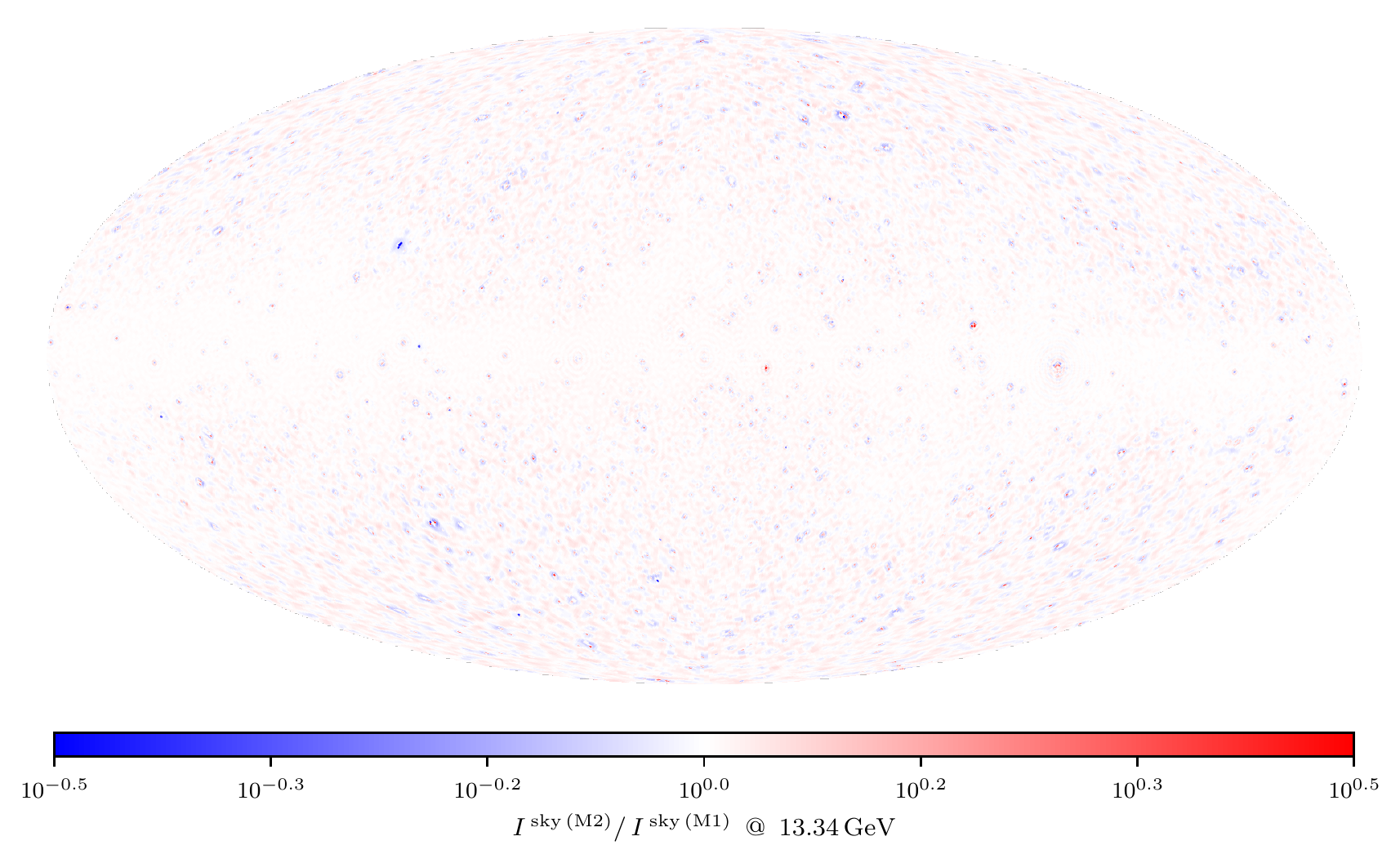}%
\end{minipage}%
\begin{minipage}[t]{0.33\textwidth}%
\includegraphics[width=1\textwidth]{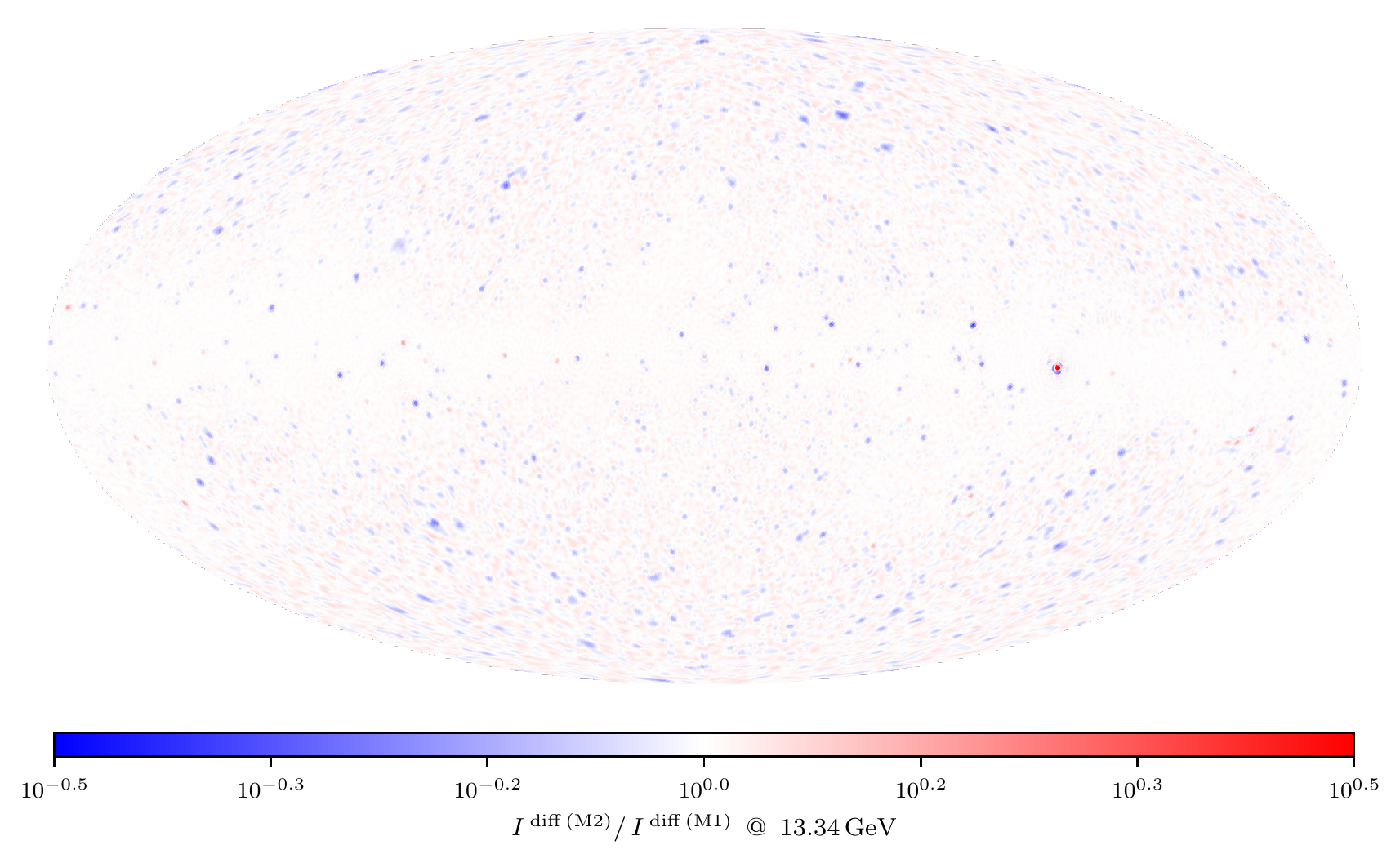}%
\end{minipage}%
\begin{minipage}[t]{0.33\textwidth}%
\includegraphics[width=1\textwidth]{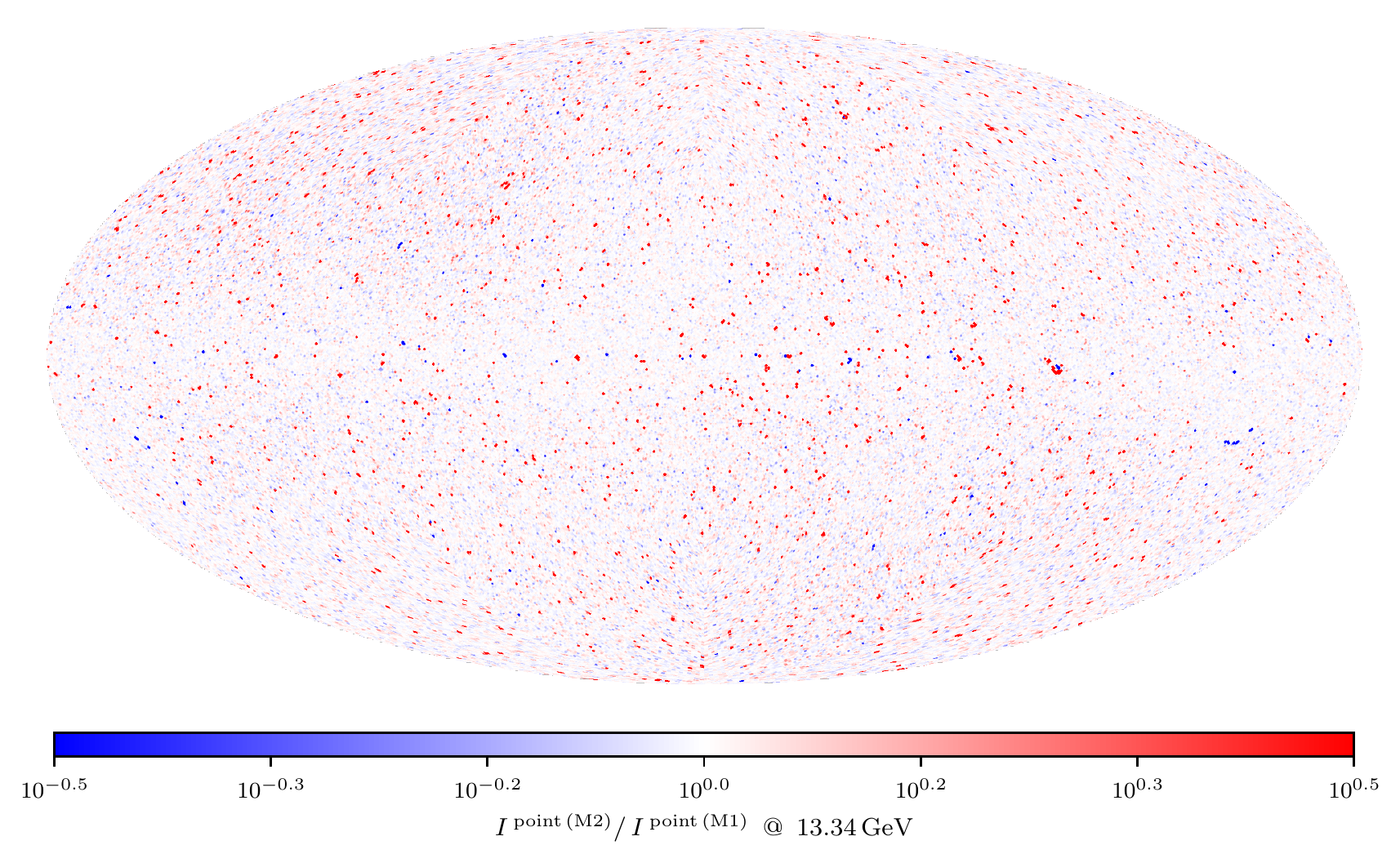}%
\end{minipage}
\begin{minipage}[t]{0.33\textwidth}%
\includegraphics[width=1\textwidth]{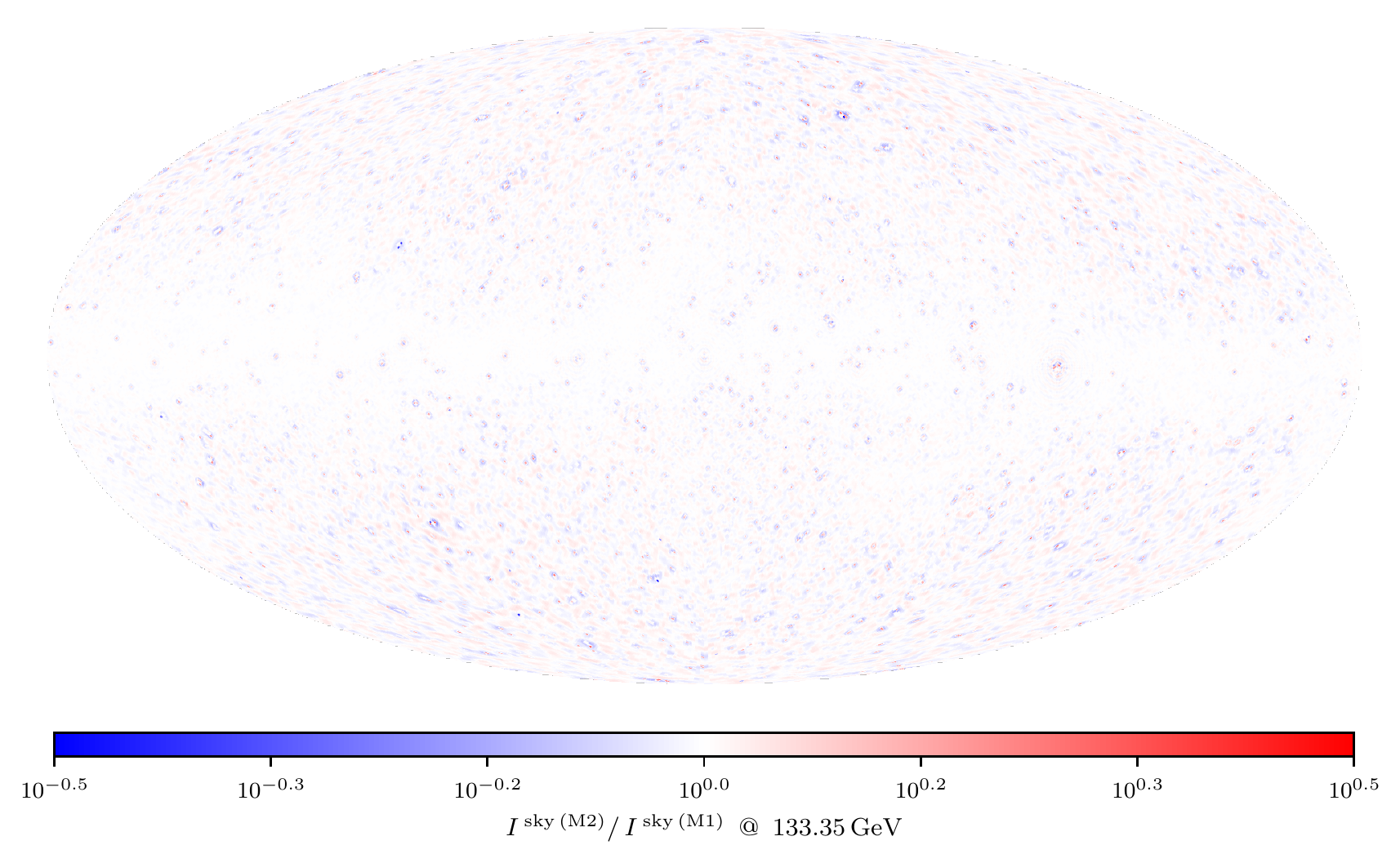}%
\end{minipage}%
\begin{minipage}[t]{0.33\textwidth}%
\includegraphics[width=1\textwidth]{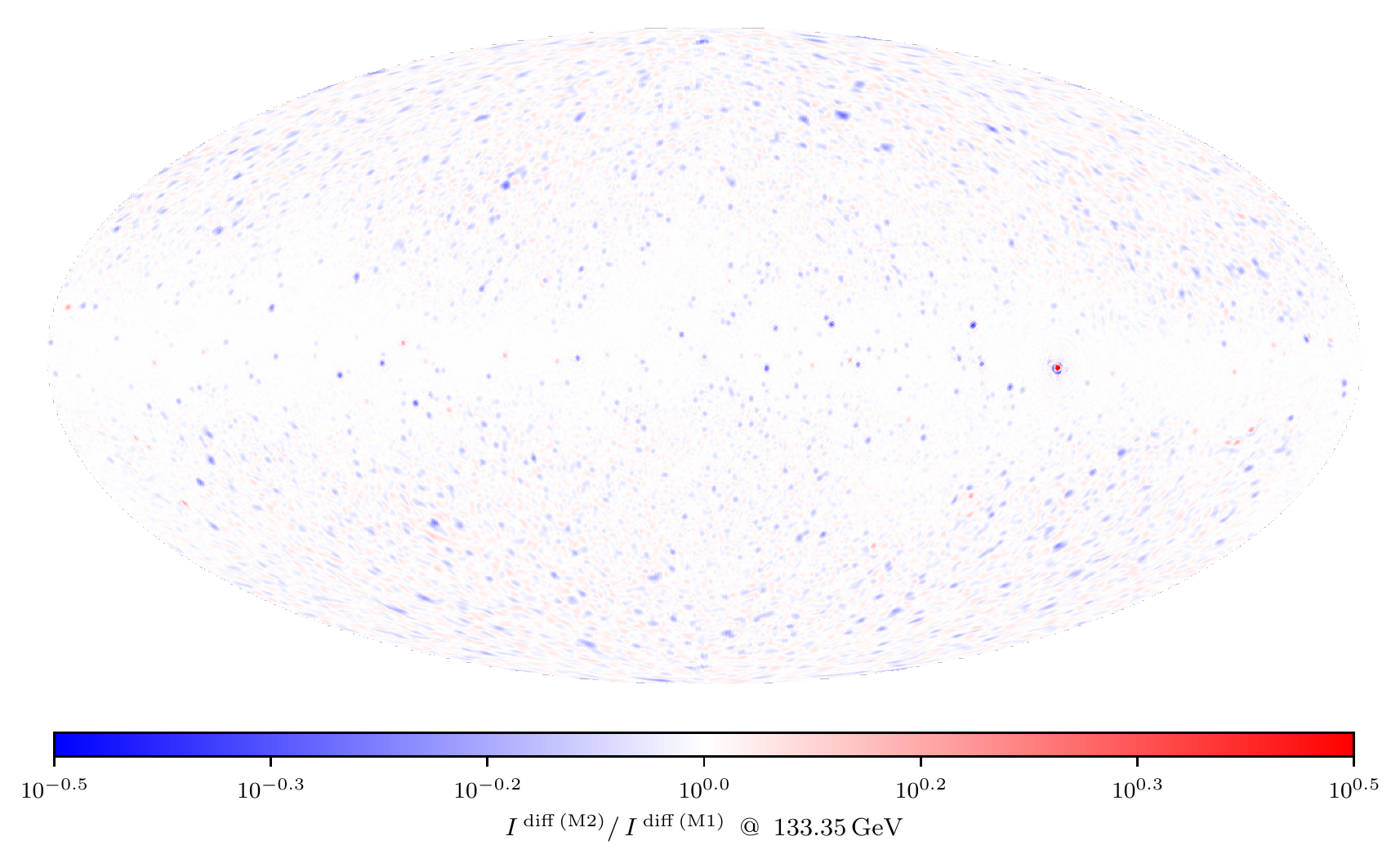}%
\end{minipage}%
\begin{minipage}[t]{0.33\textwidth}%
\includegraphics[width=1\textwidth]{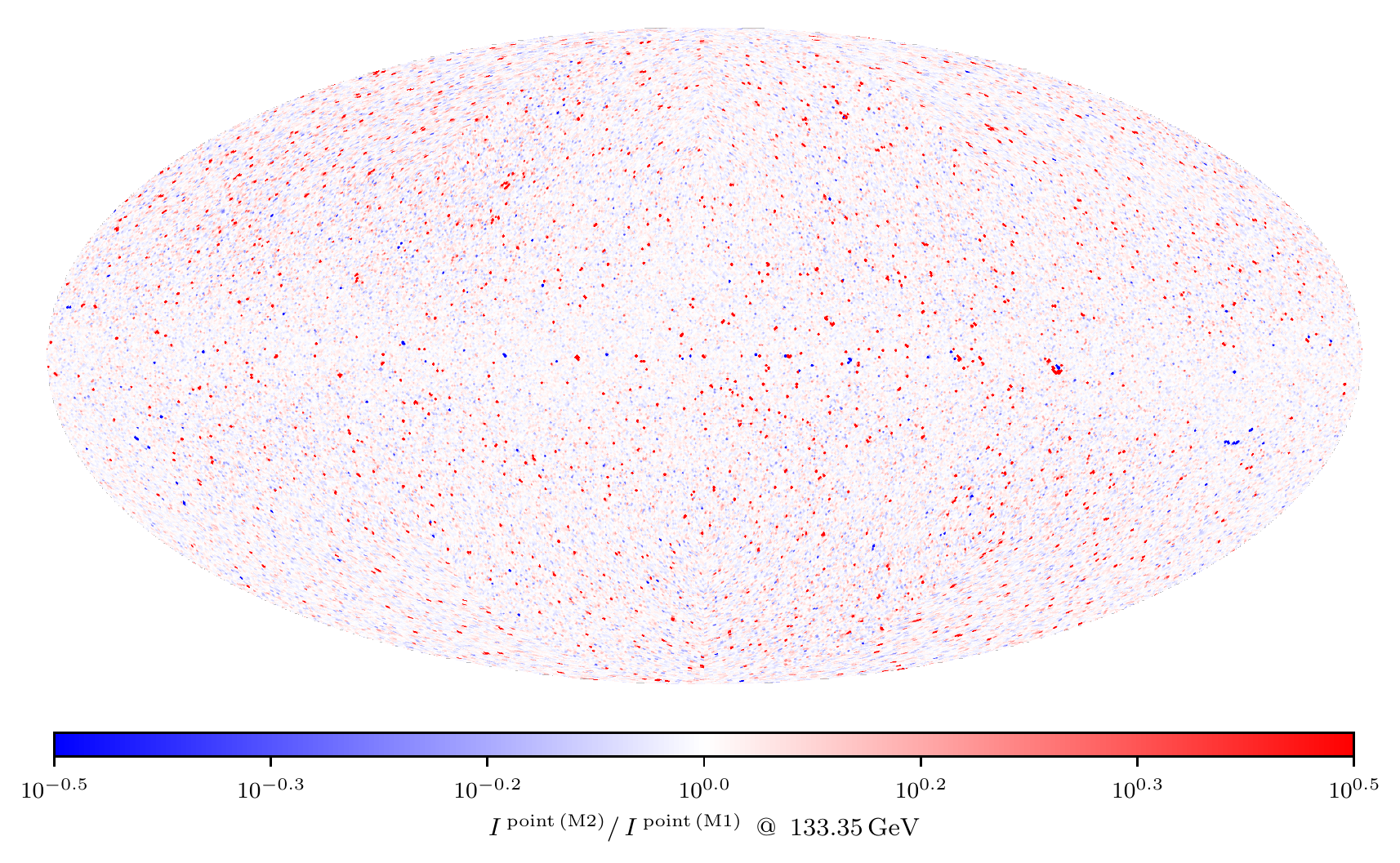}%
\end{minipage}
\caption{Ratio maps of the reconstructions based on M2 and M1.
\textbf{Left}: Ratios of total sky flux values, $I^\mathrm{\:(M2)}/\,I^\mathrm{\:(M1)}$.
\textbf{Middle}: Ratios of diffuse flux values, $I^\mathrm{\:diffuse\:sum\,(M2)}/\,I^\mathrm{\:diffuse\,(M1)}$.
\textbf{Right}: Ratios of PS flux values, $I^\mathrm{\:point\,(M2)}/\,I^\mathrm{\:point\,(M1)}$.
Energy bins: \unit[1.00--1.77]{GeV} (\textbf{top}), \unit[10.0--17.8]{GeV} (\textbf{middle}), and \unit[100--178]{GeV} (\textbf{bottom}).
The colorbars are logarithmic and shows equal fluxes as white.}
\label{fig:comp-m1-vs-m2-maps}
\end{figure*}

Figure~\ref{fig:comp-m1-vs-m2-maps} shows the ratios of reconstructed total, diffuse and point-like fluxes in a low, medium, and high energy bin.
We observe multiple systematic deviations between the reconstructions:
First, the low energy ratio maps for the total sky and diffuse emission have a bluish tint,
corresponding to a 20\% reduction in isotropic emissions from M1 to M2.
In the higher energy bins shown, this effect is not present.
Second, there appears to be a shift of emission from the diffuse to the point-like emission component from M1 to M2,
marked by corresponding blue spots in the diffuse and red pixels PS maps.
This effect is visible in all energy bins shown.
The only counter-example is in the location of Vela, where M2 found more PS flux than M1.
Third, the PS maps show dim PSs found in one but not both reconstructions,
visible as a background of slightly blue and red pixels.

% ---- fig: ps m1 vs m2 scatterplot
\begin{figure}
\resizebox{\hsize}{!}{\includegraphics{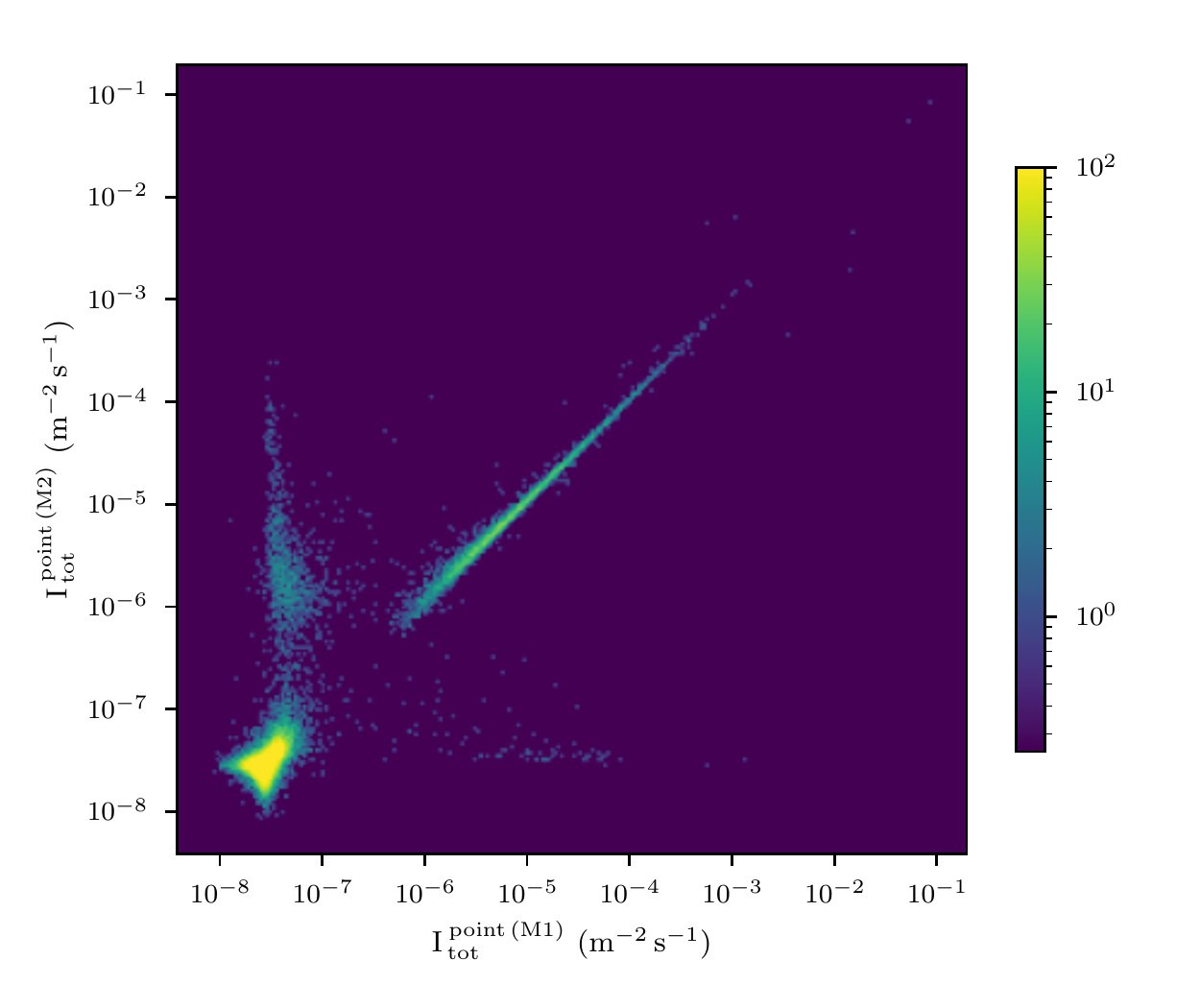}}
\caption{2D histogram of M1 and M2 PS pixel brightness values $I^\mathrm{\:point}_\mathrm{tot}$.
Both brightness values are binned on identical logarithmic scales.
The histogram counts are shown with a logarithmic color scale.
We note that emission from extended objects might be distributed into multiple PS pixels as we do not model them explicitly.}
\label{fig:comp-m1-vs-m2-ps-brightness-scatterplot}
\end{figure}

To analyze the repeatability of the PS detection further,
Fig.~\ref{fig:comp-m1-vs-m2-ps-brightness-scatterplot}
shows a scatter plot of the PS pixel brightness values found with the two models.
This confirms that some PSs are switched-on in one reconstruction but remain switched-off in the other,
where they sit at the high-brightness end of the deactivated source population.
The figure shows very good agreement of PS brightness values for all switched-on source pixels
over five orders of magnitude,
with strong deviations only in a few very bright sources, which we discuss in the following section.
The PSs additionally activated in M2 with respect to M1
contribute between ${10}^{-6}$ and \unit[${10}^{-4}$]{$\mathrm{m}^{-2}\,\mathrm{s}^{-1}$},
changing their brightness by a factor of ${10}^{1.5}$ to ${10}^{3.5}$ in the process.

% ---- fig: m1 and m2 vs fermi template
\begin{figure*}
\centering
\noindent
\begin{minipage}[t]{0.32\textwidth}%
\includegraphics[width=1\textwidth]{graphics_v2/v2_wo_template_v01_u03_FB-comp-with-fermi-diffuse-template-0.78-GeV-ratio_map-51.pdf_300dpi.pdf}%
\end{minipage}\ %
\begin{minipage}[t]{0.32\textwidth}%
\includegraphics[width=1\textwidth]{graphics_v2/v2_wo_template_v01_u03_FB-comp-with-fermi-diffuse-template-13.89-GeV-ratio_map-51.pdf_300dpi.pdf}%
\end{minipage}\ %
\begin{minipage}[t]{0.32\textwidth}%
\includegraphics[width=1\textwidth]{graphics_v2/v2_wo_template_v01_u03_FB-comp-with-fermi-diffuse-template-247.03-GeV-ratio_map-51.pdf_300dpi.pdf}%
\end{minipage}
\begin{minipage}[t]{0.32\textwidth}%
\includegraphics[width=1\textwidth]{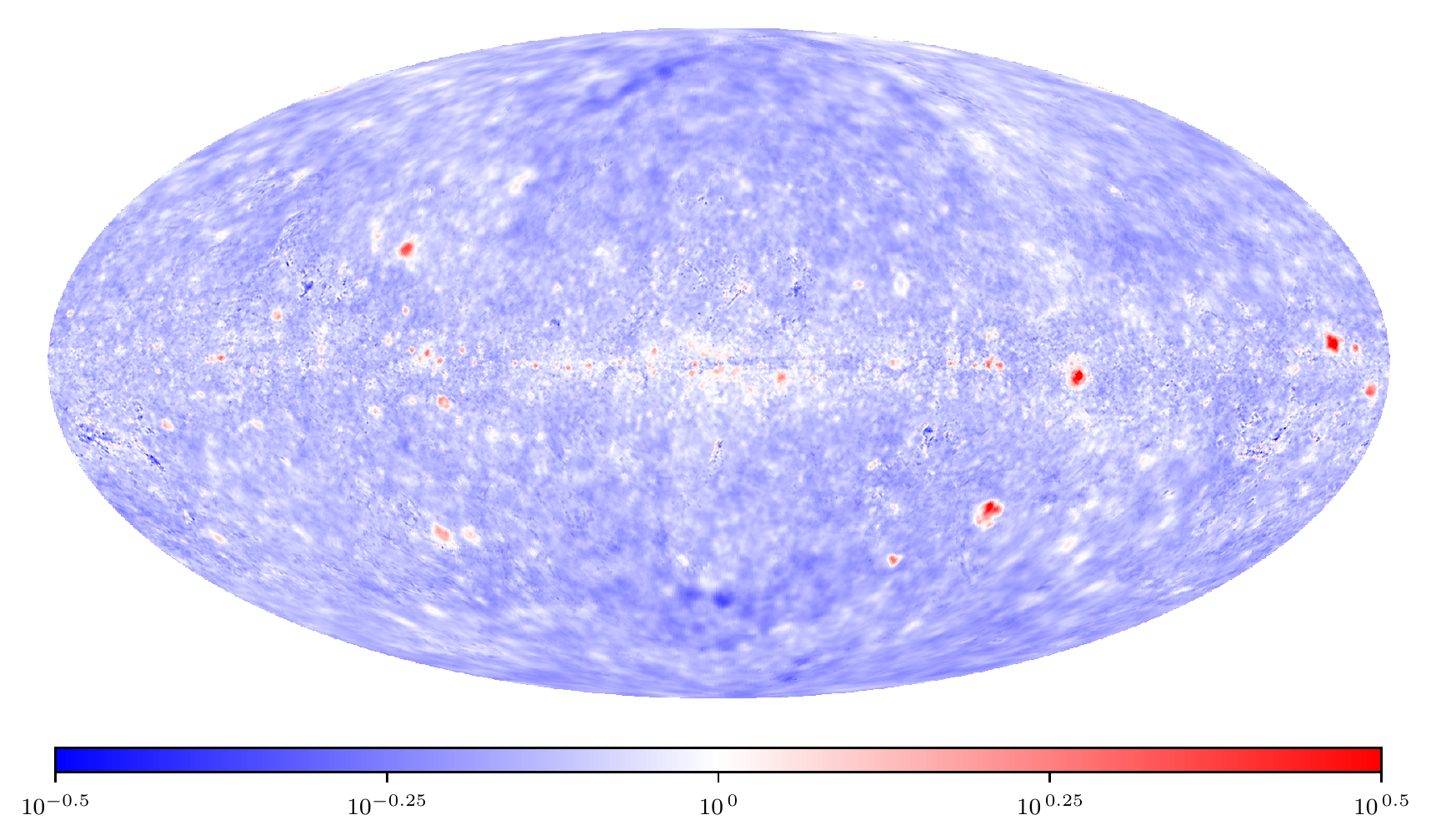}%
\end{minipage}\ %
\begin{minipage}[t]{0.32\textwidth}%
\includegraphics[width=1\textwidth]{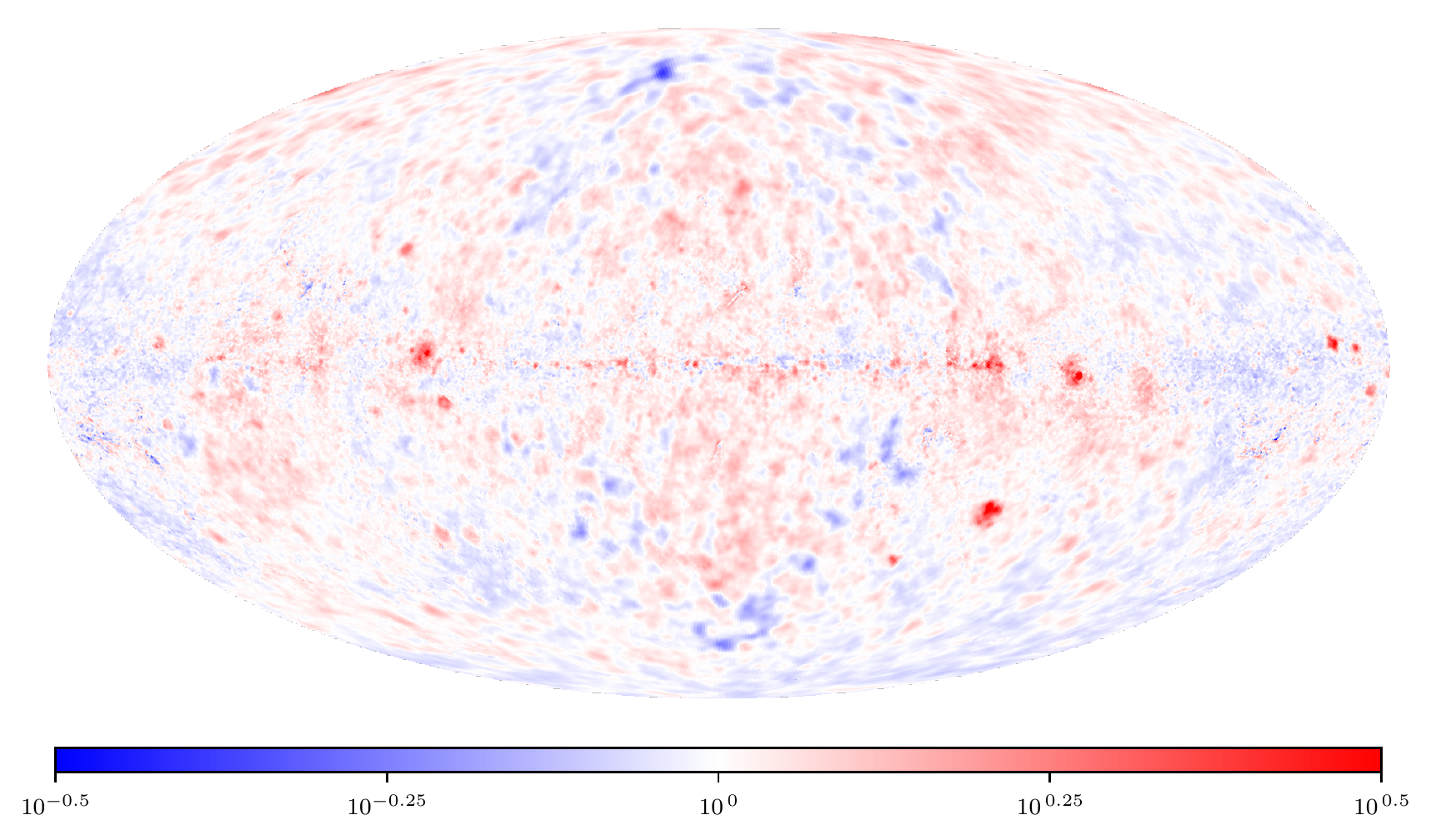}%
\end{minipage}\ %
\begin{minipage}[t]{0.32\textwidth}%
\includegraphics[width=1\textwidth]{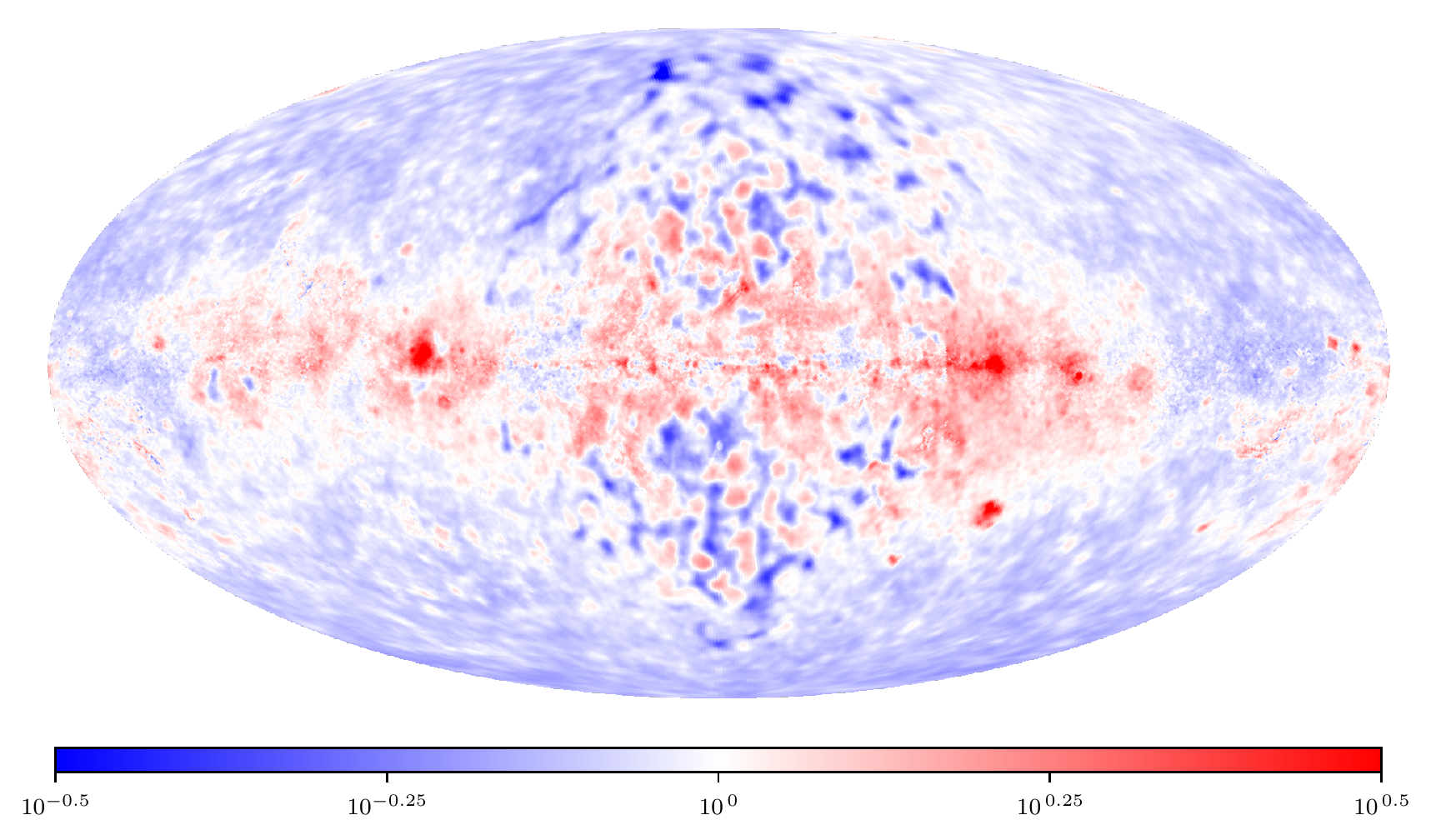}%
\end{minipage}
\caption{Ratio maps of our diffuse emission reconstructions and the \textit{Fermi} diffuse and isotropic emission templates
for the M1 (\textbf{top row}) and M2 (\textbf{bottom row}) reconstructions.
The maps show our diffuse reconstructions divided by the values predicted by the \textit{Fermi} templates in selected energy bins on a logarithmic color scale.
Numbers larger than $10^0$ indicate we have reconstructed more flux than the template predicts, and vice versa.
Energy bins: \unit[0.56--1.00]{GeV} (\textbf{left}), \unit[10.0--17.8]{GeV} (\textbf{middle}), and \unit[178--316]{GeV} (\textbf{right}).}
\label{fig:comp-m1-m2-vs-fermi}
\end{figure*}

The observed shift of emission from the diffuse to the PS component in the M2 reconstruction
also affects the agreement of this reconstruction with the \textit{Fermi} diffuse emission template.
Figure~\ref{fig:comp-m1-m2-vs-fermi} shows flux ratio maps for our diffuse reconstructions
and the \textit{Fermi} template in selected energy bins.
The flux ratio maps of M1 (top row) and M2 (bottom row) show that the reconstructions
deviate from the \textit{Fermi} templates in similar ways,
as expected based on the comparison between the diffuse emission reconstructions above.
However, there are some noteworthy differences:
In the \unit[0.56--1.00]{GeV} energy bin, we observe a flux under-prediction in the M2 map,
corresponding to the uniform reduction in isotropic background found with M2 in this bin.
In the \unit[10.0--17.8]{GeV} energy bin, the M2 flux ratio map shows less small-scale structure than the M1 maps,
related to the shift of flux from the diffuse emissions
to the PS component.
In the \unit[178--316]{GeV} energy bin, the pattern of deviations from the \textit{Fermi} templates
is very similar in the two reconstructions.

% ---- fig: ps m1 vs m2 map
%\begin{figure}
%\resizebox{\hsize}{!}{\includegraphics{graphics_v2/ps-luminosity-comparison-ratio-skyplot-v2_wo_template_v01_u03_FB-step-51-vs-v2_with_template_v01_u02_FB-step-50_300dpi.pdf}}
%\caption{Ratio map of M2 and M1 PS pixel brightness values.}
%\label{fig:comp-m1-m2-ps-brightness-map}
%\end{figure}

% ---- fig: region spectra M1 vs M2-{d, nd}

\section{Discussion} \label{sec:discussion}

\subsection{Quality of fit} \label{sec:disc-qof}
As we claim to demonstrate data-driven imaging, we first want to discuss the
achieved quality of fit. Judging from average reduced $\chi^2$ statistics,
our reconstructions are able to explain the observed data well.

However, we do note clear signs of instrument response mismodeling.
The residuals obtained with both sky models show PSF mismodeling artifacts
(see Fig.~\ref{fig:m1-residual-maps} and~\ref{fig:m2-residual-maps}).
They indicate opposing PSF width errors for \texttt{FRONT}
and \texttt{BACK} events, which we take as evidence against a systematic error
in our PSF implementation as the source of this mismodeling.
Artifacts of the mismodeling can also be found in the thermal dust template
modification map of M2 (Fig.~\ref{fig:m2-dust-modification-field-low-e}),
where we observe an unphysical over-sharpening of the Galactic disk.
Further, we observe evidence of EDF mismodeling (in the low-energy limit)
in the energy spectra of bright PSs (Fig.~\ref{fig:m1-ps-spectra}).
These IRF modeling errors are unfortunate, as they limit the fidelity of the acquired
reconstructions, but they do not invalidate the demonstrated imaging approach.

The residual histograms reveal a slight bias of our models in the
high energy limit through their ramped distribution
(see the bottom-row panels of Fig.~\ref{fig:m1-residual-histograms}).
This is a symptom of information density asymmetry between low, medium,
and high energy data bins, driven by the vastly different photon counts
in the different energy bins (see Fig.~\ref{fig:data-exposure-corrected-spectral-plot})
and our spectral modeling of the sky flux,
which assumes spectral continuity on logarithmic scales.
This leads to information propagation between the energy bins,
a central feature of spatio-spectral reconstructions.
Both discussed limitations should be taken into consideration
when deriving claims from the reconstructions presented.

Nevertheless, we achieved high-quality gamma-ray sky maps,
as shown by the comparison with the \textit{Fermi} diffuse emission template.

\subsection{Agreement with the \textit{Fermi} diffuse emission templates} \label{sec:disc-agreement-with-fermi}
Our analyses show a strong quantitative agreement of our diffuse emission reconstructions
with the diffuse emission templates published by the \textit{Fermi} Collaboration (see the discussions of Figs.~\ref{fig:m1-vs-fermi-template},
~\ref{fig:comp-m1-m2-vs-fermi}, and~\ref{fig:m2-vs-fermi-template}).
The overall high Pearson correlation coefficient between our M1 reconstruction and the templates on log-log scale, $\left<\rho\right>=0.98$,
and the geometric flux ratio mean and standard deviation of $1.01 \pm 0.07$,
together indicate close quantitative agreement of the maps.

In the low and medium energy regime,
the deviations seem to be driven by extended emissions,
which the \textit{Fermi} diffuse emission templates exclude
but which are included in our diffuse emission reconstruction.
In the high energy limit,
where observed photon numbers are low and correspondingly the data becomes less informative,
we find the largest relative differences.
Both our models produced similar deviation patterns from the templates in this limit,
which suggests the existence of systematic errors in the creation of our maps and/or the templates.
Relevant disagreements include the level of isotropic background flux,
two large-scale smooth diffuse emission structures at low Galactic latitudes,
and a pattern of medium-scale ({5\textdegree}) deviations in a large region surrounding the GC.
We defer the analysis of these high-energy deviations to future work.

Overall, we observe a strong agreement of our reconstructions with the templates over many orders of magnitude of photon energy and flux density.
We take this as a validation of our imaging approach and consider it an independent replication of the \textit{Fermi} diffuse emission templates in the energy range of \unit[0.5--100]{GeV}, up to the spatial and spectral resolution achieved in this work.

\subsection{Template-informed imaging}
The reconstructions based on the template-informed and template-free imaging models
presented in this work broadly agree,
indicating that no strong bias was introduced by the template.
The differences found between the reconstruction demonstrate the benefits
of template-informed imaging.
In the M2 reconstructions, additional PSs were identified
that had been absorbed by the diffuse component in the M1 reconstruction,
as evidenced by the comparison of the diffuse reconstructions with the \textit{Fermi} template (see the discussion of Fig.~\ref{fig:comp-m1-m2-vs-fermi}).
This was driven by the inclusion of the template,
which eliminated the need for the diffuse modification fields $\tau^\mathrm{\:dust}$
and $\tau^\mathrm{\:nd}$ to promote structures of this scale via their
angular CPS models.
Still, the template-informed component was able to incorporate the extended
object emission, speaking to its flexibility with respect to the template
and its data-drivenness.

The template-driven imaging produced separate reconstructions of dust-related and additional diffuse emissions,
allowing us to study them individually.
Judging from the log-log best-fit slopes between the obtained diffuse maps and the dust template
(see Fig.~\ref{fig:m2-diffuse-scatterplot-dust}),
and the spectral index maps of the components (see Fig.~\ref{fig:m2-diffuse-spectral-index-maps}),
we find the separation into thermal dust template correlated and independent emissions was successful.

The dust correlated diffuse component shows a spectral hardening toward the GC,
as reported by \citet{gaggero2015hardening}, \citet{acero2016templates}, and \citet{pothast2018hardening}.
The spectral index shift from $-1.7$ to $-1.45$ is consistent with the values reported for the \unit[2--228.65]{GeV} interval studied by \citet{pothast2018hardening},
who find a spectral index hardening for the hadronic emission component from
$-2.7$ at \unit[33]{kpc} distance from the GC to
$-2.45$ at \unit[5]{kpc} distance from the GC.
We want to point out that recently, \citet{vecchiotti2022hardening} hypothesized that the hardening observed in LAT studies may be an artifact of the LAT's detection limits.
Being a data-driven analysis, our work will certainly be susceptible to such data-inherent biases.
If the hypothesis is found to be correct,
updated instrument sensitivity models should allow this effect to be corrected.

We observe multiple sites with a localized hardening of the energy spectra in the inner Galaxy,
both in the M1 diffuse spectral index maps (see Fig.~\ref{fig:m1-diffuse-spectral-index-map}),
but more clearly in the M2 $I^\mathrm{\:dust}$ spectral index map
(see the left panel of Fig.~\ref{fig:m2-diffuse-spectral-index-maps}).
These sites have an extension of a few degrees and
as they have no counterpart in the dust modification map
$\tau^\mathrm{\:dust}$ (see Fig.~\ref{fig:m2-dust-modification-field-low-e}),
we hypothesize that these structures show sites of CR injections into the ISM.
\citet{abdollahi2022crinjection} test 4FGL sources for CR production via neutral pion decay emission.
Their Fig.~6 shows a dense population of CR injecting sources in the region where we observe the spectral index aberrations.
The local spectral hardening could also originate
from PS contamination in the diffuse maps.
However, on visual inspection of Fig.~\ref{fig:m2-sf-plots} and~\ref{fig:m1-sf-plots},
we find no apparent PS contamination in the foci of the hypothesized outflow structures.
Because of this and the extension of the observed structures, we disfavor the PS contamination hypothesis.

The a priori dust-independent diffuse emission map (see Fig.~\ref{fig:m2-sky-maps} and~\ref{fig:m2-sf-plots})
shows a number of interesting features.
We observe a constellation of ``clumpy'' structures surrounding the FBs
and also finer clumpy structure within the FBs.
These may all be caused by PS contamination,
similar to the structures deleted from the diffuse reconstruction in M2, but both models robustly
incorporated these emissions into the diffuse maps.
\citet{balaji2018wavelet} report finding only little small-scale structure in FBs,
suggesting the clumpy structures observed by us
are either contamination or outside the bubbles.
As our focus is on presenting the employed methodology,
we defer the detailed study of the obtained $I^\mathrm{\:nd}$ map to future publications.

Regarding the potential limitations of our template-informed reconstructions, we have two remarks:
First, we note that the template-informed component $I^\mathrm{\:dust}$ reconstructed fluxes to one order of magnitude below the reconstructed isotropic background.
The structures found in this limit have to be considered entirely template-driven,
as even large relative changes to their flux density values would not significantly change the observed total flux in the corresponding pixels and so are not constrained by the likelihood.

Second, given that our template-informed model has a free second diffuse component that could in principle also absorb dust-connected emissions,
we cannot guarantee that the ``templating errors'' of
the thermal dust map with respect to the true dust-associated gamma-ray emission are entirely corrected by the modification field $\tau^\mathrm{\:dust}$, as assumed throughout this work.
However, the extended emission structures included into the template-informed component demonstrated sufficient flexibility of the modification field to adopt even strong deviations from the dust template, supporting the presumption of nearly complete correction.

Perspectively, replacing templates with models of the gamma-ray production throughout the Galaxy
and including data from other modalities currently used for the creation of the templates directly into the reconstruction
holds the promise of producing globally consistent reconstructions of all involved physical quantities.
\citet{hutschenreuter2023disentangling} demonstrate this in a joint reconstruction of
quantities related to the Galactic Faraday sky.
Based on pulsar dispersion measure and distance data, extragalactic
Faraday rotation measure data, and \textit{Planck} free-free and H-$\alpha$
observations, they reconstruct the Galactic dispersion measure,
line-of-sight parallel integrated magnetic field, and the emission
measure-dispersion measure path length
$\mathrm{L}_{\mathrm{DM}^2/\mathrm{EM}}$.
Performing joint reconstructions of observed gamma-ray emissions,
volumetric CR production density, transport, and interaction target densities
in the Galaxy is the subject of ongoing research.

\subsection{Point sources}
In our presentation of reconstruction results, we show a number of PS emission analyses.
For their reception, it should be noted that the PS model employed is reductionist in nature,
not resolving individual PSs, but cumulative PS fluxes from the sky regions spanned by the pixels
and has a fixed pixel brightness distribution prior.
These modeling simplifications were made for the sake of implementational simplicity,
but much more fine-grained PS analyses are
available (see for example the source catalogs procured by the \textit{Fermi} Collaboration).
Also, our model is not able to resolve sub-detection-threshold PS properties, contrary to methods from the NPTF and 1pPDF families.

For the PS brightness distribution, we find a similar shape as the \textit{Fermi} Collaboration (see the discussion of Fig.~\ref{fig:m1-ps-source-count-distribution}).
Between ${10}^{-5}$ and \unit[$5\cdot{10}^{-4}$]{$\mathrm{m}^{-2}\,\mathrm{s}^{-1}$}, the source brightness count distributions of our reconstruction\textit{s} follow power laws
of indices $-2.26$ (M1) and $-2.30$ (M2),
deviating from the prior mean of $-2.5$.
Previous works reported that the gamma-ray PS count distribution
for high latitudes ($|\mathrm{b}| \le 30\deg$) is better
described by a (multiply) broken power law:
\citet{malyshevhogg2011} find a spectral index of $n_\mathrm{1} = 2.31$ below and $n_\mathrm{2}=1.54$ above
$\mathrm{S}_\mathrm{b}=2.9\cdot10^{-5}\:\mathrm{m}^{-2}\:\mathrm{s}^{-1}$
for the energy range of \unit[1.0--300]{GeV},
while \citet{zechlin2016unveiling} find a spectral index of $n_\mathrm{1}=3.1$ below and $n_\mathrm{2}=1.97$ above
$\mathrm{S}_\mathrm{b}=2.1 \cdot 10^{-4}\:\mathrm{m}^{-2}\:\mathrm{s}^{-1}$
for the energy range of \unit[1.0--10]{GeV}.
This suggests including broken power-law parametric source brightness distribution models in future analyses
to elucidate to which degree the brightness distributions observed by us are data- and prior-driven.
However, given that the inferred PS brightness distribution strongly deviates from the prior distribution
(single power law with low brightness cutoff),
we assume it to be mostly data-driven for the activated source pixels.

The analysis of PS consistency between our two models (Fig.~\ref{fig:comp-m1-vs-m2-ps-brightness-scatterplot})
shows that most PSs are reconstructed consistently,
but this breaks down in the high-flux limit.
We believe this is an effect of extended sources, specifically close pulsar wind nebulae,
which produce both PS flux via their central pulsar and extended emission from the jet and surrounding medium.
These objects are not well representable by our emission component models,
making different splits of the flux equally well-fitting.
Because of this, different combinations of PS
and diffuse emission are reconstructed,
depending on the general state of the emission components
(for example, learned angular power spectra).
An explicit modeling of extended emission objects could
alleviate this problem and lead to more stable bright PS reconstructions.
However, for a priori unknown extended structures,
this will still fail.
The automated treatment of extended objects found superimposed on the sky in a Bayesian framework is ongoing research.

\subsection{Caveats for studies of the \unit[1--5]{GeV} energy range}
As laid out in the introduction, the GCE is subject of long-standing interest
in the gamma-ray astronomy community.
Its emission lies mostly in the \unit[1--5]{GeV} band,
making this energy range particularly relevant for down-stream analyses of the maps we provide.

In Sect.~\ref{sec:results-m1}, we discuss that we have reduced data-drivenness
in the \unit[0.56--1.0]{GeV} and \unit[1.0--1.77]{GeV} energy bins,
because the strong EDF of the LAT at energies below \unit[1]{GeV} prevents us
from using the data bin corresponding to the lowest reconstructed energy
in our likelihood.
This makes the reconstructions in the lowest two energy bins especially vulnerable to IRF mismodeling,
and warrants a discussion of the dependability of the reconstructions in the corresponding energy range.

For the PS spectra reconstructed based on M1 (Fig.~\ref{fig:m1-ps-spectra}),
we observe spectral artifacts for bright PSs
in the lowest two energy bins,
leading us to recommend excluding them from analyses.
For the M1 diffuse emission reconstruction,
comparison with the \textit{Fermi} diffuse emission templates
shows good agreement in these low energy bins (Fig.~\ref{fig:m1-vs-fermi-template}),
suggesting they are not affected as strongly by spectral artifacts.
For the M2 diffuse emission reconstruction, a difference in
isotropic background flux with respect to the \textit{Fermi} templates (see Fig.~\ref{fig:m2-vs-fermi-template})
and the M1 diffuse reconstruction (see Fig.~\ref{fig:comp-m1-vs-m2-maps})
is present in the lowest energy bin,
which could be caused by IRF mismodeling in conjunction with the described ``under-informedness.''
We therefore recommend caution when analyzing the two lowest energy bins of the M2 diffuse reconstructions.

\subsection{Cosmic ray background contamination}

Our reconstructions are based on the \texttt{\detokenize{P8R3_SOURCE}} event class,
while analyses of the diffuse emissions usually use more restrictive event classes
(\texttt{\detokenize{P8R3_ULTRACLEAN}} and \texttt{\detokenize{P8R3_ULTRACLEANVETO}}).
This means the data set we use has a higher level of CR contamination
than in comparable analyses,
which might bias our reconstruction to find more isotropic gamma-ray background than truly exists.
However, the higher photon count in the \texttt{\detokenize{P8R3_SOURCE}} event class compared to the
more commonly used restrictive event classes also makes our
analysis more sensitive to faint emission sources, which we optimized for in selecting the event class.
Different trade-offs between faint emission sensitivity and CR background susceptibility can be explored
by repeating the reconstructions with the other event classes and cuts provided by the \textit{Fermi} Collaboration.

\subsection{Potential improvements}

As already mentioned above, the presented imaging models can be improved in a number of ways.

First, we demonstrate the template-informed imaging using only one template for simplicity.
However, generally, the use of additional emission templates leads to improvements in reconstruction fidelity
and enables further separation of the emission components,
as demonstrated by the full body of work on template-based analyses.
To facilitate searches for DMA or decay,
corresponding emission models can be added.

Second, improvements to the modeling of PS
and extended emissions would lead to higher fidelity reconstructions of all emission components.
For the former, techniques such as iterative charted refinement \citep[ICR;][]{edenhofer2022icr} hold promise.
Iterative charted refinement can facilitate the construction of $\tau^\mathrm{\:c}$ fields with locally increased resolution
and allows local deviations in the correlation structure,
which are necessary for a correct modeling of extended object emission.

Third, we did not include a separate component for the isotropic diffuse background, which forced our diffuse emission component to absorb it.
In the template-informed imaging run, this limited the dynamic range of the template-free component $I^\mathrm{\:nd}$.
Including a separate isotropic emission component (low parameter count model)
would allow the non-isotropic diffuse emissions components
to learn more specific correlation structures for their emissions.

Fourth, improvements to the accuracy of the instrument response model would increase the fidelity of the obtained results, as is true for all similar analysis methods.

Fifth, the studied energy range, as well as the spatial and spectral resolution
are currently limited by computational constraints,
specifically, the computational efficiency and scalability of the numerical PSF operator implementation.
Improvements in this regard would enable higher resolution reconstructions
and make an expansion of the reconstructed energy range to lower energy bins feasible,
where the PSF becomes increasingly wide.
This would allow a more robust study of the \unit[1--5]{GeV} energy range and the GCE than the current reconstruction affords.

Further potential for improvement lies in the inference methodology itself.
\citet{frank2021geovi} recently published Geometric Variational Inference, a generalization of MGVI,
which improves the fidelity of the estimated posterior samples with respect to MGVI for non-Gaussian posterior distributions.
At the cost of additional computational resources, such methods can increase the fidelity of the reconstructed sky maps.
We leave the realization of this to future work.

\section{Conclusions} \label{sec:conclusions}

We present a template-independent spatio-spectral imaging approach for the gamma-ray sky,
based on a Bayesian framework.
We demonstrate its capabilities on data produced by the \textit{Fermi} LAT,
showing a template-free and a template-informed reconstruction of the gamma-ray sky.
In the latter case, we include the \textit{Planck} \unit[545]{GHz} thermal dust emission map
as a tracer of sites of interaction between CR protons and ISM protons in our sky model.
The reconstructions are highly data-driven,
giving them high sensitivity to unexpected emissions by construction.
This is achieved through the use of Gaussian processes for modeling the
emission components constituting the gamma-ray sky,
and in the case of the template-informed component,
for modeling deviations of the respective emission component from the template.
We used an instrument response model based on the in-flight-calibrated IRFs
published by the \textit{Fermi} Collaboration
and fully modeled the Poissonian statistics of the emissions.
With our instrument and sky models,
we achieve an overall good quality of fit,
but also observe evidence of instrument response mismodeling.

The imaging approach we present extends existing concepts for template-based imaging
to the fully data-driven limit
while still relying on classical hierarchical generative models for the inference.
This way, we retain the direct interpretability of our model parameters,
in contrast to DL-based methods.
Based on hierarchical generative models,
characteristic spatial and spectral correlation scales of the individual emission components
are detected automatically and used to image them more accurately.
The work lays out a recipe for more complex sky models based on the presented methods,
allowing the inclusion of further templates, emission components,
and more complex modeling of the existing subcomponents.

We have presented and analyzed the results of our reconstructions,
which include individual spatio-spectral maps for all assumed emission components,
summary statistics that are part of their generative models
(such as spatial and spectral power spectra),
and an uncertainty quantification of any resulting quantities in terms of self-consistent posterior samples.
Comparing our diffuse reconstructions with the sum of the \textit{Fermi} diffuse emission templates \texttt{gll\_iem\_v07} and \texttt{iso\_P8R3\_SOURCE\_V3\_v1},
we find strong quantitative agreement, with pixel-wise flux ratios in the range $0.5$--$2$,
a geometric mean flux ratio of $1.01$,
and an average Pearson correlation between the two maps on log-log scale of $\rho=0.98$.
Comparing our diffuse reconstructions with the \textit{Planck} \unit[545]{GHz} thermal dust emission map,
we find a strong linear scaling on log-log scale for the dust-informed emission component ($\alpha=0.94$)
and a weak linear scaling on log-log scale for the second diffuse component of the template-informed reconstruction ($\alpha=0.28$),
indicating a successful flux separation in the template-informed model.
We show and discuss spectral index maps of reconstructed emission components,
where we find a spectral hardening of hadronic emissions
from -1.7 in the Galactic anticenter to -1.4 in the GC,
consistent with literature values\footnote{Energy spectrum indices are given for energy-bin-integrated fluxes $I_\mathrm{ij}$ throughout this work.
To convert the stated numbers to spectral indices of differential flux $\Phi(x,E)$, subtract $\mathtt{1}$
(see Sect.~\ref{sec:methods-general-signal-def}).}.
We analyzed the retrieved pixel-wise PS fluxes
and find a population of activated PS flux pixels between ${10}^{-6}$ and \unit[${10}^{-3}$]{$\mathrm{m}^{-2}\mathrm{s}^{-1}$},
with luminosity function power-law indices of $-2.26$ and $-2.30$ between
${10}^{-5}$ and \unit[$5\cdot{10}^{-5}$]{$\mathrm{m}^{-2}\mathrm{s}^{-1}$} for our two reconstructions, respectively.
For these PS fluxes, we find energy spectral indices\footnotemark[10]
in the range $\left[-2.15, -0.35\right]$.
For the diffuse emission components, we find empirical angular power spectra with spectral indices between $-2.63$ and $-2.48$
and empirical EPSs with spectral indices between $-1.77$ and $-1.65$.

We believe the presented imaging approach is a valuable addition to existing analysis methods
and will help such methods reach their full potential.
% -------

\begin{acknowledgements}
Jakob~Knollm{\"u}ller acknowledges support by the \emph{Excellence Cluster ORIGINS}
which is funded by the \emph{Deutsche Forschungsgemeinschaft} (DFG, German Research Foundation)
under Germany{\textquoteright}s Excellence Strategy – EXC-2094-390783311.
Philipp~Arras acknowledges the financial support by the
\emph{German Federal Ministry of Education and Research (BMBF)} under grant
05A17PB1 (Verbundprojekt D-MeerKAT).
Philipp~Frank acknowledges funding through the \emph{German Federal Ministry of Education and Research (BMBF)}
for the project ErUM-IFT: Informationsfeldtheorie für Experimente an Großforschungsanlagen (Förderkennzeichen: 05D23EO1).
We thank the referee for their helpful and detailed feedback on our manuscript.
\end{acknowledgements}

\bibliographystyle{bibtex/aa}
\bibliography{ref_v2}

%\clearpage{}

\begin{appendix}

\section{Generative models of the $\gamma$-ray sky} \label{sec:appendix-generative-models}

\subsection{General remarks}

For Bayesian inference, it is convenient to express the sky components
in terms of generative models. A generative model is a description of a stochastic process that is able to generate random realizations following a desired target statistic.
Following \citet{knollmueller2018encoding},
we constructed our generative models to take
a vector of uncorrelated random variables $\xi=(\xi_{1},\xi_{2},\ldots)$,
here assumed to be standard normal distributed, with 
\begin{equation}
\mathcal{P}(\xi)=\prod_{j}\mathcal{\mathcal{{N}}}(\xi_{j}\vert0,1)=\prod_{j}\left[\exp(-\xi_{j}^{2}/2)/\sqrt{2\pi}\right],
\end{equation}
and to apply a transformation $T$ to it to turn $\xi$
into a potential signal vector $s=T(\xi)$
including all quantities of interest that are reconstructed from the observations.
The transformation $T$
has to be chosen such that the required signal statistics $\mathcal{P}(s)$
result from the transformed Gaussian statistics $\mathcal{P}(\xi)$
via
\begin{equation}
\left.\mathcal{P}(s)\right|_{s=T(\xi)}=\mathcal{P}(\xi)\left\Vert \frac{\partial T(\xi)}{\partial\xi}\right\Vert ^{-1}.
\end{equation}
Such a transformation exists for any hierarchical probabilistic model
over continuous quantities and can be accessed via inverse transform
sampling \citep{knollmueller2018encoding}.

The latent parameters $\xi$ are our effective parameterization
of the model and are to be learned from the data. The physical signal
$s$ corresponding to a recovered $\xi$ can always be obtained via $s=T(\xi)$.

For the models presented here, we make heavy use of the generative
correlated field model introduced by \citet[][Sect.~3.4]{arras2021comparison}
and \citet[][Methods]{arras_variable_2022},
summarized in \citet[][Appendix B]{frank2021geovi}.
It allows us to translate prior knowledge on the spatial and spectral
correlation functions of modeled quantities into appropriate
transformation functions
$T$.

To this end, we represent the spatial and spectral correlation functions
of the modification fields $\tau^\mathrm{\:c}$
via power spectra in the corresponding harmonic spaces,
the SH space with coordinates $k=(\ell,m)$ for the spatial dimensions and the 1D Fourier space with coordinate $q$ for the log-energy dimension.
The harmonic representation of $\tau^\mathrm{\:c}$ in this space is defined via
\begin{equation}
\tau^\mathrm{\:c}(x,y)=\sum_{\ell m}Y_{\ell m}(x)\,\int\frac{\mathrm{d}q}{2\pi}\,\mathrm{e}^{-iqy}\,\hat{\tau}^\mathrm{\:c}(k,q),
\end{equation}
where $Y$ denotes the SH functions.
Thanks to the assumed separability of the correlation functions in spatial
and spectral directions, the corresponding power spectra are separable
as well. Thus, for components c and c', we have
\begin{equation}
\langle\hat{\tau}^\mathrm{\:c}(k,q)\;\hat{\tau}^\mathrm{\:*\,c'}(k',q')\rangle_{(\tau)}=2\pi\,\delta_{cc'}\delta_{\ell\ell'}\delta_{mm'}\delta(q-q')\,\hat{C}^\mathrm{\:c}(\ell)\,\hat{D}^\mathrm{\:c}(q).
\label{eq:correlation-structure}
\end{equation}
Here, $\hat{C}^\mathrm{\:c}(\ell)$ is the spatial APS of $\tau^\mathrm{\:c}$.
It is assumed to be isotropic and therefore does not depend on
$m$.
$\hat{D}^\mathrm{\:c}(q)$ is the corresponding log-energy Fourier power spectrum of $\tau^\mathrm{\:c}$.
$\hat{\tau}^\mathrm{\:*}$ denotes the complex conjugate of $\hat{\tau}$.
The $2\pi$ in Eq.~\ref{eq:correlation-structure} results from
the adopted Fourier convention for the log-energy space,
\begin{eqnarray}
f(q) & = & \int \mathrm{d}y\,\exp(iqy)\,f(y)\\
f(y) & = & \frac{1}{2\pi} \int \mathrm{d}q\,\exp(-iqy)\,f(q).
\end{eqnarray}

The correlated field model generates harmonic space samples of the
field $\tau^\mathrm{\:c}$ by drawing independent unit Gaussian random numbers
$\xi^\mathrm{\:c}(k,q)$ and multiplying these point-wise with the amplitude
spectra according to 
\begin{equation}
\hat{\tau}^\mathrm{\:c}(k,q)=\sqrt{\hat{C}^\mathrm{\:c}(\ell) \hat{D}^\mathrm{\:c}(q)}\;\xi^\mathrm{\:c}(k,q) =: A^\mathrm{\:c}_\mathrm{\tau}(k,q)\;\xi^\mathrm{\:c}(k,q).
\end{equation}
Realizations of $\tau^\mathrm{\:c}(x,y)$ can then be obtained via the inverse
harmonic transforms to the original $(x,y)$ coordinates as given above.
By construction, these exhibit the requested correlations in spatial and log-energy dimension.

Since a priori we do not know the exact power and corresponding amplitude spectra, the correlated
field model represents them through further generative processes.
By specifying their prior statistics, we encoded our prior knowledge on the spatial and spectral correlation functions.
We show the construction here for the angular power spectra $\hat{C}(\ell)$,
but the $\hat{D}(q)$ are constructed analogously.
The CPSs are modeled similar to the gamma-ray emission models, as modified power laws in the harmonic modes, 
\begin{equation}
\hat{C}(\ell) = {|\ell|}^{\,\beta} \cdot e^{\kappa\,\left(\ln|\ell|\right)}.
\end{equation}
Here, $\beta$ specifies the power-law index and $\kappa$ models the deviations from a pure power law in $|\ell|$ on log-log scale.
We placed a Gaussian prior on the power-law index $\beta$
with prior mean $\mu_\beta$ and prior standard deviation $\sigma_\beta$.
The deviations, $\kappa$, were modeled via an integrated Wiener process (see Eqs.~20 to 24 of \citealt[]{arras2021comparison} for a precise definition).
This allows us to specify separate priors on the deviation amplitude on log-log scale and on the spectral roughness.

In the PS spectrum model $f_\mathrm{spec}$, the energy spectrum fluctuations $\epsilon_\mathrm{ij}$ were built using the same integrated Wiener process formulation as we use for $\kappa$.

\subsection{Prior parameters of the sky models}

\begin{figure*}
\centering
\begin{tikzpicture}[
  x=10.5mm,
  % y=7mm,
  % node distance=30mm,
  % on grid,
  every node/.style={draw, anchor=west, thick},
  arrow/.style={-{Stealth}, thick},
  normalleft/.style={draw, anchor=west, minimum width=49mm, align=center, minimum height=2em, fill=white},
  normalright/.style={draw, anchor=east, minimum width=49mm, align=center, minimum height=2em, fill=white},
  normalcenter/.style={draw, anchor=mid, minimum width=49mm, align=center, minimum height=2em, fill=white},
  bigcenter/.style={draw, anchor=west, minimum width=105mm, align=center, minimum height=5em, fill=white},
  smallleft/.style={draw, anchor=west, minimum width=25mm, align=center, minimum height=2em, fill=white},
  smallright/.style={draw, anchor=east, minimum width=25mm, align=center, minimum height=2em, fill=white},
  smallcenter/.style={draw, anchor=mid, minimum width=25mm, align=center, minimum height=2em, fill=white},
  ]

\node [normalcenter, ultra thick] (sky_1) at (4.25, 10.5) {\textbf{M1:} $\;I = I^\mathrm{\:point} + I^\mathrm{\:diff}$};

\draw[arrow] (2.3,9.3) -- (sky_1); %(4.1,10.15); % I_point_1 -> sky_1
\draw[arrow] (6.3, 9.3) -- (sky_1); %(5.55,10.15); % I_diffuse_1 -> sky_1

\draw[arrow] (1.7,5.70) -- (1.70, 8.65); % I_point_0 -> I_point_1
\draw[arrow] (3.7,8.) -- (3, 8.65); % f_point -> I_point_1

\draw[arrow] (3.775, 6.85) -- (3.775, 7.40); % alpha_point -> f_point

\node [smallleft] (I_point_1) at (0, 9) {$I^\mathrm{\:point}_\mathrm{ij} = I^\mathrm{\:point}_\mathrm{tot,\:j} \cdot \frac{f_\mathrm{spec,\:ij}}{\sum_\mathrm{i} f_\mathrm{spec,\:ij}}$};
\node [smallcenter, fill=white] (f_point) at (3.75, 7.75) {$f_\mathrm{spec,\:ij} = {10}^{\:\alpha^\mathrm{\:point}_\mathrm{j} \cdot y_\mathrm{i} \,+\, \epsilon_\mathrm{ij}}$};
\node [smallcenter, fill=white] (alpha_point) at (3.8, 6.5) {$\alpha^\mathrm{\:point}_\mathrm{j} = \mu_\text{ap} + \sigma_\text{ap} \: \xi_\text{ap}^i$};
\node [smallleft, fill=white] (I_point_1_zero) at (0, 5.25) {$I^\mathrm{\:point}_\mathrm{tot,\,j} \sim \textsc{InvGamma}\left(I^\mathrm{\:point}_\mathrm{tot,\,j}\:|\:\alpha_{\text{I}\Gamma}, \beta_{\text{I}\Gamma}\right)$};
\node [smallright] (I_diffuse_1) at (9, 9) {$I^\mathrm{\:diff}_\mathrm{ij} = I^\mathrm{\:diff}_{0} \cdot \left(\frac{E_i}{E_0}\right)^{\alpha^\mathrm{\:diff}} \cdot {10}^{\,\tau^\mathrm{\:diff}_\mathrm{ij}}$};

% -----------------------------------------------------------------------------------------

\node [normalcenter, ultra thick] (sky_2) at (13, 10.5) {\textbf{M2:} $\;I = I^\mathrm{\:point} + I^\mathrm{\:diff}$};

\node [smallleft] (I_point_2) at (10, 9) {$I^\mathrm{\:point}_\mathrm{ij}$};
\node [smallright] (I_diffuse_2) at (15.5, 9) {$I^\mathrm{\:diff} = I^\mathrm{\:dust} + I^\mathrm{\:nd}$};
\draw[arrow] (I_point_2) -- (sky_2);
\draw[arrow] (I_diffuse_2) -- (sky_2); % I_diffuse_2 -> sky_2

\node [smallright] (I_diff_nd) at (16, 7.5) {$I^\mathrm{\:nd}_\mathrm{ij} \;\mathrel{\hat{=}}\; I^\mathrm{\:diff}_\mathrm{ij}$};
\draw[arrow] (I_diff_nd) -- (I_diffuse_2);

% --------------------------------------

\node [normalright, densely dashed] (rhos) at (16, 5.375) {$\tau = \textsc{F}^{-1}(A_\tau\;\xi_\tau^\text{ex})$\\
{\tiny $\tau \in \{\tau^\mathrm{\:diff},\tau^\mathrm{\:dust},\tau^\mathrm{\:nd}\}$}};

\draw[arrow, dashed] (rhos) -- (8.6, 8.55); % I_diffuse_1
\draw[arrow, dashed] (rhos) -- (I_diff_nd);

\node [smallleft] (I_diffuse_dust) at (7.75, 7.5) {$I^\mathrm{\:dust}_\mathrm{ij} = I^\mathrm{\:dust}_{0} \cdot \left(\frac{E_i}{E_0}\right)^{\alpha^\mathrm{\:dust}} \cdot {10}^{\,\tau^\mathrm{\:dust}_\mathrm{ij}}\cdot \nicefrac{I^\mathrm{\:T}_\mathrm{\:j}}{\langle I^\mathrm{\:T}_\mathrm{j} \rangle_\mathrm{geom}}$};
\draw[arrow] (I_diffuse_dust) -- (I_diffuse_2);
\draw[arrow, dashed] (rhos) -- (12.4, 7.07); % I_diffuse_dust

%\node[smallleft, densely dashed](epsilon_kappa) at (7, 5.375) {$\epsilon\;\;\;\;\;\;\;\;\;\;\;\;\;\kappa$};
%\draw[arrow, dashed] (7.3, 5.75) -- (5.25, 7.85); % epsilon_kappa -> f_spec
%\draw[arrow, dashed] (9.4, 5.375) -- (11.3, 5.375); % epsilon_kappa -> rhos
\end{tikzpicture}
\caption{Structure of the main generative hierarchical Bayesian models used
in the inference. The initial, abstract, and Gaussian-distributed
degrees of freedom, $\xi$, are mostly left out for the sake of clarity.
The dashed component indicates a set of
instances of the Gaussian correlated field model introduced
by \citet[][Sect.~3.4]{arras2021comparison}.
\textbf{Left:} Template-free model M1 with diffuse and point-like fluxes $I^\mathrm{\:diff}$ and $I^\mathrm{\:point}$.
\textbf{Right:} Template-informed model M2 with the additional template-informed diffuse flux component $I^\mathrm{\:dust}$.
$I^\mathrm{\:nd}$ is identical in construction to $I^\mathrm{\:diff}$, but has different prior parameter values.
$I^\mathrm{\:point}_\mathrm{ij}$ is defined identically in both models.}
\label{fig:generative-model-structure}
\end{figure*}
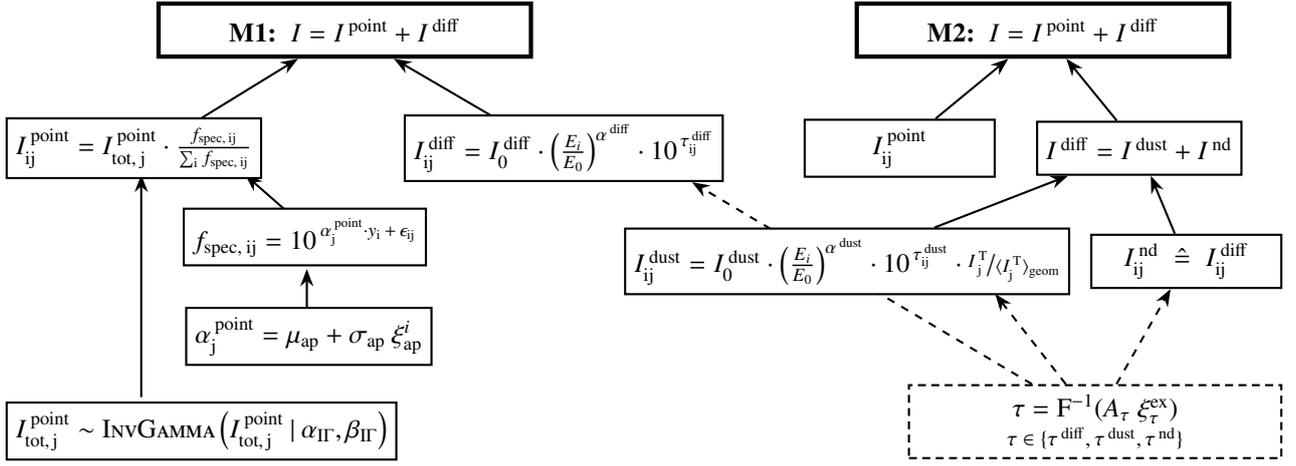

\begin{table}
\centering
\begin{tabular}{|| c | c ||}
\hline
\textbf{Parameter} & \textbf{Value}\\
\hline
\hline
$\alpha_{\text{I}\Gamma}$ & $1.5$\\
\hline
$\beta_{\text{I}\Gamma}$ & \unit[$5\cdot{10}^{-4}$]{$(\mathrm{m}^{2}\,\mathrm{s}\,\mathrm{sr})^{-1}$}\\
\hline
$\epsilon_\mathrm{flex}$ & $\mathfrak{m}=0.5,\;\mathfrak{s}=0.25$\\
\hline
$\epsilon_\mathrm{asp}$ & $\mathfrak{m}=0.05,\;\mathfrak{s}=0.05$\\
\hline
$\mu_\mathrm{ap}$ & $-1.25$\\
\hline
$\sigma_\mathrm{ap}$ & $0.3$\\
\hline
\multicolumn{2}{c}{}\\
\end{tabular}
\caption{Prior parameters of the PS emission model in M1 and M2.}
\label{tab:ps-prior-parameters}
\end{table}

\begin{table}
\centering
\begin{tabular}{|| c | c ||}
\hline
\textbf{Parameter} & \textbf{Value}\\
\hline
\hline
\multicolumn{2}{||c||}{\rule{0pt}{3ex}$I^\mathrm{\:diff}\;$ (M1)}\\
\hline
$I^\mathrm{\:diff}_0$ & \unit[${10}^{-2}$]{$\mathrm{m}^{-2}\,\mathrm{s}^{-1}\,\mathrm{sr}^{-1}$}\\
\hline
$\alpha^\mathrm{\:diff}$ & $-1.466$\\
\hline
$\hat{C}_\beta$ & $\mu_\beta = -3.0,\;\sigma_\beta = 0.25$\\
\hline
$\tau^\mathrm{\:diff}_\mathrm{fluct\,x}$ & $\mathfrak{m}=0.75,\;\mathfrak{s}=0.187$\\
\hline
$\hat{D}_\gamma$ & $\mu_\gamma = -3.5,\;\sigma_\gamma = 0.25$\\
\hline
$\tau^\mathrm{\:diff}_\mathrm{fluct\,y}$ & $\mathfrak{m}=0.1,\;\mathfrak{s}=0.05$\\
\hline
\multicolumn{2}{||c||}{\rule{0pt}{3ex}$I^\mathrm{\:nd}\;$ (M2)}\\
\hline
$I^\mathrm{\:nd}_0$ & $I^\mathrm{\:diff}_0 / 2$\\
\hline
$\alpha^\mathrm{\:nd}$ & $-1.25$\\
\hline
$\hat{C}_\beta$ & $\mu_\beta = -3.0,\;\sigma_\beta = 0.25$\\
\hline
$\tau^\mathrm{\:nd}_\mathrm{fluct\,x}$ & $\mathfrak{m}=0.75,\;\mathfrak{s}=0.187$\\
\hline
$\hat{D}_\gamma$ & $\mu_\gamma = -3.5,\;\sigma_\gamma = 0.25$\\
\hline
$\tau^\mathrm{\:nd}_\mathrm{fluct\,y}$ & $\mathfrak{m}=0.1,\;\mathfrak{s}=0.05$\\
\hline
\multicolumn{2}{||c||}{\rule{0pt}{3ex}$I^\mathrm{\:dust}\;$ (M2)}\\
\hline
$I^\mathrm{\:dust}_0$ & $I^\mathrm{\:diff}_0 / 2$\\
\hline
$\alpha^\mathrm{\:dust}$ & $-1.65$\\
\hline
$\hat{C}_\beta$ & $\mu_\beta = -3.5,\;\sigma_\beta = 0.25$\\
\hline
$\tau^\mathrm{\:dust}_\mathrm{fluct\,x}$ & $\mathfrak{m}=0.25,\;\mathfrak{s}=0.125$\\
\hline
$\hat{D}_\gamma$ & $\mu_\gamma = -3.0,\;\sigma_\gamma = 0.25$\\
\hline
$\tau^\mathrm{\:dust}_\mathrm{fluct\,y}$ & $\mathfrak{m}=0.1,\;\mathfrak{s}=0.025$\\
\hline
\multicolumn{2}{||c||}{\rule{0pt}{3ex}Common parameters}\\
\hline
$\mathrm{Std}[\langle\tau^\mathrm{\:c}\rangle]$ & $\mathfrak{m}=0.5,\;\mathfrak{s}=0.05$\\
\hline
$\hat{C}\kappa_\mathrm{flex}$, $\hat{D}\kappa_\mathrm{flex}$ & $\mathfrak{m}=0.1,\;\mathfrak{s}=0.1$\\
\hline
$\hat{C}\kappa_\mathrm{asp}$, $\hat{D}\kappa_\mathrm{asp}$ & $\mathfrak{m}=0.1,\;\mathfrak{s}=0.05$\\
\hline
\multicolumn{2}{c}{}\\
\end{tabular}
\caption{Prior parameters of the diffuse emission components of M1 and M2.}
\label{tab:diff-prior-parameters}
\end{table}

Figure~\ref{fig:generative-model-structure} shows the hierarchical structure of the employed sky models M1 and M2,
while Table~\ref{tab:ps-prior-parameters} and~\ref{tab:diff-prior-parameters} provide the corresponding prior parameters.
In the following, we give additional background information to help with their interpretation.

Single parameter values indicate the parameter is a constant in the model and cannot be modified during reconstruction.
Parameter values typeset as tuples indicate the model parameter is assumed to be either normal or log-normal distributed a priori.
If $\mu$ and $\sigma$ values are given, we chose a normal distribution with mean $\mu$ and standard deviation $\sigma$ as the parameter prior.
If  $\mathfrak{m}$ and $\mathfrak{s}$ values are given,
a log-normal prior distribution in accordance with \citet[Eq.~16 and Eq.~17]{arras2021comparison} is assumed.

A practical demonstration of how the correlated field parameters affect the generated field samples
can be found at the NIFTy documentation page\footnote{\url{https://ift.pages.mpcdf.de/nifty/user/getting_started_4_CorrelatedFields.html}}
and the source code of this demonstration is published at the NIFTy code repository\footnote{
\url{https://gitlab.mpcdf.mpg.de/ift/nifty/-/blob/ebd57b33300c631a3bf4c8d6bebe2e5a2c0ed368/demos/getting_started_4_CorrelatedFields.py}}.
We used the ``\texttt{add\_fluctuations}'' call of the correlated field model.
The model parameters for the APS and EPS
$\beta$ and $\gamma$ correspond to its ``\texttt{loglogavgslope}''  parameter,
while $\kappa_\mathrm{flex}$ and $\kappa_\mathrm{asp}$ correspond to the ``\texttt{flexibility}''  and ``\texttt{asperity}'' parameters.
The correlated field standard deviation in the energy dimension $\tau^\mathrm{\:c}_\mathrm{fluct\,y}$
and the spatial dimension $\tau^\mathrm{\:c}_\mathrm{fluct\,x}$
is set via the \texttt{fluctuations} parameter in the \texttt{add\_fluctuations} function.
The a priori standard deviation of the offset from zero of the $\tau$ fields is parameterized by the $\textrm{Std}[\langle\tau^\mathrm{\:c}\rangle]$ parameter,
which corresponds to the ``\texttt{offset\_std}'' parameter of the ``\texttt{CorrelatedFieldMaker}.''

\FloatBarrier
\section{Instrument response model implementation} \label{sec:appendix-instrument-response-model}

\subsection{Overview}

The \textit{Fermi} Collaboration provides high-quality IRF models for the LAT, obtained
via Monte Carlo simulations of the instrument in action and validated with online
measurements.

\subsection{Point spread function}

The PSF probabilistically describes how the
\textit{Fermi} LAT misclassifies the origin direction of recorded gamma-ray
photons.
Of the \textit{Fermi} LAT IRF components,
the convolution with the PSF was the hardest to model numerically.
Usually, convolutions can be efficiently performed by multiplication
in a corresponding harmonic space. However, the \textit{Fermi} LAT PSF is not
band-limited in SH space, which makes it unfit
for this approach. We therefore chose to numerically represent the
PSF convolution operation with a sparse matrix application. For this,
we calculate the contribution of each pixel to every other pixel under
convolution with the PSF. To quantify this contribution, the PSF needed
to be integrated over the receiving pixel areas, once for each source
pixel. We perform this integration using Monte Carlo methods, specifically
direct sampling from the angular deviation distribution described
by the PSF. The drawn samples were binned into the respective sky pixels
and the count normalized by the total sample count. This integration
method yields the highest accuracy in the pixels receiving the most
flux under the PSF convolution, while still capturing the broad-tailed
nature of the PSF. The PSF application matrix was sparsified by discarding
entries with contributions below $10^{-7}$ times the maximum contribution.

% ---- fig: psf-flux-fractions ----
\begin{figure*}
\centering
\noindent
\begin{minipage}[t]{0.49\textwidth}%
\includegraphics[width=1\textwidth]{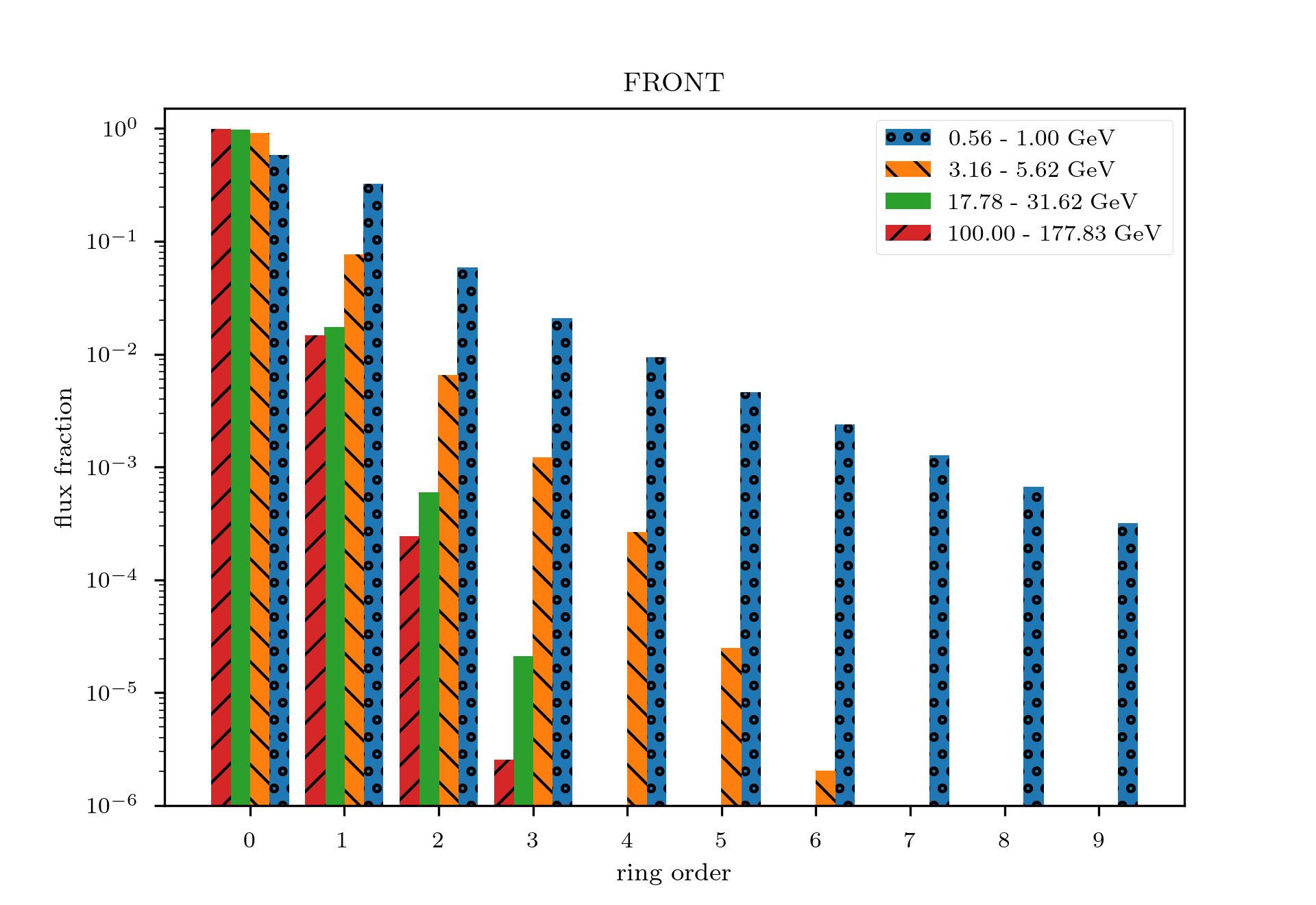}%
\end{minipage}%
\begin{minipage}[t]{0.49\textwidth}%
\includegraphics[width=1\textwidth]{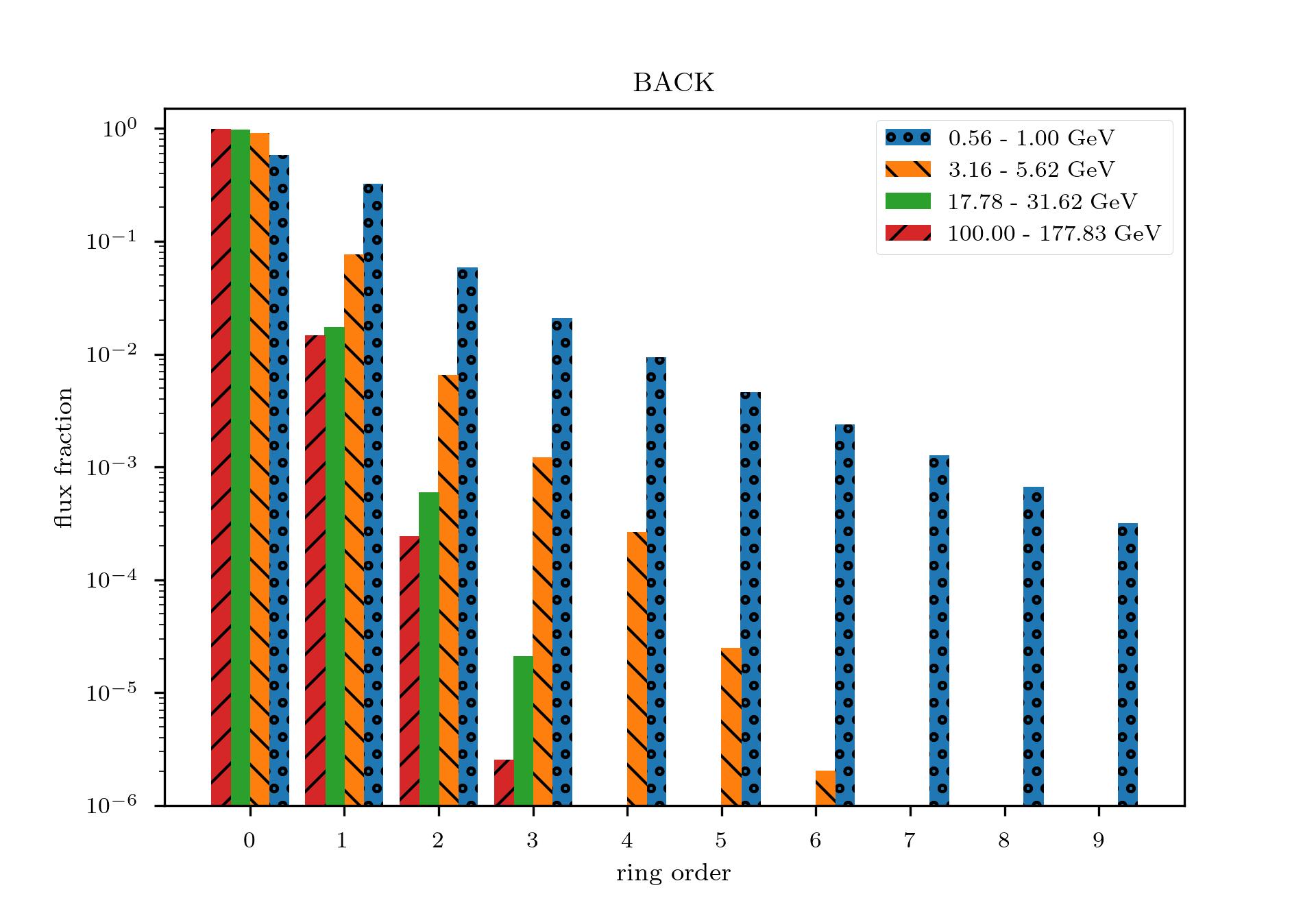}%
\end{minipage}
\caption{Flux fractions distributed into concentric rings around the origin pixels by the PSF
of the $\cos(\theta) \in \left(0.9,\,1.0\right]$ incidence direction bin
for \texttt{FRONT} and \texttt{BACK} events and selected energy bins.
Ring order 0 corresponds to the origin pixel itself, ring order 1 to the eight pixels surrounding it,
and so forth.
Table~\ref{tab:ring-angles} provides the average $\delta\theta$ values of the pixels in the rings.
Flux fractions are shown on a logarithmic scale from ${10}^{-6}$ to $1.0$.}
\label{fig:psf-flux-fractions}
\end{figure*}

To evaluate how the PSF modeling interacts with the spatial discretization,
we analyzed how the PSF matrices distribute flux into concentric rings
of pixels around selected source pixels.
For this, we calculated how much flux is retained in the source pixel itself,
in the eight pixels surrounding it, in the 26 pixels surrounding these, and further on.
We call these groups of pixels rings of order zero, one, two, and higher.
Figure~\ref{fig:psf-flux-fractions} shows the flux fractions distributed into
the respective rings for \texttt{FRONT} and \texttt{BACK} events
in selected energy bins.
The figure shows the expected patterns.
With increasing bin energy, the instrument point spread becomes weaker.
This is captured by our PSF model matrices, which leave increasing
amounts of flux in the source pixel for increasing energy
(see the ring order zero bars in both panels).
For the two highest energy bins shown, only an original flux fraction on the order of
one percent is distributed away from the source pixel for \texttt{FRONT} events.
For \texttt{BACK} events, in the same energy bins on the order of 10\% of the flux is
redistributed by the PSF model matrix.
\begin{table*}
\centering
\begin{tabular}{|| c || c | c | c | c | c | c | c | c | c | c ||}
\hline
\textbf{Ring order} & 0 & 1 & 2 & 3 & 4 & 5 & 6 & 7 & 8 & 9 \\
\hline
\textbf{Average $\delta\theta$} & $0.0^\circ$ & $0.57^\circ$ & $1.09^\circ$ & $1.62^\circ$ & $2.18^\circ$ & $2.73^\circ$ & $3.29^\circ$ & $3.87^\circ$ & $4.45^\circ$ & $5.07^\circ$ \\
\hline
\multicolumn{2}{c}{}\\
\end{tabular}
\caption{Average $\delta\theta$ values of the pixels belonging
to the rings presented in Fig.~\ref{fig:psf-flux-fractions}.}
\label{tab:ring-angles}
\end{table*}
Table~\ref{tab:ring-angles} provides the average $\delta\theta$ values
of the pixels belonging to the rings.
The $\delta\theta$ values of the pixels are calculated by finding the angle between the vectors
pointing to the center of the source pixel and the center of the receiving pixel
from the center of the sphere.

\subsection{Energy dispersion function}

The EDF probabilistically describes how the
\textit{Fermi} LAT misclassifies the energy of recorded gamma-ray photons.
Similar to the PSF modeling, the effect of the convolution with the EDF of the LAT was modeled numerically using a matrix.
Since the EDF describes how specific photon energies are misclassified,
a double integral needs to be calculated to estimate the contributions
of individual energy bins toward each other: First, integration of
the true photon energy over the source energy bin and second, integration
of the recorded photon energy over the target energy bin. Since the
photon flux density is not constant within the energy bins, in the
first integral a weighting factor has to be considered. We estimated
this weighting factor from a power-law fit to the raw photon counts.

% ---- fig: EDFs ----
\begin{figure*}
\centering
\noindent
\begin{minipage}[t]{0.49\textwidth}%
\includegraphics[width=1\textwidth]{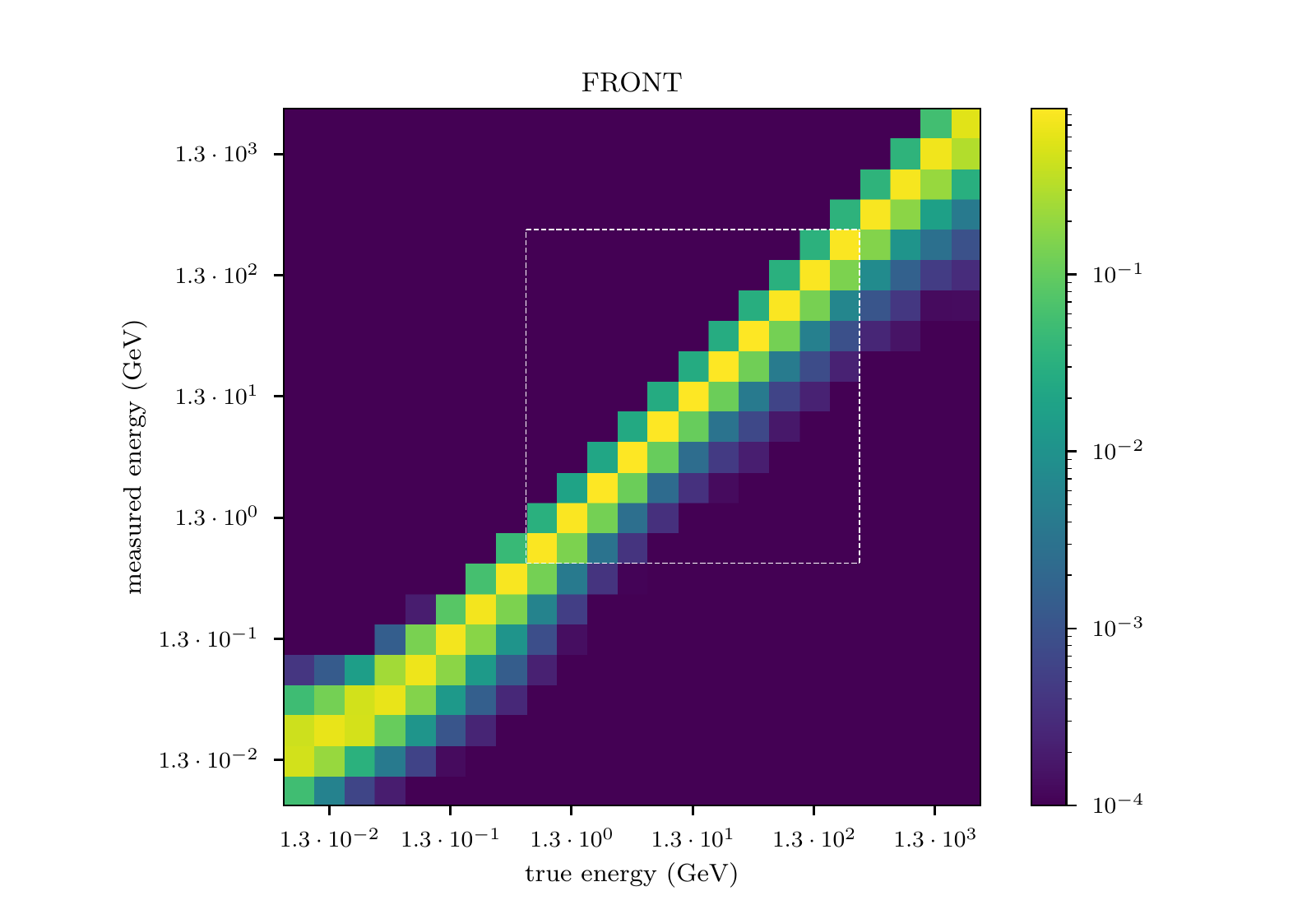}%
\end{minipage}%
\begin{minipage}[t]{0.49\textwidth}%
\includegraphics[width=1\textwidth]{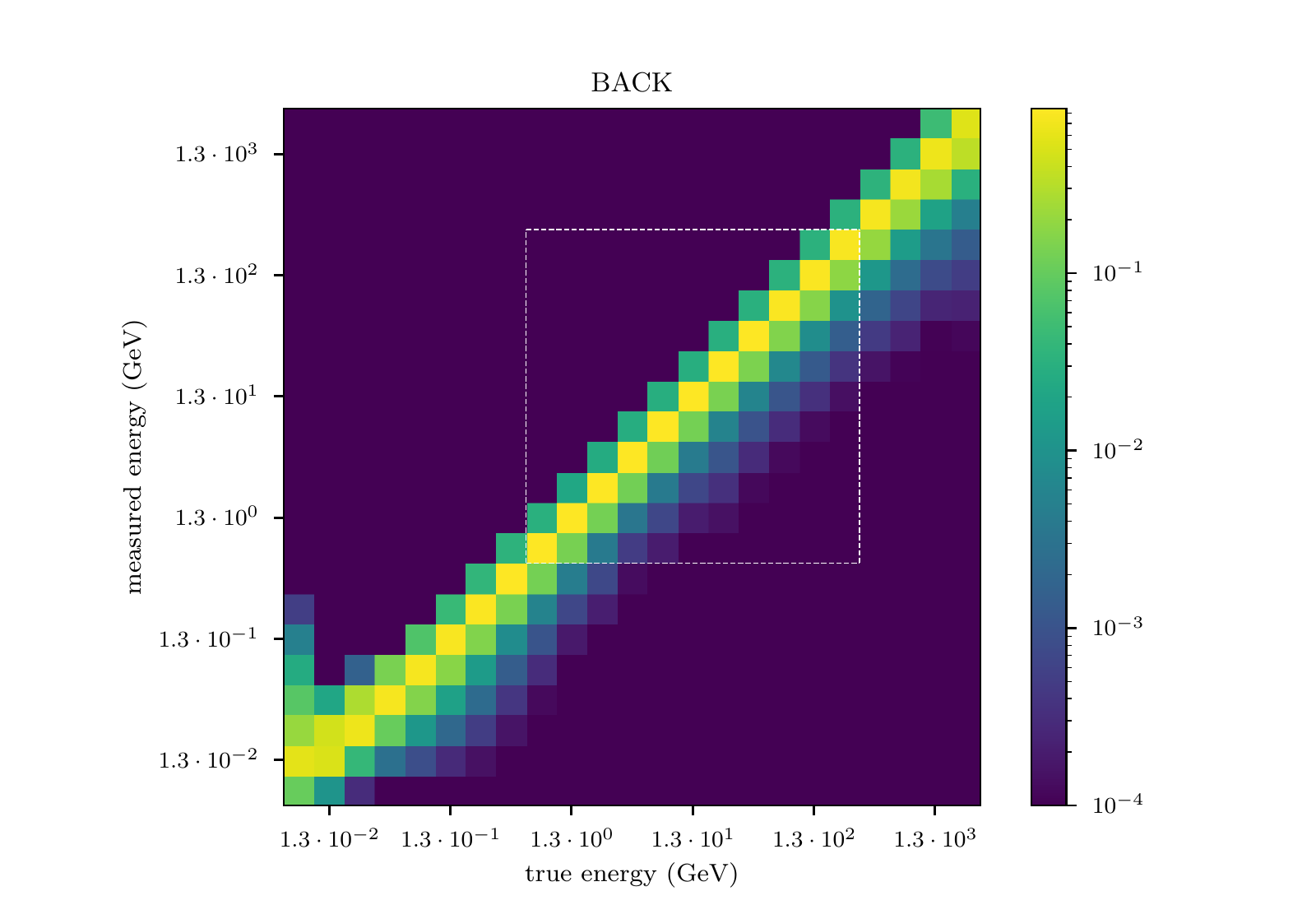}%
\end{minipage}
\caption{Energy dispersion matrices of the $\cos(\theta) \in \left(0.9,\,1.0\right]$ incidence direction bin
for \texttt{FRONT} and \texttt{BACK} events.
True and measured energies are shown on logarithmic scales.
The figure shows how flux from each true energy is distributed to the
observed energy bins by the measurement process
and has a logarithmic color scale.
The white rectangle marks the EDF matrix entries used in this work,
which were selected based on the studied energy range.}
\label{fig:edf-matrix}
\end{figure*}

Figure~\ref{fig:edf-matrix} shows the distribution of flux between energy bins
prescribed by the EDF matrix employed in our model.
The \textit{Fermi} Collaboration published a similar figure entitled ``Detector Response Matrix''
on the website describing the usage of the Pass 8 energy dispersion characterization\footnote{
\url{https://fermi.gsfc.nasa.gov/ssc/data/analysis/documentation/Pass8_edisp_usage.html}}.

\subsection{Exposure maps}

Exposure maps were calculated based on the spacecraft files provided
with the event tables by the \textit{Fermi} Collaboration
and take into account the data cuts we describe in Sect.~\ref{sec:methods-dataset}.
To get total exposure values for all data bins,
first we calculated exposure maps for each \unit[30]{s} time slice specified in the spacecraft files,
and afterward took their sum along the temporal axis as the total exposure.
To calculate the exposure values for a time slice,
we first calculated a binary ``data bin reachability mask''
based on the instrument pointing, the data bins' incidence direction window,
the Earth albedo cut for the corresponding time frame, and the zenith cut.
Second, for all data bins included by the reachability mask
we set the exposure value to the active observation time of the LAT in the corresponding time slice (as specified by the spacecraft files).
All data bins ``unreachable'' during the time slice were set to have \unit[0]{s} of exposure.
The resulting time slice exposure maps were added up for all included mission weeks to form the total exposure maps.

\subsection{Effective area}

The effective area of the instrument is provided by the \textit{Fermi} Collaboration
at much higher resolution than the other IRFs. For the sake of simplicity, we
averaged the values over the coarser incidence direction and energy
bins employed in the PSF parameter table and used these averages in our instrument response
model.

\FloatBarrier
\section{Color coding of multi-energy sky maps} \label{sec:appendix-color-coding}
Throughout this article, we show several sky maps in which energy information is color-coded.
In these maps, we simulate how a non-color-blind human would perceive the visualized gamma-ray fluxes
if the photon energies were shifted into the visible light spectrum range.
This presentation is optimized for a natural perception of spectral variations.

To obtain high spectral contrast and a natural appearance of the resulting maps,
we performed the following preprocessing steps before mapping to perceived colors:
First, we multiplied the spatio-spectral flux map with an energy-dependent scaling function
to compensate for flux density decrease toward high energies
(see the middle panel of Fig.~\ref{fig:data-exposure-corrected-mf-plot}).
Second, we transformed the values to a logarithmic scale.
Third, we clipped the logarithmic values to a predefined range
(corresponding to the visualized dynamic range)
and rescaled them to lie in the $[0, 1]$ interval.
Fourth, we applied a gamma correction with exponent $1.675$.

For the perceived color mapping,
we then mapped the reconstructed gamma-ray energies to visible light wavelengths (log-linearly)
and for each pixel used the CIE 1931 model of human color perception to determine the chromaticity and luminosity
a human would perceive if looking into a light source of the corresponding visible light spectrum.
The corresponding perceived color values of the pixels were then embedded in the sRGB color space \citep{ICE_sRGB} for rendering.

As a last step, we applied a color saturation correction.
For this, we calculated the gray value $L$ of each image pixel from the sRGB values,
\begin{equation}
L = 0.2989 r \:+\: 0.5870 g \:+\: 0.1140 b
,\end{equation}
and the color difference of each pixel from its gray value,
\begin{equation}
\left(\delta r,\, \delta g,\, \delta b\right) = \left(r - L,\, g - L,\, b - L\right).
\end{equation}
The final sRGB pixel values were then set to the color-saturation-enhanced
\begin{equation}
\left(r',\, g',\, b'\right) \,=\, 2.2 \cdot \left(\delta r,\, \delta g,\, \delta b\right)\, + L.
\end{equation}

The bottom panel of Fig.~\ref{fig:data-exposure-corrected-mf-plot}
shows the resulting mapping of mono-energetic gamma-ray fluxes
(after energy-dependent scaling) to sRGB values.
Figure~\ref{fig:m2-mf-plot-desaturated} shows selected M2 reconstruction
results rendered without application of the color saturation enhancement.

\FloatBarrier
\section{Additional plots} \label{sec:appendix-plots}

This section contains additional figures that were excluded from the main text to save space,
but that we nevertheless want to include for completeness.

% ---- fig: m1 sf diffuse plots
\begin{figure*}
\centering
\noindent
\begin{minipage}[t]{0.32\textwidth}%
\includegraphics[width=1\textwidth]{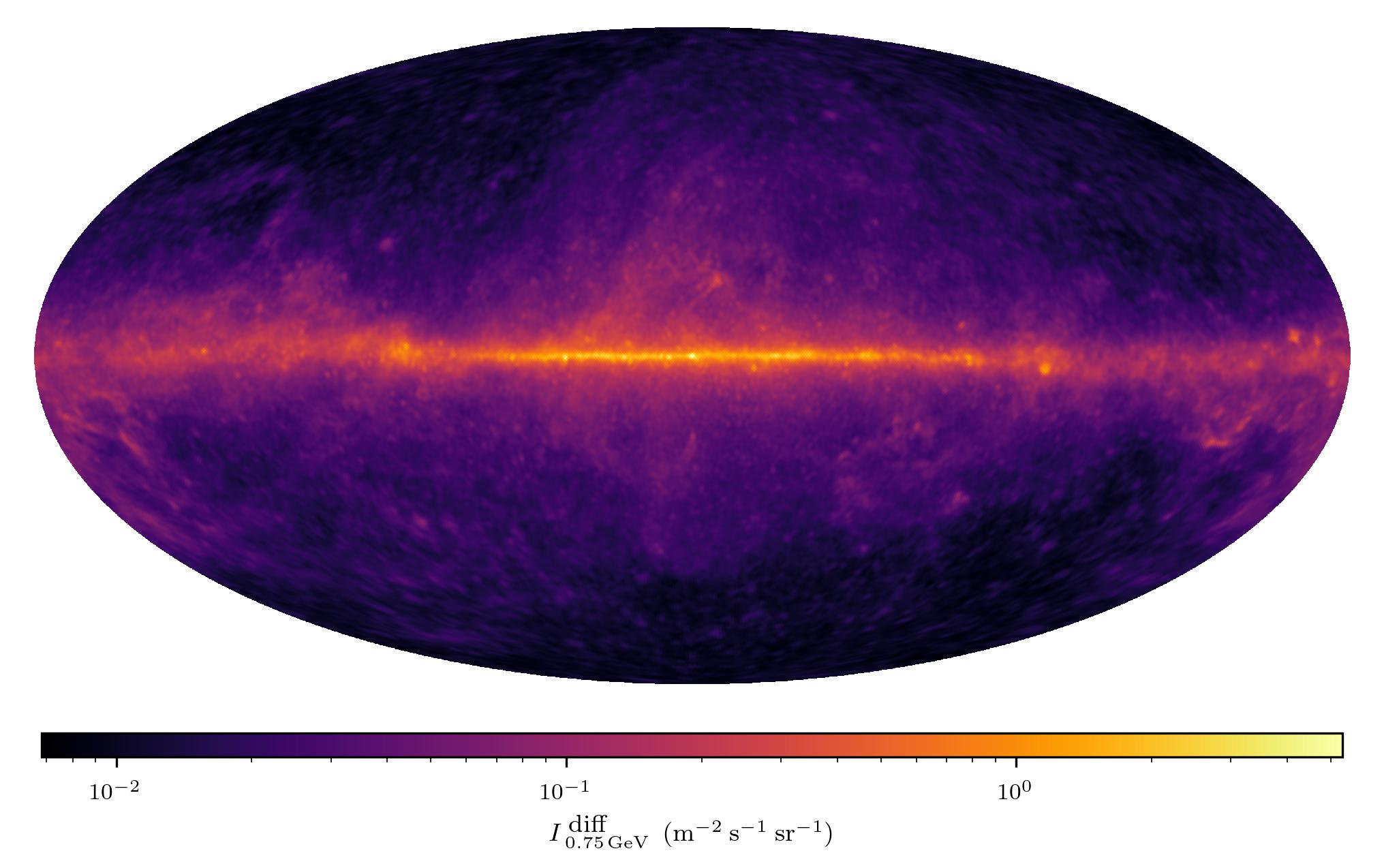}%
\end{minipage}\ %
\begin{minipage}[t]{0.32\textwidth}%
\includegraphics[width=1\textwidth]{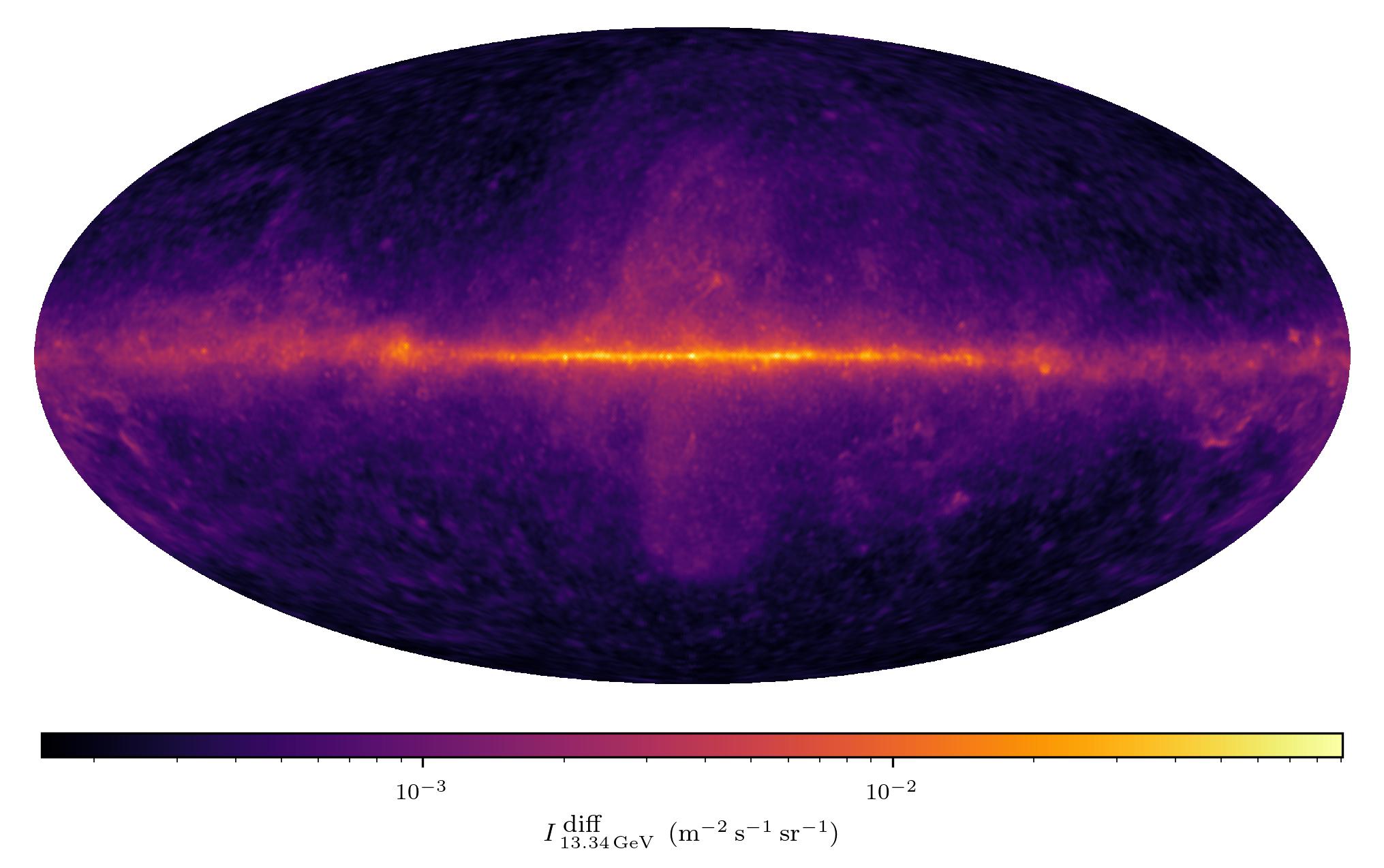}
\end{minipage}\ %
\begin{minipage}[t]{0.32\textwidth}%
\includegraphics[width=1\textwidth]{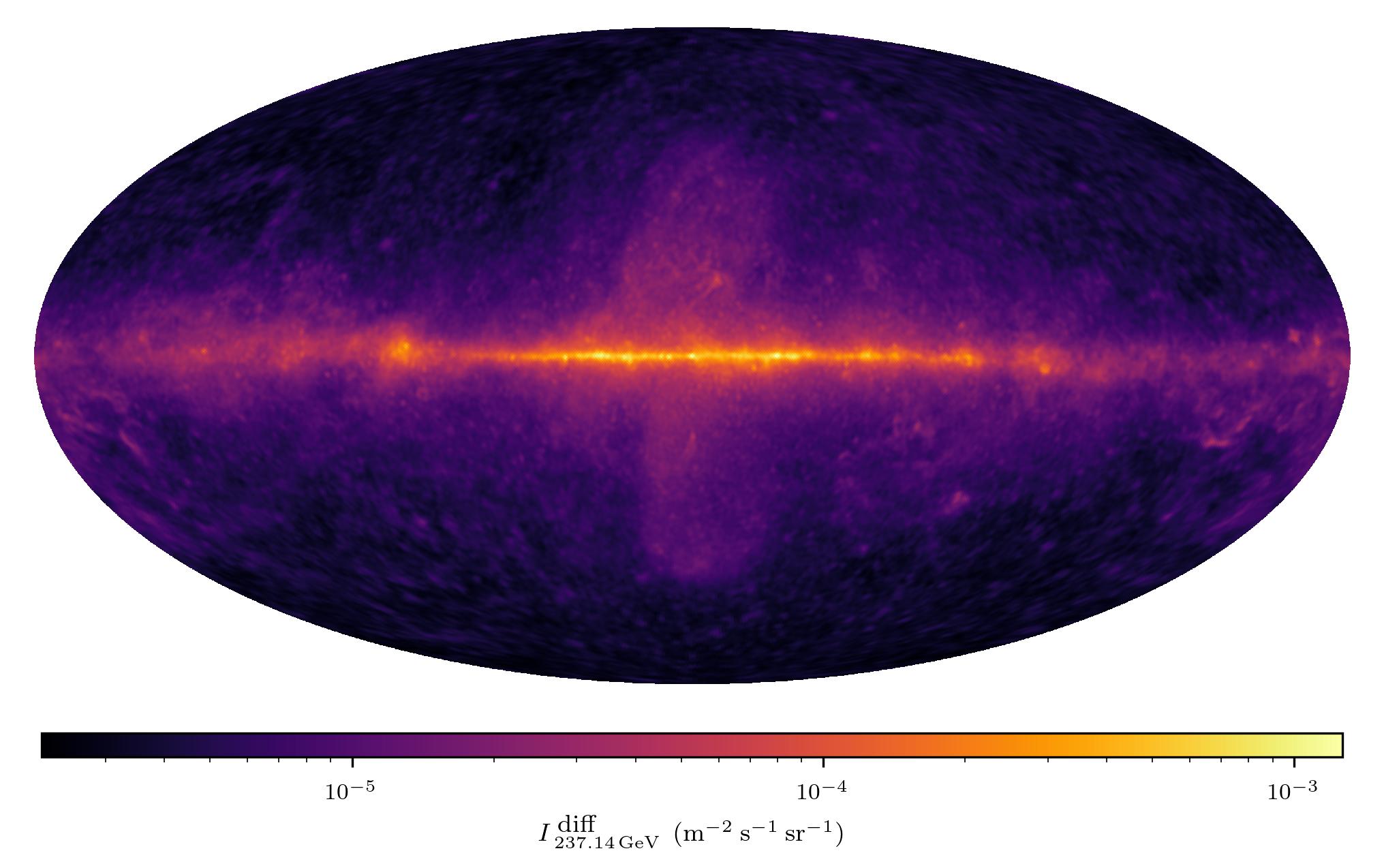}
\end{minipage}
\caption{Single energy plots of the M1 diffuse reconstruction with a logarithmic color scale.
Energy bins: \unit[0.56--1.00]{GeV} (\textbf{left}), \unit[10.0--17.8]{GeV} (\textbf{middle}), and \unit[178--316]{GeV} (\textbf{right}).
All panels have individual brightness scales and visualize the full dynamic range of their respective maps.
}
\label{fig:m1-sf-plots}
\end{figure*}

Figure~\ref{fig:m1-sf-plots} displays the M1 diffuse emission reconstruction in the \unit[0.56--1.00]{GeV}, \unit[10.0--17.8]{GeV}, and \unit[178--316]{GeV} energy bins.
The progression of maps shows the transition from a sky dominated by hadronic emission in the \unit[1]{GeV} energy regime
to one increasingly dominated by leptonic emissions in the \unit[100]{GeV} energy regime,
where the FBs and Galactic disk are visible as the most prominent structures.
The corresponding M2 diffuse emission reconstruction maps are shown in Fig.~\ref{fig:m2-sf-plots}.

% ---- fig: data histogram for selected bins
\begin{figure*}
\centering
\noindent
\includegraphics[width=1\textwidth]{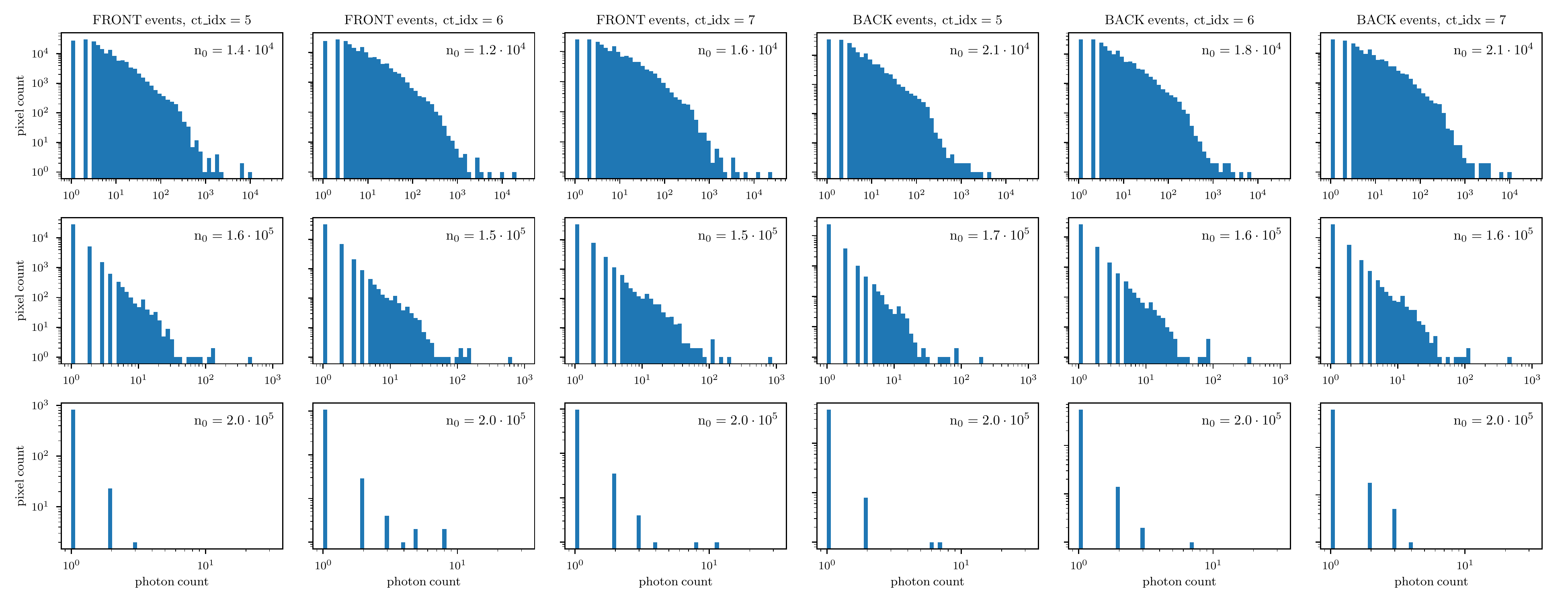}
\caption{Histograms of data photon counts for all response bins with the energies
\unit[1.00--1.78]{GeV} (\textbf{top}), \unit[10.0--17.8]{GeV} (\textbf{middle}), and \unit[178--316]{GeV} (\textbf{bottom}).
Photon counts and histogram pixel counts are shown on logarithmic scales.
The number of pixels with zero observed photons, $\mathrm{n}_{0}$, is displayed as text in each panel.
The apparent photon count sparsity in the $<$ \unit[10]{counts} regime is caused by the sparsity of integer numbers on the
logarithmic scale in that regime.
Figure~\ref{fig:m1-residual-histograms} shows the residual histograms of the selected bins.}
\label{fig:data-histograms}
\end{figure*}

Figure~\ref{fig:data-histograms} shows the empirical distribution of photon counts in all response bins considered in this paper
for the energy bins \unit[1.00--1.78]{GeV}, \unit[10.0--17.8]{GeV}, and \unit[178--316]{GeV}.
The counts in the \unit[1.00--1.78]{GeV} energy bin follow a broken power law
with an index of approximately -1.35 below the break at \unit[$1.5\cdot{10}^2$]{ph}
and indices between -2.5 and -2.0 above the break.
The existence of the break is consistent in all data response bins of this energy.
In the \unit[10.0--17.8]{GeV} energy bin the photon counts also follow a power law with an index of approximately -3.1 without an apparent break.
The power-law behavior seems to also hold for the \unit[178--316]{GeV} energy bins.
Here, we refrain from calculating approximate power-law exponents as the low overall photon counts lead to
high variability between the different response bins.

% ---- fig: m2 residual maps
\begin{figure*}
\centering
\noindent
\begin{minipage}[t]{0.33\textwidth}%
\includegraphics[width=1\textwidth]{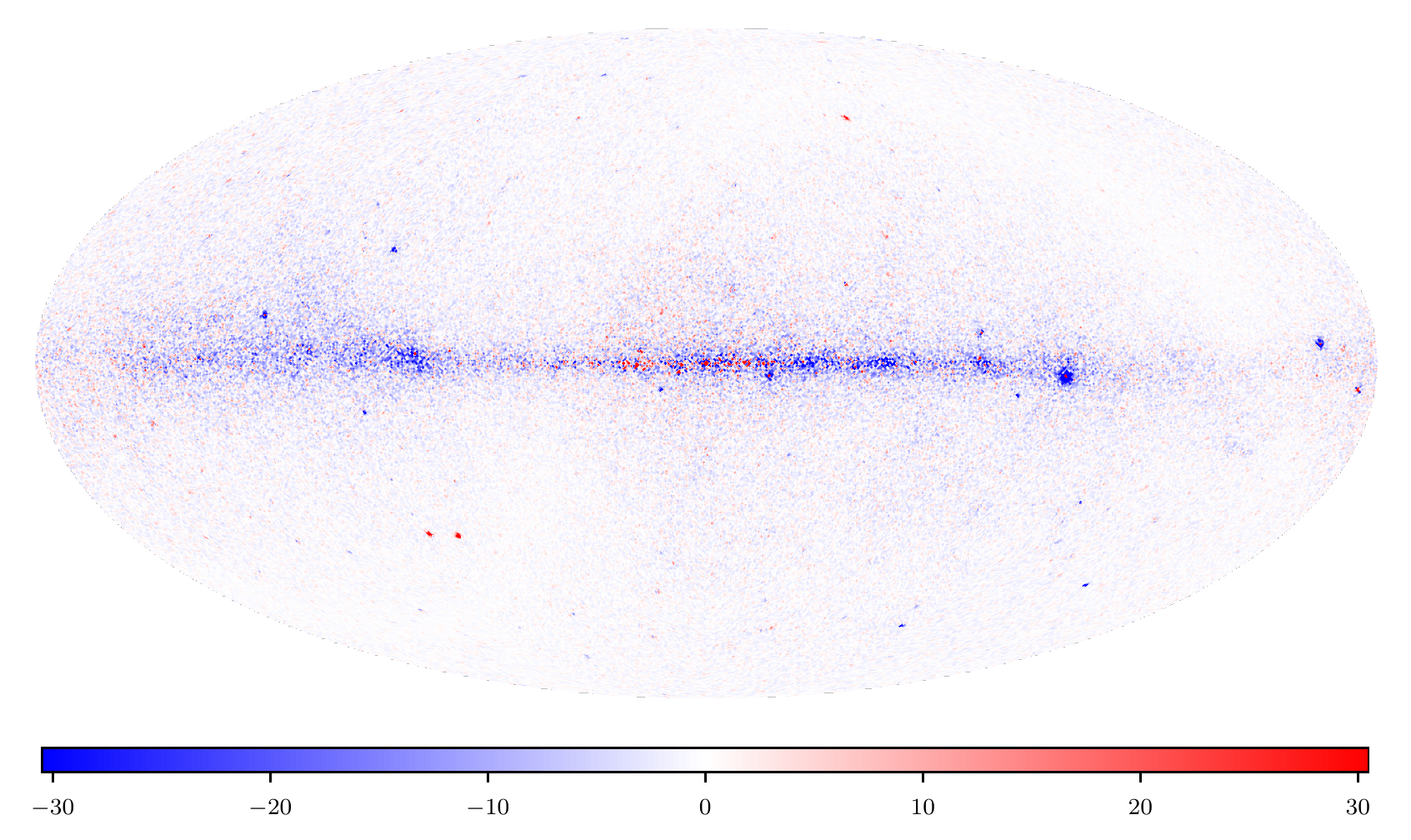}%
\end{minipage}%
\begin{minipage}[t]{0.33\textwidth}%
\includegraphics[width=1\textwidth]{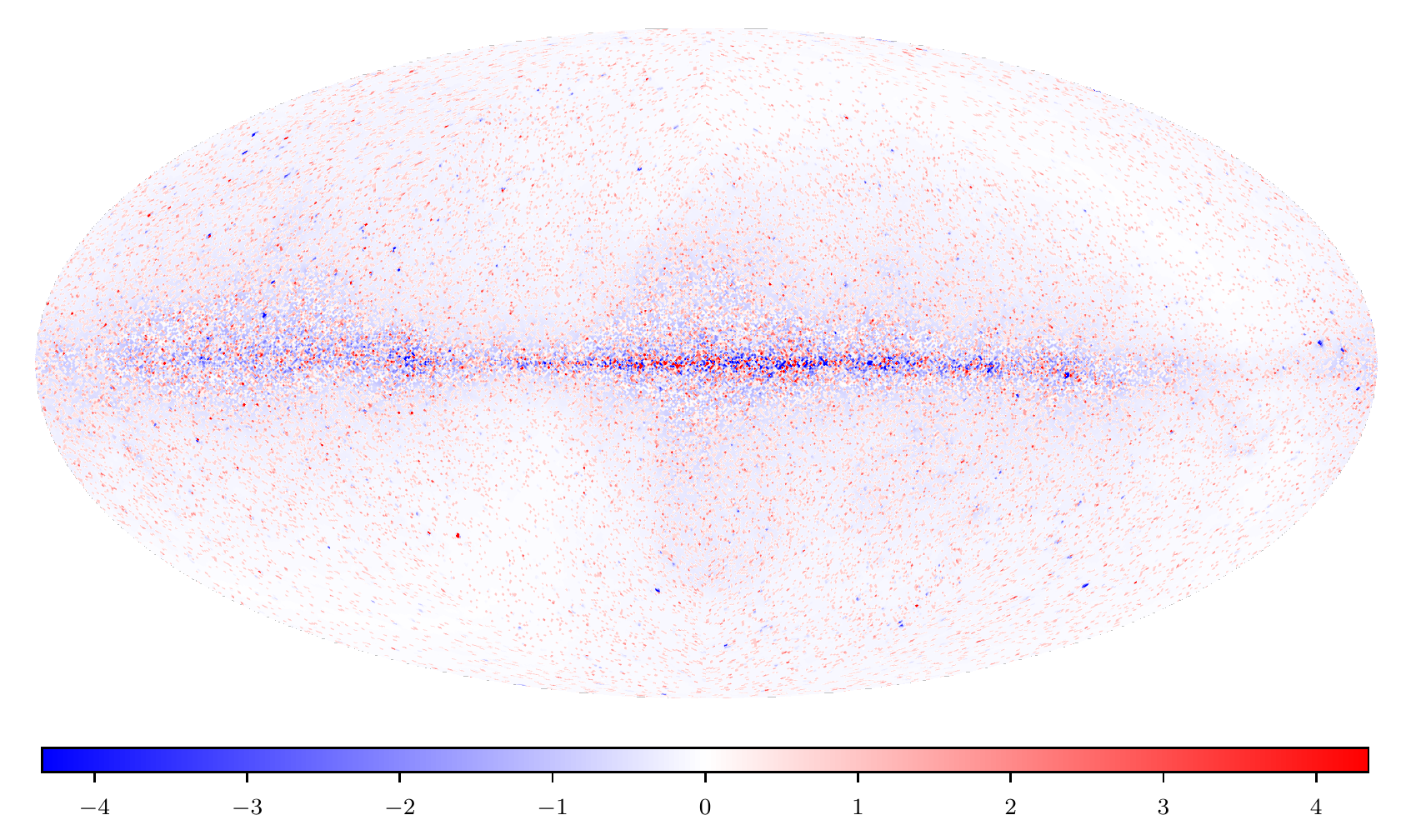}%
\end{minipage}%
\begin{minipage}[t]{0.33\textwidth}%
\includegraphics[width=1\textwidth]{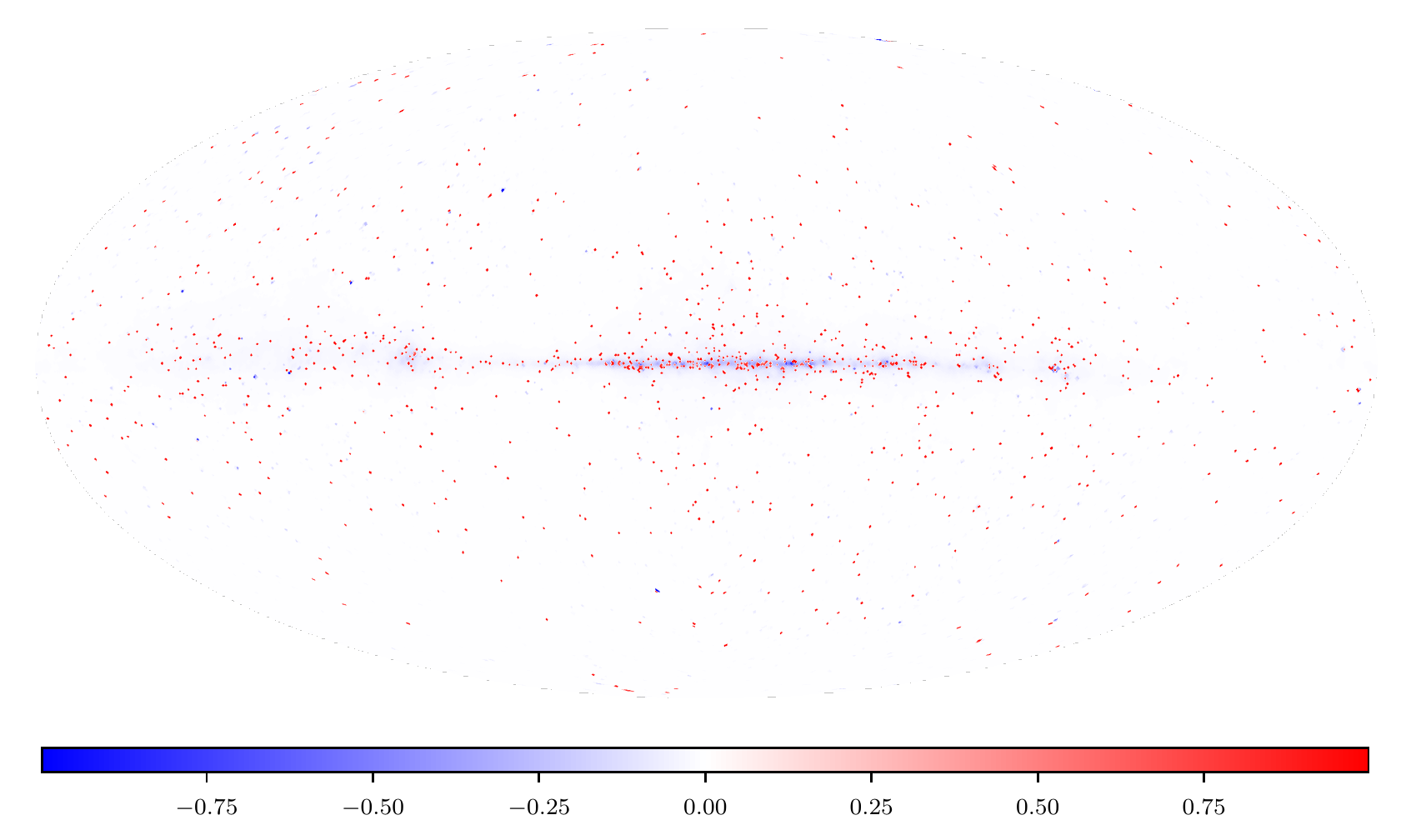}%
\end{minipage}

\begin{minipage}[t]{0.33\textwidth}%
\includegraphics[width=1\textwidth]{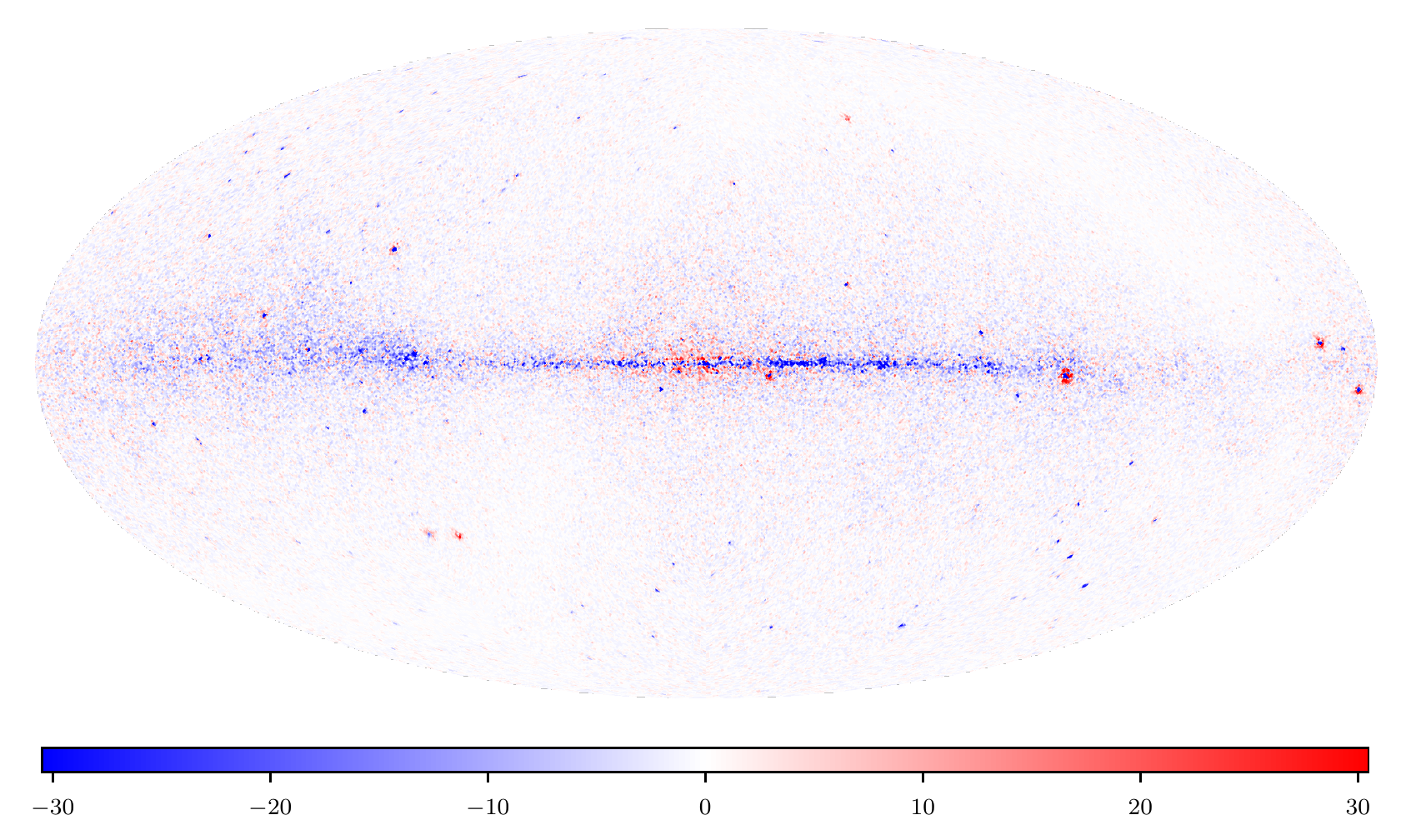}%
\end{minipage}%
\begin{minipage}[t]{0.33\textwidth}%
\includegraphics[width=1\textwidth]{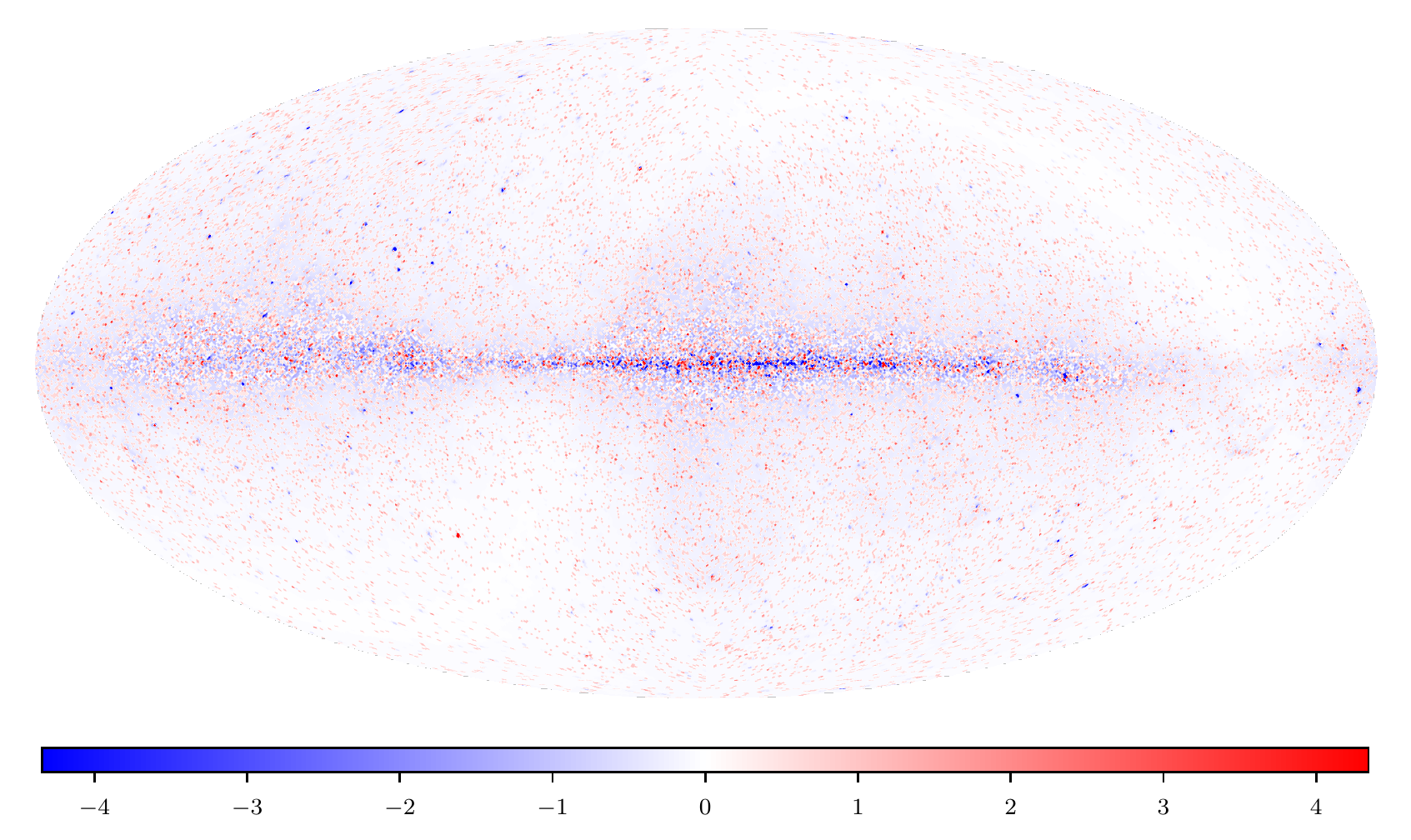}%
\end{minipage}%
\begin{minipage}[t]{0.33\textwidth}%
\includegraphics[width=1\textwidth]{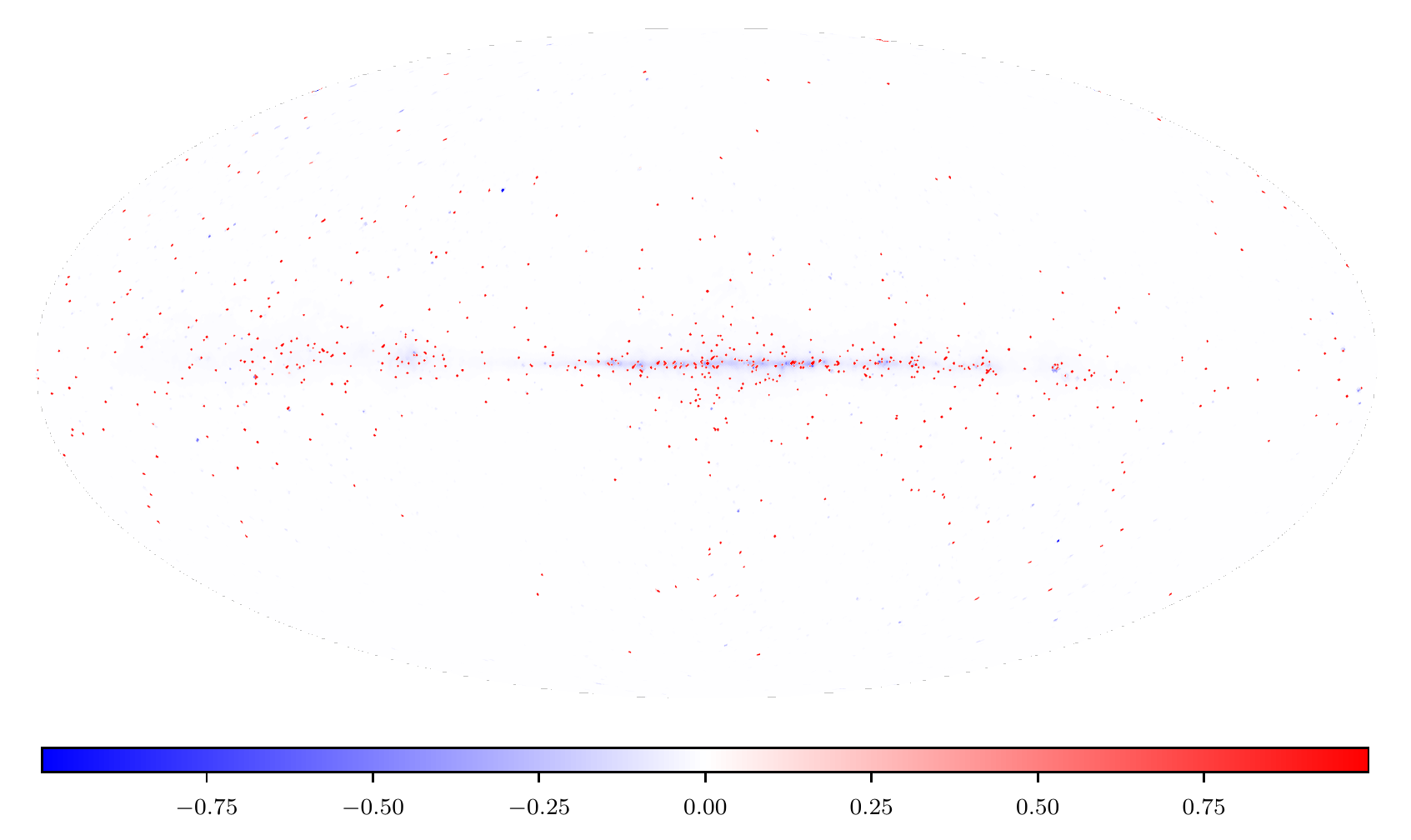}%
\end{minipage}
\caption{Selected residual maps for the M2 reconstruction shown for the data bin with $\cos(\theta_\mathrm{w'}) \in (0.9,1.0]$.
\textbf{Top}: \texttt{FRONT} events.
\textbf{Bottom}: \texttt{BACK} events.
Energy bins: \unit[1.00--1.78]{GeV} (\textbf{left}), \unit[10.0--17.8]{GeV} (\textbf{middle}), and \unit[178--316]{GeV} (\textbf{right}).
The maps are clipped to the 99.9th percentile of the residual amplitude in their respective energy bin.
}
\label{fig:m2-residual-maps}
\end{figure*}

Figure~\ref{fig:m2-residual-maps} shows selected residuals of the reconstruction based on the M2 (template-informed) model.
Similar to the residuals of the M1 reconstruction (see Fig.~\ref{fig:m1-residual-maps}),
the M2 residuals are largest in the Galactic plane and around extended sources such as Vela.
We observe the same inversion of residuals between \texttt{FRONT} and \texttt{BACK} events around the brightest PSs,
and a general agreement with the residuals of the M1 reconstruction.

% ---- fig: m2 dust modification maps
\begin{figure*}
\centering
\noindent
\begin{minipage}[t]{0.32\textwidth}%
\includegraphics[width=1\textwidth]{graphics_v2/v2_with_template_v01_u02_FB-rec-dust_modification-sample_gmean-e_idx-9-mollweide-50_300dpi.jpg}%
\end{minipage}\ %
\begin{minipage}[t]{0.32\textwidth}%
\includegraphics[width=1\textwidth]{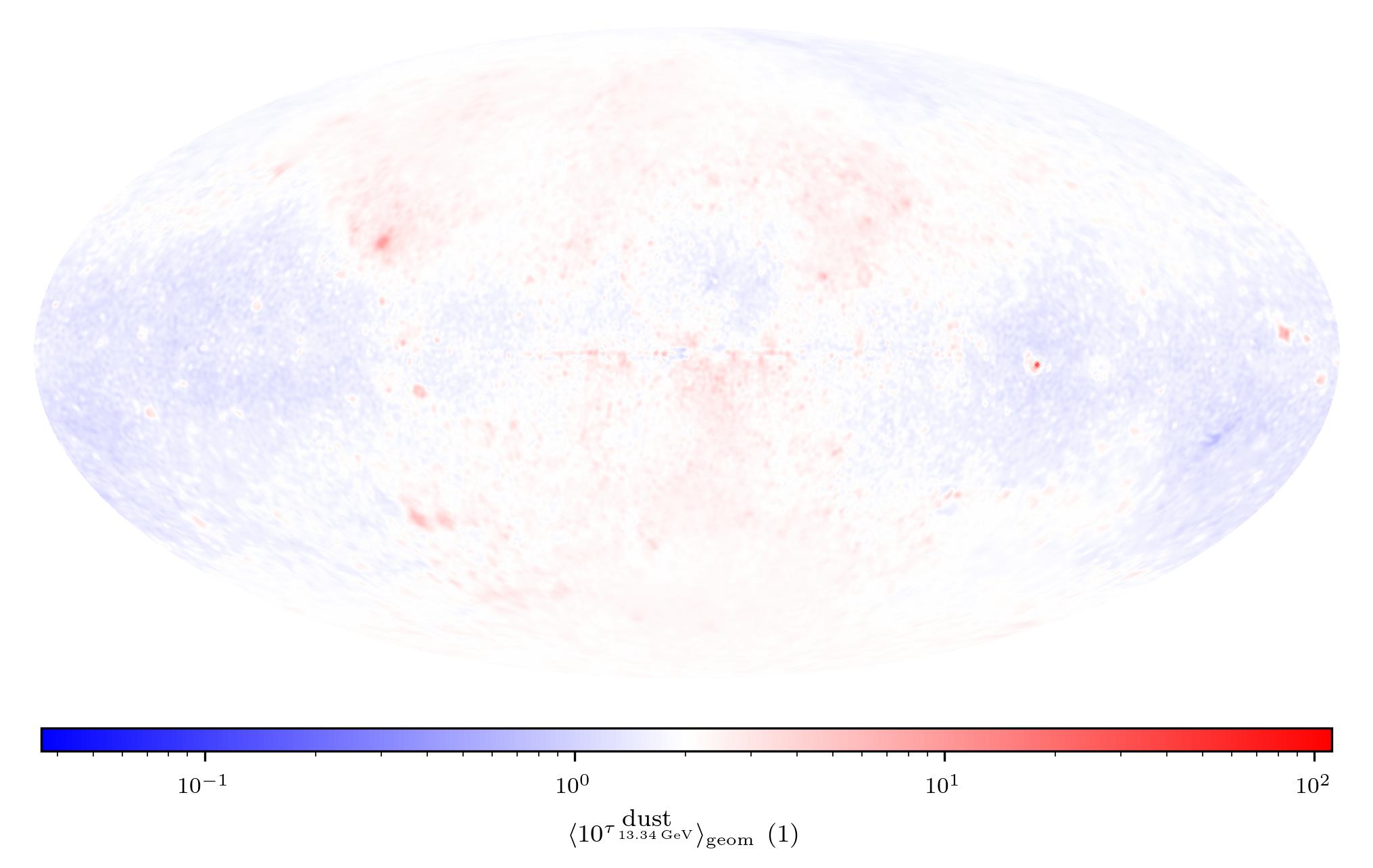}%
\end{minipage}\ %
\begin{minipage}[t]{0.32\textwidth}%
\includegraphics[width=1\textwidth]{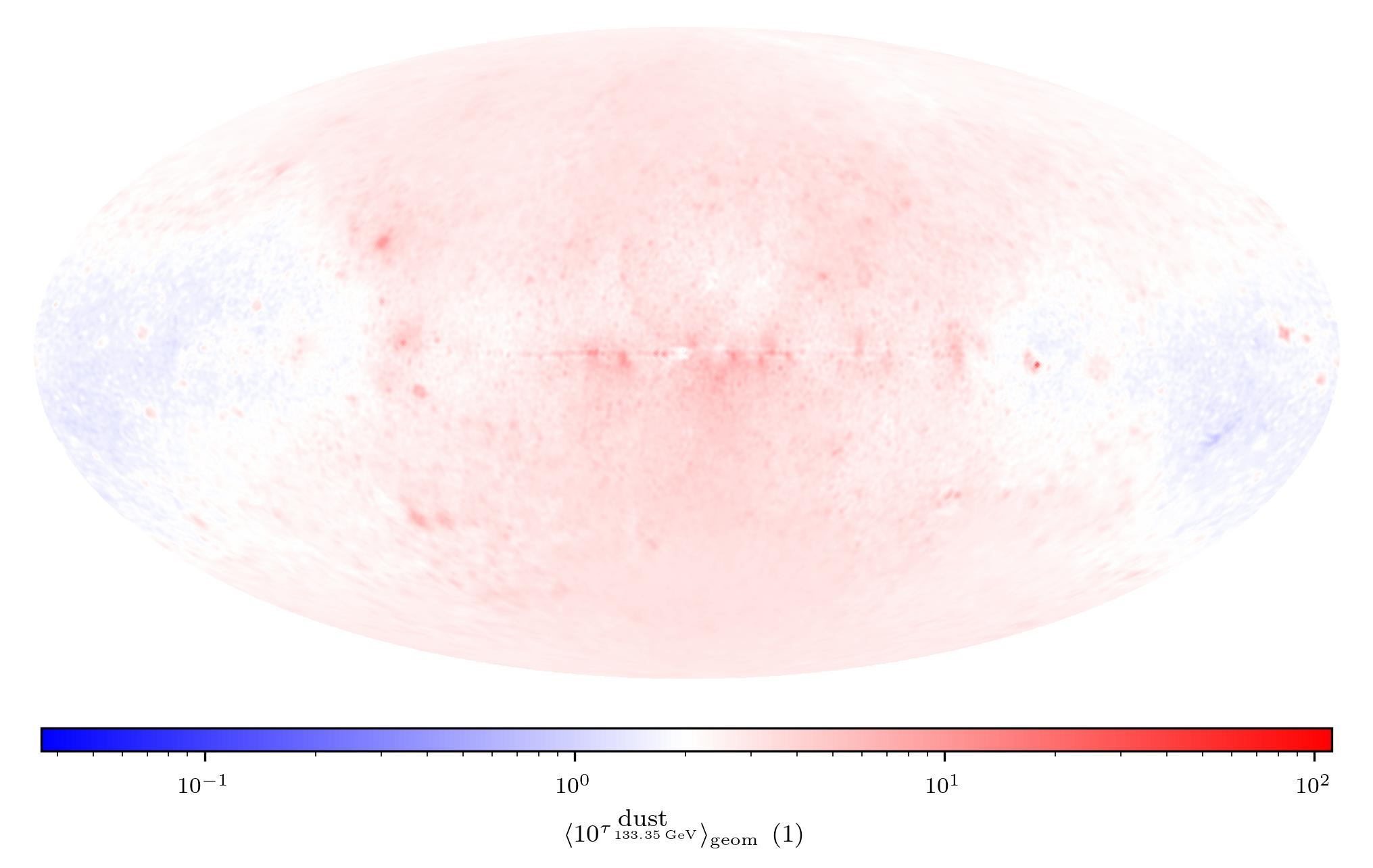}%
\end{minipage}
\begin{minipage}[t]{0.32\textwidth}%
\includegraphics[width=1\textwidth]{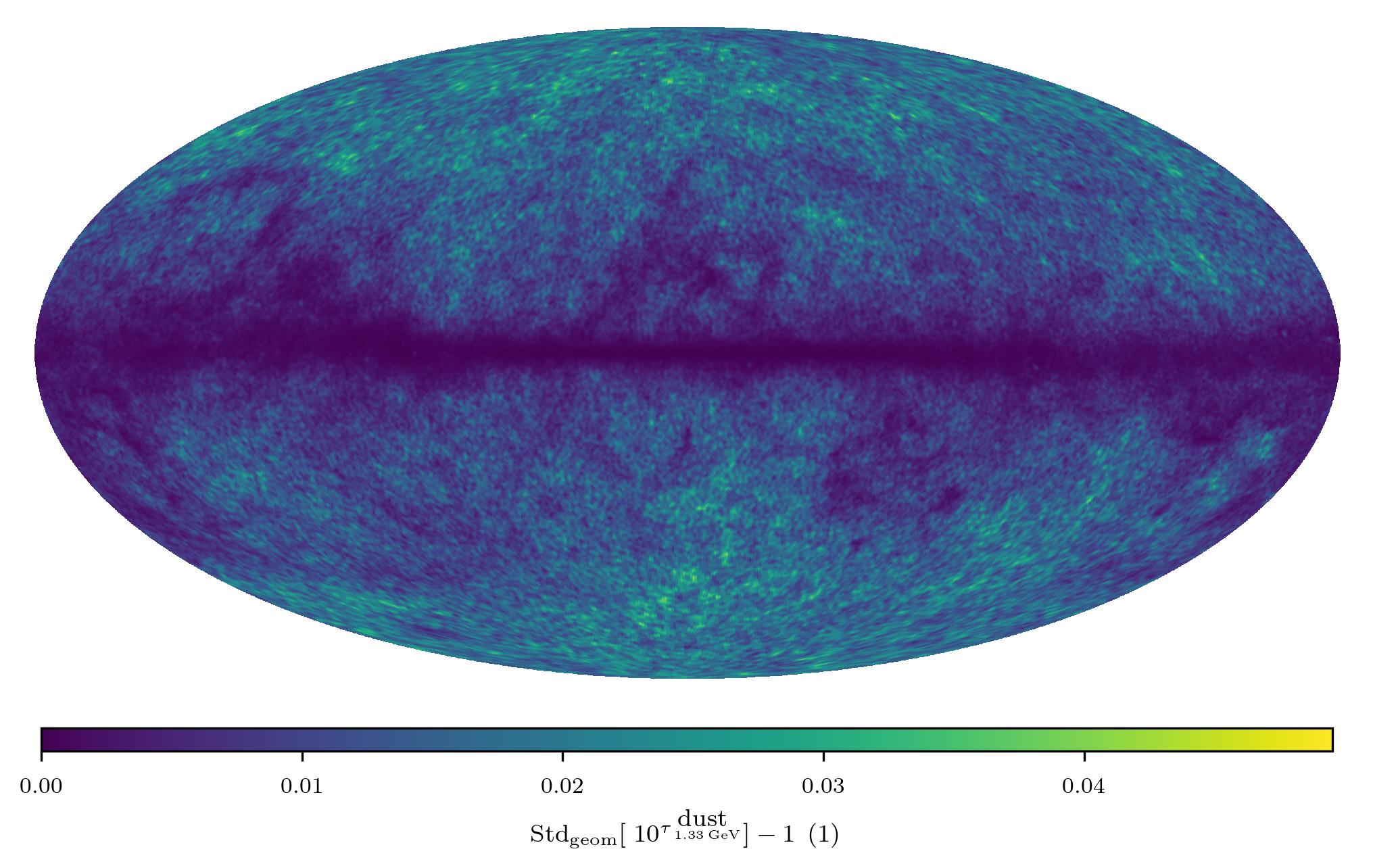}%
\end{minipage}\ %
\begin{minipage}[t]{0.32\textwidth}%
\includegraphics[width=1\textwidth]{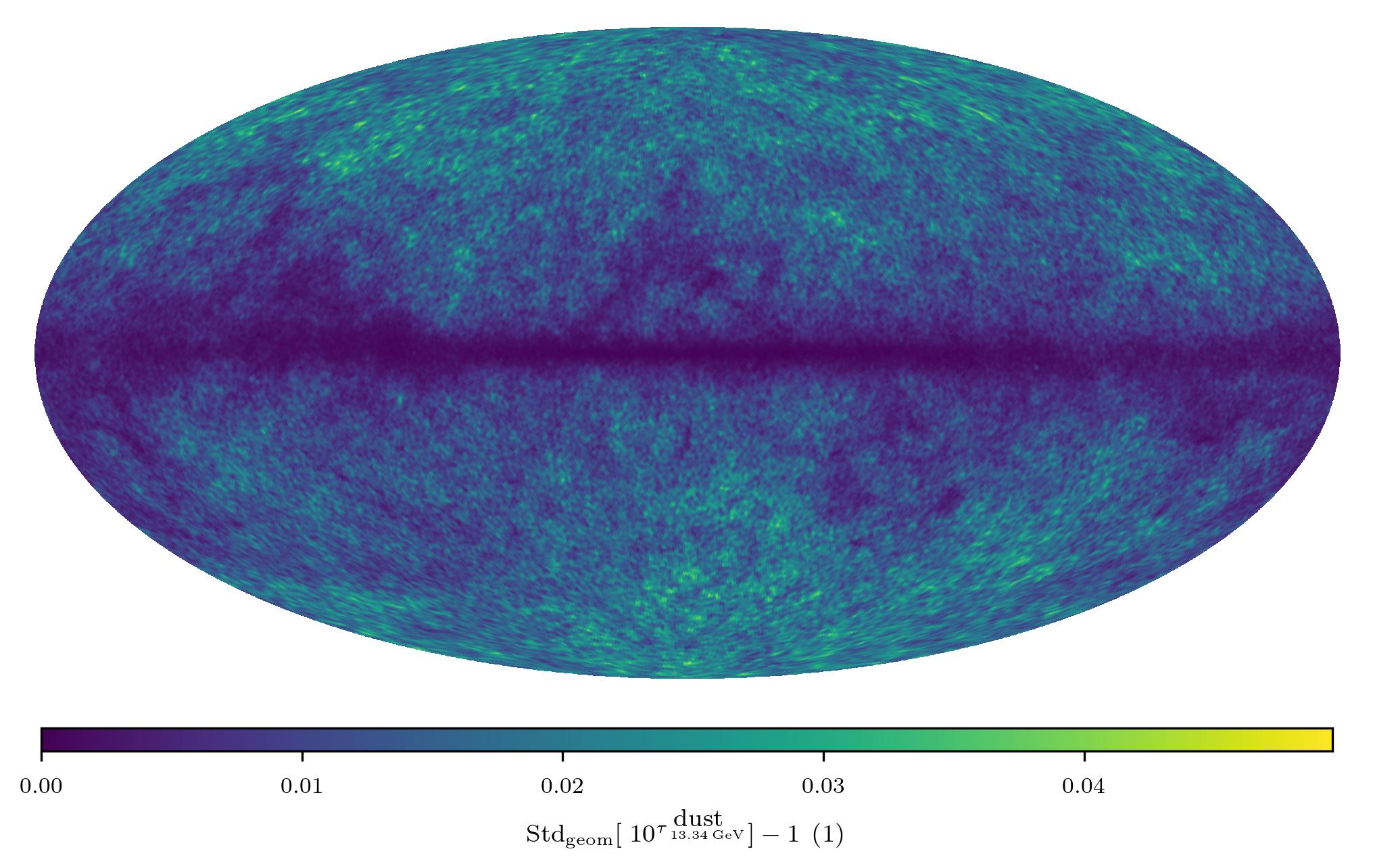}%
\end{minipage}\ %
\begin{minipage}[t]{0.32\textwidth}%
\includegraphics[width=1\textwidth]{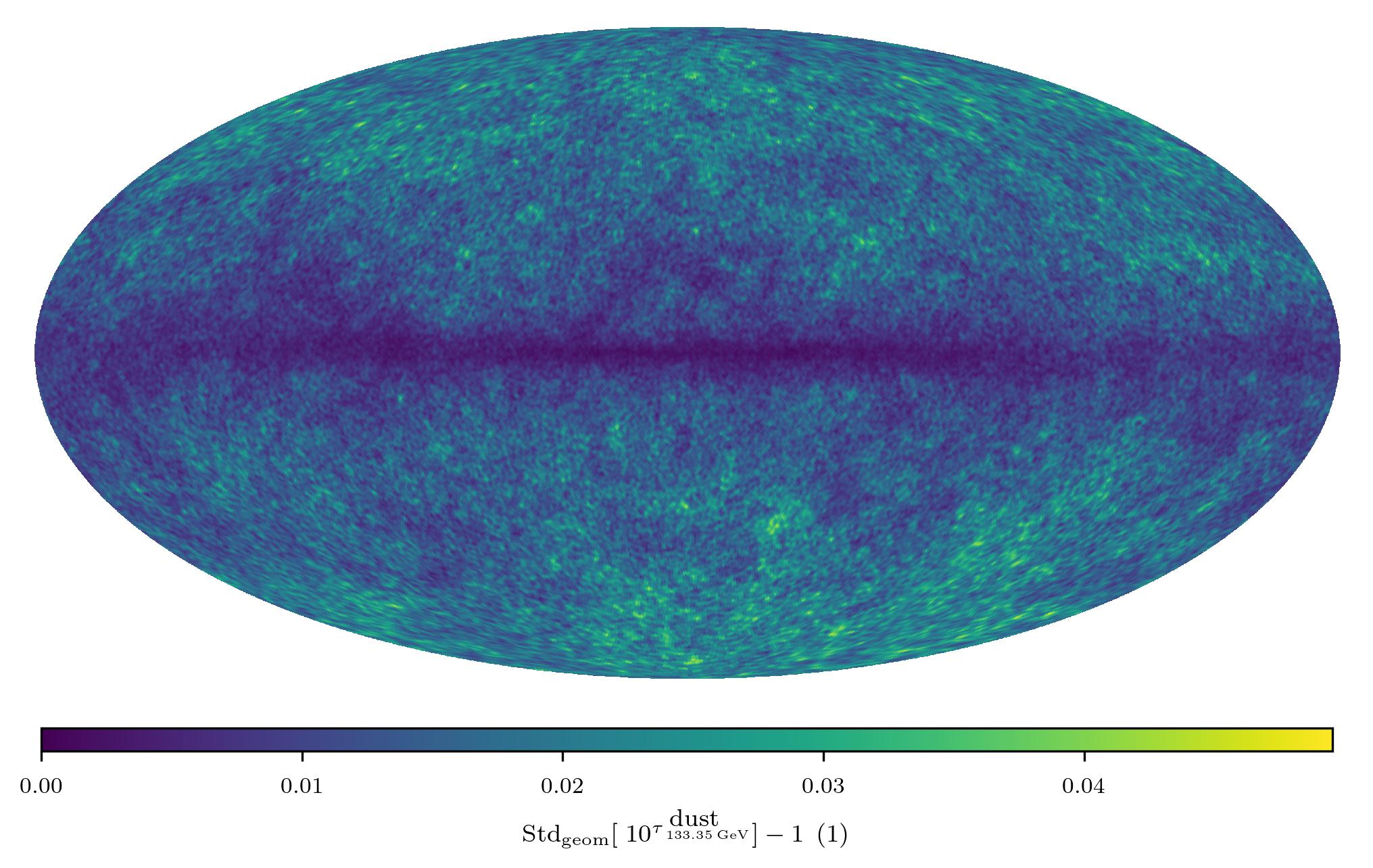}%
\end{minipage}
\caption{Posterior statistics of the M2 thermal dust template modification field for selected energy bins.
\textbf{Top row}: Posterior mean of the multiplicative template modification fields
on logarithmic color scales.
\textbf{Bottom row}: Corresponding posterior multiplicative uncertainty of the modification fields
on linear color scales.
Energy bins: \unit[1.0--1.78]{GeV} (\textbf{left}), \unit[10.0--17.8]{GeV} (\textbf{middle}), and \unit[100--178]{GeV} (\textbf{right}).}
\label{fig:m2-dust-modification-field-all-e}
\end{figure*}

Figure~\ref{fig:m2-dust-modification-field-all-e} shows the modification field $\tau^\mathrm{\:dust}$ learned for the \textit{Planck} thermal dust emission template
in the M2 reconstruction run and the corresponding multiplicative uncertainty maps
in the energy bins \unit[1.0--1.78]{GeV}, \unit[10.0--17.8]{GeV}, and \unit[100--178]{GeV}.
Toward higher energies, the average of the field increases slightly,
indicating that the reconstruction preferred a harder spectrum than the
prior encoded,
which specifies a base power-law spectral index of $\alpha^\mathrm{\:dust} = -1.65$.
The multiplicative uncertainty appears to be stable across the energy bands shown,
but with regions of low multiplicative uncertainty in the \unit[1.00--1.78]{GeV} and \unit[10.0--17.8]{GeV} energy bins
where the dust emissions predict the hadronic gamma-ray emissions to a high degree of accuracy.

% ---- fig: m2 vs fermi template ----
\begin{figure*}
\centering
\noindent
\begin{minipage}[t]{0.65\textwidth}%
\includegraphics[width=1\textwidth]{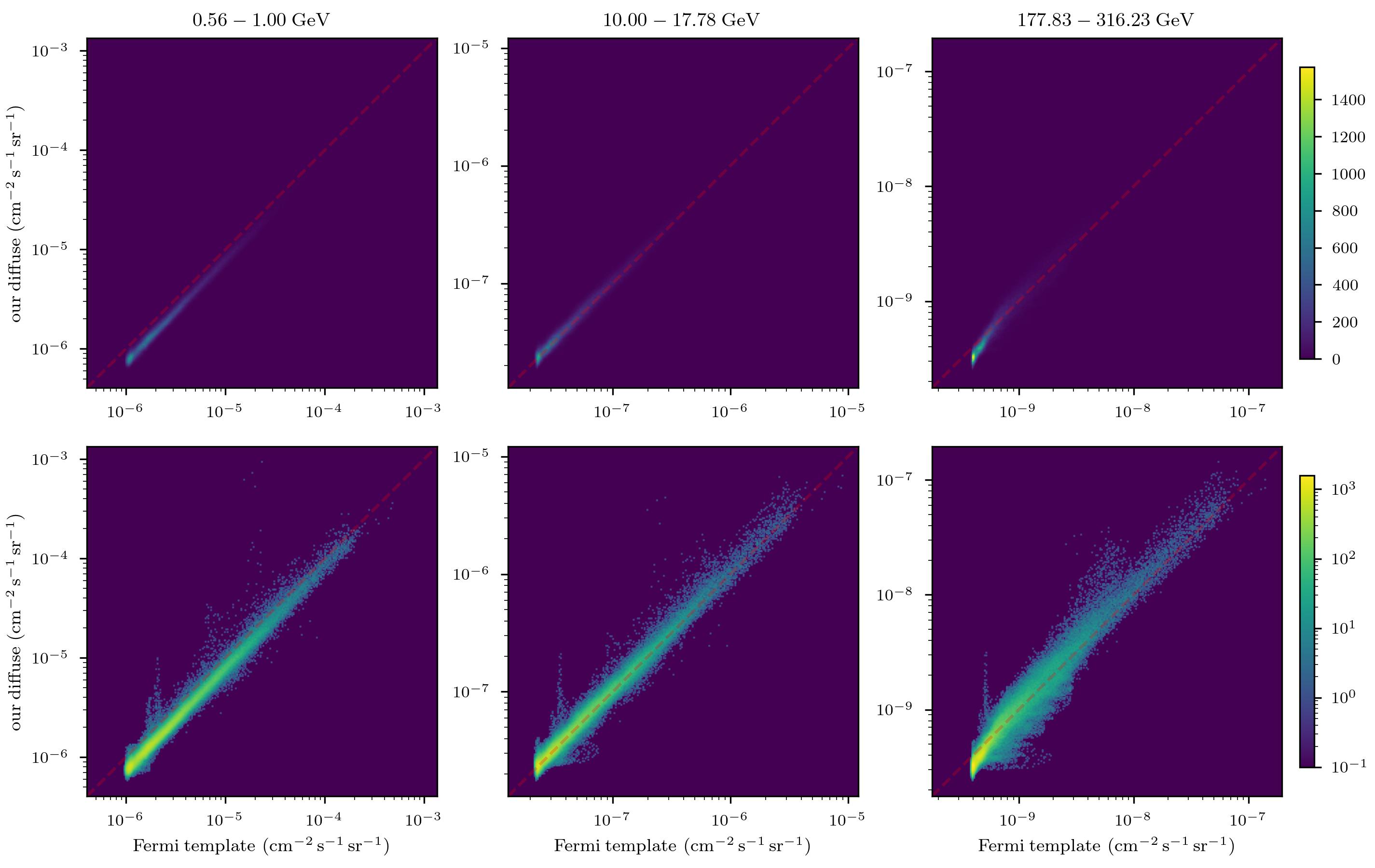}%
\end{minipage}\ %
\begin{minipage}[t]{0.25\textwidth}%
\vspace{-165pt}
\includegraphics[width=1\textwidth]{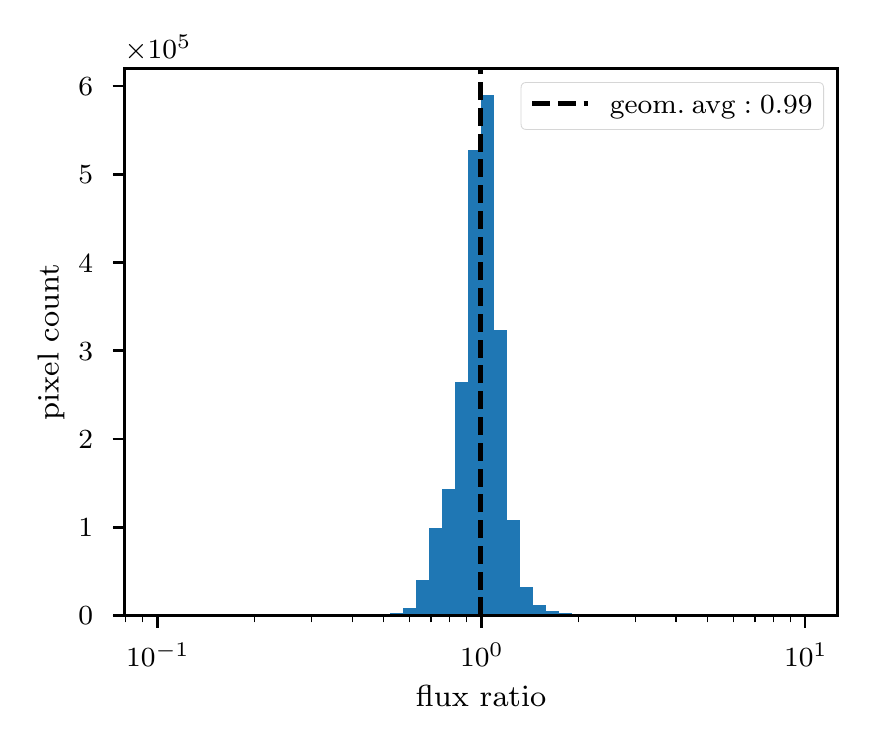}
\end{minipage}
\caption{
Comparison of the M2 diffuse reconstruction with the diffuse emission templates published by the \textit{Fermi} Collaboration (diffuse foreground: \texttt{gll\_iem\_v07}, isotropic background: \texttt{iso\_P8R3\_SOURCE\_V3\_v1}).
\textbf{Left}: 2D histograms of the diffuse flux found by our M2 reconstruction vs. the diffuse emission templates for the energy bins
\unit[0.56--1.00]{GeV}, \unit[10.0--17.8]{GeV}, and \unit[178--316]{GeV}.
The histogram bins are logarithmically spaced in both fluxes.
The top row shows the histogram counts with a linear color scale,
while the bottom row shows them with a logarithmic color scale.
The dashed red line marks perfect agreement.
\textbf{Right}: Histogram of flux ratios (our diffuse reconstruction divided by the corresponding voxel values predicted by the template) for all spatio-spectral bins.
Flux ratios are binned and displayed on a logarithmic scale.
Numbers larger than $10^0$ indicate we reconstruct more flux than the template predicts.}
\label{fig:m2-vs-fermi-template}
\end{figure*}

Figure~\ref{fig:m2-vs-fermi-template} shows the 2D histograms of fluxes in the \textit{Fermi} diffuse emission templates and our M2 diffuse emission reconstruction.
As already known from the comparison of the M1 and M2 reconstructions, the M2 reconstruction under-predicts diffuse emission in the lowest energy bin.
This effect is visible in the top-left panel of Fig.~\ref{fig:m2-vs-fermi-template} as an offset between the dotted red line and the line of high histogram counts.
The corresponding panels for the M1 diffuse reconstruction are provided in Fig.~\ref{fig:m1-vs-fermi-template}.

% ---- fig: m2 ps source count distribution
\begin{figure}
\resizebox{\hsize}{!}{\includegraphics{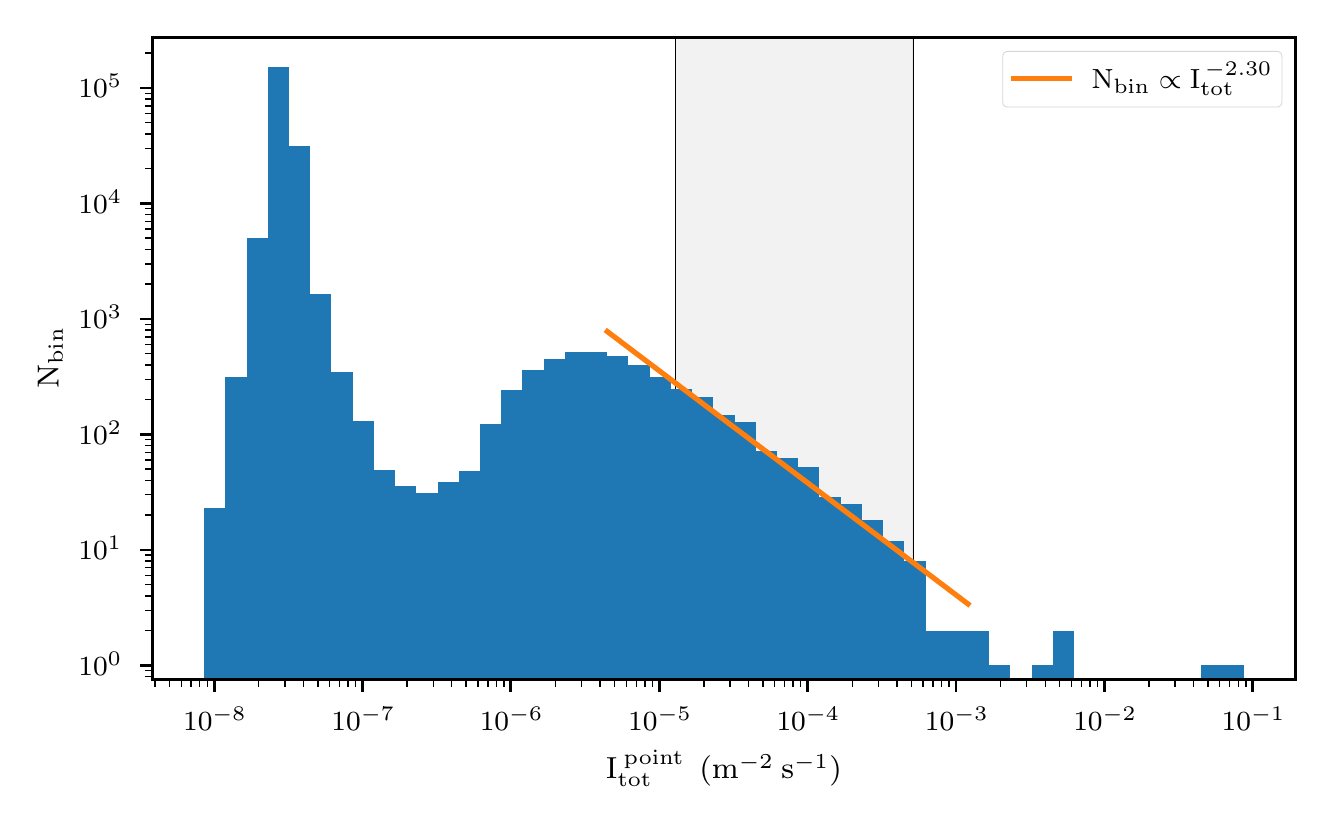}}
\caption{PS pixel count histogram for the M2 reconstruction on log-log scale.
The orange line shows a power-law fit to the brightness distribution in the brightness range highlighted in gray.
Flux values below \unit[$5\cdot{10}^{-7}$]{$\mathrm{m}^{-2}\,\mathrm{s}^{-1}$}
should be regarded as non-detections.
The distribution function in this regime is prior-driven.}
\label{fig:m2-ps-source-count-distribution}
\end{figure}

Figure~\ref{fig:m2-ps-source-count-distribution} shows the SCD found for the PS pixels
in the M2 reconstruction.
It follows a similar shape as the SCD found with M1 (see Fig.~\ref{fig:m1-ps-source-count-distribution}).
The M2 SCD follows a slightly steeper power law in the highlighted range.
This might be an effect of the increased PS flux contribution in the M2 reconstruction compared with the M1 reconstruction.
Point source pixels not or only slightly participating in the M1 reconstruction took on flux in the M2 reconstruction,
increasing the ratio of low-flux but active PS pixels to high-flux PS pixels and thereby changing the SCD power-law index.

% ---- fig: m2 desaturated spatio-spectral plot
\begin{figure}
\begin{minipage}[t]{\hsize}%
\includegraphics[width=1\textwidth]{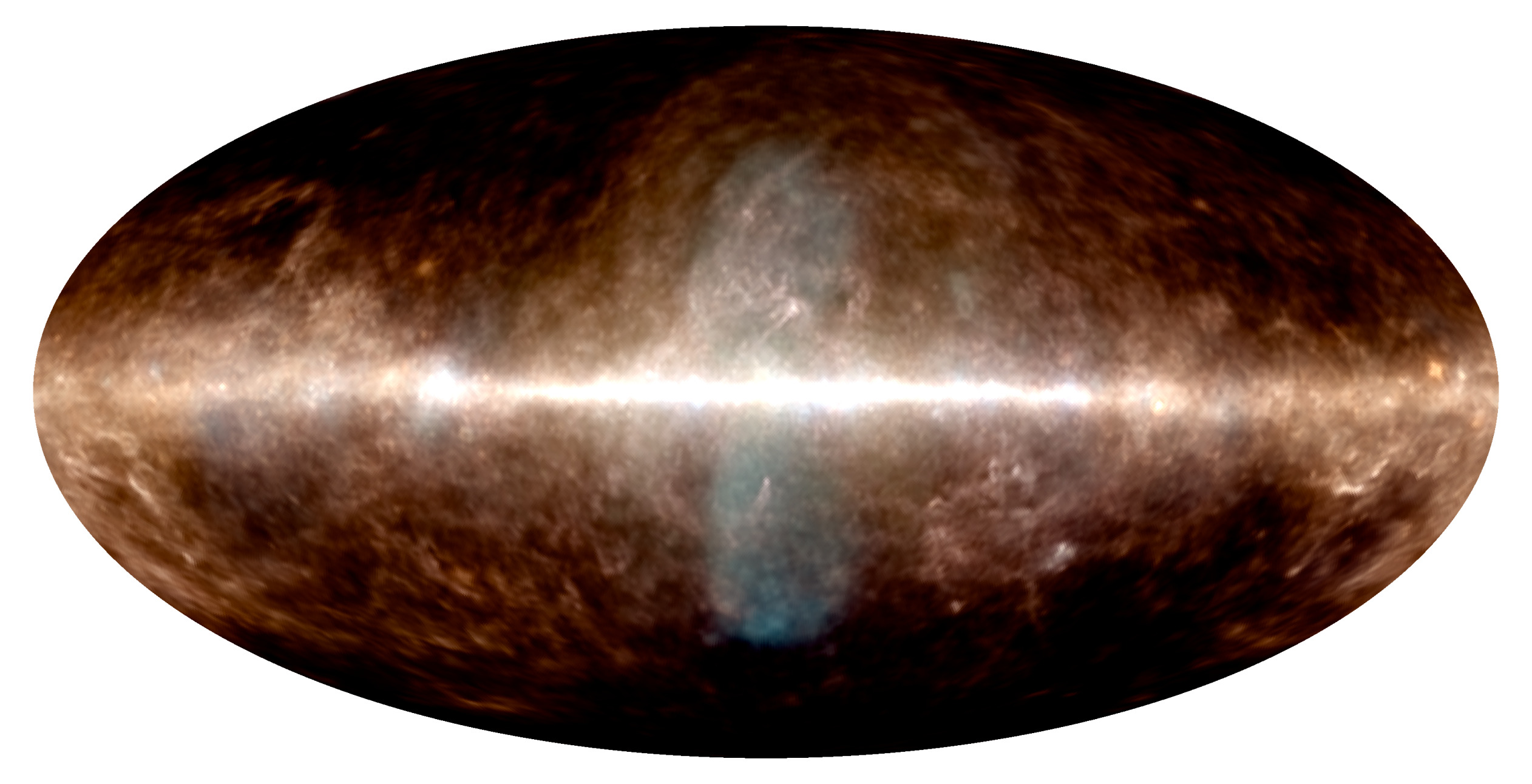}%
\end{minipage}
\begin{minipage}[t]{\hsize}%
\includegraphics[width=1\textwidth]{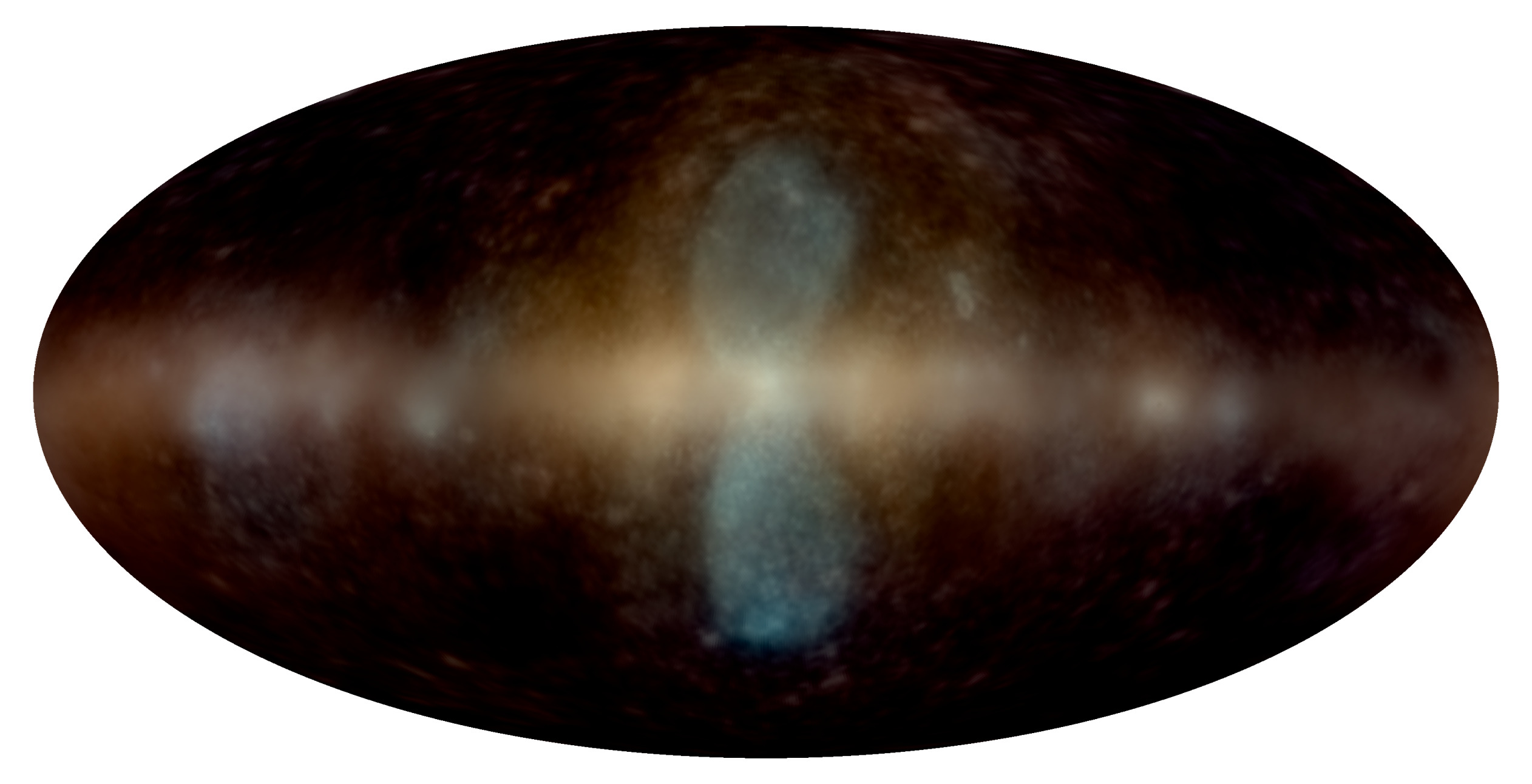}%
\end{minipage}
\begin{minipage}[t]{\hsize}%
\vspace{0.0cm}
\includegraphics[width=1\textwidth]{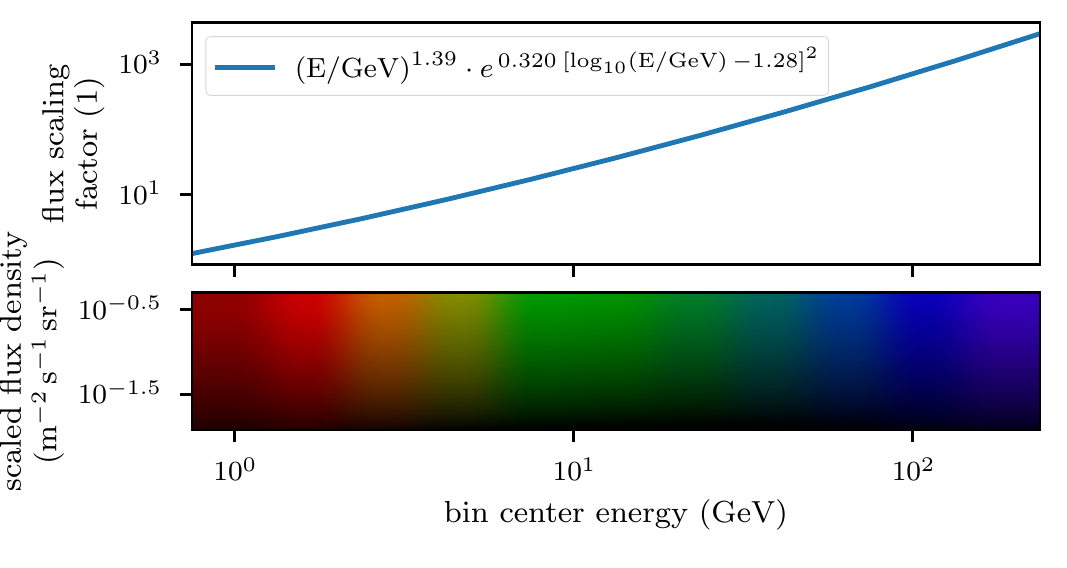}%
\end{minipage}
\caption{
Spatio-spectral plotting without color saturation enhancement.
The sky maps in the top and middle panel are based on the same color rendering as
Fig.~\ref{fig:data-exposure-corrected-mf-plot},~\ref{fig:m1-sky-maps}, and~\ref{fig:m2-sky-maps},
but were produced without the color saturation enhancement
described in Appendix~\ref{sec:appendix-color-coding}.
\textbf{Top:} Total diffuse emission reconstruction based on M2.
\textbf{Middle:} Template-free diffuse emission $I^\mathrm{\:nd}$ reconstruction based on model M2.
\textbf{Bottom}: Energy-dependent flux scaling factor (identical to Fig.~\ref{fig:data-exposure-corrected-mf-plot})
and mapping of photon energies and scaled flux densities to perceived colors without the color saturation enhancement.
\label{fig:m2-mf-plot-desaturated}
}
\end{figure}

Figure~\ref{fig:m2-mf-plot-desaturated} shows the spatio-spectral emission maps
for the total diffuse and template-free diffuse emission component
of the M2 reconstruction without the color saturation enhancement
described in Appendix~\ref{sec:appendix-color-coding} and the corresponding color mapping.
These natural saturation maps are better suited to show fine structures
than the saturation-enhanced maps displayed in Fig.~\ref{fig:m2-sky-maps},
but because of their lower color saturation do not show spectral variations as clearly.

\end{appendix}
\end{document}